\RequirePackage{amsmath} 
\documentclass[longauth,traditabstract]{aa}

\usepackage[nonamebreak]{natbib}
\usepackage[stable]{footmisc}
\usepackage{amsmath}
\usepackage{amssymb}
\usepackage{txfonts}
\usepackage{placeins}
\usepackage{natbib}
\bibpunct{(}{)}{;}{a}{}{,} %
\usepackage{graphicx}
\usepackage{epstopdf}
\usepackage{ifthen}
\usepackage[table,usenames,dvipsnames]{xcolor}
\definecolor{linkcolor}{rgb}{0.6,0,0}
\definecolor{citecolor}{rgb}{0,0,0.75}
\definecolor{urlcolor}{rgb}{0.12,0.46,0.7}
\usepackage[breaklinks, colorlinks, urlcolor=urlcolor, linkcolor=linkcolor,citecolor=citecolor,pdfencoding=auto]{hyperref}
\hypersetup{linktocpage}
\usepackage{overpic}
\usepackage{fixltx2e}
\usepackage{rotating}

\usepackage{booktabs}
\usepackage{enumitem}

\usepackage{etoolbox}
\makeatother

\usepackage[utf8]{inputenc} 

\def\setsymbol#1#2{\expandafter\def\csname #1\endcsname{#2}}
\def\getsymbol#1{\csname #1\endcsname}

\def\Planck{\textit{Planck}}

\newbox\tablebox    \newdimen\tablewidth
\def\leaderfil{\leaders\hbox to 5pt{\hss.\hss}\hfil}
\def\endPlancktable{\tablewidth=\columnwidth 
    $$\hss\copy\tablebox\hss$$
    \vskip-\lastskip\vskip -2pt}
\def\endPlancktablewide{\tablewidth=\textwidth 
    $$\hss\copy\tablebox\hss$$
    \vskip-\lastskip\vskip -2pt}
\def\tablenote#1 #2\par{\begingroup \parindent=0.8em
    \abovedisplayshortskip=0pt\belowdisplayshortskip=0pt
    \noindent
    $$\hss\vbox{\hsize\tablewidth \hangindent=\parindent \hangafter=1 \noindent
    \hbox to \parindent{$^#1$\hss}\strut#2\strut\par}\hss$$
    \endgroup}
\def\doubleline{\vskip 3pt\hrule \vskip 1.5pt \hrule \vskip 5pt}

\def\L2{\ifmmode L_2\else $L_2$\fi}

\def\DeltaT{\ifmmode \Delta T\else $\Delta T$\fi}
\def\deltat{\ifmmode \Delta t\else $\Delta t$\fi}
\def\fknee{\ifmmode f_{\rm knee}\else $f_{\rm knee}$\fi}
\def\Fmax{\ifmmode F_{\rm max}\else $F_{\rm max}$\fi}
\def\solar{\ifmmode{\rm M}_{\mathord\odot}\else${\rm M}_{\mathord\odot}$\fi}
\def\Msolar{\ifmmode{\rm M}_{\mathord\odot}\else${\rm M}_{\mathord\odot}$\fi}
\def\Lsolar{\ifmmode{\rm L}_{\mathord\odot}\else${\rm L}_{\mathord\odot}$\fi}
\def\inv{\ifmmode^{-1}\else$^{-1}$\fi}
\def\mo{\ifmmode^{-1}\else$^{-1}$\fi}
\def\sup#1{\ifmmode ^{\rm #1}\else $^{\rm #1}$\fi}
\def\expo#1{\ifmmode \times 10^{#1}\else $\times 10^{#1}$\fi}
\def\,{\thinspace}
\def\lsim{\mathrel{\raise .4ex\hbox{\rlap{$<$}\lower 1.2ex\hbox{$\sim$}}}}
\def\gsim{\mathrel{\raise .4ex\hbox{\rlap{$>$}\lower 1.2ex\hbox{$\sim$}}}}
\let\lea=\lsim

\def\simprop{\mathrel{\raise .4ex\hbox{\rlap{$\propto$}\lower 1.2ex\hbox{$\sim$}}}}
\def\deg{\ifmmode^\circ\else$^\circ$\fi}
\def\pdeg{\ifmmode $\setbox0=\hbox{$^{\circ}$}\rlap{\hskip.11\wd0 .}$^{\circ}
          \else \setbox0=\hbox{$^{\circ}$}\rlap{\hskip.11\wd0 .}$^{\circ}$\fi}
\def\arcs{\ifmmode {^{\scriptstyle\prime\prime}}
          \else $^{\scriptstyle\prime\prime}$\fi}
\def\arcm{\ifmmode {^{\scriptstyle\prime}}
          \else $^{\scriptstyle\prime}$\fi}
\newdimen\sa  \newdimen\sb
\def\parcs{\sa=.07em \sb=.03em
     \ifmmode \hbox{\rlap{.}}^{\scriptstyle\prime\kern -\sb\prime}\hbox{\kern -\sa}
     \else \rlap{.}$^{\scriptstyle\prime\kern -\sb\prime}$\kern -\sa\fi}
\def\parcm{\sa=.08em \sb=.03em
     \ifmmode \hbox{\rlap{.}\kern\sa}^{\scriptstyle\prime}\hbox{\kern-\sb}
     \else \rlap{.}\kern\sa$^{\scriptstyle\prime}$\kern-\sb\fi}
\def\ra[#1 #2 #3.#4]{#1\sup{h}#2\sup{m}#3\sup{s}\llap.#4}
\def\dec[#1 #2 #3.#4]{#1\deg#2\arcm#3\arcs\llap.#4}
\def\deco[#1 #2 #3]{#1\deg#2\arcm#3\arcs}
\def\rra[#1 #2]{#1\sup{h}#2\sup{m}}

\def\dots{\relax\ifmmode \ldots\else $\ldots$\fi}
\def\WHzsr{\ifmmode $W\,Hz\mo\,sr\mo$\else W\,Hz\mo\,sr\mo\fi}
\def\mHz{\ifmmode $\,mHz$\else \,mHz\fi}
\def\GHz{\ifmmode $\,GHz$\else \,GHz\fi}
\def\mKs{\ifmmode $\,mK\,s$^{1/2}\else \,mK\,s$^{1/2}$\fi}
\def\muKs{\ifmmode \,\mu$K\,s$^{1/2}\else \,$\mu$K\,s$^{1/2}$\fi}
\def\muKRJs{\ifmmode \,\mu$K$_{\rm RJ}$\,s$^{1/2}\else \,$\mu$K$_{\rm RJ}$\,s$^{1/2}$\fi}
\def\muKHz{\ifmmode \,\mu$K\,Hz$^{-1/2}\else \,$\mu$K\,Hz$^{-1/2}$\fi}
\def\MJysr{\ifmmode \,$MJy\,sr\mo$\else \,MJy\,sr\mo\fi}
\def\MJysrmK{\ifmmode \,$MJy\,sr\mo$\,mK$_{\rm CMB}\mo\else \,MJy\,sr\mo\,mK$_{\rm CMB}\mo$\fi}
\def\microns{\ifmmode \,\mu$m$\else \,$\mu$m\fi}

\def\muK{\ifmmode \,\mu$K$\else \,$\mu$\hbox{K}\fi}
\def\microK{\ifmmode \,\mu$K$\else \,$\mu$\hbox{K}\fi}
\def\muW{\ifmmode \,\mu$W$\else \,$\mu$\hbox{W}\fi}
\def\kms{\ifmmode $\,km\,s$^{-1}\else \,km\,s$^{-1}$\fi}
\def\kmsMpc{\ifmmode $\,\kms\,Mpc\mo$\else \,\kms\,Mpc\mo\fi}

\providecommand{\sorthelp}[1]{}

\newcommand{\onesig}[1]{(68\,\%,~\text{#1})}
\newcommand{\twosig}[1]{(95\,\%,~\text{#1})}

\newcommand{\leftparbox}[2]{\parbox{#1}{\begin{flushleft} #2 \end{flushleft}}}
\newcommand{\oneonesig}[4][5cm]{
\begin{equation}
\left.
  #2 \quad\mbox{\text{\leftparbox{#1}{(68\,\%,~#3)#4}}}
  \right.
\end{equation}
}
\newcommand{\onetwosig}[4][5cm]{
\begin{equation}
\left.
  #2 \quad\mbox{\text{\leftparbox{#1}{(95\,\%,~#3)#4}}}
  \right.
\end{equation}
}
\newcommand{\twoonesig}[4][\pbwidth]{
\begin{equation}
\left.
 \begin{aligned}
#2 \\ #3
 \end{aligned}
\ \right\} \ \ \mbox{\text{\leftparbox{#1}{68\,\%,~#4}}}
\end{equation}
}
\newcommand{\twotwosig}[4][\pbwidth]{
\begin{equation}
\left.
 \begin{aligned}
#2 \\ #3
 \end{aligned}
\ \right\} \ \ \mbox{\text{\leftparbox{#1}{95\,\%,~#4}}}
\end{equation}
}
\newcommand{\threeonesig}[5][\pbwidth]{
\begin{equation}
\left.
 \begin{aligned}
#2 \\ #3 \\ #4
 \end{aligned}
\ \right\} \ \ \mbox{\text{\leftparbox{#1}{68\,\%,~#5}}}
\end{equation}
}
\newcommand{\threetwosig}[5][\pbwidth]{
\begin{equation}
\left.
 \begin{aligned}
#2 \\ #3 \\ #4
 \end{aligned}
\ \right\} \ \ \mbox{\text{\leftparbox{#1}{95\,\%,~#5}}}
\end{equation}
}

\newenvironment{unindentedlist}{
 \begin{list}{{$\bullet$}}{
  \setlength\partopsep{0pt}
  \setlength\parskip{0pt}
  \setlength\parsep{0pt}
  \setlength\topsep{0pt}
  \setlength\itemsep{0pt}
  \setlength{\itemindent}{\leftmargin}
  \setlength{\leftmargin}{0pt}
 }
}{
 \end{list}
}

\newcommand{\sroll}{\texttt{SRoll}}

\let\oldell\ell
\renewcommand{\ell}{\texorpdfstring{\oldell}{ℓ}}

\newcommand{\camspec}{{\tt CamSpec}}
\newcommand{\plik}{{\tt Plik}}
\newcommand{\pliklite}{{\tt plik\_lite}}

\newcommand{\commander}{{\tt Commander}}

\newcommand{\smica}{{\tt SMICA}}
\newcommand{\simall}{{\tt SimAll}}

\newcommand{\CFHTLENS}{{CFHTLenS}}
\newcommand{\KIDS}{{KiDS}}

\newcommand{\effchisquare}{\chi^2_{\rm eff}}
\newcommand{\dchisquare}{\Delta\effchisquare}

\newcommand{\mksym}[1]{\ifmmode {\rm #1}\else #1\fi}
\newcommand{\dataplus}{\allowbreak+}

\newcommand{\BAO}{\mksym{BAO}}
\newcommand{\lensing}{\mksym{lensing}}

\newcommand{\TT}{\mksym{TT}}
\newcommand{\TE}{\mksym{TE}}
\newcommand{\EE}{\mksym{EE}}
\newcommand{\TTTEEE}{\mksym{TT,TE,EE}}
\newcommand{\planckTTonly}{\planck\ \TT}
\newcommand{\planckTTTEEEonly}{\planck\ \TTTEEE}

\newcommand{\lowE}{\mksym{lowE}}

\newcommand{\simallEE}{\simall}
\newcommand{\parthenope}{{\tt PArthENoPE}}

\newcommand{\planckTT}{\planckTTonly\dataplus\lowE}
\newcommand{\planckTE}{\planck\ \TE\dataplus\lowE}
\newcommand{\planckEE}{\planck\ \EE\dataplus\lowE}

\newcommand{\planckall}{\planckTTTEEEonly\dataplus\lowE}
\newcommand{\planckTTBAO}{\planckTT\dataplus\BAO}
\newcommand{\planckTTlensing}{\planckTT\dataplus\lensing}
\newcommand{\planckallBAO}{\planckall\dataplus\BAO}
\newcommand{\planckalllensing}{\planckall\dataplus\lensing}

\newcommand{\shortpol}{\TTTEEE}
\newcommand{\shortTT}{\TT\dataplus\lowE}
\newcommand{\shortall}{\TTTEEE\dataplus\lowE}

\newcommand{\wzero}{w_0}
\newcommand{\lnAs}{\ln(10^{10} A_{\rm s})}
\newcommand{\As}{A_{\rm s}}
\newcommand{\Astau}{\As e^{-2\tau}}

\newcommand{\ns}{n_{\rm s}}
\newcommand{\lcdm}{\texorpdfstring{{$\rm{\Lambda CDM}$}}{ΛCDM}}
\newcommand{\boldlcdm}{\texorpdfstring{$\boldsymbol{\Lambda}$CDM}{ΛCDM}}

\newcommand{\rzerotwo}{r_{0.002}}
\newcommand{\nszerotwo}{n_{{\rm s},0.002}}
\newcommand{\Alens}{A_{\rm L}}

\newcommand{\thetaMC}{\theta_{\rm MC}}
\newcommand{\nrun}{d \ns / d\ln k}
\newcommand{\nrunrun}{d^2 \ns / d\ln k^2}

\newcommand{\zre}{z_{\text{re}}}
\newcommand{\yhe}{Y_{\text{P}}}
\newcommand{\ypbbn}{Y_{\text{P}}^{\rm BBN}}
\newcommand{\taun}{\tau_{\rm n}}

\newcommand{\nnu}{N_{\rm eff}}
\newcommand{\neff}{\nnu}
\newcommand{\mnu}{\sum m_\nu}
\newcommand{\sumnu}{\sum m_\nu}
\newcommand{\mnusterile}{m_{\nu,\, \mathrm{sterile}}^{\mathrm{eff}}}
\newcommand{\meffsterile}{\mnusterile}

\newcommand{\msthermal}{m_{\rm sterile}^{\rm thermal}}

\newcommand{\rdrag}{r_{\rm drag}}

\newcommand{\rstar}{r_{\ast}}

\newcommand{\thetastar}{\theta_{\ast}}

\newcommand{\DVBAO}{D_{\rm V}}

\newcommand{\DM}{D_{\rm M}}

\providecommand{\lea}{\la}

\providecommand{\text}[1]{\rm{#1}}

\newcommand{\Mpc}{\text{Mpc}}

\newcommand{\Hunit}{~\text{km}~\text{s}^{-1} \Mpc^{-1}}

\providecommand{\muK}{\mu\rm{K}}

\newcommand{\mpl}{m_{\text{Pl}}}
\newcommand{\eV}{\,\text{eV}}
\newcommand{\MeV}{\,\text{MeV}}

\providecommand{\Omk}{\Omega_K}

\providecommand{\Omb}{\Omega_{\mathrm{b}}}
\providecommand{\Omc}{\Omega_{\mathrm{c}}}
\providecommand{\Omm}{\Omega_{\mathrm{m}}}
\providecommand{\omb}{\omega_{\mathrm{b}}}
\providecommand{\omc}{\omega_{\mathrm{c}}}

\providecommand{\CAMB}{{\tt camb}}
\providecommand{\COSMOMC}{{\tt CosmoMC}}

\providecommand{\LCDM}{{$\rm{\Lambda CDM}$}}
\providecommand{\COSMOREC}{{\tt CosmoRec}}
\providecommand{\HYREC}{{\tt HyRec}}
\providecommand{\RECFAST}{{\tt recfast}}
\providecommand{\HALOFIT}{{\tt halofit}}

\newcommand{\begm}{\begin{pmatrix}}
\newcommand{\enm}{\end{pmatrix}}

\newcommand\ba{\begin{eqnarray}}
\newcommand\ea{\end{eqnarray}}
\newcommand\bea{\begin{eqnarray}}
\newcommand\eea{\end{eqnarray}}

\newcommand\be{\begin{equation}}
\newcommand\ee{\end{equation}}

\newcommand{\clo}{\mathcal{O}}
\newcommand{\clp}{\mathcal{P}}

\newcommand{\clr}{\mathcal{R}}

\def\pmb#1{\setbox0=\hbox{#1}%
    \kern-.025em\copy0\kern-\wd0
    \kern.05em\copy0\kern-\wd0
    \kern-.025em\raise.0433em\box0}

\providecommand{\simlt}{\lea}

\def\p2Y{\;_2Y}
\def\m2Y{\;_{-2}Y}
\def\beglet{
  \addtocounter{equation}{1}%
  \setcounter{parentequation}{\value{equation}}%
  \setcounter{equation}{0}%
  \def\theequation{\arabic{parentequation}\alph{equation}}%
  \ignorespaces
}
\def\endlet{
  \setcounter{equation}{\value{parentequation}}%
  \def\theequation{\arabic{equation}}%
}
\providecommand{\beglet}{\begin{subequations}}
\providecommand{\endlet}{\end{subequations}}

\newcommand{\neutron}{{\rm n}}

\newcommand{\calibC}{\ensuremath{c}}
\newcommand{\paramsI}{\citetalias{planck2013-p11}}
\newcommand{\paramsII}{\citetalias{planck2014-a15}}

\defcitealias{planck2013-p11}{PCP13}
\defcitealias{planck2014-a15}{PCP15}
\defcitealias{planck2016-l06}{PCP18}

\defcitealias{pb2015}{BKP}

\defcitealias{planck2013-p08}{PPL13}
\defcitealias{planck2014-a13}{PPL15}
\defcitealias{planck2016-l05}{PPL18}

\newcommand{\likeI}{\citetalias{planck2013-p08}}
\newcommand{\likeII}{\citetalias{planck2014-a13}}
\newcommand{\likeIII}{\citetalias{planck2016-l05}}

\defcitealias{planck2016-l08}{PL2018}
\defcitealias{planck2014-a17}{PL2015}
\defcitealias{planck2013-p12}{PL2013}
\newcommand{\PlanckLensThree}{\citetalias{planck2016-l08}}
\newcommand{\PlanckLensTwo}{\citetalias{planck2014-a17}}

\newcommand{\PLA}{\citetalias{PLA}}
\defcitealias{PLA}{PLA}

\defcitealias{planck2014-a16}{PDE15}
\newcommand{\PDEII}{\citetalias{planck2014-a16}} \newcommand{\rep}[1]{#1}

\newcommand{\planck}{\Planck}
\newcommand{\WMAP}{WMAP}

\newcommand{\BK}{BK15}

\def\aap{{A\&A}}
\def\apj{{ApJ}}
\def\mnras{{MNRAS}}
\def\apjs{{ApJS}}
\def\prd{{PRD}}
\def\prl{{PRL}}
\def\jcap{{JCAP}}

\newcommand{\Alenssec}{Sect.~\ref{sec:Alens}}

\newcommand{\isdraft}[1]{}

\begin{document}

\title{\vglue -10mm\Planck\ 2018 results. VI. Cosmological parameters}
\titlerunning{Cosmological parameters}
\authorrunning{Planck Collaboration}

\author{\small
Planck Collaboration: N.~Aghanim\inst{54}
\and
Y.~Akrami\inst{15, 57, 59}
\and
M.~Ashdown\inst{65, 5}
\and
J.~Aumont\inst{95}
\and
C.~Baccigalupi\inst{78}
\and
M.~Ballardini\inst{21, 41}
\and
A.~J.~Banday\inst{95, 8}
\and
R.~B.~Barreiro\inst{61}
\and
N.~Bartolo\inst{29, 62}
\and
S.~Basak\inst{85}
% ...
}
\institute{\small
AIM, CEA, CNRS, Universit\'{e} Paris-Saclay, Universit\'{e} Paris-Diderot, Sorbonne Paris Cit\'{e}, F-91191 Gif-sur-Yvette, France\goodbreak
\and
APC, AstroParticule et Cosmologie, Universit\'{e} Paris Diderot, CNRS/IN2P3, CEA/lrfu, Observatoire de Paris, Sorbonne Paris Cit\'{e}, 10, rue Alice Domon et L\'{e}onie Duquet, 75205 Paris Cedex 13, France\goodbreak
\and
African Institute for Mathematical Sciences, 6-8 Melrose Road, Muizenberg, Cape Town, South Africa\goodbreak
\and
Aix Marseille Univ, CNRS, CNES, LAM, Marseille, France\goodbreak
\and
Astrophysics Group, Cavendish Laboratory, University of Cambridge, J J Thomson Avenue, Cambridge CB3 0HE, U.K.\goodbreak
\and
Astrophysics \& Cosmology Research Unit, School of Mathematics, Statistics \& Computer Science, University of KwaZulu-Natal, Westville Campus, Private Bag X54001, Durban 4000, South Africa\goodbreak
\and
CITA, University of Toronto, 60 St. George St., Toronto, ON M5S 3H8, Canada\goodbreak
\and
CNRS, IRAP, 9 Av. colonel Roche, BP 44346, F-31028 Toulouse cedex 4, France\goodbreak
\and
Cahill Center for Astronomy and Astrophysics, California Institute of Technology, Pasadena CA,  91125, USA\goodbreak
\and
California Institute of Technology, Pasadena, California, U.S.A.\goodbreak
\and
Centre for Theoretical Cosmology, DAMTP, University of Cambridge, Wilberforce Road, Cambridge CB3 0WA, U.K.\goodbreak
\and
Computational Cosmology Center, Lawrence Berkeley National Laboratory, Berkeley, California, U.S.A.\goodbreak
\and
DTU Space, National Space Institute, Technical University of Denmark, Elektrovej 327, DK-2800 Kgs. Lyngby, Denmark\goodbreak
\and
D\'{e}partement de Physique Th\'{e}orique, Universit\'{e} de Gen\`{e}ve, 24, Quai E. Ansermet,1211 Gen\`{e}ve 4, Switzerland\goodbreak
\and
D\'{e}partement de Physique, \'{E}cole normale sup\'{e}rieure, PSL Research University, CNRS, 24 rue Lhomond, 75005 Paris, France\goodbreak
\and
Departamento de Astrof\'{i}sica, Universidad de La Laguna (ULL), E-38206 La Laguna, Tenerife, Spain\goodbreak
\and
Departamento de F\'{\i}sica, Universidad de Oviedo, C/ Federico Garc\'{\i}a Lorca, 18 , Oviedo, Spain\goodbreak
\and
Department of Astrophysics/IMAPP, Radboud University, P.O. Box 9010, 6500 GL Nijmegen, The Netherlands\goodbreak
\and
Department of Mathematics, University of Stellenbosch, Stellenbosch 7602, South Africa\goodbreak
\and
Department of Physics \& Astronomy, University of British Columbia, 6224 Agricultural Road, Vancouver, British Columbia, Canada\goodbreak
\and
Department of Physics \& Astronomy, University of the Western Cape, Cape Town 7535, South Africa\goodbreak
\and
Department of Physics and Astronomy, University College London, London WC1E 6BT, U.K.\goodbreak
\and
Department of Physics and Astronomy, University of Sussex, Brighton BN1 9QH, U.K.\goodbreak
\and
Department of Physics, Gustaf H\"{a}llstr\"{o}min katu 2a, University of Helsinki, Helsinki, Finland\goodbreak
\and
Department of Physics, Princeton University, Princeton, New Jersey, U.S.A.\goodbreak
\and
Department of Physics, University of California, Berkeley, California, U.S.A.\goodbreak
\and
Department of Physics, University of California, One Shields Avenue, Davis, California, U.S.A.\goodbreak
\and
Department of Physics, University of California, Santa Barbara, California, U.S.A.\goodbreak
\and
Dipartimento di Fisica e Astronomia G. Galilei, Universit\`{a} degli Studi di Padova, via Marzolo 8, 35131 Padova, Italy\goodbreak
\and
Dipartimento di Fisica e Scienze della Terra, Universit\`{a} di Ferrara, Via Saragat 1, 44122 Ferrara, Italy\goodbreak
\and
Dipartimento di Fisica, Universit\`{a} La Sapienza, P. le A. Moro 2, Roma, Italy\goodbreak
\and
Dipartimento di Fisica, Universit\`{a} degli Studi di Milano, Via Celoria, 16, Milano, Italy\goodbreak
\and
Dipartimento di Fisica, Universit\`{a} degli Studi di Trieste, via A. Valerio 2, Trieste, Italy\goodbreak
\and
Dipartimento di Fisica, Universit\`{a} di Roma Tor Vergata, Via della Ricerca Scientifica, 1, Roma, Italy\goodbreak
\and
European Space Agency, ESAC, Planck Science Office, Camino bajo del Castillo, s/n, Urbanizaci\'{o}n Villafranca del Castillo, Villanueva de la Ca\~{n}ada, Madrid, Spain\goodbreak
\and
European Space Agency, ESTEC, Keplerlaan 1, 2201 AZ Noordwijk, The Netherlands\goodbreak
\and
Gran Sasso Science Institute, INFN, viale F. Crispi 7, 67100 L'Aquila, Italy\goodbreak
\and
HEP Division, Argonne National Laboratory, Lemont, IL 60439, USA\goodbreak
\and
Haverford College Astronomy Department, 370 Lancaster Avenue, Haverford, Pennsylvania, U.S.A.\goodbreak
\and
Helsinki Institute of Physics, Gustaf H\"{a}llstr\"{o}min katu 2, University of Helsinki, Helsinki, Finland\goodbreak
\and
INAF - OAS Bologna, Istituto Nazionale di Astrofisica - Osservatorio di Astrofisica e Scienza dello Spazio di Bologna, Area della Ricerca del CNR, Via Gobetti 101, 40129, Bologna, Italy\goodbreak
\and
INAF - Osservatorio Astronomico di Padova, Vicolo dell'Osservatorio 5, Padova, Italy\goodbreak
\and
INAF - Osservatorio Astronomico di Trieste, Via G.B. Tiepolo 11, Trieste, Italy\goodbreak
\and
INAF, Istituto di Radioastronomia, Via Piero Gobetti 101, I-40129 Bologna, Italy\goodbreak
\and
INAF/IASF Milano, Via E. Bassini 15, Milano, Italy\goodbreak
\and
INFN - CNAF, viale Berti Pichat 6/2, 40127 Bologna, Italy\goodbreak
\and
INFN, Sezione di Bologna, viale Berti Pichat 6/2, 40127 Bologna, Italy\goodbreak
\and
INFN, Sezione di Ferrara, Via Saragat 1, 44122 Ferrara, Italy\goodbreak
\and
INFN, Sezione di Milano, Via Celoria 16, Milano, Italy\goodbreak
\and
INFN, Sezione di Roma 1, Universit\`{a} di Roma Sapienza, Piazzale Aldo Moro 2, 00185, Roma, Italy\goodbreak
\and
INFN, Sezione di Roma 2, Universit\`{a} di Roma Tor Vergata, Via della Ricerca Scientifica, 1, Roma, Italy\goodbreak
\and
IUCAA, Post Bag 4, Ganeshkhind, Pune University Campus, Pune 411 007, India\goodbreak
\and
Imperial College London, Astrophysics group, Blackett Laboratory, Prince Consort Road, London, SW7 2AZ, U.K.\goodbreak
\and
Institut d'Astrophysique Spatiale, CNRS, Univ. Paris-Sud, Universit\'{e} Paris-Saclay, B\^{a}t. 121, 91405 Orsay cedex, France\goodbreak
\and
Institut d'Astrophysique de Paris, CNRS (UMR7095), 98 bis Boulevard Arago, F-75014, Paris, France\goodbreak
\and
Institut f\"{u}r Theoretische Teilchenphysik und Kosmologie, RWTH Aachen University, D-52056 Aachen, Germany\goodbreak
\and
Institute Lorentz, Leiden University, PO Box 9506, Leiden 2300 RA, The Netherlands\goodbreak
\and
Institute of Astronomy, University of Cambridge, Madingley Road, Cambridge CB3 0HA, U.K.\goodbreak
\and
Institute of Theoretical Astrophysics, University of Oslo, Blindern, Oslo, Norway\goodbreak
\and
Instituto de Astrof\'{\i}sica de Canarias, C/V\'{\i}a L\'{a}ctea s/n, La Laguna, Tenerife, Spain\goodbreak
\and
Instituto de F\'{\i}sica de Cantabria (CSIC-Universidad de Cantabria), Avda. de los Castros s/n, Santander, Spain\goodbreak
\and
Istituto Nazionale di Fisica Nucleare, Sezione di Padova, via Marzolo 8, I-35131 Padova, Italy\goodbreak
\and
Jet Propulsion Laboratory, California Institute of Technology, 4800 Oak Grove Drive, Pasadena, California, U.S.A.\goodbreak
\and
Jodrell Bank Centre for Astrophysics, Alan Turing Building, School of Physics and Astronomy, The University of Manchester, Oxford Road, Manchester, M13 9PL, U.K.\goodbreak
\and
Kavli Institute for Cosmology Cambridge, Madingley Road, Cambridge, CB3 0HA, U.K.\goodbreak
\and
Kavli Institute for the Physics and Mathematics of the Universe (Kavli IPMU, WPI), UTIAS, The University of Tokyo, Chiba, 277- 8583, Japan\goodbreak
\and
Laboratoire d'Oc{\'e}anographie Physique et Spatiale (LOPS), Univ. Brest, CNRS, Ifremer, IRD, Brest, France\goodbreak
\and
Laboratoire de Physique Subatomique et Cosmologie, Universit\'{e} Grenoble-Alpes, CNRS/IN2P3, 53, rue des Martyrs, 38026 Grenoble Cedex, France\goodbreak
\and
Laboratoire de Physique Th\'{e}orique, Universit\'{e} Paris-Sud 11 \& CNRS, B\^{a}timent 210, 91405 Orsay, France\goodbreak
\and
Lawrence Berkeley National Laboratory, Berkeley, California, U.S.A.\goodbreak
\and
Low Temperature Laboratory, Department of Applied Physics, Aalto University, Espoo, FI-00076 AALTO, Finland\goodbreak
\and
Max-Planck-Institut f\"{u}r Astrophysik, Karl-Schwarzschild-Str. 1, 85741 Garching, Germany\goodbreak
\and
Mullard Space Science Laboratory, University College London, Surrey RH5 6NT, U.K.\goodbreak
\and
NAOC-UKZN Computational Astrophysics Centre (NUCAC), University of KwaZulu-Natal, Durban 4000, South Africa\goodbreak
\and
National Centre for Nuclear Research, ul. L. Pasteura 7, 02-093 Warsaw, Poland\goodbreak
\and
Physics Department, Shahid Beheshti University, Velenjak, Tehran 19839, Iran\goodbreak
\and
Purple Mountain Observatory, No. 8 Yuan Hua Road, 210034 Nanjing, China\goodbreak
\and
SISSA, Astrophysics Sector, via Bonomea 265, 34136, Trieste, Italy\goodbreak
\and
San Diego Supercomputer Center, University of California, San Diego, 9500 Gilman Drive, La Jolla, CA 92093, USA\goodbreak
\and
School of Chemistry and Physics, University of KwaZulu-Natal, Westville Campus, Private Bag X54001, Durban, 4000, South Africa\goodbreak
\and
School of Physical Sciences, National Institute of Science Education and Research, HBNI, Jatni-752050, Odissa, India\goodbreak
\and
School of Physics and Astronomy, Cardiff University, Queens Buildings, The Parade, Cardiff, CF24 3AA, U.K.\goodbreak
\and
School of Physics and Astronomy, Sun Yat-sen University, 2 Daxue Rd, Tangjia, Zhuhai, China\goodbreak
\and
School of Physics and Astronomy, University of Nottingham, Nottingham NG7 2RD, U.K.\goodbreak
\and
School of Physics, Indian Institute of Science Education and Research Thiruvananthapuram, Maruthamala PO, Vithura, Thiruvananthapuram 695551, Kerala, India\goodbreak
\and
School of Physics, The University of New South Wales, Sydney NSW 2052, Australia\goodbreak
\and
Simon Fraser University, Department of Physics, 8888 University Drive, Burnaby BC, Canada\goodbreak
\and
Sorbonne Universit\'{e}, Institut Lagrange de Paris (ILP), 98 bis Boulevard Arago, 75014 Paris, France\goodbreak
\and
Sorbonne Universit\'{e}, Observatoire de Paris, Universit\'{e} PSL, \'{E}cole normale sup\'{e}rieure, CNRS, LERMA, F-75005, Paris, France\goodbreak
\and
Sorbonne Universit\'{e}, UMR7095, Institut d'Astrophysique de Paris, 98 bis Boulevard Arago, F-75014, Paris, France\goodbreak
\and
Space Research Institute (IKI), Russian Academy of Sciences, Profsoyuznaya Str, 84/32, Moscow, 117997, Russia\goodbreak
\and
Space Science Data Center - Agenzia Spaziale Italiana, Via del Politecnico snc, 00133, Roma, Italy\goodbreak
\and
Space Sciences Laboratory, University of California, Berkeley, California, U.S.A.\goodbreak
\and
The Oskar Klein Centre for Cosmoparticle Physics, Department of Physics, Stockholm University, AlbaNova, SE-106 91 Stockholm, Sweden\goodbreak
\and
Universit\'{e} de Toulouse, UPS-OMP, IRAP, F-31028 Toulouse cedex 4, France\goodbreak
% ...
}
\date{\vglue -1.5mm \today\vglue -5mm}

\abstract{\vglue -3mm
We present cosmological parameter results from the final full-mission \planck\ measurements of the cosmic microwave background (CMB) anisotropies, combining information from the temperature and polarization maps and the lensing reconstruction. Compared to the 2015 results, improved measurements of large-scale polarization allow the reionization optical depth to be measured with higher precision, leading to significant gains in the precision of other correlated parameters. Improved modelling of the small-scale polarization leads to more robust constraints on many parameters, with residual modelling uncertainties estimated to affect them only at the $0.5\,\sigma$ level. We find good consistency with the standard spatially-flat 6-parameter
\lcdm\ cosmology having a power-law spectrum of adiabatic scalar perturbations (denoted ``base \lcdm'' in this paper), from polarization, temperature, and lensing, separately and in combination. A combined analysis gives
dark matter density $\Omc h^2 = 0.120\pm 0.001$,
baryon density $\Omb h^2 = 0.0224\pm 0.0001$,
scalar spectral index $\ns = 0.965\pm 0.004$, and optical depth $\tau = 0.054\pm 0.007$ (in this abstract we quote $68\,\%$ confidence regions on measured parameters and $95\,\%$ on upper limits). The angular acoustic scale is measured to $0.03\,\%$ precision, with $100\theta_*=1.0411\pm 0.0003$.
These results are only weakly dependent on the cosmological model and remain stable, with somewhat increased errors, in many commonly considered extensions.
Assuming the base-\lcdm\ cosmology, the inferred (model-dependent) late-Universe parameters are: Hubble constant
$H_0 = (67.4\pm 0.5)\Hunit$; matter density parameter $\Omm = 0.315\pm 0.007$;
and matter fluctuation amplitude $\sigma_8 = 0.811\pm 0.006$.
We find no compelling evidence for extensions to the base-\lcdm\ model. Combining with baryon acoustic oscillation (BAO) measurements (and considering single-parameter extensions) we constrain the effective extra relativistic degrees of freedom to be $\nnu = 2.99\pm 0.17$, in agreement with the Standard Model prediction $\nnu = 3.046$, and find that the neutrino mass is tightly constrained to $\mnu < 0.12\,\eV$.
The CMB spectra continue to prefer higher lensing amplitudes than predicted in base \lcdm\ at over $2\,\sigma$, which pulls some parameters that affect the lensing amplitude away from the \lcdm\ model; however, this is not supported by the
lensing reconstruction or (in models that also change the background geometry) BAO data. The joint constraint with BAO measurements on spatial curvature is consistent with a flat universe, $\Omk = 0.001\pm 0.002$. Also combining with Type Ia supernovae (SNe), the dark-energy equation of state parameter is measured to be
$\wzero = -1.03\pm 0.03$, consistent with a cosmological constant.
We find no evidence for deviations from a purely power-law primordial spectrum, and combining with data from BAO, BICEP2, and Keck Array data, we place a limit on the tensor-to-scalar ratio $r_{0.002} < 0.06$.
Standard big-bang nucleosynthesis predictions for the helium and deuterium abundances for the base-\lcdm\ cosmology are in excellent agreement with observations.
The \planck\ base-\lcdm\ results are in good agreement with BAO, SNe, and some galaxy lensing observations, but in slight tension with the Dark Energy Survey's combined-probe results including galaxy clustering (which prefers lower fluctuation amplitudes or matter density parameters), and in significant, $3.6\,\sigma$, tension with local measurements of the Hubble constant (which prefer a higher value). Simple model extensions that can partially resolve these tensions are not favoured by the \planck\ data.
}


\maketitle
\clearpage

\tableofcontents

\section{Introduction} \label{sec:intro}
Since their discovery \citep{Smoot92}, temperature anisotropies in the
cosmic microwave background (CMB) have become one of the most powerful
ways of studying cosmology and the physics of the early Universe. This
paper reports the final results on cosmological parameters from the
Planck Collaboration.\footnote{\Planck\
(\url{https://www.esa.int/Planck}) is a project of the European Space
Agency (ESA) with instruments provided by two scientific consortia
funded by ESA member states and led by Principal Investigators from
France and Italy, telescope reflectors provided through a
collaboration between ESA and a scientific consortium led and funded
by Denmark, and additional contributions from NASA (USA).}  Our first
results were presented
in \citet[][hereafter \paramsI]{planck2013-p11}. These were based on
temperature ($TT$) power spectra and CMB lensing measurements from the
first 15.5 months of \Planck\ data combined with the Wilkinson
Microwave Anisotropy Probe (\WMAP) polarization likelihood at
multipoles $\ell \le 23$ \citep{Bennett:2012zja} to constrain the
reionization optical depth
$\tau$. \citet[][hereafter \paramsII]{planck2014-a15} reported results
from the full \Planck\ mission (29 months of observations with the
High Frequency Instrument, HFI), with substantial improvements in the
characterization of the \Planck\ beams and absolute calibration
(resolving a difference between the absolute calibrations of \WMAP\
and \Planck).  The focus of \paramsII, as in \paramsI, was on
temperature observations, though we reported preliminary results on
the high-multipole $TE$ and $EE$ polarization spectra. In addition, we
used polarization measurements at low multipoles from the Low
Frequency Instrument (LFI) to constrain the value of $\tau$.

Following the completion of \paramsII, a concerted effort by
the \Planck\ team was made to reduce systematics in the HFI
polarization data at low multipoles.  First results were presented
in \citet{planck2014-a10}, which showed evidence for a lower value of
the reionization optical depth than in the 2015 results.
Further improvements to the HFI
polarization maps prepared for the 2018 data release are described
in \citet{planck2016-l03}. In this paper, we constrain $\tau$ using a
new low-multipole likelihood constructed from these maps. The
improvements in HFI data processing since \paramsII\ have very
little effect on the $TT$, $TE$, and $EE$ spectra at high
multipoles. However, this paper includes characterizations of the
temperature-to-polarization leakage and relative calibrations of the
polarization spectra enabling us to produce a combined \TTTEEE\
likelihood that is of sufficient fidelity to be used to test cosmological
models (although with some limitations, which will be described in detail
in the main body of this paper). The focus of this paper, therefore,
is to present updated cosmological results from \Planck\ power spectra and CMB
lensing measurements using temperature and polarization.

\paramsI\ showed that the \Planck\ data were remarkably consistent with a
spatially-flat \LCDM\ cosmology with purely adiabatic, Gaussian
initial fluctuations, as predicted in simple inflationary models.  We
refer to this model, which can be specified by six parameters, as
``base'' \LCDM\ in this paper. Note that in the base \LCDM\ cosmology we
assume a single minimal-mass neutrino eigenstate.  We investigated a grid of one- and
two-parameter extensions to the base-\LCDM\ cosmology (varying, for
example, the sum of neutrino masses, effective number of relativistic degrees of freedom
$\neff$, spatial curvature $\Omega_K$, or dark-energy equation of state
$w_0$), finding no statistically significant preference for any
departure from the base model. These conclusions were reinforced using
the full \planck\ mission data in \paramsII.

The analyses reported in \paramsI\ and \paramsII\ revealed some
discrepancies (often referred to as ``tensions'')
with non-\Planck\ data in the context of \lcdm\ models (e.g., \rep{distance-ladder} measurements
of the Hubble constant and determinations of the present-day amplitude of
the fluctuation spectrum), including other CMB
experiments \citep{Story:2013}. As a result, it is important to test
the fidelity of the \Planck\  data as thoroughly as
possible.  First, we would like to emphasize that where it has been
possible to compare data between different experiments at the map
level (therefore eliminating cosmic variance), they have been found to
be consistent within the levels set by instrument noise, apart from
overall differences in absolute calibration; comparisons between WMAP
and \Planck\ are described by \citet{Huang:2018xle}, between the Atacama
Cosmology Telescope (ACT) and \Planck\ by \citet{Louis:2014}, and
between the South Pole Telescope (SPT) and \Planck\ by \citet{Hou:2018}.
There have also been claims of internal inconsistencies
in the \Planck\ $TT$ power spectrum between
frequencies \citep{Spergel:2013rxa}
and between the \lcdm\ parameters obtained from low and high multipoles
\citep{Addison:2015wyg}. In addition,
the \Planck\ $TT$ spectrum preferred more lensing than expected in the
base-\LCDM\ model (quantified by the phenomenological $\Alens$ parameter
defined in Sect.~\ref{subsec:CMBlensing}) at moderate statistical
significance, raising the question of whether there are unaccounted for
systematic effects lurking within the \Planck\ data. These issues were
largely addressed in \citet{planck2014-a13}, \paramsII, and in an
associated paper, \citet{planck2016-LI}. We revisit these issues in this
paper at the cosmological parameter level, using consistency with the \planck\ polarization spectra as an
additional check. Since 2013, we have improved the absolute calibration
(fixing the amplitudes of the power spectra), added
 \Planck\ polarization, full-mission \Planck\ lensing , and produced a new low-multipole polarization
likelihood from the \Planck\ HFI. Nevertheless,  the key parameters of the
base-\LCDM\ model reported in this paper,  agree to better than $1\,\sigma_{2013}$\footnote{Here
$\sigma_{2013}$ is the standard deviation quoted on parameters
in \paramsI.} with those determined from the nominal mission temperature
data in \paramsI, with the exception of $\tau$ (which is lower in the
2018 analysis by $1.1\,\sigma_{2013}$). The cosmological parameters
from \Planck\ have remained remarkably stable since the first data
release in 2013.

The results from \planck\ are in very good agreement with simple
single-field models of inflation \citep{planck2013-p17,
planck2014-a24}. We have found no evidence for primordial
non-Gaussianity \citep{planck2013-p09a,planck2014-a19}, setting
stringent upper limits. Nor have we found any evidence for
isocurvature perturbations or cosmic defects \citep[see \paramsII\
and][]{planck2014-a24}. \Planck, together with
Bicep/Keck \citep{pb2015} polarization measurements, set tight limits
on the amplitude of gravitational waves generated during
inflation. These results are updated in this paper and in the
companion papers, describing more comprehensive tests of inflationary
models \citep{planck2016-l10} and primordial non-Gaussianity
\citep{planck2016-l09}. The \planck\ results require adiabatic, Gaussian
initial scalar fluctuations, with a red-tilted spectrum. The upper
limits on gravitational waves then require flat inflationary potentials,
which has stimulated new developments in inflationary model building
\citep[see e.g.,][and references therein]{Ferrara:2013, Kallosh:2013,
Galante:2015, Akrami:2018}. Some authors \citep{Ijjas:2013, Ijjas:2016}
have come to a very different conclusion, namely that the
\Planck/Bicep/Keck results require special initial conditions and therefore
 disfavour inflation.
This controversy lies firmly in the theoretical
domain \citep[see e.g.,][]{Guth:2014,Linde:2017}, since observations of the
CMB constrain only a limited number of $e$-folds during inflation, not the
initial conditions.  Post \Planck, inflation remains a viable and attractive
mechanism for accounting for the structure that we see in the Universe.

The layout of this paper is as follows. Section~\ref{sec:model}
describes changes to our theoretical modelling since \paramsII\ and
summarizes the likelihoods used in this paper. More comprehensive
descriptions of the power-spectrum likelihoods are given
in \citet{planck2016-l05}, while the 2018 \Planck\ CMB lensing
likelihood is described in detail
in \citet{planck2016-l08}. Section~\ref{sec:lcdm} discusses the
parameters of the base-\LCDM\ model, comparing parameters derived from
the \Planck\ $TT$, $TE$, and $EE$ power spectra. Our best estimates of the
 base-\LCDM\ cosmological parameters are derived from the full \Planck\
TT,TE,EE likelihood combined with \Planck\ CMB lensing and an HFI-based
low-multipole polarization likelihood to constrain $\tau$.  We compare
the \Planck\ $TE$ and $EE$ spectra with power spectra measured from recent
ground-based experiments in Sect.~\ref{sec:highell}.

The \Planck\ base-\LCDM\ cosmology is compared with external data sets
in Sect.~\ref{sec:datasets}. CMB power spectrum measurements suffer
from a ``geometric degeneracy'' \citep[see][]{Efstathiou:1999} which
limits their ability to constrain certain extensions to the base cosmology
(for example, allowing $\Omega_K$ or $w_0$ to vary).  \Planck\ lensing
measurements partially break the geometric degeneracy, but it is
broken very effectively with the addition of baryon acoustic oscillation
(BAO) measurements from
galaxy surveys. As in \paramsI\ and \paramsII\ we use BAO measurements
as the primary external data set to combine with \Planck.  We adopt
this approach for two reasons. Firstly, BAO-scale determinations are
relatively simple geometric measurements, with little scope for bias
from systematic errors. Secondly, the primary purpose of this paper
is to present and emphasize the \Planck\ results. We therefore make
minimal use of external data sets in reporting our main results, rather
than combining with many different data sets. Exploration of multiple
data sets can be done by others using the Monte Carlo Markov chains and
\Planck\ likelihoods released through the Planck Legacy Archive
(\PLA).\footnote{\url{https://pla.esac.esa.int}} Nevertheless,
Sect.~\ref{sec:datasets} presents a comprehensive survey of the
consistency of the \Planck\ base-\LCDM\ cosmology with different types
of astrophysical data, including Type 1a supernovae, redshift-space
distortions, galaxy shear surveys, and galaxy cluster counts. These
data sets are consistent with the
\Planck\ base-\LCDM\ cosmology with, at worst, moderate tensions at about the
$2.5\,\sigma$ level. \rep{Distance-ladder} measurements of the Hubble constant, $H_0$,
are an exception, however. The latest measurement from \citet{Riess:2019cxk}
is discrepant with the \Planck\ base-\LCDM\ value for $H_0$ at about the
\rep{$4.4\,\sigma$} level. This large discrepancy, and its possible implications for cosmology, is discussed in Sect.~\ref{sec:hubble}.

Section~\ref{sec:internalconsistency} investigates the internal
consistency of the \Planck\ base-\LCDM\ parameters, presenting
additional tests using the $TE$ and $EE$ spectra, \rep{as well as a discussion
of systematic uncertainties.}  Results from our main
grid of parameter constraints on one- or two-parameter extensions to
the base-\LCDM\ cosmology are presented in Sect.~\ref{sec:maingrid}. That
section also includes discussions of more complex models of dark
energy and modified gravity \citep[updating the results presented
in][]{planck2014-a16}, primordial nucleosynthesis, reionization,
recombination, and dark matter
annihilation. Section~\ref{sec:conclusions} summarizes our main
conclusions.

\section{Methodology and likelihoods}\label{sec:model}
\subsection{Theoretical model}

The definitions, methodology, and notation used in this paper largely
follow those adopted in the earlier Planck Collaboration papers dealing with
cosmological parameters (\paramsI, \paramsII).
Our baseline assumption is the \lcdm\ model with purely adiabatic scalar primordial perturbations with a power-law spectrum. We assume three neutrinos species, approximated as two massless states and a single massive neutrino of mass $m_\nu = 0.06\eV$.
We put flat priors on the baryon density $\omb\equiv \Omb h^2$, cold dark matter density $\omc\equiv \Omc h^2$, an approximation to the observed angular size of the sound horizon at recombination $\thetaMC$, the reionization optical depth $\tau$, the initial super-horizon amplitude of curvature perturbations $\As$ at $k=0.05\,{\Mpc}^{-1}$, and the primordial spectral index $\ns$.
Other parameter definitions, prior limits, and notation are described explicitly in table~1 of \paramsI; the only change is that we now take the amplitude prior to be flat in $\log \As$ over the range $1.61 <\log (10^{10}\As) <3.91$ (which makes no difference to \planck\ results, but is consistent with the range used for some external data analyses).

Changes in our physical modelling compared with \paramsII\ are as follows.
\begin{itemize}
 \item For modelling the small-scale nonlinear matter power spectrum, and calculating the effects of CMB lensing,
 we use the \HALOFIT\ technique \citep{Smith:2002dz} as before,
but now replace the \citet{Takahashi:2012em} approach with {\tt HMcode}, the
fitting method of \citet{Mead:2015yca,Mead:2016zqy}, as implemented in \CAMB\
\citep{Lewis:1999bs}.
\item For each model in which the fraction of baryonic mass in helium $\yhe$ is
{\it not} varied independently of other parameters, the value is now set using an updated big-bang
nucleosynthesis (BBN) prediction by interpolation on a grid of values calculated using version 1.1 of the \parthenope\ BBN code
\citep[][version 2.0 gives identical results]{Pisanti:2007hk}. We now use a fixed fiducial neutron decay-constant value of
$\tau_\neutron = 880.2\,{\rm s}$, neglecting uncertainties.
Predictions from \parthenope\ for the helium mass fraction ($\yhe\approx 0.2454$, nucleon fraction $\ypbbn\approx 0.2467$ from \planck\ in \lcdm) are lower than those from the code of~\citet{Pitrou:2018cgg} for the same value of
$\tau_\neutron$ by $\Delta\yhe \approx 0.0005$; however, other parameter results would be consistent to well within $0.1 \sigma$.
See Sect.~\ref{sec:BBN} for further discussion of BBN parameter uncertainties and code variations.
\end{itemize}

Building upon many years of theoretical effort, the computation of CMB power
spectra and the related likelihood functions has now become highly efficient and robust.
Our main results are based upon the lensed CMB power spectra computed
with the August 2017 version of the \CAMB\footnote{\url{https://camb.info}}
Boltzmann code \citep{Lewis:1999bs} and parameter constraints are based on the July 2018 version of {\tt CosmoMC}\footnote{\url{https://cosmologist.info/cosmomc/}} \citep{Lewis:2002ah,Lewis:2013hha}.
We have checked that there is very good consistency between these results and equivalent results computed  using the {\tt class} Boltzmann code~\citep{Blas:2011rf} and {\tt MontePython} sampler~\citep{Audren:2012wb,Brinckmann:2018cvx}.
Marginalized densities, limits, and contour plots are generated using updated adaptive kernel density estimates (with corrections for boundary and smoothing biases) as calculated using the {\tt getdist} package\footnote{\url{https://getdist.readthedocs.io/}} (also part of {\tt CosmoMC}), which improves average accuracy for a given number of posterior samples compared to the version used in our previous analyses.

A few new derived parameters have been added to the output of the {\tt CosmoMC} chains to allow comparisons and combinations with external data sets. A full description of all parameters is provided in
the tables presented in the Explanatory Supplement \citep{planck2016-ES}, and parameter chains are available on the \PLA.

\subsection{Power spectra and likelihoods}\label{sec:likelihood}
Since the 2015 \Planck\ data release, most of the effort on the low-level data
processing has been directed to improving the fidelity of the polarization data
at low multipoles. The first results from this effort were reported
in \citet{planck2014-a10} and led to a new determination of the reionization
optical depth, $\tau$.  The main results presented in this paper are
based on the 2018 HFI maps produced with the \sroll\ mapmaking algorithm
described in detail in \cite{planck2016-l03}, supplemented with LFI data
described in \citet{planck2016-l02}.

Because \Planck-HFI measures polarization by differencing the
signals measured by polarization-sensitive bolometers (PSBs), a number of
instrumental effects need to be controlled to achieve high precision
in the absolute calibrations of each detector. These include: effective
gain variations arising from nonlinearities in the
analogue-to-digital electronics and thermal fluctuations; far-field
beam characterization, including long bolometer time constants; and
differences in detector bandpasses. The \sroll\ mapmaking solution
for the 100--353\,GHz channels minimizes map residuals between all HFI
detectors at a given frequency, using absolute calibrations based on
the orbital dipole, together with a bandpass-mismatch model
constructed from spatial templates of the foregrounds and a
parametric model characterizing the remaining systematics. We refer
the reader to \citet{planck2016-l03} for details of the
implementation of \sroll. The fidelity of the \sroll\ maps can be
assessed using various null tests (e.g., splitting the data by
half-mission, odd-even surveys, and different detector combinations) and by the
consistency of the recovered Solar dipole solution. These tests are
described in \cite{planck2016-l03} and demonstrate that the Solar
dipole calibration is accurate to about one part in $10^4$ for the three
lowest-frequency HFI channels. Large-scale intensity-to-polarization leakage,
caused by calibration mismatch in the \sroll\ maps, is then reduced
to levels $\la 10^{-6} \mu{\rm K}^2$ at $\ell > 3$.

The low-multipole polarization likelihood used in this paper is based on the
\sroll\ polarization maps and series of end-to-end simulations that are
used to characterize the noise properties and remaining biases in the
\sroll\ maps. This low-multipole likelihood is summarized in
Sect.~\ref{sec:lowl}\ and is described in more  detail in \citet{planck2016-l05}.

As in previous \Planck\ papers, the baseline likelihood is a hybrid,
patching together a low-multipole likelihood at $\ell<30$ with a
Gaussian likelihood constructed from pseudo-cross-spectrum estimates
at higher multipoles. Correlations between the low and high multipoles
are neglected. In this paper, we have used two independent high-multipole
\TTTEEE\ likelihoods.\footnote{We use roman letters, such as \TTTEEE, to refer
to particular likelihood combinations, but use italics, such as $TT$, when
discussing power spectra more generally.}
The \plik\ likelihood, which is adopted as the baseline
in this paper, is described in Sect.~\ref{sec:plik}, while the \camspec\
likelihood is described in Sect.~\ref{sec:camspec}
and Appendix \ref{appendix:camspec}. These two likelihoods are in very good
agreement in TT, but show small differences in TE and EE, as described
below and in the main body of this paper.  Section~\ref{sec:lensing}
summarizes the \Planck\ CMB lensing likelihood, which is described in
greater detail in \citet{planck2016-l08}.

Before summarizing the high-multipole likelihoods, we make a few
remarks concerning the 2018 \sroll\ maps.  The main aim of the \sroll\
processing is to reduce the impact of systematics at low multipoles
and hence the main differences between the 2015 and 2018 HFI maps are
at low multipoles.  Compared to the 2015 HFI maps, the \sroll\ maps
eliminate the last 1000 HFI scanning rings (about 22 days of
observations) because these were less thermally stable than the rest
of the mission. \sroll\ uses higher resolution maps to determine the
destriping offsets compared to the 2015 maps, leading to a reduction of
about 12\,\% in the noise levels at 143\,GHz \citep[see figure~10
of][]{planck2016-l03}. A tighter requirement on the reconstruction
of $Q$ and $U$ values at each pixel leads to more missing pixels in the
2018 maps compared to 2015.  These and other changes to the
2018 \Planck\ maps have very little impact on the temperature and
polarization spectra at high multipoles (as will be demonstrated
explicitly in Fig.~\ref{fig:systematics} below).

There are, however, data-processing effects that need to be accounted
for to create an unbiased temperature+polarization likelihood at high multipoles
from the \sroll\ maps. In simplified form, the power absorbed by a detector at time $t$ on the sky is
\begin{equation}
P(t) \hspace{-0.8mm} = \hspace{-0.8mm} G\left\{I + \rho\left[ Q\cos 2(\psi(t) + \psi_0)+U \sin 2(\psi(t) + \psi_0) \right] \right\} + n(t), \ \ \label{bol}
\end{equation}
where $I$, $Q$, and $U$ are the beam-convolved Stokes parameters seen by
the detector at time $t$, $G$ is the effective gain (setting the
absolute calibration), $\rho$ is the detector polarization efficiency,
$\psi(t)$ is the roll angle of the satellite, $\psi_0$ is the detector
polarization angle, and $n(t)$ is the noise. For a perfect
polarization-sensitive detector, $\rho=1$, while for a perfect
unpolarized detector, $\rho = 0$. The polarization efficiencies and
polarization angles for the HFI bolometers were measured on the ground
and are reported in \cite{rosset2010}. For polarization-sensitive
detectors the ground-based measurements of polarization angles were
measured to an accuracy of approximately $1^\circ$ and the polarization
efficiencies to a quoted accuracy of 0.1--0.3\,\%. The \sroll\
mapmaking algorithm assumes the ground-based measurements of
polarization angles and efficiencies, which cannot be separated
because they are degenerate with each other.  Errors in the
polarization angles induce leakage from $E$ to $B$ modes, while errors
in the polarization efficiencies lead to gain mismatch between $I$, $Q$ and $U$.
 Analysis of the \Planck\ $TB$ and $EB$ spectra (which should
be zero in the absence of parity-violating physics) reported
in \citet{planck2016-l03}, suggest errors in the polarization angles of
$\la0.5^\circ$, within the error estimates reported in \cite{rosset2010}.
However, systematic errors in the polarization efficiencies are found to be
several times larger than the \cite{rosset2010} determinations
(which were limited to characterizations of the feed and detector sub-assemblies
and did not characterize the system in combination with the telescope)
leading to effective calibration offsets in the
polarization spectra. These polarization efficiency differences, which are
detector- and hence frequency-dependent, need to be calibrated to
construct a high-multipole likelihood. To give some representative numbers,
the \citet{rosset2010} ground-based measurements estimated polarization
efficiencies for the PSBs, with typical values of 92--96\,\% at 100\,GHz,
83--93\,\% at 143\,GHz, and 94--95\,\% at 217\,GHz (the three frequencies
used to construct the high-multipole polarization likelihoods). From the
\sroll\ maps, we find evidence of systematic errors in the polarization
efficiencies of order 0.5--1\,\% at 100 and 217\,GHz and up to 1.5\,\%
at 143\,GHz.  Differences between the main beams of the PSBs introduce
temperature-to-polarization leakage at high multipoles. We use the
{\tt QuickPol} estimates of the temperature-polarization beam transfer
function matrices, as described in \citet{quickpolHivon}, to correct
for temperature-to-polarization leakage.  Inaccuracies in the
corrections for effective polarization efficiencies and
temperature-to-polarization leakage are the main contributors to systematic
errors in the \Planck\ polarization spectra at high multipoles.

In principle, $B$-mode polarization spectra
contain information about lensing and primordial tensor modes. However,
for \planck,  $B$-mode polarization spectra are strongly noise dominated on
all angular scales. Given the very limited information contained in the \Planck\
$B$-mode spectra (and the increased complexity involved) we
do not include  $B$-mode power
spectra in the likelihoods; however, for an estimate of the lensing $B$-mode
power spectrum see \citealt{planck2016-l08}, hereafter \PlanckLensThree.

\label{sec:likelihoods}

\subsubsection{The baseline \plik\ likelihood}
\label{sec:plik}
The \plik\ high-multipole likelihood \citep[described in detail
in][hereafter PPL18]{planck2016-l05} is a Gaussian approximation to
the probability distributions of the $TT$, $EE$, and $TE$ angular
power spectra, with semi-analytic covariance matrices calculated
assuming a fiducial cosmology. It includes multipoles in the range
$30\le\ell\le 2508$ for $TT$ and $30\le\ell\le 1996$ for $TE$ and $EE$, and
is constructed from half-mission cross-spectra measured from
 the \mbox{100-,} 143-, and 217-GHz HFI frequency maps.

The TT likelihood uses four half-mission cross-spectra, with different multipole cuts to avoid multipole regions where noise dominates due to the limited resolution of the beams and
to ensure foreground contamination is correctly handled by our foreground model: $100\times100$
 ($\ell=30$--1197); $143\times143$ ($\ell=30$--1996); $143\times217$
 ($\ell=30$--2508); and $217\times217$ ($\ell=30$--2508). The TE and
 EE likelihoods also include the $100\times143$ and $100\times217$
 cross-spectra to improve the signal-to-noise ratio, and have
 different multipole cuts: $100\times100$ ($\ell=30$--$999$);
 $100\times143$ ($\ell=30$--999); $100\times217$ ($\ell=505$--999);
 $143\times143$ ($\ell=30$--1996); $143\times217$ ($\ell=505$--1996);
 and $217\times217$ ($\ell=505$--1996).  The 100-, 143-, and 217-GHz
 intensity maps are masked to reduce Galactic dust, CO, extended
 sources, and point-source contamination (a different point-source mask
 is used at each frequency), as well as badly-conditioned/missing
 pixels, effectively retaining 66, 57, and 47\,\% of the sky after
 apodization, respectively (see equation~10 in \paramsII\ for a definition of the effective sky fraction). The apodization is applied to reduce the
 mask-induced correlations between modes, and reduces the effective
 sky fraction by about $10\,\%$ compared to the unapodized masks.  The 100-, 143-, and 217-GHz maps in
 polarization are masked only for Galactic contamination and
 badly-conditioned or missing pixels, effectively retaining 70, 50, and
 41\,\% of the sky after apodization, respectively.

The baseline likelihood uses the different frequency power spectra without coadding them,
 modelling the foreground and instrumental effects with nuisance parameters that
 are marginalized over at the parameter estimation level, both in
 temperature and in polarization.  To reduce the size of the
 covariance matrix and data vector, the baseline \plik\ likelihood uses binned
 band powers, which give an excellent approximation to the unbinned likelihood
 for smooth theoretical power spectra. Unbinned versions of the likelihoods are
 also available and provide almost identical results to the binned spectra for all of the
theoretical models considered in our main parameter grid (Sect. \ref{sec:grid}).

\begin{figure*}
\begin{center}
\includegraphics{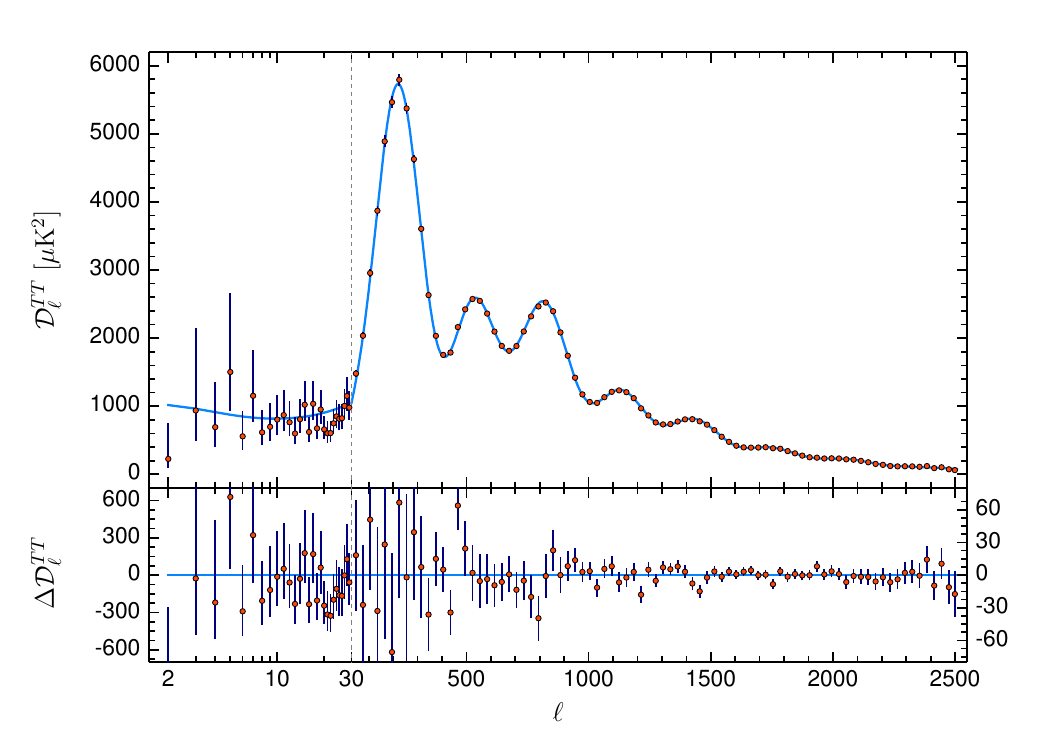}
\end{center}
\caption{
\Planck\ 2018 temperature power spectrum. At multipoles $\ell \geq 30$ we show the frequency-coadded temperature spectrum computed from the \plik\ cross-half-mission likelihood, with foreground and other nuisance parameters fixed to a best fit assuming the base-\lcdm\ cosmology. In the multipole range $2\leq\ell\leq 29$, we plot the power spectrum estimates from the {\tt Commander} component-separation algorithm, computed over 86\,\% of the sky. The base-\lcdm\ theoretical spectrum best fit to the \planckalllensing\ likelihoods is plotted in light blue in the upper panel. Residuals with respect to this model are shown in the lower panel. The error bars show $\pm 1\,\sigma$ diagonal uncertainties, including cosmic variance (approximated as Gaussian) and not including uncertainties in the foreground model at $\ell\ge 30$.
Note that the vertical scale changes at $\ell=30$, where the horizontal axis switches from logarithmic to linear.
}
\label{fig:coadded}
\end{figure*}

\begin{figure*}
\begin{center}
\includegraphics[width=0.85\textwidth]{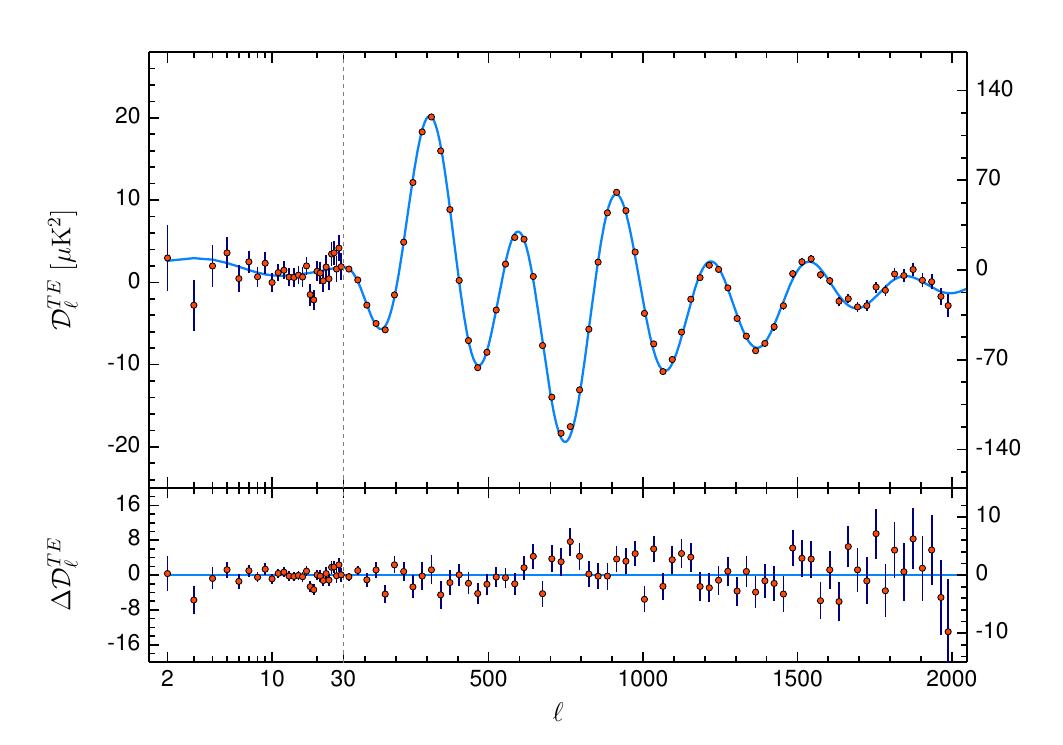}
\includegraphics[width=0.85\textwidth]{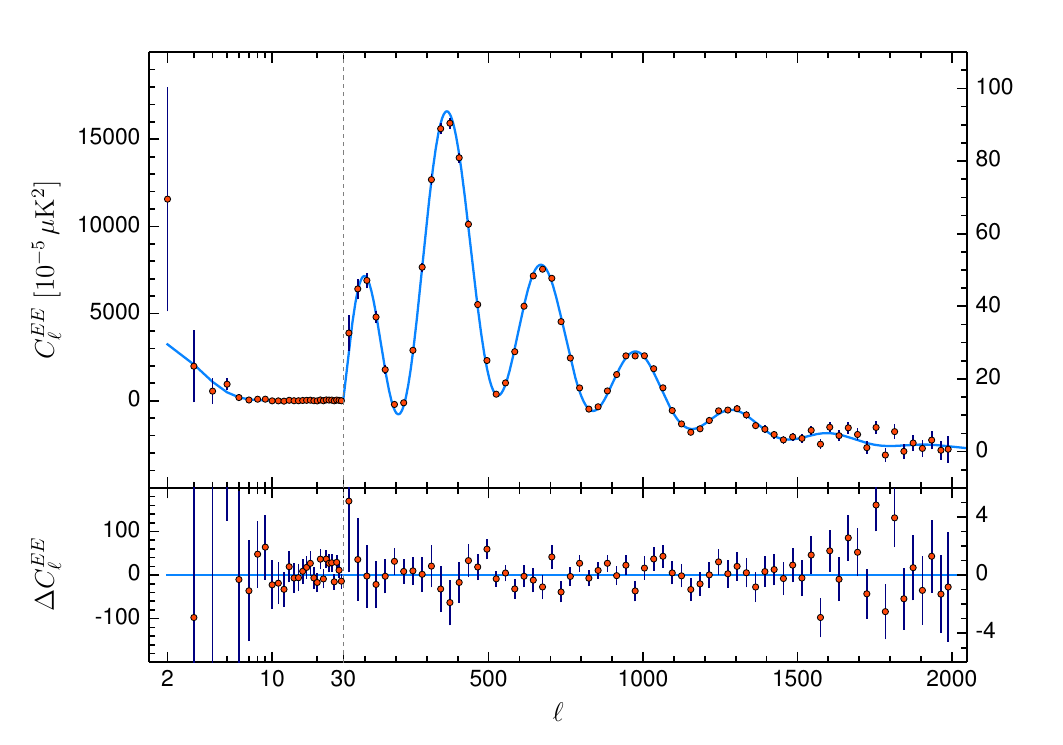}
\end{center}
\caption{
\Planck\ 2018 $TE$ (top) and $EE$ (bottom) power spectra. At multipoles $\ell \geq 30$ we show the coadded frequency spectra computed from the \plik\ cross-half-mission likelihood with foreground and other nuisance parameters fixed to a best fit assuming the base-\lcdm\ cosmology. In the multipole range $2\leq\ell\leq 29$, we plot the power spectra estimates from the \simall\ likelihood (though only the $EE$ spectrum is used in the baseline parameter analysis at $\ell\le 29$). The best-fit base-\lcdm\ theoretical spectrum fit to the \planckalllensing\ likelihood is plotted in light blue in the upper panels. Residuals with respect to this model are shown in the lower panels. The error bars show Gaussian $\pm 1\sigma$ diagonal uncertainties including cosmic variance. Note that the vertical scale changes at $\ell=30$, where the horizontal axis switches from logarithmic to linear. \label{fig:coaddedpol}
}
\end{figure*}

The major changes with respect to the 2015 \plik\ likelihood are the following.
\begin{unindentedlist}
\item \textit{Beams}. In 2015, the effective beam window functions were calculated assuming the same average sky fraction at all frequencies. In this new release, we apply beam window functions  calculated for the specific sky fraction retained at each frequency. The impact on the spectra is small, at the level of approximately 0.1\,\% at $\ell=2000$.
\item \textit{Dust modelling in $TT$}. The use of intensity-thresholded point-source masks modifies the power spectrum of the Galactic dust emission, since such masks include point-like bright Galactic dust regions. Because these point-source masks are frequency dependent, a different dust template is constructed from the 545-GHz maps for each power spectrum used in the likelihood. This differs from the approach adopted in 2015, which used a Galactic dust template with the same shape at all frequencies. As in 2015, the Galactic dust amplitudes are then left free to vary, with priors determined from cross-correlating the frequency maps used in the likelihood with the 545-GHz maps.
These changes produce small correlated shifts in the dust, cosmic infrared
background (CIB), and point-source amplitudes, but have negligible
impact on cosmological parameters.
\item \textit{Dust modelling in $TE$ and $EE$}. Dust amplitudes in $TE$ are varied with
Gaussian priors as in 2015, while in $EE$ we fix the dust amplitudes to the values
obtained using the cross-correlations with 353-GHz maps, for the reasons detailed in \likeIII.
The choice of fixing the dust amplitudes in $EE$ has a small impact (of the order of $0.2\,\sigma$) on the base-\LCDM\ results when combining into the full ``\TTTEEE,'' \plik\ likelihood because $EE$ has lower statistical power compared to $TT$ or $TE$; however, dust modelling in $EE$ has a greater effect when parameters are estimated from $EE$ alone (e.g., fixing the dust amplitude in $EE$ lowers $\ns$ by $0.8\,\sigma$, compared to allowing the dust amplitude to vary.)
\item \textit{Correction of systematic effects in the polarization spectra}. In the 2015 \Planck\ analysis,
small differences in the inter-frequency comparisons of $TE$ and $EE$
foreground-corrected polarization power spectra were identified and
attributed to systematics such as temperature-to-polarization leakage
 and polarization efficiencies, which had not been
characterized adequately at the time. For the $2018$ analysis we have applied the following corrections to the \plik\ spectra.
\begin{itemize}
\item \textit{Beam-leakage correction}. The $TE$ and $EE$ pseudo-spectra are corrected for temperature-to-polarization leakage caused by beam mismatch, using polarized beam matrices  computed with
the \texttt{QuickPol} code described in \citet{quickpolHivon}. The beam-leakage correction template is calculated using fiducial theoretical spectra computed from the best-fit \LCDM\ cosmology fitted to the $TT$ data,
together with \texttt{QuickPol} estimates of the
HFI polarized beam transfer-function matrices. This template is then included in our data model. The correction for beam leakage has a larger impact on $TE$ than on $EE$. For base-\LCDM\ cosmology, correcting for the leakage induces shifts of $\la 1\,\sigma$ when constraining parameters with \TTTEEE, namely $+1.1\,\sigma$ for $\omb$, $-0.7\,\sigma$ for $\omc$, $+0.7\,\sigma$ for $\theta_{\rm MC}$, and $+0.5\,\sigma$ for $\ns$, with smaller changes for other parameters.

\item \textit{Effective polarization efficiencies}.
We estimate the effective polarization efficiencies of the \sroll\ maps by comparing the frequency polarization power spectra to
fiducial spectra computed from the best-fit base-\LCDM\ model determined from the temperature data.
The details and limitations of this procedure are described in \likeIII\ and briefly summarized further below. Applying these polarization efficiency estimates, we find relatively small shifts to the base-\LCDM\ parameters determined from the
\TTTEEE\ likelihood, with the largest shifts in $\omb$ ($+0.4\,\sigma$), $\omc$ ($+0.2\,\sigma$), and $\ns$ ($+0.2\,\sigma$).
The parameter shifts are small because the
polarization efficiencies at different frequencies partially average out in the
coadded $TE$ spectra (see also Fig.~\ref{fig:systematics}, discussed in
Sect.~\ref{sec:lcdm}).

\item \textit{Correlated noise in auto-frequency cross-spectra and subpixel effects.}
The likelihood is built using half-mission cross-spectra to avoid noise biases from auto-spectra. However, small residual correlated noise contributions may still be present. The pixelization of the maps introduces an additional noise term because
the centroid of the ``hits'' distribution of the detector samples in each pixel does not necessarily lie at the pixel centre.
The impact of correlated noise is evaluated using the end-to-end simulations described in \citet{planck2016-l03},
 while the impact of subpixel effects is estimated with analytic calculations.
Both effects are included in the \plik\ data model, but have negligible impact on cosmological parameters.
\end{itemize}
\end{unindentedlist}
\vskip 2pt
Of the systematic effects listed above, correction for the polarization efficiencies has the largest uncertainty.
We model these factors as effective polarization calibration parameters $\calibC^{EE}_{\nu}$, defined at the power spectrum level for a frequency spectrum $\nu\times\nu$.\footnote{Thus, the polarization efficiency for a cross-frequency spectrum $\nu\times\nu'$ in, e.g., $EE$ is $\sqrt{\calibC^{EE}_{\nu}\times \calibC^{EE}_{\nu'}}$.}
To correct for errors in polarization efficiencies and large-scale beam-transfer
function errors, we recalibrate the $TE$ and $EE$ spectra against a
fiducial theoretical model to minimize
\beglet
\begin{eqnarray}
\chi^2 = (C^{\rm D} - \tens{G} C^{\rm Th})\tens{M}^{-1}(C^{\rm D} - \tens{G} C^{\rm Th}),  \label{chi1}
\end{eqnarray}
with respect to the $\calibC^{EE}_{\nu}$ parameters contained in the diagonal calibration matrix $\tens{G}$ with elements
\begin{eqnarray}
\tens{G}_{i,i} &=& \left( \frac{1}{\sqrt{\calibC^{XX}_{\nu}\calibC^{YY}_{\nu'}}} + \frac{1}{\sqrt{\calibC^{XX}_{\nu'}\calibC^{YY}_{\nu}}} \right)_{i,i} \,, \label{eq:caldef}
\end{eqnarray}
\endlet
where the index $i=1,N$ runs over the multipoles $\ell$ and frequencies $\nu\times\nu'$ of the spectra contained in the $C^{\rm D}$ data vector of dimension $N$; $C^{\rm D}$ contains the $C_\ell$ frequency spectra either for $XY=TE$ or $XY=EE$, fit separately. In Eq.~\eqref{chi1}, $\tens{M}$ is the covariance matrix for the appropriate spectra included in the fit, while the $\calibC^{TT}_{\nu}$ temperature calibration parameters are fixed. We perform the fit only using multipoles $\ell=200\mbox{--}1000$ to minimize the impact of inaccuracies in the foreground modelling or noise, and we test the stability of the results by fitting either one frequency spectrum or all the frequency spectra at the same time.
The recalibration is computed with respect to a fiducial model vector $C^{\rm Th}$ because the \Planck\ polarization
spectra are noisy and it is not possible to inter-calibrate the spectra to a precision of better than $1\,\%$ without invoking a reference model. The fiducial theoretical spectra $C^{\rm Th}_{\ell}$ contained in $C^{\rm Th}$ are derived from the best-fit temperature data alone, assuming the base-\LCDM\ model, adding the beam-leakage model and fixing the Galactic dust amplitudes to the central values of the priors obtained from using the $353$-GHz maps.
This is clearly a model-dependent procedure, but given that we fit over a restricted range of multipoles, where the
$TT$ spectra are measured to cosmic variance, the resulting polarization calibrations are insensitive to small changes in the
underlying cosmological model.

In principle, the polarization efficiencies found by fitting the $TE$
spectra should be consistent with those obtained from $EE$.  However,
the polarization efficiency at $143\times 143$, $\calibC^{EE}_{143}$,
derived from the $EE$ spectrum is about $2\,\sigma$ lower than that
derived from $TE$ (where the $\sigma$ is the uncertainty of the $TE$
estimate, of the order of $0.02$). This difference may be a
statistical fluctuation or it could be a sign of residual systematics
that project onto calibration parameters differently in $EE$ and
$TE$. We have investigated ways of correcting for effective
polarization efficiencies: adopting the estimates from $EE$ (which are
about a factor of 2 more precise than $TE$) for both the $TE$ and $EE$
spectra (we call this the ``map-based'' approach); or applying
independent estimates from $TE$ and $EE$ (the ``spectrum-based''
approach). In the baseline \plik\ likelihood we use the map-based
approach, with the polarization efficiencies fixed to the efficiencies
obtained from the fits on $EE$:
$\left(\calibC^{EE}_{100}\right)_{\mathrm{EE\, fit}}=1.021$;
$\left(\calibC^{EE}_{143}\right)_{\mathrm{EE\,fit}}=0.966$; and
$\left(\calibC^{EE}_{217}\right)_{\mathrm{EE\,
fit}}=1.040$. The \camspec\ likelihood, described in the next section,
uses spectrum-based effective polarization efficiency corrections,
leaving an overall temperature-to-polarization calibration free to
vary within a specified prior.

 The use of spectrum-based polarization efficiency estimates (which essentially differs by applying to $EE$ the
efficiencies given  above, and to $TE$ the efficiencies obtained fitting the $TE$ spectra, $\left(\calibC^{EE}_{100}\right)_{\mathrm{TE \,fit}}=1.04$, $\left(\calibC^{EE}_{143}\right)_{\mathrm{TE \,fit}}=1.0$, and $\left(\calibC^{EE}_{217}\right)_{\mathrm{TE\,fit}}=1.02$), also has a small, but non-negligible impact on cosmological parameters. For example, for the \LCDM\ model, fitting the
 \plik\ \TTTEEE+lowE likelihood, using spectrum-based polarization efficiencies, we find small shifts
in the base-\LCDM\ parameters compared with ignoring spectrum-based polarization efficiency corrections entirely; the largest of these
shifts are $+0.5\,\sigma$ in $\omb$, $+0.1\,\sigma$ in $\omc$, and $+0.3\,\sigma$ in $\ns$ (to be compared to $+0.4\,\sigma$ in $\omb$, $+0.2\,\sigma$ in $\omc$, and $+0.2\,\sigma$ in $\ns$ for the map-based case).
Furthermore, if we introduce the phenomenological $\Alens$ parameter
(discussed in much greater detail in \Alenssec), using the baseline \TTTEEE+lowE likelihood gives
 $\Alens=1.180\pm0.065$,  differing from unity by $2.7\sigma$ (the value of $\Alens$ is unchanged with respect to the case where we ignore polar efficiencies entirely, $1.180\pm 0.065$). Switching to spectrum-based polarization efficiency corrections
changes this estimate to $\Alens=1.142\pm 0.066$ differing from unity by $2.1 \sigma$. Readers of this paper should therefore
not over-interpret the \Planck\ polarization results and should be aware of the
sensitivity of these results to small changes in the specific choices and assumptions
made in constructing the polarization likelihoods, which are not accounted for in the likelihood error model. To emphasize this
point, we also give results from the \camspec\ likelihood (see, e.g., Table~\ref{table:default}), described in the next section, which has been constructed independently of \plik. We also note that if we apply the \camspec\ polarization masks
and spectrum-based polarization efficiencies in the  \plik\ likelihood, then the cosmological parameters from the
two likelihoods are in close agreement.

The coadded 2018 \plik\ temperature and polarization power spectra and residuals with respect to the base-\LCDM\ model are shown in Figs.~\ref{fig:coadded} and \ref{fig:coaddedpol}.

\subsubsection{The \camspec\ likelihood}\label{sec:camspec}

The \camspec\ temperature likelihood was used as the baseline for the
first analysis of cosmological parameters from \Planck, reported
in \paramsI, and was described  in \likeI. \rep{A detailed description of
\camspec\ and its generalization to polarization is given in \cite{Efstathiou:2019}.}
 For \paramsII,
the \camspec\ temperature likelihood was unaltered from that adopted
in \likeI, except that we used half-mission cross-spectra instead of
detector-set cross-spectra and made minor modifications to the
foreground model. For this set of papers, the \camspec\ temperature
analysis uses identical input maps and masks as \plik\ and is
unaltered from \paramsII, except for the following details.

\noindent
$\bullet$ In previous versions we used half-ring difference maps (constructed
from the first and second halves of the scanning rings within each pointing
period) to estimate noise. In this release we have used differences between
maps constructed from odd and even rings. The use of odd-even differences makes
almost no difference to the temperature analysis, since the temperature spectra
that enter the likelihood are signal dominated over most of the multipole range.
However, the odd-even noise estimates give higher noise levels than half-ring
difference estimates at multipoles ${\la}\,500$ (in qualitative agreement
with end-to-end simulations), and this improves the $\chi^2$ of the
polarization spectra. This differs from the \plik\ likelihood, which 
uses the half-ring difference maps to estimate the noise levels, together with a correction
to compensate for correlated noise, as described in \likeIII.

\noindent
$\bullet$ In \paramsII, we used power-spectrum templates for the CIB
from the halo models described in \cite{planck2013-pip56}. The
overall amplitude of the CIB power spectrum  at 217\,GHz was allowed to vary as one of the ``nuisance'' parameters in the likelihood, but the relative amplitudes
at $143\times217$ and $143\times143$  were fixed to the values given by the model. In the
2018 analysis, we retain the template shapes from \cite{planck2013-pip56}, but allow
free amplitudes at $217\times217$, $143\times 217$, and $143\times 143$. The CIB
is ignored at 100\,GHz. We made these changes to the 2018 \camspec\ likelihood
to reduce any source of systematic bias associated with the specific model of \cite{planck2013-pip56},
since this model is uncertain at low frequencies and fails
to match {\it Herschel}-SPIRE measurements \citep{Viero:2013} of the CIB anisotropies at 350 and $500\,\mu$m for $\ell \ga 3000$
\citep{Mak:2017}. This change was implemented to see whether it had any impact on the value of the lensing parameter $\Alens$
(see \Alenssec); however,  it has a negligible effect on $\Alens$ or on other cosmological parameters. The
 \plik\ likelihood retains the 2015 model for the CIB.
\noindent

\noindent
$\bullet$ In \paramsII\ we used a single functional form for the Galactic
dust power spectrum template, constructed by computing differences of
$545\times 545$ power spectra determined using different masks. The
dust template was then rescaled to match the dust amplitudes at lower
frequencies for the masks used to form the likelihood. In the
2018 \camspec\ likelihood we use dust templates computed from the
$545\times 545$ spectra, using  masks with exactly the same point-source holes
as those used to compute the $100\times 100$, $143 \times 143$,
$143\times 217$, and $217 \times 217$ power spectra that are used in
the likelihood. The \plik\ likelihood adopts a similar approach and the
\camspec\ and \plik\ dust templates are in very good agreement.

In forming the temperature likelihood, we apply multipole cuts to the
temperature spectra as follows: $\ell_{\rm min}=30$, $\ell_{\rm max} =
1200$ for the $100\times 100$ spectrum; $\ell_{\rm min}=30$,
$\ell_{\rm max} = 2000$ for the $143\times 143$ spectrum; and $\ell_{\rm
min}=500$, $\ell_{\rm max} = 2500$ for $143\times 217$ and $217\times
217$.  As discussed in previous papers, the $\ell_{\rm min}$ cuts
applied to the $143\times 217$ and $217\times 217$ spectra are imposed to
reduce any potential systematic biases arising from Galactic dust at
these frequencies. A foreground model is included in computing the
covariance matrices, assuming that foregrounds are isotropic and
Gaussian. This model underestimates the contribution of Galactic dust
to the covariances, since this component is anisotropic on the
sky.  However, dust always makes a very small contribution to the
covariance matrices in the \camspec\ likelihood. \cite{Mak:2017}
describe a simple model to account for the Galactic dust contributions
to covariance matrices.

It is important to emphasize that these changes to the 2018 \camspec\
TT likelihood are largely cosmetic and have very little impact on
cosmological parameters. This can be assessed by comparing
the \camspec\ TT results reported in this paper with those in \paramsII. The
main changes in cosmological parameters from the TT likelihood come
from the tighter constraints on the optical depth, $\tau$, adopted in
this paper.

In polarization, \camspec\ uses a different methodology to \plik. In
temperature, there are a number of frequency-dependent foregrounds at
high multipoles that are described by a physically motivated parametric model containing
``nuisance'' parameters.  These nuisance parameters are
sampled, along with cosmological parameters, during Markov chain Monte
Carlo (MCMC) exploration
of the likelihood.  The TT likelihood is therefore a power-spectrum-based
component-separation tool and it is essential to retain
cross-power spectra for each distinct frequency combination. For the
\Planck\ $TE$ and $EE$ spectra, however, Galactic dust is by far the dominant
foreground contribution. At the multipoles and sensitivities
accessible to \planck, polarized point sources make a negligible
contribution to the foreground \citep[as verified by ACTPol and
SPTpol;][]{Louis2016, Henning2017}, so the only foreground that
needs to be subtracted is polarized Galactic dust emission. As
described in \paramsII, we subtract polarized dust emission from each $TE$/$ET$
and $EE$ spectrum using the 353-GHz half-mission maps. This is done in
an analogous way to the construction of 545-GHz-cleaned temperature
maps described in \paramsII\ and Appendix \ref{appendix:camspec}.
Since the 353-GHz maps are
noisy at high multipoles we use the cleaned spectra at multipoles
$\le 300$ and extrapolate the dust model to higher multipoles by
fitting power laws to the dust estimates at lower multipoles.

The polarization spectra are then corrected for temperature-to-polarization
leakage and effective polarization efficiencies as described below,
assuming a fiducial theoretical power spectrum.  The corrected $TE$/$ET$ spectra and $EE$
spectra for all half-mission cross-spectra constructed from 100-,
143-, and 217-GHz maps are then coadded to form a single $TE$ spectrum
and a single $EE$ spectrum for the \camspec\ likelihood. The
polarization part of the \camspec\ likelihood therefore contains no
nuisance parameters other than overall calibration factors $c_{TE}$
and $c_{EE}$ for the $TE$ and $EE$ spectra. Since the \camspec\ likelihood uses coadded $TE$ and $EE$ spectra,
we do not need to bin the spectra to form a \TTTEEE\ likelihood. 
The polarization masks used in \camspec\ are based on 353--143\,GHz polarization maps that are
degraded in resolution and thresholded on $P=(Q^2 +U^2)^{1/2}$.  
The default \camspec\ polarization mask used for the 2018 analysis
preserves a fraction $f_{\rm sky}=57.7\,\%$ and is apodized to give an effective
sky fraction (see equation~10 of \paramsII) of $f^{\rm W}_{\rm sky} = 47.7\,\%$. We use the same
polarization mask for all frequencies. The \camspec\ polarization masks differ from 
those used in the \plik\ likelihood, which uses intensity-thresholded masks in polarization
(and therefore a larger effective sky area in polarization, as described in the previous section).

To construct covariance matrices, temperature-to-polarization leakage
corrections, and effective polarization efficiencies, we need to adopt
a fiducial model. For the 2018 analysis, we adopted the best-fit
\camspec\ base-\LCDM\ model from \paramsII\ to construct a likelihood
from the 2018 temperature maps. We then ran a minimizer on the TT
likelihood, imposing a prior of $\tau = 0.05 \pm 0.02$, and the best-fit
base-\LCDM\ cosmology was adopted as our fiducial model.  To deal with
temperature-to-polarization leakage, we used the {\tt QuickPol} polarized beam matrices to
compute corrections to the $TE$ and $EE$ spectra assuming the fiducial
model. The temperature-to-polarization leakage corrections are
relatively small for $TE$ spectra (although they have some impact on cosmological
parameters, consistent with the behaviour of the \plik\ likelihood
described in the previous section), but are negligible for $EE$ spectra.

To correct for effective polarization efficiencies (including large-scale transfer functions 
arising from errors in the polarized beams) we recalibrated each $TE$, $ET$, and $EE$ spectrum against the
fiducial model spectra by minimizing
\begin{equation}
\chi^2 = \sum_{\ell_1\ell_2} (C^{\rm D}_{\ell_1} - \alpha_{\rm P} C^{\rm Th}_{\ell_1})\tens{M}^{-1}_{{\ell_1}{\ell_2}}(C^{\rm D}_{\ell_2} - \alpha_{\rm P} C^{\rm Th}_{\ell_2}), \label{chi0}
\end{equation}
with respect to $\alpha_{\rm P}$, where $C^{\rm D}_\ell$ is the beam-corrected data spectrum ($TE$, $ET$, or $EE$) 
corrected for temperature-to-polarization leakage,  $\tens{M}$ is the covariance matrix for the appropriate spectrum, and the sums extend over
$200 \le \ell \le 1000$.  We calibrate each $TE$ and $EE$ spectrum individually, rather than computing map-based polarization calibrations. Although there is a good correspondence between spectrum-based calibrations and map-based calibrations, we find evidence for some differences, particularly for the $143\times143$ $EE$ spectrum in agreement with the \plik\ analysis. Unlike \plik,  we adopt 
spectrum-based calibrations of polarization efficiencies in preference to map-based calibrations.

As in temperature, we apply multipole cuts to the polarization
spectra prior to coaddition in order to reduce sensitivity to dust subtraction, beam estimation, and noise modelling. For $TE/ET$ spectra we use: $\ell_{\min}=30$ and $\ell_{\rm max}=1200$ for the $100\times100$, $100\times 143$ and $100\times 217$ spectra; 
$\ell_{\min}=30$ and $\ell_{\rm max}=2000$ for $143\times 143$ and $143\times 217$; and $\ell_{\min}=500$ and $\ell_{\rm max}=2500$ for the $217\times 217$ cross-spectrum. For $EE$, we use: $\ell_{\min}=30$ and $\ell_{\rm max}=1000$ for $100\times100$; $\ell_{\min}=30$ and $\ell_{\rm max}=1200$ for $100\times143$; $\ell_{\min}=200$  and $\ell_{\rm max}=1200$ for $100\times 217$; $\ell_{\min}=30$ and $\ell_{\rm max}=1500$ for $143\times143$; $\ell_{\min}=300$ and $\ell_{\rm max}=2000$ for $143\times217$; and $\ell_{\min}=500$ and $\ell_{\rm max}=2000$
for $217\times 217$. Since dust is subtracted from the polarization spectra, we do not include a dust model in the polarization
covariance matrices. Note that at low multipoles,  $\ell \la 300$, Galactic dust dominates over the CMB signal in $EE$ at all
frequencies. We experimented with different polarization masks and different multipole cuts and found stable results from the
\camspec\ polarization likelihood.

To summarize, for the $TT$ data \plik\ and \camspec\ use very similar
methodologies and a similar foreground model, and the power spectra used in
the likelihoods only differ in the handling of missing pixels. As a result,
there is close agreement between the two temperature likelihoods. In
polarization, different polarization masks are applied and different methods
are used for correcting Galactic dust, effective polarization calibrations,
and temperature-to-polarization leakage.
In addition, the polarization covariance matrices differ at low multipoles.
As described in Appendix \ref{appendix:camspec}, the two codes give
similar results in polarization for base \LCDM\ and most of the
extensions of \LCDM\ considered in this paper, and there would be no
material change to most of the science conclusions in this paper were one
to use the \camspec\ likelihood in place of \plik. However, in cases where there
are differences that could have an impact on the scientific interpretation
(e.g., for $A_{\rm L}$, $\mnu$, and $\Omega_K$) we show results from both
codes. This should give the reader an impression of the sensitivity of the
science results to different methodologies and choices made in constructing the
polarization blocks of the high-multipole likelihoods.

\subsubsection{The low-$\ell$ likelihood}\label{sec:lowl}

The HFI low-$\ell$ polarization likelihood is based on the full-mission HFI 100-GHz and 143-GHz Stokes $Q$ and $U$ low-resolution maps, cleaned through a template-fitting procedure using LFI 30-GHz \citep{planck2016-l02} and HFI 353-GHz maps,\footnote{The polarized synchrotron component is fitted only at 100\,GHz, being negligible at 143\,GHz. For the polarized dust component, following the prescription contained in \citet{planck2016-l03}, the low-$\ell$ HFI polarization likelihood uses the 353-GHz polarization-sensitive-bolometer-only map.} which are used as tracers of polarized synchrotron and thermal dust, respectively (for details about the cleaning procedure see \likeIII).
Power spectra are calculated based on a quadratic maximum-likelihood estimation of the cross-spectrum between the 100- and 143-GHz data, and the multipole range used spans $\ell=2$ to $\ell=29$.

We only use the $EE$ likelihood (``lowE'') for the main parameter results in this paper.
The likelihood code, called \simall, is based on the power spectra. It is constructed using an extension of the {\tt SimBaL} algorithm presented in \citet{planck2014-a10}, using  300 end-to-end simulations characterizing the HFI noise and residual systematics \citep[see][for details]{planck2016-l03}  to build an empirical probability distribution of the $EE$ spectra
(ignoring the off-diagonal correlations). \rep{The $TE$ spectrum at low multipoles does not provide tight
constraints compared to $EE$ because of cosmic variance. However, \likeIII\ discusses the $TE$ spectra
at low multipoles constructed by cross-correlating the \commander\ component-separated map
with the 100- and 143-GHz maps. The $TE$ spectra show excess variance compared to simulations
at low multipoles, most notably at $\ell\,{=}\,5$ and at $\ell\,{=}\,18$ and $19$, for reasons that are not understood.
No attempt has been made to fold in \commander\ component-separation errors in the statistical analysis. We have therefore excluded the $TE$ spectrum at low multipoles (with the added benefit of simplifying the construction of the \simall\  likelihood). Little information is lost by discarding the $TE$ spectrum. Evidently, further work is
required to understand the behaviour of $TE$ at low multipoles; however, as discussed in~\likeIII, the $\tau$
constraint derived from $TE$ to $\ell_{\rm max}\,{=}\,10$ ($\tau = 0.051 \pm 0.015$) is consistent with results
derived from the \simall\ EE likelihood summarized below.}

Using the \simall\ likelihood combined with the low-$\ell$ temperature \commander\ likelihood \citep[see][]{planck2016-l04}, varying $\ln(10^{10}\As)$ and $\tau$, but fixing other cosmological parameters to those of a fiducial base-\LCDM\ model (with parameters very close to those
of the baseline \LCDM\ cosmology in this paper), \likeIII\ reports the optical depth measurement\footnote{The corresponding marginalized amplitude parameter is $\ln(10^{10}\As) = 2.924\pm 0.052$, which gives $\As$ about 10\,\% lower than the value obtained from the joint fits in Sect.~\ref{sec:lcdm}. The $\tau$ constraints quoted here are lower than the joint results, since the small-scale power has a preference for higher $\As$ (and hence higher $\tau$ for the well-measured $\As e^{-2\tau}$ combination) at high multipoles, related to the
preference for more lensing discussed in Sect.~\ref{sec:internalconsistency}.}
\oneonesig[3cm]{\tau = 0.0506\pm 0.0086}{lowE}{.}
This is significantly tighter than the LFI-based constraint used in the 2015 release ($\tau = 0.067\pm 0.022$), and differs by about half a sigma from the result of~\citet{planck2014-a10} ($\tau = 0.055\pm 0.009$).
The latter change is driven mainly by the removal of the last 1000 scanning rings in the 2018 \sroll\ maps,
higher variance in the end-to-end simulations, and differences in the 30-GHz map used as a synchrotron tracer \citep[see appendix~A of][]{planck2016-l02}.
The impact of the tighter optical depth measurement on cosmological parameters compared to the 2015 release is discussed in Sect.~\ref{sec:paramchanges}. The error model in the final likelihood does not fully include all modelling
uncertainties and differences between likelihood codes, but the different approaches lead to estimates of $\tau$ that
are consistent within their respective $1\,\sigma$ errors.

In addition to the default \simall\ \lowE\ likelihood used in this paper, the LFI polarization likelihood has also been updated for the 2018 release, as described in detail in \likeIII. It gives consistent results to \simall, but with larger errors ($\tau = 0.063\pm 0.020$); we give a more detailed comparison of the various $\tau$ constraints in Sect.~\ref{sec:reionization}.

The low-$\ell$ temperature likelihood is based on maps from the
\commander\ component-separation algorithm, as discussed in detail
in~\citet{planck2016-l04}, with a Gibbs-sample-based Blackwell-Rao
likelihood that accurately accounts for the non-Gaussian shape of the
posterior at low multipoles, as in 2015. The CMB maps that are used
differ in several ways from the 2015 analysis. Firstly, since the 2018
analysis does not produce individual bolometer maps (since it is optimized to
reduce large-scale polarization systematics) the number of foreground
components that can be constrained is reduced compared to 2015. The 2018 \commander\ analysis
only fits the CMB, a single general low-frequency power-law component,
thermal dust, and a single CO component with spatially constant line
ratios between 100, 217, and 353\,GHz.  Secondly, the 2018 analysis is  based
only on \planck\ data and so does not including the WMAP and Haslam 408-MHz maps.
Finally, in order to be conservative with respect to CO emission, the
sky fraction  has been reduced to 86\,\% coverage, compared to
93\,\% in 2015. The net effect is a small increase in errors, and the
best-fit data points are correspondingly slightly more scattered compared
to 2015.
The (arbitrary) normalization of the \commander\ likelihood was also changed, so that a
theory power spectrum equal to the best-fit power spectrum points will, by definition,
 give $\effchisquare=0$.

\subsubsection{Likelihood notation}

Throughout this paper, we adopt the following labels for likelihoods:
(i) \planckTT\ denotes the combination of the high-$\ell$ TT likelihood at
multipoles $\ell\ge 30$, the low-$\ell$ temperature-only \commander\ likelihood,
and the low-$\ell$ EE likelihood from \simall;
(ii) labels such as \planck\ TE+lowE denote the
TE likelihood at $\ell\ge 30$ plus the low-$\ell$ EE \simall\ likelihood;
and (iii) \planckall\
denotes the combination of the combined likelihood using $TT$, $TE$,
and $EE$ spectra at $\ell\ge 30$, the low-$\ell$ temperature \commander\ likelihood, and the low-$\ell$ \simall\ EE likelihood.  For brevity we sometimes drop the ``\planck'' qualifier where it should be clear, and unless otherwise stated high-$\ell$ results are based on the \plik\ likelihood.  $TE$ correlations at $\ell \le 29$ are not included in any of the results
presented in this paper.

\subsubsection{Uncertainties on cosmological parameters}

\rep{

To maximize the accuracy of the results, various choices can be made in the construction of the high-multipole likelihoods. Examples of these are the sky area, noise models, multipole ranges, frequencies, foreground parameterization, and priors, as detailed for this release of \Planck\ data in \likeIII. The cosmological parameters and their uncertainties depend on these options. It is therefore necessary to test the sensitivity of the results with respect to such choices.
In particular, when removing or adding independent information (e.g., by lifting or adding priors, or by measuring parameters from different multipole ranges), we {\it do expect\/} cosmological parameters to shift. The crucial question, however, is whether these are in agreement with statistical expectations.
If they are consistent with being statistical excursions, then the noise model, along with foreground and instrumental nuisance parameters (e.g., polarization efficiencies), may be a consistent representation of the data. In this case, the uncertainties quoted in this paper should accurately describe the combined noise and sample variance due to finite data. Different choices of sky area, multipole range, etc., will produce changes in the parameters, but they will be adequately described by the quoted uncertainties.
On the other hand, if the shifts {\it do not\/} agree with statistical expectations, they might be an indication of unmodelled systematic effects.

In \likeIII\ we discuss a series of tests indicating the overall robustness of our results. 
Internal to the \plik\ likelihood code, we consider the CMB spectra, errors, and resulting parameters as we vary the input data, $\ell$ range, sky area, etc. We also consider the effect of known sources of systematic uncertainty, such as high-frequency oscillations in the raw time-ordered data and temperature-to-polarization leakage. We further test the baseline likelihood using extensive simulations; these tests demonstrate the solidity of our results.
As a specific example, when lifting all priors on nuisance parameters (such as calibration and foregrounds), the posterior mean on the number of relativistic species $\nnu$ shifts upwards by about $1\,\sigma$. We quantify in \likeIII\ that this is statistically not anomalous, since lifting priors reduces information and, as a consequence, error bars also increase. 

Only in a small number of areas, do such tests show mild internal disagreements at the level of spectra and parameters. 
One example is the higher than expected $\chi^2$ of the \plik\ TE frequency-likelihood, which can be traced back to a small mismatch between the different cross-frequency spectra. When we co-add the foreground-cleaned frequency TE  spectra into one CMB spectrum (which is less sensitive to such a mismatch), the related $\chi^2$ is in better agreement with expectations. A second example is the choice of polarization-efficiency corrections, which has a small impact on the final results and is further discussed below.

We have also compared the results from the \plik\ likelihood with those obtained with \camspec\ in Sects.~\ref{sec:plik} and \ref{sec:camspec} and Appendix~\ref{appendix:camspec}, as well as in \likeIII\ \citep[see also][]{Efstathiou:2019}. 
Some of the likelihood choices (e.g., sky area and multipole range) will give different detailed results within the expected sample variance. 
Others, such as the models for noise (bias-corrected half-ring difference for \plik\ versus odd-even rings for \camspec) and polarization efficiency, may give a hint of residual systematic uncertainties. 
If we restrict ourselves to temperature, the \plik\ and \camspec\ likelihoods are in excellent accord, with most parameters agreeing to better than $0.5\,\sigma$ ($0.2\,\sigma$ on the \LCDM\ model). 
On the other hand, we find indications (discussed in more detail in \likeIII) that the polarization efficiencies of the frequency-channel maps differ when measured in the $TE$ or $EE$ spectra, and the \plik\ and \camspec\ likelihoods have explored different choices of polarization efficiency corrections. This and polarization-noise modelling may be responsible for differences in the details of the resulting polarization spectra and parameters.

For the base-\LCDM\ model, the results from \plik\ and \camspec\ for the \TTTEEE\ likelihoods are in good agreement (see Table~\ref{table:default}), again with most parameters agreeing to better than $0.5\,\sigma$. We also find differences between the \plik\ and \camspec\ TTTEEE likelihoods for some extended models, especially for the single-parameter extensions with $\Alens$ (at $0.7\,\sigma$) and $\Omega_{\rm K}$ (at $0.5\,\sigma$); these differences are discussed in Sects.~\ref{sec:Alens} and \ref{sec:curv}, respectively, where we show results for both likelihoods. For both $\Alens$ and $\Omega_{\rm K}$, the \plik\ \TTTEEE\ likelihood pulls away from the base-\LCDM\ model with a slightly higher significance than the \camspec\ \TTTEEE\ likelihood. The is due, at least in part, to the choice of how to model polarization efficiencies, as discussed in \likeIII. For the $\Omk$ case, for example, the $\Delta \chi^2$ between the \LCDM\ and \LCDM+$\Omk$ models for \shortall\ is $\Delta \chi^2=11$, of which $8.3$ $\Delta\chi$ points are due to the improvement of the \plik\ \TTTEEE\ likelihood. Using spectrum-based polarization efficiencies, instead of map-based ones\footnote{As explained in Sections~\ref{sec:plik} and \ref{sec:camspec}, the ``map-based'' approach applies the same polarization efficiency corrections estimated from $EE$ to both the $TE$ and $EE$ spectra, while the ``spectrum-based'' approach applies  independent estimates obtained from $TE$ and $EE$ to the $TE$ and $EE$ spectra, respectively.} reduces that total difference to $\Delta \chi^2=5.2$, of which $\Delta \chi^2=4.6$ is due to the \plik\ likelihood. This is in agreement with the $\Delta \chi^2$ value obtained for these models by \camspec, which uses spectrum-based polarization efficiencies, with $\Delta \chi^2=4.3$.

Other details of choices in the likelihood functions impact the difference in parameters; however, these comprise both expected statistical fluctuations (due to differing raw data cuts and sky coverage) and possible residual systematic errors. For both extended models the \Planck\ TTTEEE likelihoods are usually combined with other data to break parameter degeneracies. For these parameters, the addition of either \Planck\ lensing or BAO data overwhelms any differences between the \plik\ and \camspec\ likelihoods and so we find almost identical results.

In this paper we therefore do not explicitly model an increase in error bars due to these residual systematic errors --- any such characterization would inevitably be incomplete, and it would also be impossible to give the necessary probabilistic characterization required for meaningful quantitative error bars. Instead our best-fit values, posterior means, errors and limits should (as always) be considered as conditional on the cosmological model and our best knowledge of the \Planck\ instruments and astrophysical foregrounds, as captured by the baseline likelihoods.
}

\subsection{The CMB lensing likelihood}\label{sec:lensing}

The CMB photons that arrive here today traverse almost the entire observable Universe. Along the way their paths are deflected by gradients in the gravitational potentials associated with inhomogeneities in the Universe \citep{Blanchard87}.
The dominant effects \citep[e.g.,][]{Lewis:2006fu,Hanson:2009kr} are a smoothing of
the acoustic peaks, conversion of $E$-mode polarization to $B$-mode
polarization, and generation of a connected 4-point function, each of which
can be measured in high angular resolution, low-noise observations, such
as those from \Planck.

\Planck\ was the first experiment to measure the lensing signal to sufficient
precision for it to become important for the determination of cosmological parameters, providing sensitivity to parameters that affect the late-time expansion, geometry, and clustering \citep[hereafter PL2013]{planck2013-p12}. In \citet[hereafter PL2015]{planck2014-a17} the \planck\ lensing reconstruction was
improved by including polarization information.
The \Planck\ lensing measurement is still the most significant detection of
CMB lensing to date.
In this final data release we report a measurement of the power spectrum of the lensing potential, $C_L^{\phi\phi}$, from the 4-point function, with a precision of around $2.6\,\%$ on the amplitude, as discussed in detail in \PlanckLensThree.
We demonstrate the robustness of the reconstruction to a variety of tests over lensing multipoles $8\le L \le 400$,
and conservatively restrict the likelihood to this range to reduce the impact of possible systematics. Compared to 2015, the multipole range is extended from $L_{\rm min}=40$ down to $L_{\rm min}=8$, with other analysis changes mostly introducing random fluctuations in the band powers, due to improvements in the noise modelling and the somewhat different mixture of frequencies being used in the foreground-cleaned \smica\ maps \citep[see][]{planck2016-l04}. The signal-to-noise per multipole is almost the same as in 2015, which, combined with the wider multipole range, makes the likelihood just slightly more powerful than in 2015. 
CMB lensing can provide complementary information to the \planck\ CMB power spectra, since it
it probes much lower redshifts, including $z\la 2$, when dark energy becomes important.
The lensing effect depends on the propagation of photons on null geodesics, and hence depends on  the background geometry and Weyl potential (the combination of scalar metric perturbations that determines the Weyl spacetime curvature tensor; see e.g.~\cite{Lewis:2006fu}).

\begin{figure}[tbp!]
\begin{center}
\includegraphics[angle=0]{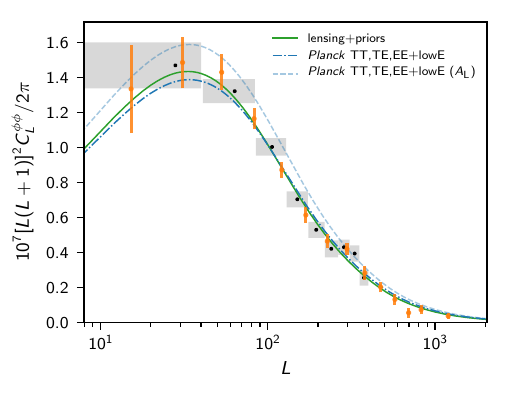}
\end{center}
\caption {CMB lensing-potential power spectrum, as measured by \planck\ (see \PlanckLensThree\ for
a detailed description of this measurement). Orange points show the full range of scales reconstructed with a logarithmic binning, while grey bands show the error and multipole range of the conservative band powers used for the likelihood, with black points showing the average multipole of the band weight. The solid line shows the best \lcdm\ fit to the conservative points alone, and the dot-dashed line shows the prediction from the best fit to the \planck\ CMB power spectra alone. The dashed line shows the prediction from the best fit to the CMB power spectra when the lensing amplitude $\Alens$ is also varied ($\Alens=1.19$ for the best-fit model; see \Alenssec\ for a detailed discussion of $\Alens$).
}
\label{fig:lenspower}
\end{figure}

We approximate the lensing likelihood as Gaussian in the estimated band powers, making perturbative corrections for the small
dependence of band powers on the cosmology, as described in \PlanckLensTwo. We neglect correlations between the 2- and 4-point functions, which are negligible at \planck\ sensitivity
\citep{Schmittfull:2013uea,Peloton:2016kbw}.
As in \PlanckLensTwo, band powers at multipoles $L>400$ are less robust than over $8\le L \le 400$,
with some evidence for a curl-test failure, and possibly also systematic differences between individual frequencies
that we were unable to resolve. Multipoles at $L<8$ are very sensitive to the large mean-field correction on these scales, and hence are sensitive to the fidelity of the simulations used to estimate the mean field. As described above, our baseline cosmological results therefore conservatively use only the
multipole range $8\le L \le 400$.

\begin{figure}[tbp!]
\begin{center}
\includegraphics[width=\columnwidth]{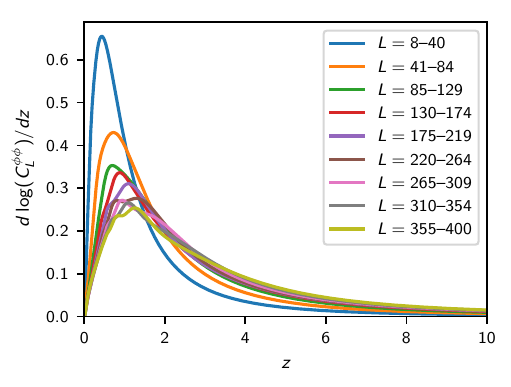}
\end{center}
\caption{Contributions to the conservative CMB lensing band powers (see text and Fig. \ref{fig:lenspower}) as a function of redshift in the base-\LCDM\ model (evaluated here, and only here, using the Limber approximation~\citep{LoVerde:2008re} on all scales). Multipole ranges of the corresponding band powers are shown in the legend.
}
\label{fig:lenskernel}
\end{figure}

The \Planck\ measurements of $C_L^{\phi\phi}$ are plotted in Fig.~\ref{fig:lenspower},
where they are compared to the predicted spectrum from the best-fitting
base-\lcdm\ model of Sect.~\ref{sec:lcdm}, and Fig.~\ref{fig:lenskernel} shows the corresponding broad redshift ranges that contribute to the lensing band powers in the \lcdm\ model.
Fig.~\ref{fig:lenspower} shows that the lensing data are in excellent agreement with the predictions inferred from the CMB power spectra in the base-\lcdm\ model ($\chi^2_{\rm eff} = 8.9$ for 9 binned conservative band-power measurements, $\chi^2_{\rm eff} =14.0$ for 14 bins over the full multipole range; we discuss agreement in extensions to the \lcdm\ model in more detail below). The lensing data prefer lensing power spectra that are slightly tilted towards less power on small scales compared to the best fit to the CMB power
spectra. This small tilt pulls joint constraints a small fraction of an error bar towards
parameters that give a lower lensing amplitude on small scales. Parameter results from the full multipole range would be a little tighter and largely consistent with the conservative band powers, although preferring slightly lower fluctuation amplitudes (see \PlanckLensThree).

As described in detail in \PlanckLensThree, the lensing likelihood (in combination with some weak priors) can alone provide \lcdm\ parameter constraints that are competitive with current galaxy lensing and clustering, measuring
\oneonesig[4cm]{\sigma_8 \Omm^{0.25} = 0.589\pm 0.020}{\planck\ lensing}{.\label{equ:lens1}}
Combined with BAO (see Sect.~\ref{sec:BAO} below) and a baryon density prior to break the main degeneracy between $H_0$, $\Omm$, and $\sigma_8$ (described in \PlanckLensTwo), individual parameters $H_0$, $\Omm$, and $\sigma_8$ can also separately be constrained to a precision of a few percent.
We use $\Omb h^2=0.0222\pm0.0005$ \citep[motivated by the primordial deuterium abundance measurements of][see also Sect.~\ref{sec:BBN}]{Cooke:2017cwo}, which gives
\threeonesig[3.5cm]{H_0 &= 67.9^{+1.2}_{-1.3}\Hunit,}{\sigma_8 &= 0.811\pm 0.019,}{\Omm &= 0.303^{+0.016}_{-0.018},}
 {\text{lensing+BAO\label{lensingBAOconstraint}}. }
The constraints of Eq.~(\ref{equ:lens1}) and (\ref{lensingBAOconstraint}) in are in very good agreement with the estimates derived from the \Planck\ power spectra and are independent of how the \Planck\ power spectra depend on the cosmological model
at high multipoles. This is a strong test of the internal consistency of the \Planck\ data. The \Planck\ lensing constraints in Eqs.~\eqref{equ:lens1} and \eqref{lensingBAOconstraint}, and the consistency of these results with the
\Planck\ power spectrum likelihoods,
should be borne in mind when comparing \Planck\ results
with other astrophysical data (e.g., direct measurements of $H_0$ and
galaxy shear surveys, see Sect.~\ref{sec:datasets}).

In this paper, we focus on joint constraints with the main \planck\ power spectrum results, where the lensing power spectrum tightens measurements of the fluctuation amplitude and improves constraints on extended models, especially when allowing for spatial curvature.

 A peculiar feature of the  \planck\ TT likelihood, reported in \paramsI\ and \paramsII, is the favouring of
high values for the lensing consistency parameter $\Alens$ (at about $2.5\,\sigma$). This result
is discussed in detail in \Alenssec. It is clear from Fig.~\ref{fig:lenspower}, however, that the \planck\ lensing likelihood prefers values of $\Alens$ close to unity and cosmological parameters that are close to those of the best-fit base-\LCDM\ parameters derived from the \planckalllensing\ likelihood (i.e., without allowing $\Alens$ to vary).

\label{subsec:CMBlensing}

\section{Constraints on base \boldlcdm}\label{sec:lcdm}

\begin{figure*}[htbp!]
\begin{center}
\includegraphics[width=18cm]{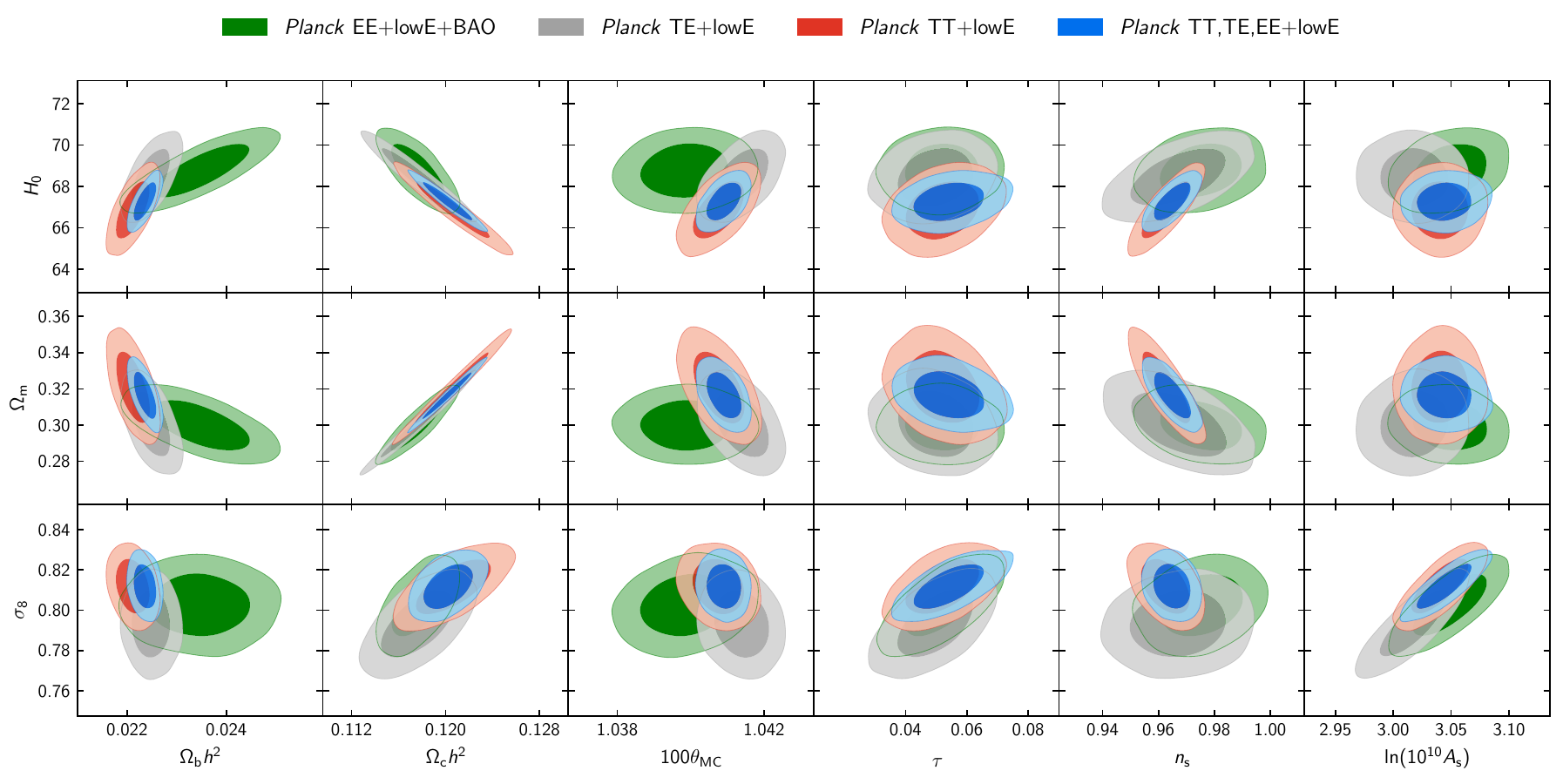}
\end{center}
\vspace{-3mm}
\caption {Constraints on parameters of the base-\lcdm\ model from the separate \planck\ $EE$, $TE$, and $TT$ high-$\ell$ spectra combined with low-$\ell$ polarization (\lowE), and, in the case of $EE$ also with BAO (described in Sect.~\ref{sec:BAO}), compared to the joint result using \planckall.
Parameters on the bottom axis are our sampled MCMC parameters with flat priors, and parameters on the left axis are derived parameters (with $H_0$ in $\Hunit$).
Contours contain $68\,\%$ and $95\,\%$ of the probability.
}
\label{fig:polrectangle}
\end{figure*}

The \planck\ measurement of seven acoustic peaks in the CMB temperature power spectrum allows cosmological parameters to be constrained extremely accurately. In previous papers, we have focussed on parameters derived from the $TT$ power spectrum. The $TE$ and $EE$ polarization spectra provide a powerful consistency check on the underlying model and also help to break some partial parameter degeneracies. The goal of this section is to explore the consistency of cosmological parameters of the base-\LCDM\ cosmology determined from $TT$, $TE$, and $EE$ spectra
and to present results from the combinations of these spectra,
which are significantly more precise that those determined using $TT$
alone.

Figure~\ref{fig:polrectangle} shows 2-dimensional marginalized
constraints on the six MCMC sampling parameters of the base-\lcdm\ model
used to  explore the parameter posteriors, plotted against the following
derived parameters: the Hubble constant $H_0$,
late-time clustering amplitude $\sigma_8$ and
matter density parameter $\Omm$ (defined including a 0.06-eV mass
neutrino).  Table~\ref{table:default} gives individual parameter
constraints using our baseline parameter combination
\planckalllensing. These represent the legacy results on the
cosmological \LCDM\ parameters from the \Planck\ satellite, and are
currently the most precise measurements coming from a single CMB
experiment. We give the best-fit values, as well as the
marginalized posterior mean values, along with the corresponding 68\,\%
probability intervals. Table~\ref{table:default} also quantifies the
small changes in parameters that are found when using the \plik\ and \camspec\
high-$\ell$ polarization analyses described in Sect.~\ref{sec:likelihoods} and
Appendix~\ref{appendix:camspec}.  Table~\ref{LCDMcompare} gives
marginalized parameter constraints from the various CMB spectra,
individually and without CMB lensing, including a wider variety of
derived parameters of physical interest.

We now discuss in more detail the parameters that are most directly
measured by the data and how these relate to constraints on individual
parameters of more general interest.

\begin{table*}[htbp!]
\caption{
Base-\lcdm\ cosmological parameters from \planckalllensing. Results for the parameter best fits, marginalized means and $68\,\%$ errors from our default analysis using the \plik\ likelihood are given in the first two numerical columns. The  \camspec\ likelihood results give some idea of the remaining modelling uncertainty in the high-$\ell$ polarization, though parts of the small shifts are due to slightly different sky areas in polarization. The ``Combined'' column give the average of the \plik\ and \camspec\ results, assuming equal weight. The combined errors are from the equal-weighted probabilities, hence including some uncertainty from the systematic difference between them; however, the differences between the high-$\ell$ likelihoods are so small that they have little effect on the 1$\,\sigma$ errors. The errors do not include modelling uncertainties in the lensing and low-$\ell$ likelihoods or other  modelling errors (such as temperature foregrounds) common to both high-$\ell$ likelihoods. A total systematic uncertainty of around $0.5\,\sigma$ may be more realistic, and values should not be overinterpreted beyond this level. The best-fit values give a representative model that is an excellent fit to the baseline likelihood, though models nearby in
the  parameter space  may have very similar likelihoods.
The first six parameters here are the ones on which we impose
flat priors and use as sampling parameters; the remaining parameters are derived from the first six. Note that $\Omm$ includes the contribution from one neutrino with a mass of 0.06\,eV.  The quantity $\thetaMC$ is an approximation to the acoustic scale angle, while $\thetastar$ is the full numerical result.
\label{table:default}
}
\vskip -4mm
\newdimen\tblskip \tblskip=5pt
\setbox\tablebox=\vbox{
 \newdimen\digitwidth
 \setbox0=\hbox{\rm 0}
 \digitwidth=\wd0
 \catcode`*=\active
 \def*{\kern\digitwidth}
 \newdimen\signwidth
 \setbox0=\hbox{+}
 \signwidth=\wd0
 \catcode`!=\active
 \def!{\kern\signwidth}
 \newdimen\dpwidth
 \setbox0=\hbox{.}
 \dpwidth=\wd0
 \catcode`?=\active
 \def?{\kern\dpwidth}
\halign{\hbox to 1.2in{#\leaderfil}\tabskip 1em&
\hfil#\hfil& \hfil#\hfil\tabskip 4em& \hfil#\hfil\tabskip 1em& \hfil#\hfil&
\hfil#\hfil\tabskip=0pt\cr
\noalign{\doubleline}
\omit\hfil Parameter\hfil& \plik\ best fit& \plik\ [1]& \camspec\ [2]& ($[2]-[1]$)/$\sigma_1$& Combined\cr
\noalign{\vskip 3pt\hrule\vskip 3pt}
$\Omega_{\mathrm{b}}h^2$& $0.022383$& $0.02237\pm0.00015$& $0.02229\pm0.00015$& $-0.5$& $0.02233\pm0.00015$\cr
$\Omega_{\mathrm{c}}h^2$& $0.12011 $& $0.1200\pm0.0012$& $0.1197\pm0.0012$& $-0.3$& $0.1198\pm0.0012$\cr
$100\theta_{\mathrm{MC}}$& $1.040909$& $1.04092\pm0.00031$& $1.04087\pm0.00031$& $-0.2$& $1.04089\pm0.00031$\cr
$\tau$& $0.0543$& $0.0544\pm0.0073$& $0.0536^{+0.0069}_{-0.0077}$& $-0.1$& $0.0540\pm0.0074$\cr
$\ln(10^{10}A_\mathrm{s})$& $3.0448$& $3.044\pm0.014$& $3.041\pm0.015$& $-0.3$& $3.043\pm0.014$\cr
$n_\mathrm{s}$& $0.96605$& $0.9649\pm0.0042$& $0.9656\pm0.0042$& $+0.2$& $0.9652\pm0.0042$\cr
\noalign{\vskip 3pt\hrule\vskip 4pt}
$\Omega_{\mathrm{m}} h^2$ & $ 0.14314 $  &  $ 0.1430\pm 0.0011 $  &  $ 0.1426\pm 0.0011 $  &  $ -0.3 $  &  $ 0.1428\pm 0.0011 $ \cr
$H_0 \,[\Hunit]$& $67.32$& $67.36\pm0.54*$& $67.39\pm0.54*$& $+0.1$& $67.37\pm0.54*$\cr
$\Omega_{\mathrm{m}}$& $0.3158$& $0.3153\pm0.0073$& $0.3142\pm0.0074$& $-0.2$& $0.3147\pm0.0074$\cr
$\mathrm{Age}\, [\mathrm{Gyr}]$& $13.7971$& $13.797\pm0.023*$& $13.805\pm0.023*$& $+0.4$& $13.801\pm0.024*$\cr
$\sigma_8$& $0.8120$& $0.8111\pm0.0060$& $0.8091\pm0.0060$& $-0.3$& $0.8101\pm0.0061$\cr
$S_8\equiv \sigma_8 (\Omm/0.3)^{0.5}$& $0.8331$& $0.832\pm0.013$& $0.828\pm0.013$& $-0.3$& $0.830\pm0.013$\cr
$z_{\mathrm{re}}$& $7.68$& $7.67\pm0.73$& $7.61\pm0.75$& $-0.1$& $7.64\pm0.74$\cr
$100\theta_\ast$& $1.041085$& $1.04110\pm0.00031$& $1.04106\pm0.00031$& $-0.1$& $1.04108\pm0.00031$\cr
$r_{\mathrm{drag}} \,[{\rm Mpc}]$& $147.049$& $147.09\pm0.26**$& $147.26\pm0.28**$& $+0.6$& $147.18\pm0.29**$\cr
\noalign{\vskip 3pt\hrule\vskip 3pt}}}
\endPlancktablewide
\end{table*}

\begin{table*}
\begin{center}

\caption{Parameter $68\,\%$ intervals for the base-\lcdm\ model from
\planck\ CMB power spectra, in combination with CMB lensing reconstruction and BAO. The top group of six rows are the base parameters, which are sampled in the MCMC analysis with flat priors. The middle group lists derived parameters. The bottom three rows show the temperature foreground amplitudes $f^{TT}_{\ell = 2000}$ for the corresponding frequency spectra (expressed
as the contribution to  $D^{TT}_{\ell=2000}$ in units of $(\mu{\rm K})^2$).
In all cases the helium mass fraction used is predicted by BBN
(posterior mean $\yhe\approx 0.2454$, with theoretical uncertainties in the BBN predictions dominating over the \Planck\ error
on $\Omb h^2$). The reionization redshift mid-point $\zre$ and optical depth $\tau$ here assumes a simple $\tanh$ model (as discussed in the text) for the reionization of hydrogen and simultaneous first reionization of helium.
Our baseline results are based on \planckalllensing\ (as also given in Table~\ref{table:default}).
}
\label{LCDMcompare}
\begingroup
\openup 5pt
\newdimen\tblskip \tblskip=5pt
\nointerlineskip
\vskip -4mm
\scriptsize
\setbox\tablebox=\vbox{
    \newdimen\digitwidth
    \setbox0=\hbox{\rm 0}
    \digitwidth=\wd0
    \catcode`"=\active
    \def"{\kern\digitwidth}
    \newdimen\signwidth
    \setbox0=\hbox{+}
    \signwidth=\wd0
    \catcode`!=\active
    \def!{\kern\signwidth}
\halign{
\hbox to 0.9in{$#$\leaderfil}\tabskip=1.5em&$#$\hfil&$#$\hfil&$#$\hfil&$#$\hfil&$#$\hfil&\hfil$#$\hfil\tabskip=0pt\cr
\noalign{\doubleline}
\multispan1\hfil \hfil&\multispan1\hfil \shortTT\hfil&\multispan1\hfil TE+\lowE\hfil&\multispan1\hfil EE+\lowE\hfil&\multispan1\hfil \shortall\hfil&\multispan1\hfil \shortall+\lensing\hfil&\multispan1\hfil \shortall+\lensing+BAO\hfil\cr
\noalign{\vskip -3pt}
\omit\hfil Parameter\hfil&\omit\hfil 68\% limits\hfil&\omit\hfil 68\% limits\hfil&\omit\hfil 68\% limits\hfil&\omit\hfil 68\% limits\hfil&\omit\hfil 68\% limits\hfil&\omit\hfil 68\% limits\hfil\cr
\noalign{\vskip 3pt\hrule\vskip 5pt}
\Omega_{\mathrm{b}} h^2&0.02212\pm 0.00022&0.02249\pm 0.00025&0.0240\pm 0.0012&0.02236\pm 0.00015&0.02237\pm 0.00015&0.02242\pm 0.00014\cr
\Omega_{\mathrm{c}} h^2&0.1206\pm 0.0021&0.1177\pm 0.0020&0.1158\pm 0.0046&0.1202\pm 0.0014&0.1200\pm 0.0012&0.11933\pm 0.00091\cr
100\theta_{\mathrm{MC}}&1.04077\pm 0.00047&1.04139\pm 0.00049&1.03999\pm 0.00089&1.04090\pm 0.00031&1.04092\pm 0.00031&1.04101\pm 0.00029\cr
\tau&0.0522\pm 0.0080&0.0496\pm 0.0085&0.0527\pm 0.0090&0.0544^{+0.0070}_{-0.0081}&0.0544\pm 0.0073&0.0561\pm 0.0071\cr
\ln(10^{10} A_\mathrm{s})&3.040\pm 0.016&3.018^{+0.020}_{-0.018}&3.052\pm 0.022&3.045\pm 0.016&3.044\pm 0.014&3.047\pm 0.014\cr
n_\mathrm{s}&0.9626\pm 0.0057&0.967\pm 0.011&0.980\pm 0.015&0.9649\pm 0.0044&0.9649\pm 0.0042&0.9665\pm 0.0038\cr
\noalign{\vskip 5pt\hrule\vskip 3pt}
H_0\,[{\rm km}\,{\rm s}^{-1}\,{\rm Mpc}^{-1}]&66.88\pm 0.92&68.44\pm 0.91&69.9\pm 2.7&67.27\pm 0.60&67.36\pm 0.54&67.66\pm 0.42\cr
\Omega_\Lambda&0.679\pm 0.013&0.699\pm 0.012&0.711^{+0.033}_{-0.026}&0.6834\pm 0.0084&0.6847\pm 0.0073&0.6889\pm 0.0056\cr
\Omega_{\mathrm{m}}&0.321\pm 0.013&0.301\pm 0.012&0.289^{+0.026}_{-0.033}&0.3166\pm 0.0084&0.3153\pm 0.0073&0.3111\pm 0.0056\cr
\Omega_{\mathrm{m}} h^2&0.1434\pm 0.0020&0.1408\pm 0.0019&0.1404^{+0.0034}_{-0.0039}&0.1432\pm 0.0013&0.1430\pm 0.0011&0.14240\pm 0.00087\cr
\Omega_{\mathrm{m}} h^3&0.09589\pm 0.00046&0.09635\pm 0.00051&0.0981^{+0.0016}_{-0.0018}&0.09633\pm 0.00029&0.09633\pm 0.00030&0.09635\pm 0.00030\cr
\sigma_8&0.8118\pm 0.0089&0.793\pm 0.011&0.796\pm 0.018&0.8120\pm 0.0073&0.8111\pm 0.0060&0.8102\pm 0.0060\cr
S_8\equiv \sigma_8(\Omega_{\rm m}/0.3)^{0.5}&0.840\pm 0.024&0.794\pm 0.024&0.781^{+0.052}_{-0.060}&0.834\pm 0.016&0.832\pm 0.013&0.825\pm 0.011\cr
\sigma_8 \Omega_{\mathrm{m}}^{0.25}&0.611\pm 0.012&0.587\pm 0.012&0.583\pm 0.027&0.6090\pm 0.0081&0.6078\pm 0.0064&0.6051\pm 0.0058\cr
z_{\mathrm{re}}&7.50\pm 0.82&7.11^{+0.91}_{-0.75}&7.10^{+0.87}_{-0.73}&7.68\pm 0.79&7.67\pm 0.73&7.82\pm 0.71\cr
10^9 A_{\mathrm{s}}&2.092\pm 0.034&2.045\pm 0.041&2.116\pm 0.047&2.101^{+0.031}_{-0.034}&2.100\pm 0.030&2.105\pm 0.030\cr
10^9 A_{\mathrm{s}} e^{-2\tau}&1.884\pm 0.014&1.851\pm 0.018&1.904\pm 0.024&1.884\pm 0.012&1.883\pm 0.011&1.881\pm 0.010\cr
\mathrm{Age}\,[\mathrm{Gyr}]&13.830\pm 0.037&13.761\pm 0.038&13.64^{+0.16}_{-0.14}&13.800\pm 0.024&13.797\pm 0.023&13.787\pm 0.020\cr
z_\ast&1090.30\pm 0.41&1089.57\pm 0.42&1087.8^{+1.6}_{-1.7}&1089.95\pm 0.27&1089.92\pm 0.25&1089.80\pm 0.21\cr
r_\ast\,[\mathrm{Mpc}]&144.46\pm 0.48&144.95\pm 0.48&144.29\pm 0.64&144.39\pm 0.30&144.43\pm 0.26&144.57\pm 0.22\cr
100\theta_\ast&1.04097\pm 0.00046&1.04156\pm 0.00049&1.04001\pm 0.00086&1.04109\pm 0.00030&1.04110\pm 0.00031&1.04119\pm 0.00029\cr
z_{\mathrm{drag}}&1059.39\pm 0.46&1060.03\pm 0.54&1063.2\pm 2.4&1059.93\pm 0.30&1059.94\pm 0.30&1060.01\pm 0.29\cr
r_{\mathrm{drag}}\,[\mathrm{Mpc}]&147.21\pm 0.48&147.59\pm 0.49&146.46\pm 0.70&147.05\pm 0.30&147.09\pm 0.26&147.21\pm 0.23\cr
k_{\mathrm{D}}\,[\mathrm{Mpc}^{-1}]&0.14054\pm 0.00052&0.14043\pm 0.00057&0.1426\pm 0.0012&0.14090\pm 0.00032&0.14087\pm 0.00030&0.14078\pm 0.00028\cr
z_{\mathrm{eq}}&3411\pm 48&3349\pm 46&3340^{+81}_{-92}&3407\pm 31&3402\pm 26&3387\pm 21\cr
k_{\mathrm{eq}}\,[\mathrm{Mpc}^{-1}]&0.01041\pm 0.00014&0.01022\pm 0.00014&0.01019^{+0.00025}_{-0.00028}&0.010398\pm 0.000094&0.010384\pm 0.000081&0.010339\pm 0.000063\cr
100\theta_{\rm{s,eq}}&0.4483\pm 0.0046&0.4547\pm 0.0045&0.4562\pm 0.0092&0.4490\pm 0.0030&0.4494\pm 0.0026&0.4509\pm 0.0020\cr
\noalign{\vskip 5pt\hrule\vskip 3pt}
f_{2000}^{143}&31.2\pm 3.0&&&29.5\pm 2.7&29.6\pm 2.8&29.4\pm 2.7\cr
f_{2000}^{143\times217}&33.6\pm 2.0&&&32.2\pm 1.9&32.3\pm 1.9&32.1\pm 1.9\cr
f_{2000}^{217}&108.2\pm 1.9&&&107.0\pm 1.8&107.1\pm 1.8&106.9\pm 1.8\cr
\noalign{\vskip 5pt\hrule\vskip 3pt}
} 
} 
\endPlancktable
\endgroup
\end{center}
\end{table*}

\subsection{Acoustic scale}

The acoustic oscillations in $\ell$ seen in the CMB power spectra
correspond to a sharply-defined acoustic angular scale on the sky,
given by $\thetastar\equiv \rstar/\DM$ where $\rstar$ is the comoving
sound horizon at recombination quantifying the distance the
photon-baryon perturbations can influence, and $\DM$ is the comoving
angular diameter distance\footnote{The quantity $\DM$ is $(1+z)D_{\rm
 A}$, where $D_{\rm A}$ is the usual angular diameter distance.}
that maps this distance into an angle on the sky. \planck\ measures
\oneonesig[4cm]{100\theta_* = 1.04097 \pm 0.00046}{\planckTT}{,}
corresponding to a precise 0.05\,\% measurement of the angular scale
$\theta_* = (0\pdeg59643\pm 0\pdeg00026)$. The angular scales of the
peaks in the polarization spectrum and cross-spectrum are different,
since the polarization at recombination is sourced by quadrupolar
flows in the photon fluid, which are out of phase with the density
perturbations. The polarization spectra can, however, be used to measure
the same acoustic scale parameter, giving a stringent test on the
assumption of purely adiabatic perturbation driving the oscillations.
From the polarization spectra we find
\beglet
\begin{eqnarray}
100\theta_* &=& 1.04156 \pm 0.00049 \quad\onesig{\planckTE}, \\
100\theta_* &=& 1.04001 \pm 0.00086 \quad\onesig{\planckEE},
\end{eqnarray}
\endlet
in excellent agreement with the
temperature measurement. The constraint from $TE$ is of similar
precision to that from $TT$: although the polarization data are much
noisier, the $TE$ and $EE$ spectra have more distinct acoustic peaks,
which helps improve the signal-to-noise ratio of the
acoustic scale measurement. Using the
combined likelihood we find: \oneonesig[4cm]{100\theta_* = 1.04109 \pm
 0.00030}{\shortall}{,} a measurement with 0.03\,\%
precision.\footnote{Doppler aberration due to the Earth's motion means
  that $\theta_*$ is expected to vary over the sky at the $10^{-3}$
  level; however, averaged over the likelihood masks, the expected
  bias for \Planck\ is below $0.1 \,\sigma$.}

 Because of  its simple geometrical interpretation, $\theta_*$ is measured very robustly and almost independently of the cosmological model (see Table~\ref{tab:base_extensions}). It is the CMB analogue of the transverse baryon acoustic oscillation scale $\rdrag/\DM$ measured  from galaxy surveys, where $\rdrag$ is the comoving sound horizon at the end of the baryonic-drag epoch  (see Sect.~\ref{sec:BAO}).
 In \lcdm, the CMB constraint can be expressed as a tight 0.04\,\%-precision relation between $\rdrag\,h$ and $\Omm$ as
 \oneonesig[3cm]{\left(\frac{\rdrag h}{\rm Mpc}\right)\left(\frac{\Omm}{0.3}\right)^{0.4} = 101.056\pm 0.036}{\shortall}{.}
  The sound horizon $\rdrag$ depends primarily on the matter, baryon, and radiation densities, which for fixed observed CMB temperature today,\footnote{We take $T_0 = 2.7255{\rm K}$~\citep{Fixsen:2009ug}, with the $\pm 0.0006 {\rm K}$ error having negligible impact on results.} gives a 0.05\,\% constraint on the combination
  \oneonesig[2.5cm]{\Omm^{0.3}h(\Omb h^2)^{-0.16}=0.87498\pm 0.00052}{\shortall}{.}
Marginalizing out the dependence on the baryon density, the remaining degeneracy between the matter density and Hubble parameters is well approximated by a constraint on the parameter combination $\Omm h^3$~\citep{Percival:2002gq}. We find a $0.3\,\%$ constraint from \Planck:
\oneonesig[4cm]{\Omm h^3 = 0.09633 \pm 0.00029}{\shortall}{,\label{ommhthree}}
 corresponding to an anti-correlation between the matter density $\Omm h^2$ and the Hubble parameter. This correlation can also be seen in Fig.~\ref{fig:polrectangle} as an anti-correlation between the dark-matter density $\Omc h^2$ and $H_0$, and a corresponding positive correlation between $\Omc h^2$ and $\Omm$.

\subsection{Hubble constant and dark-energy density}

The degeneracy between $\Omm$ and $H_0$ is not exact, but the constraint on these parameters individually is
substantially less precise than Eq.~\eqref{ommhthree}, giving
\twoonesig[2.5cm]{H_0 &= (67.27 \pm 0.60) \, \Hunit,}{\Omm &= 0.3166 \pm 0.0084,}{\shortall. \label{equ:hOparams}}
It is important to emphasize that the values given in Eq.~\eqref{equ:hOparams} assume the base-\LCDM\ cosmology with minimal neutrino mass. These estimates are highly model dependent and this needs to be borne in mind when comparing with other measurements, for example the direct measurements of $H_0$ discussed in Sect.~\ref{sec:hubble}. The values in Eq.~\eqref{equ:hOparams} are in very good
agreement with the independent constraints of Eq.~\eqref{lensingBAOconstraint} from \Planck\ CMB lensing+BAO. Including
 CMB lensing sharpens the determination of $H_0$ to a $0.8\,\%$ constraint:
\oneonesig[3cm]{H_0=(67.36 \pm 0.54)\Hunit}{\shortall\dataplus\lensing}{.}
This value is our ``best estimate'' of $H_0$ from \Planck, assuming the base-\LCDM\ cosmology.

Since we are considering a flat universe in this section,
a constraint on $\Omm$ translates directly into a constraint on the dark-energy density parameter, giving
\oneonesig{\Omega_\Lambda=0.6847 \pm 0.0073}{\shortall\dataplus\lensing}{.}
In terms of a physical density, this corresponds to $\Omega_\Lambda h^2 = 0.3107\pm 0.0082$, or cosmological constant $\Lambda = (4.24\pm 0.11) \times10^{-66}\eV^2
=(2.846\pm 0.076)\times 10^{-122}\,m_{\rm Pl}^2$ in natural units (where $m_{\rm Pl}$ is the Planck mass).

\subsection{Optical depth and the fluctuation amplitude}
\label{subsec:amplitudes}

\begin{figure}
\begin{center}
\includegraphics[width=\columnwidth]{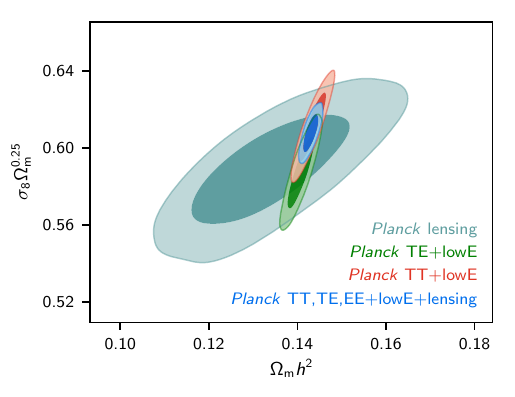}
\end{center}
\caption {
Base-\lcdm\ $68\,\%$ and $95\,\%$ marginalized constraint contours for the matter density and $\sigma_8 \Omm^{0.25}$, a fluctuation amplitude parameter that is well constrained by the CMB-lensing likelihood. The \planck\ TE, TT, and lensing likelihoods all overlap in a consistent region of parameter space, with the
combined likelihood substantially reducing the allowed parameter space.
}
\label{fig:lensingommh2}
\end{figure}

Since the CMB fluctuations are linear up to lensing corrections, and the lensing corrections are largely oscillatory, the average observed CMB power spectrum amplitude scales nearly proportionally with the primordial comoving curvature power spectrum amplitude $\As$ (which we define at the pivot scale $k_0=0.05\,{\Mpc}^{-1}$). The sub-horizon CMB anisotropies are however scattered by free electrons that are present after reionization, so the observed amplitude actually scales with $\As e^{-2\tau}$, where $\tau$ is the reionization optical depth (see Sect.~\ref{sec:reionization} for further discussion of reionization constraints).
This parameter combination is therefore well measured, with the $0.6\,\%$ constraint
\oneonesig[3.5cm]{\As e^{-2\tau} = (1.884 \pm 0.012)\times 10^{-9}}{\shortall}{.}
In this final \planck\ release the optical depth is well constrained by the large-scale polarization measurements from the \Planck\ HFI, with the joint constraint
\oneonesig{\tau = 0.0544^{+0.0070}_{-0.0081}}{\shortall}{.}
Assuming simple $\tanh$ parameterization of the ionization
fraction,\footnote{\label{footnote:tanh}For reference, the ionization fraction $x_{\rm e}=n_{\rm e}/n_{\rm H}$ in the tanh model
is assumed to have the redshift dependence~\citep{Lewis:2008wr}:
$$
x_e = {1+ n_{\rm He}/n_{\rm H} \over 2} \left [ 1 + {\rm tanh} \left ( {y(z_{\rm re}) - y(z) \over \Delta y} \right ) \right ], \nonumber
$$
where $y(z) = (1 + z)^{3/2}$, $\Delta y = \frac{3}{2}(1+z_{\rm re})^{1/2} \Delta z$, with $\Delta z = 0.5$.  Helium is assumed to be singly ionized with hydrogen at $z\gg 3$, but at lower redshifts we add the very small contribution from the second reionization of helium with a similar tanh transition at $z=3.5$.
}
 this implies a mid-point redshift of reionization
\oneonesig{\zre = 7.68 \pm 0.79}{\shortall}{,}
and a one-tail upper limit of $\zre < 9.0$ ($95\,\%$).
This is consistent with observations of high-redshift quasars that suggest the Universe was fully reionized by $z\,{\approx}\,6$ \citep{bouwens2015}. We do not include the astrophysical constraint that $\zre \ga 6.5$ in our default parameter results, but if required results including this prior are part of the published tables on the Planck Legacy Archive (\PLA).
A more detailed discussion of reionization histories consistent with \planck\ and results from other \planck\ likelihoods\ is deferred to Sect.~\ref{sec:reionization}.

The measurement of the optical depth breaks the $\As e^{-2\tau}$ degeneracy, giving a $1.5\,\%$ measurement of the primordial amplitude:
\oneonesig[4cm]{\As = (2.101^{+0.031}_{-0.034})\times 10^{-9}}{\shortall}{. \label{equ:amp}}
Since the optical depth is reasonably well constrained, degeneracies with other
cosmological parameters contribute to the error in Eq.~\eqref{equ:amp}.
 From the temperature spectrum alone there is a significant degeneracy between $\As e^{-2\tau}$ and $\Omm h^2$, since for fixed $\theta_*$,
larger values of these parameters will increase and decrease the small-scale power, respectively.
This behaviour is mitigated in our joint constraint with polarization because the polarization spectra have a different dependence on $\Omm h^2$; polarization is generated by causal sub-horizon quadrupole scattering at recombination, but the temperature spectrum has multiple sources and is also sensitive to non-local redshifting effects as the photons leave the last-scattering surface \citep[see, e.g.,][for further discussion]{Galli:2014kla}.

Assuming the \lcdm\ model, the \planck\ CMB parameter amplitude constraint can be converted into a fluctuation amplitude at the present day,
conventionally quantified by the $\sigma_8$ parameter.
The CMB lensing reconstruction power spectrum also constrains the late-time fluctuation amplitude more directly, in combination with the matter density.
Figure~\ref{fig:lensingommh2} shows constraints on the matter density and amplitude parameter combination $\sigma_8 \Omm^{0.25}$ that is well measured by the CMB lensing spectrum (see \PlanckLensTwo\ for details).
There is good consistency between the temperature, polarization, and lensing constraints here, and using their combination significantly reduces the allowed parameter space. In terms of the late-time fluctuation amplitude parameter $\sigma_8$ we find the combined result
\oneonesig{\sigma_8=0.8111 \pm 0.0060}{\planckalllensing}{.}
Measurements of galaxy clustering, galaxy lensing, and clusters can also measure $\sigma_8$, and we discuss consistency of these constraints within the \lcdm\ model in more detail in Sect.~\ref{sec:datasets}.

\begin{figure*}[htbp!]
\begin{center}
\includegraphics[width=18cm]{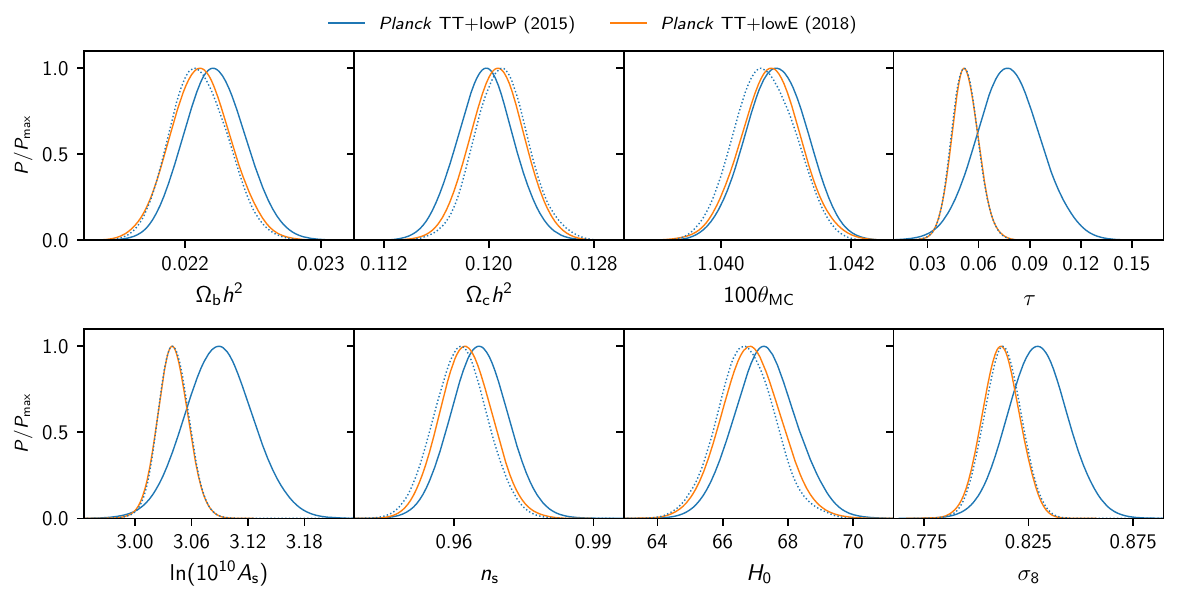}
\vspace{1cm}
\includegraphics[width=18cm]{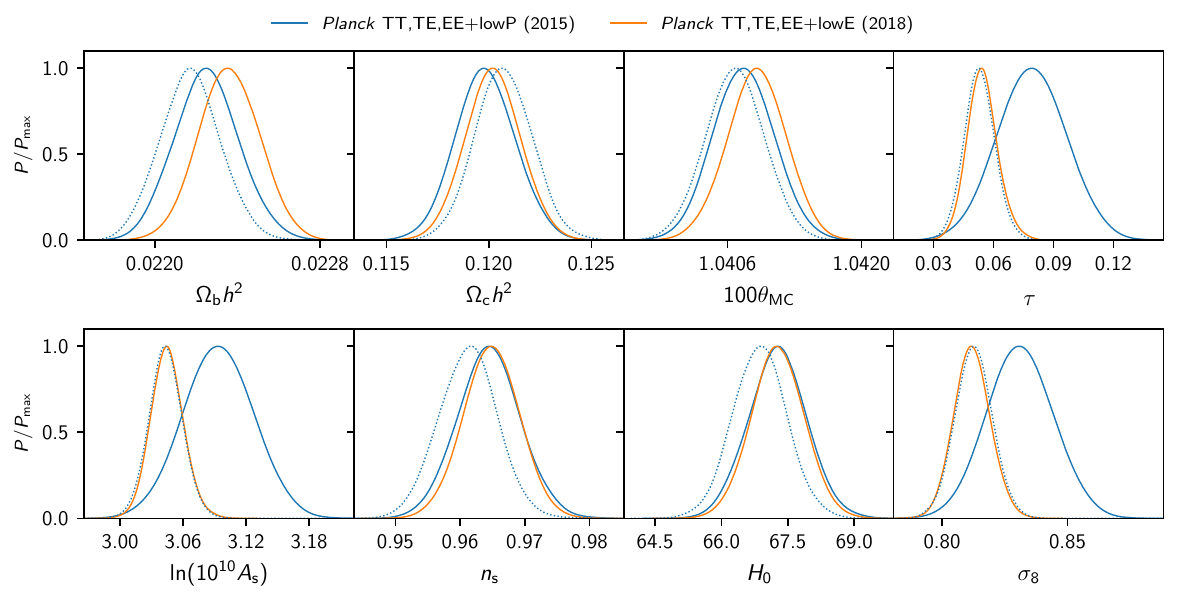}
\end{center}
\vspace{-3mm}
\caption {Comparison between the 2015 and 2018 marginalized \LCDM\ parameters. Dotted lines show the 2015 results, replacing the 2015 ``lowP'' low-$\ell$ polarization likelihood with the new 2018 ``lowE'' \simallEE\ likelihood, isolating the impact of the change in the low-$\ell$ polarization likelihood
(and hence the constraints on $\tau$).
}
\label{fig:comparison_2015_2018}
\end{figure*}

\subsection{Scalar spectral index}

The scale-dependence of the CMB power spectrum constrains the slope of
the primordial scalar power spectrum, conventionally
parameterized by the power-law index $\ns$, where $\ns=1$ corresponds
to a scale-invariant spectrum.  The matter and baryon densities also
affect the scale-dependence of the CMB spectra, but in a way that
differs from a variation in $n_s$, leading to relatively mild degeneracies
between these parameters. Assuming that the primordial power spectrum is an
exact power law we find
\oneonesig{\ns = 0.9649 \pm 0.0042}{\planckalllensing}{, \label{equ:n_s}}
which is $8\,\sigma$ away from
scale-invariance ($\ns=1$), confirming the red tilt of the spectrum at high
significance in \lcdm.  Section~\ref{sec:early}
and~\cite{planck2016-l10} discuss the implications of this result for models of
inflation and include constraints on models with primordial tensor modes
and a scale-dependent scalar spectral index.

\begin{figure*}[htbp!]
\begin{center}
\includegraphics[width=18cm]{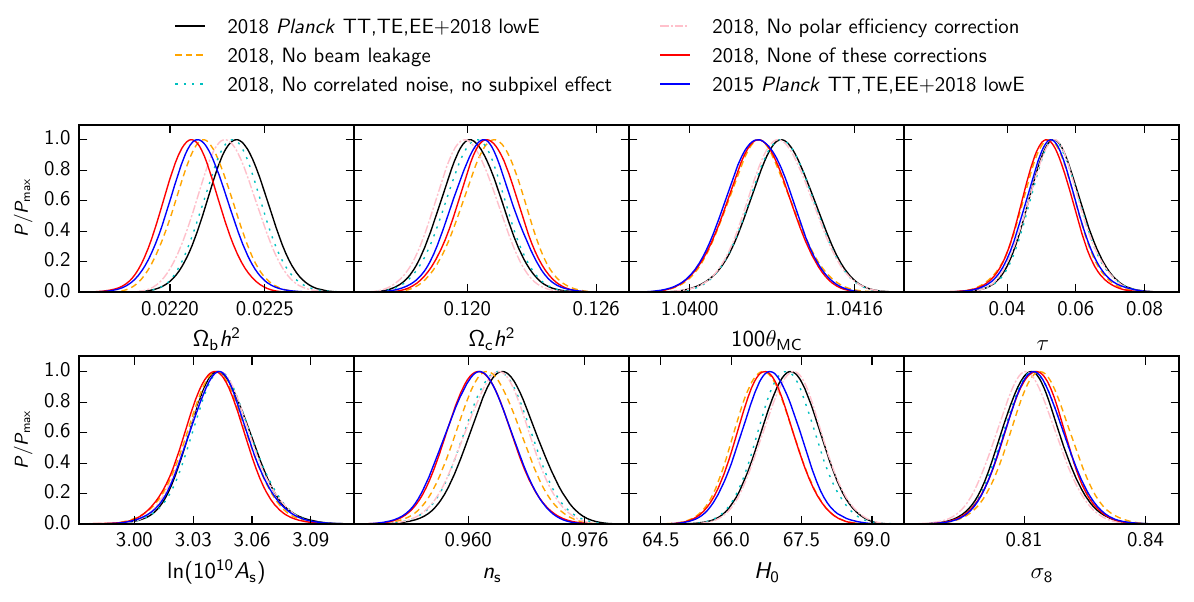}
\end{center}
\vspace{-3mm}
\caption {Impact of corrections for systematic effects on 2018 marginalized \LCDM\ parameters from \planckall. The plot shows the baseline results (black solid line), and the baseline result excluding corrections for various effects: beam leakage (dashed orange); polarization efficiencies (dot-dashed pink); and subpixel effects and correlated noise (dotted cyan). The impact of not including any of these corrections is shown in solid blue, and agree fairly well with the 2015 results if the 2015 low-$\ell$ polarization likelihood is replaced with 2018 lowE likelihood (2015 \planckTTTEEEonly\dataplus 2018 \lowE). This shows that corrections for polarization
systematics account for most of the small changes between the 2015 and 2018 results that are not caused by the change in optical depth.
}
\label{fig:comparison_2015_neweffects}

\begin{center}
\includegraphics[width=17cm]{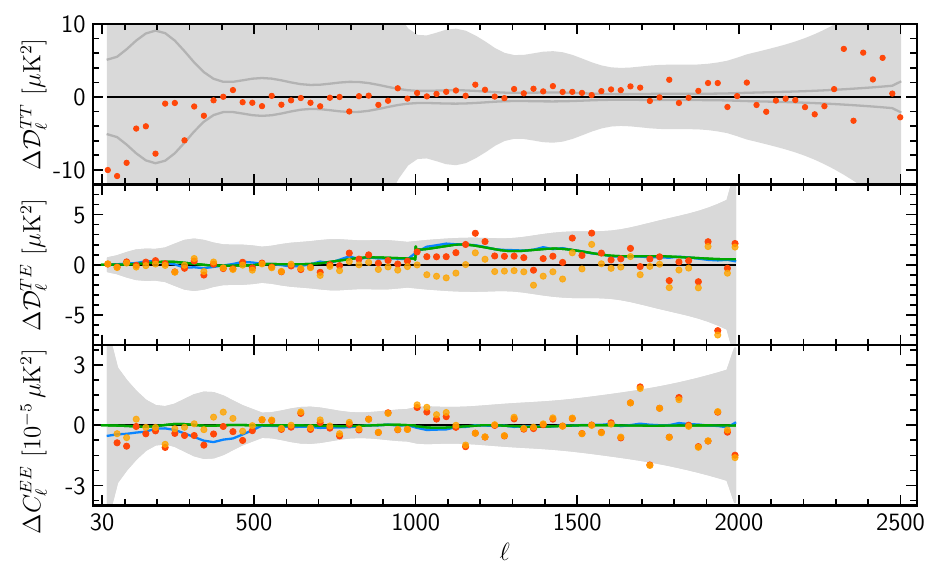}
\end{center}
\vspace{-3mm}
\caption {Differences between the 2018 and 2015 coadded power spectra at high $\ell$ in $TT$, $TE$, and $EE$ from top to bottom (red points). The 2015 $TT$ spectrum has been recalibrated by a factor of 1.00014. For $TE$ and $EE$, the orange points show the same differences but without applying the polar-efficiency and beam-leakage corrections to the 2018 spectra. This shows that the differences between the two data releases in polarization are caused mainly by these two effects. Finally, the green line shows the coadded beam-leakage correction, while the blue line shows the sum of the beam-leakage and polar-efficiency corrections. The grey band shows the $\pm1\,\sigma$ errors of the 2018 power spectra (for $TT$, the grey line also shows error bars scaled down by a factor of 10).
}
\label{fig:systematics}
\end{figure*}

\subsection{Matter densities}

The matter density can be measured from the CMB spectra using the scale-dependence of the amplitude, since for fixed $\theta_*$ a larger matter density reduces the small-scale CMB power. The matter density also affects the amount of lensing in the CMB spectra and the amplitude of the CMB-lensing reconstruction spectrum.
The matter density is well constrained to be
\oneonesig[3.5cm]{\Omm h^2= 0.1430 \pm 0.0011}{\planckalllensing}{.}
The matter mostly consists of cold dark matter, with density constrained at the percent level:
\oneonesig[3.5cm]{\Omc h^2= 0.1200 \pm  0.0012}{\planckalllensing}{.}
Changes in the baryon density affect the spectrum in characteristic ways, modifying the relative heights of the even and odd acoustic peaks, due to the effect of baryons on the depth of first and subsequent acoustic (de)compressions. Despite
comprising less than a sixth of the total matter content, the baryon effects
on the power spectra are sufficiently distinctive that the baryon-density
parameter is measured at sub-percent level accuracy with \Planck:
\oneonesig[3.5cm]{\Omb h^2 =0.02237 \pm 0.00015}{\planckalllensing}{.}
There is a partial degeneracy with $\ns$, which can also affect the relative heights of the first few peaks. This is most evident in $TE$, but is
reduced in $TT$ because of the larger range of scales that are measured by \planck\ with low noise.

\subsection{\rep{Changes in the base-\lcdm\ parameters between the
2015 and 2018 data releases}}
\label{sec:paramchanges}

Figure~\ref{fig:comparison_2015_2018} compares the parameters of the
base-{\lcdm} model measured from the final data release with those
reported in \paramsII.  To differentiate between changes caused by the
new \lowE\ polarization likelihood, and therefore generated by the
change in the measured optical depth to reionization, we also show the
result of using the 2015 likelihoods in combination with the 2018
\lowE\ polarization likelihood at low multipoles.
Figure~\ref{fig:comparison_2015_2018} includes the results
for both \planckTT\ and \planckall.\footnote{The published 2015
  parameter constraints and chains had a small error in the priors for
  the polarization Galactic foregrounds, which was subsequently
  corrected in the published likelihoods. The impact on cosmological
  parameters was very small. Here we compare with the uncorrected 2015 chains, not
  the published 2015 likelihood.}

The main differences in \LCDM\ parameters between the 2015 and the 2018 releases are caused by the following effects.
\begin{unindentedlist}

\item{New polarization low-$\ell$ likelihood.} The use of the new HFI low-$\ell$ polarization likelihood in place of the 2015 LFI likelihood is the largest cause of shifts between the 2015 and 2018 parameters. The lowering and tightening of the constraint on $\tau$ is responsible for a $1\,\sigma$ decrease of $\lnAs$ through the $\Astau$ degeneracy. This in turn decreases the smoothing due to gravitational lensing at high multipoles, which is compensated by an increase of about $1\,\sigma$ in $\omc$. This decreases the amplitude of the first acoustic
peak, so $\ns$ shifts to a lower value by about $0.5\,\sigma$ to restore power. Further adjustments are then achieved by the changes of $\thetastar$ and $\omb$ by about $0.5\,\sigma$.

\item{Polarization corrections in the high-$\ell$ likelihood.}
As described in detail in Sect.~\ref{sec:likelihoods}, the largest changes
from 2015 are caused by corrections applied to the polarization spectra.
To isolate the causes of shifts introduced by changes in the high-$\ell$ likelihood, Fig.~\ref{fig:comparison_2015_neweffects} compares 2018 results
neglecting corrections to the polarization spectra with results from the
2015 high-$\ell$ likelihood combined with the 2018 lowE likelihood
(so that both sets of results are based on similar constraints on $\tau$).
The shift towards larger values in $\omb$ by around $1\,\sigma$ is mainly caused by
the beam-leakage correction in the TE high-$\ell$ likelihood, which is also responsible for an increase of approximately $0.5\,\sigma$ in $\ns$, compensating for the shift in $\ns$ as a result of the change in $\tau$ since
2015. The beam-leakage correction also changes $\omc$ (by $-0.7\,\sigma$) and $\thetaMC$ ($+0.7\,\sigma$). The other corrections implemented in 2018 have a smaller impact on the \LCDM\ parameters, as described in detail in \citet{planck2016-l05}.
\end{unindentedlist}

Figure~\ref{fig:systematics} presents the differences between the coadded spectra from 2018 and 2015. This plot shows the stability of the $TT$ spectra, while also demonstrating that the main differences in polarization between the 2015 and 2018 releases are caused by the 2018 corrections for polarization efficiencies and beam leakage.

\section{Comparison with high-resolution experiments}\label{sec:highell}
\begin{figure}[htbp!]
\begin{center}
\centering
\hspace{-3mm}\includegraphics[width=0.965\columnwidth]{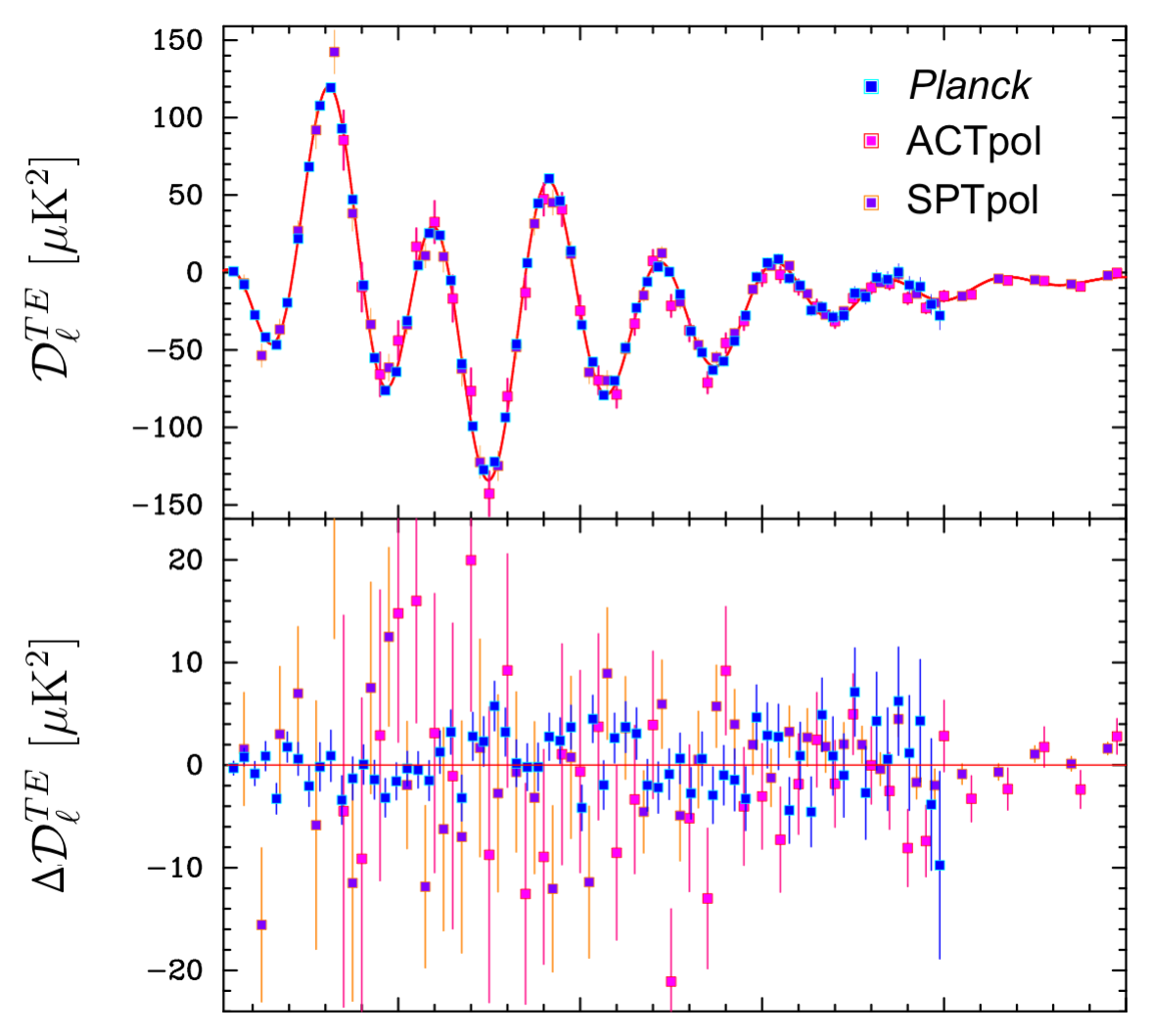} \\
\includegraphics[width=\columnwidth]{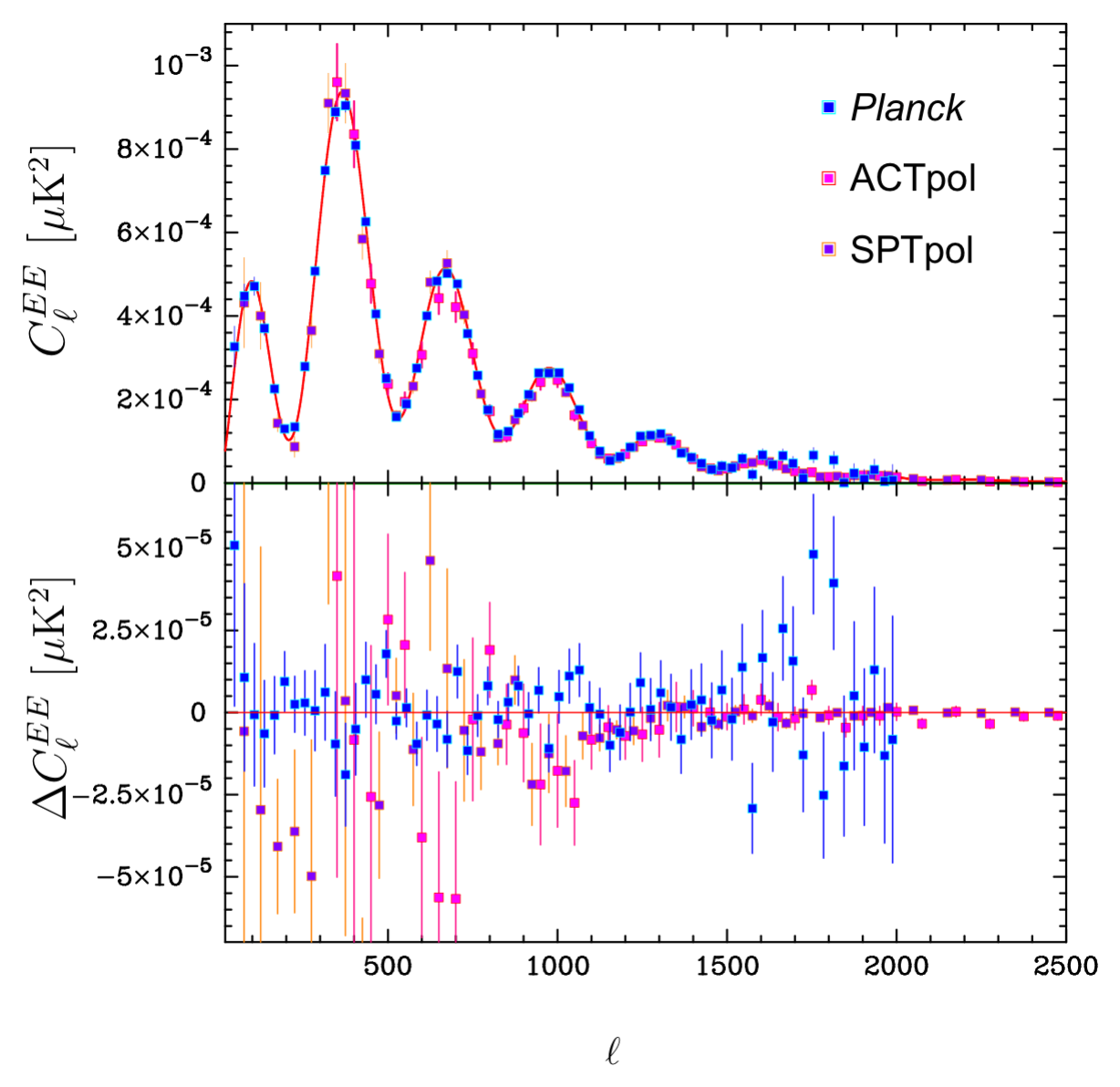}
\end{center}
\caption{Comparison of the \Planck\ \plik, ACTPol, and SPTpol $TE$ and $EE$ power spectra.
The solid lines show the best-fit base-\LCDM\ model for \planckalllensing. The lower panel in each pair of plots shows the residuals
relative to this theoretical model. The ACTPol and SPTpol $TE$ and $EE$ spectra are
as given in \citet{Louis2016} and \citet{Henning2017}, i.e., without adjusting nuisance parameters to fit the \Planck\ theoretical model. The error
bars show $\pm1\,\sigma$ uncertainties.}
\label{figure:teee_all}
\end{figure}

As discussed in \paramsI\ and \paramsII, \Planck\ $TT$ spectra are
statistically much more powerful than temperature data from current
high-resolution experiments such as the Atacama Cosmology
Telescope \citep[ACT, e.g.,][]{Das:2014} and the South Pole
Telescope \citep[SPT, e.g.,][]{Story:2013, George:2014}. As a result,
the \Planck\ temperature data dominate if they are combined with ACT
and SPT data. In \paramsII, the high-resolution temperature data were
used only to constrain low-amplitude components of the foreground model,
which are otherwise weakly constrained by \Planck\ data alone (with
very little impact on cosmological parameters). We adopt the same
approach in this paper.

Since the publication of \paramsII, \cite{Hou:2018} have performed a
direct map-based comparison of the SPT temperature data at 150\,GHz
with the \Planck\ 143-GHz maps over the same area of sky (covering
2540\,deg$^2$), finding no evidence for any systematic error in
either data set after accounting for an overall difference in
calibration. Temperature power spectrum comparisons between \Planck\
and SPT are reported in a companion paper by \citet{Aylor:2017}. They
find  cosmological parameters for base \LCDM\
derived from \Planck\ and SPT over the same patch of sky and multipole
range to be in excellent agreement. In particular, by comparing parameters
determined over the multipole range 650--2000 from both experiments, the reduction
in sample variance allows a test that is sensitive to systematic errors that could cause
shifts in parameter posteriors comparable to the widths of the \paramsII\ posteriors.
The parameters determined over the SPT sky area differ slightly, but not significantly,
 from the best-fit \LCDM\ parameters reported in \paramsII\ based on a much larger area
of sky. \cite{Aylor:2017} also find a tendency for the base-\LCDM\
parameters derived from SPT to shift as the multipole range is
increased, but at low statistical significance.

Polarization measurements have become a major focus for ground-based
CMB experiments. High resolution $TE$ and $EE$ spectra have been measured
by the ACT Polarimeter (ACTPol) and the polarization-sensitive
receiver of SPT (SPTpol).  Following two seasons
of observations, ACTPol has covered 548\,deg$^2$ along the celestial
equator at 149\,GHz with data and analysis presented in \citet{Naess2014}
and \citet{Louis2016}. The ACTPol spectra span the multipole
range $350 <\ell< 9000$. SPTpol polarization spectra from 100\,deg$^2$
in the southern hemisphere at 150\,GHz were first reported
in \citet{Crites2014} and recently extended to 500\,deg$^2$ \citep{Henning2017}. The SPTpol spectra span the multipole
range $50 < \ell < 8000$. In contrast, the \Planck\ $TE$ and $EE$ power spectra lose
statistical power at multipoles $\ga 1500$.
 The ACTPol and SPTpol spectra are compared with the \Planck\ $TE$ and $EE$
spectra in Fig.~\ref{figure:teee_all}. The polarization spectra measured from
these three very different experiments are in excellent agreement.

\begin{table*}[htbp!]
\caption{
Minimum $\chi^2$ values fitting the SPTpol spectra to the best-fit \Planck\ and SPTpol \LCDM\ cosmologies
(as described in the text). $N_{\rm b}$ gives the number of band powers in
each spectrum. The deviation of $\chi^2_{\rm min}$ from the
expectation $\langle \chi^2_{\rm min} \rangle = N_{\rm dof}$ is given by
the columns labelled $N_\sigma$, where $N_\sigma = (\chi^2_{\rm min} -
N_{\rm dof})/\sqrt{2N_{\rm dof}}$, and $N_{\rm dof} = N_{\rm b} -8$.
The last two columns give $\chi^2_{\rm p}$ for parameter
differences (Eq.~\ref{equ:highell1}) and the associated PTEs.}
\label{tab:SPTpol}
\begingroup
\vskip -4mm
\newdimen\tblskip \tblskip=5pt
\setbox\tablebox=\vbox{
 \newdimen\digitwidth
 \setbox0=\hbox{\rm 0}
 \digitwidth=\wd0
 \catcode`*=\active
 \def*{\kern\digitwidth}
 \newdimen\signwidth
 \setbox0=\hbox{+}
 \signwidth=\wd0
 \catcode`!=\active
 \def!{\kern\signwidth}
\halign{\hbox to 2.0in{#\leaderfil}\tabskip 1em&
\hfil#\hfil\tabskip 2em&
\hfil#\hfil\tabskip 0.5em& \hfil#\hfil\tabskip 2em&
\hfil#\hfil\tabskip 0.5em& \hfil#\hfil\tabskip 2em&
\hfil#\hfil\tabskip 2em& \hfil#\hfil\tabskip=0pt\cr
\noalign{\doubleline}
\omit& & \multispan2\hfil\Planck\ cosmology\hfil& \multispan2\hfil SPT cosmology\hfil& & \cr
\noalign{\vskip -5pt}
\omit& & \multispan2\hrulefill& \multispan2\hrulefill& & \cr
\noalign{\vskip 3pt}
\omit\hfil SPTpol spectrum\hfil& $N_{\rm b}$& **$\chi^2_{\rm min}$& $N_\sigma$& *$\chi^2_{\rm min}$& $N_\sigma$&$\chi^2_p$& PTE\cr
\noalign{\vskip 3pt\hrule\vskip 5pt}
$TE+EE$  & 112& **146.1& 2.91&  *137.4&   2.31& 9.85& 0.08\cr
$TE$     & *56& ***71.4& 2.38&  **70.3&   2.27& 3.38& 0.64\cr
$EE$     & *56& ***67.3& 1.96&  **61.4&   1.37& 8.21& 0.15\cr
\noalign{\vskip 5pt\hrule\vskip 3pt}}}
\endPlancktablewide
\endgroup
\end{table*}

For the base-\LCDM\ cosmology, the cosmological parameters should have
converged close to their true values by multipoles  $\sim 2000$. Since
ACTPol and SPTpol cover a much smaller sky area than \Planck\ the
errors on their $TE$ and $EE$ spectra are larger than those of \Planck\ at
low multipoles (see Fig.~\ref{figure:teee_all}). As a consequence, the
current ACTPol and SPTpol polarization constraints on the parameters of the
base-\LCDM\ cosmology are much weaker than those derived
from \Planck. The ACTPol results \citep{Louis2016} are consistent with
the \Planck\ base-\LCDM\ parameters and showed a small improvement in
constraints on extensions to the base cosmology that affect the
damping tail. Similar results were found by SPTpol,
though \citet{Henning2017} noted a $\ga 2\,\sigma$ tension with
the base-\LCDM\ model and found a trend for the parameters of the
base-\LCDM\ model to drift away from the \Planck\ solution as the
SPTpol likelihood is extended to higher multipoles.
To assess these results we have performed some tests of the consistency of
the latest \Planck\ results and the SPTpol spectra.

As a reference model for SPTpol we adopt the base-\LCDM\ parameters for
the combined $TE+EE$ fit to the SPTpol data from table~5
of \cite{Henning2017}. It is worth noting that the best-fit SPTpol cosmology is strongly excluded by
the \Planck\ $TT$ spectra and by the \Planck\ $TE+EE$ spectra. We use the \planckalllensing\ base-\LCDM\ best-fit
cosmology (as plotted in Fig.~\ref{figure:teee_all}) as a
reference model for \Planck. For each model, we ran the public version
of the SPTpol likelihood code,\footnote{Downloaded from
\url{http://pole.uchicago.edu/public/data/henning17/}\,. Note that we discovered errors
in the way that the covariances matrices were loaded for separate $TE$
and $EE$ analyses, which have been corrected in the analysis presented
here.} sampling the nuisance parameters using the same priors as
in \cite{Henning2017}. The best-fit values of $\chi^2$ are listed in
Table~\ref{tab:SPTpol}. As in \cite{Henning2017}, in assigning  significance
levels to these values, we take the number of degrees of freedom to be equal
to the number of band powers minus eight, corresponding to five cosmological parameters
($\omega_{\rm b}$, $\omega_{\rm c}$, $\theta_{\rm MC}$, $n_{\rm s}$, $\As e^{-2\tau}$)
and three nuisance parameters with flat priors.

As found by \cite{Henning2017}, the SPTpol $TE$ spectrum gives nearly
identical values of $\chi^2$ for both the SPTpol and \Planck\
cosmologies and so does not differentiate between them; however, the
$\chi^2$ values are high, at the $2.3\,\sigma$ level. The SPTpol $EE$ spectrum
provides weaker constraints on cosmological parameters than the $TE$
spectrum and is clearly better fit by the SPTpol cosmology. If the SPTpol
covariance matrix is accurate, the combined TE+EE
SPTpol data disfavour the \Planck\ \LCDM\ cosmology quite strongly and
disfavour any 6-parameter \LCDM\ cosmology.
For \LCDM\ models,  outliers distributed over a wide range of multipoles
contribute to the high $\chi^2$ values,
notably at $\ell=124$, $324$, $1874$, $2449$, and $3249$ in TE, and $\ell=1974$ and $6499$ in EE.

We can assess consistency of the parameter differences, $\pmb{$\Delta$} {\bf
p}$, between the two experiments by computing,
\begin{equation}
    \chi^2_p =    \pmb{$\Delta$} {\bf p}^{\sf T}  \tens{C}_p^{-1} \pmb{$\Delta$} {\bf p}, \label{equ:highell1}
\end{equation}
where $\tens{C}_p$ is the covariance matrix for SPTpol parameters (we
neglect the errors in the \Planck\ parameters, which are much
smaller). Values for $\chi_p^2$ are given in Table~\ref{tab:SPTpol}
together with probabilities to exceed (PTEs) computed from a $\chi^2$ distribution with five
degrees of freedom. We find no evidence for any statistically significant
inconsistency between the two sets of parameters, even for the
combined $TE+EE$ SPTpol likelihood. We also note that the parameter
$\As e^{-2 \tau}$ makes quite a large contribution to $\chi^2_p$ for
the $TE+EE$ and $EE$ spectra, but is sensitive to possible systematic
errors in the SPTpol polarization efficiency calibration \citep[which, as
discussed, is not well understood]{Henning2017}.
Varying the maximum multipole used in the SPTpol likelihood ($\ell_{\rm max}$), we find that
the parameters of the SPTpol $TE+EE$ cosmology converge by $\ell_{\rm max}=2500$;
higher multipoles do not contribute significantly to the SPTpol base-\LCDM\
solution.

\cite{Henning2017} reported a trend for the parameters of the base-\LCDM\
cosmology to change as the SPTpol likelihood is extended to higher
multipoles, which they suggested may be an indication of new
physics. However, this effect is not of high statistical significance
and cannot be tested by the \Planck\ spectra, which become less
sensitive than the SPTpol spectra at multipoles ${\ga}\,1500$. The
consistency of the base-\LCDM\ cosmology at high multipoles in polarization should become
clearer in the near future as more polarization data are
accumulated by ACTPol and SPTpol.

\section{Comparison with other astrophysical data sets}
\label{sec:datasets}

\subsection{Baryon acoustic oscillations}\label{sec:BAO}

As in \paramsI\ and \paramsII\, baryon acoustic oscillation (BAO)
measurements from galaxy redshift surveys are used as the primary
non-CMB astrophysical data set in this paper.  The acoustic scale
measured by BAOs, at around 147 {\rm Mpc}, is much larger than the
scale of virialized structures.  This separation of scales makes BAO
measurements insensitive to nonlinear physics, providing a robust
geometrical test of cosmology.  It is for this reason that BAO
measurements are given high weight compared to other non-CMB data in
this and in previous \Planck\ papers. BAO features in the galaxy power
spectrum were first detected by
\cite{Cole:05} and \cite{Eisenstein:05}. Since their discovery,
BAO measurements have improved in accuracy via a number of ambitious galaxy surveys.
As demonstrated in \paramsI\ and \paramsII\, BAO results
from galaxy surveys have been consistently in excellent agreement with
the best-fit base-\LCDM\ cosmology inferred from \Planck. More recently, the
redshift reach of BAO measurements has been increased using
 quasar redshift surveys and Lyman-$\alpha$ absorption lines detected in
quasar spectra.

\begin{figure}
\begin{center}
\includegraphics[width=85mm,angle=0]{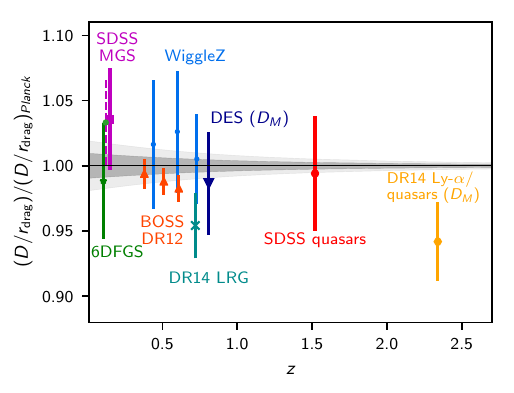}
\end{center}
\caption {
Acoustic-scale distance measurements
 divided by the corresponding mean distance ratio from \planckalllensing\ in the base-\LCDM\ model.
 The points, with their $1\,\sigma$ error bars are as follows: green star,
6dFGS \citep{Beutler:2011hx}; magenta square, SDSS MGS \citep{Ross:2014qpa};
 red triangles, BOSS DR12 \citep{Alam:2016hwk}; small blue circles, WiggleZ \citep[as analysed by][]{Kazin:14};
 large dark blue triangle, DES \citep{Abbott:2017wcz};
 cyan cross, DR14 LRG \citep{Bautista:2017wwp};
 red circle, SDSS quasars \citep{Ata:2017dya};
and orange hexagon, which shows the combined BAO constraints from
\rep{BOSS DR14 Lyman-$\alpha$ \citep{Agathe:2019vsu}
 and Lyman-$\alpha$ cross-correlation with quasars,
 as cited in \citep{Blomqvist:2019rah}.}
 The green point with magenta dashed line is the 6dFGS and MGS joint analysis result of \citet{Carter:2018vce}.
 All ratios are for the averaged distance $\DVBAO(z)$, except for DES and BOSS Lyman-$\alpha$, where the ratio plotted is $\DM$ (results for $H(z)$ are shown separately in Fig.~\ref{fig:Hz}).
 The grey bands show the $68\,\%$ and $95\,\%$ confidence ranges allowed for the ratio $\DVBAO(z)/\rdrag$ by \planckalllensing\ (bands for $\DM/\rdrag$ are very similar).
 }
\label{fig:BAO}
\end{figure}

\begin{figure*}
\vspace{0.9cm}
\begin{center}
\includegraphics[width=\hsize,angle=0]{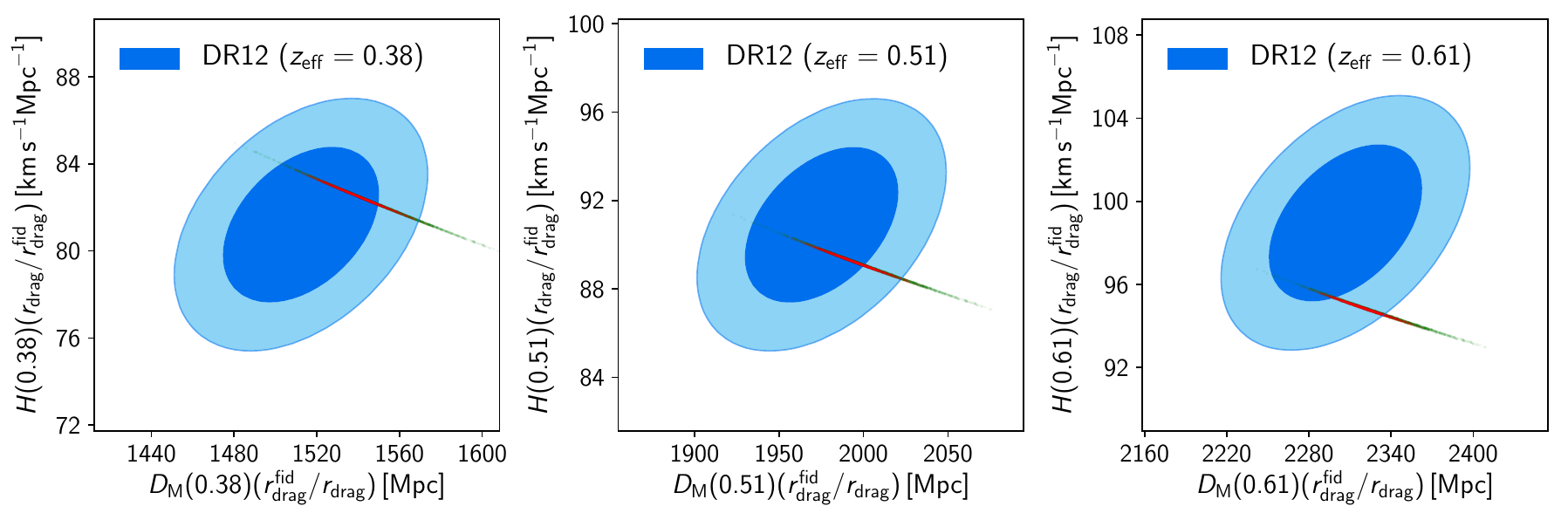}
\end{center}
\caption {Constraints on the comoving angular diameter distance
$\DM(z)$ and Hubble parameter $H(z)$ at the three central redshifts of the \citet{Alam:2016hwk}
analysis of BOSS DR12. The dark blue and light blue regions show 68\,\% and 95\,\% CL, respectively.  The fiducial sound horizon adopted
by \citet{Alam:2016hwk} is $\rdrag^{\rm fid} = 147.78\,{\rm Mpc}$.
Green points show samples from \planckTT\ chains, and red points corresponding samples from \planckall+lensing, indicating good consistency with BAOs; one can also see the shift towards slightly lower $\DM$ and higher $H$ as more CMB
data are added.
\label{BOSSprob}
\vspace{0.7cm}
}
\end{figure*}

Figure~\ref{fig:BAO} summarizes the latest BAO results, updating figure~14 of \paramsII.
This plot shows the acoustic-scale distance ratio $\DVBAO(z)/\rdrag$ measured
from surveys with effective redshift $z$,
divided by the mean acoustic-scale ratio in the base-\LCDM\ cosmology using
\planckall+lensing. Here
$\rdrag$ is the comoving sound horizon at the end of the baryon drag epoch and
$\DVBAO$ is a combination of the comoving angular diameter distance $\DM(z)$ and Hubble
parameter $H(z)$:
\begin{equation}
 \DVBAO(z) = \left [ \DM^2(z) {cz \over H(z)} \right ]^{1/3}. \label{BAO1}
\end{equation}
The grey bands in the figure show the $\pm 1\,\sigma$ and $\pm 2\, \sigma$ ranges allowed by \Planck\ in the base-\LCDM\ cosmology.

Compared to figure~14 of \paramsII, we have replaced the Baryon Oscillation Spectroscopic Survey (BOSS) LOWZ and CMASS results of \citet{Anderson:2013zyy} with the
latest BOSS data
release 12 (DR12) results summarized by \cite{Alam:2016hwk}. That paper reports
``consensus'' results on BAOs (weighting together
 different BAO analyses of BOSS DR12) reported by \cite{Ross:2016gvb}, \cite{Vargas-Magana:2016imr}, and \cite{Beutler:2016ixs} in three redshift slices with effective redshifts
$z_{\rm eff} = 0.38$, $0.51$, and $0.61$. These new measurements, shown by the
red triangles in Fig.~\ref{fig:BAO}, are in good
agreement with the \Planck\ base-\LCDM\ cosmology.

By using quasars, it has become possible to extend BAO measurements to
redshifts greater than unity. \cite{Ata:2017dya} have measured the BAO scale
$D_V$ at an effective redshift of $z_{\rm eff} = 1.52$ using a sample
of quasars from the extended Baryon Oscillation Survey (eBOSS). This
measurement is shown by the red circle in Fig.~\ref{fig:BAO} and is
 also in very good agreement with \Planck. The results of the
\citet{Ata:2017dya} analysis also agree well with other analyses of the
eBOSS quasar sample \citep[e.g.,][]{Gil-Marin:2018cgo}.

\rep{At even higher redshifts BAOs have been measured in the 
Lyman $\alpha$ spectra of quasars \citep{Delubac:2014, Font-Ribera:2014,
 Bautista:2017zgn, Bourboux:2017cbm, Agathe:2019vsu,
 Blomqvist:2019rah}. In the first preprint version of this paper, we
 compared the \Planck\ results with those from BAO features measured
 from the flux-transmission correlations of Sloan Digital Sky Survey
 (SDSS) DR12 quasars \citep{Bautista:2017zgn} and with the
 cross-correlation of the Ly\,$\alpha$ forest with SDSS
 quasars \citep{Bourboux:2017cbm}. The combined result on
 $\DM/\rdrag$ from these analyses was about $2.3\,\sigma$ lower than
 expected from the best-fit \Planck\ base-\LCDM\ cosmology. The
 \cite{Bautista:2017zgn} and \citet{Bourboux:2017cbm} analyses have been
 superseded by equivalent studies of a larger sample of SDSS DR14 quasars
 reported in \cite{Agathe:2019vsu} and \cite{Blomqvist:2019rah}.
 The combined result for $\DM/\rdrag$ from these analyses
 \citep[as quoted by][]{Blomqvist:2019rah} is plotted as the orange hexagon
 on Fig.~\ref{fig:BAO} and lies within $1.7\,\sigma$ of the \Planck\ best-fit
 model. The errors on these high-redshift BAO measurements are still quite
 large in comparison with the galaxy measurements and so we do not
 include them in our default BAO compilation.\footnote{The first preprint
 version of this paper showed that the inclusion of the \cite{Bautista:2017zgn}
 and \citet{Bourboux:2017cbm} Ly$\,\alpha$ BAO results had a minor impact
 on the parameters of the base-\LCDM\ cosmology. The impact of the more recent
 Ly$\,\alpha$ results of \cite{Agathe:2019vsu} and \cite{Blomqvist:2019rah}
 will be even lower, since they are in closer agreement with the \Planck\
 best-fit cosmology.}}

The more recent BAO analyses solve for the positions of
the BAO feature in both the line-of-sight and transverse directions (the
distortion in the transverse direction caused by the background cosmology
is sometimes called the Alcock-Paczynski effect, \citealt{Alcock:79}), leading
to joint constraints on the angular diameter distance $\DM(z_{\rm eff})$
and the Hubble parameter $H(z_{\rm eff})$.
These constraints for the BOSS DR12 analysis
are plotted in Fig.~\ref{BOSSprob}.  Samples from the \planckTT\ and \planckalllensing\
likelihood are shown in green and red, respectively, demonstrating that BAO and \planck\ polarization data with lensing
consistently pull parameters in the same direction (towards slightly lower $\Omc h^2$). We find
the same behaviour for \Planck\ when adding polarization and lensing to the $TT$ likelihood separately.
This demonstrates the remarkable consistency of the \Planck\ data, including polarization and CMB lensing
with the galaxy BAO measurements.
Evidently, the \Planck\ base-\LCDM\ parameters are in good agreement with both the
isotropized $\DVBAO$ BAO measurements plotted in Fig.~\ref{fig:BAO}, and with the anisotropic
constraints plotted in Fig.~\ref{BOSSprob}.

In this paper, we use the 6dFGS and SDSS-MGS
measurements of $\DVBAO/\rdrag$ \citep{Beutler:2011hx, Ross:2014qpa} and
the final DR12 anisotropic BAO measurements of
\citet{Alam:2016hwk}. Since the WiggleZ volume partially overlaps that of the
BOSS-CMASS sample, and the correlations have not been quantified, we do not
use the WiggleZ results in this paper. It is clear from Fig.~\ref{fig:BAO} that
the combined BAO likelihood for the lower redshift points is dominated by the BOSS measurements.

In the base-\LCDM\ model, the \Planck\ data constrain
the Hubble constant $H_0$ and matter density $\Omm$ to high precision:
\twoonesig[2.8cm]{H_0 &= (67.36 \pm 0.54) \, \Hunit,}{\Omm &=
 0.3158 \pm 0.0073,}{\shortpol\dataplus\lowE+\lensing.  \label{BAO_H01}}
With the addition of the BAO measurements, these constraints are strengthened
 to
\twoonesig[2.9cm]{H_0 &= (67.66 \pm 0.42) \, \Hunit,}{\Omm &= 0.3111 \pm 0.0056,}{\shortpol\dataplus\lowE+\lensing\dataplus\BAO. \label{BAO_H02}}
These numbers are in very good agreement with the constraints given in Eq.~\eqref{lensingBAOconstraint}, which exclude the high-multipole \Planck\ likelihood.
Section~\ref{sec:hubble} discusses the consistency of direct measurements of $H_0$ with these estimates and
Hubble parameter measurements from the line-of-sight component of BAOs at higher redshift.

As discussed above, we have excluded Ly\,$\alpha$ BAOs from our default BAO compilation. The full likelihood
for the combined Ly\,$\alpha$ and Ly\,$\alpha$-quasar cross-correlations reported in \citet{Bourboux:2017cbm} is not
yet available; nevertheless, we can get an indication of the impact of including these measurements
by assuming uncorrelated Gaussian errors on $\DM/\rdrag$ and $\rdrag H$. Adding these measurements to
\planckall\ and our default BAO compilation shifts $H_0$ higher, and $\Omm h^2$ and $\sigma_8$ lower,
by approximately $0.3\,\sigma$. The joint \planck+BAO result then gives
$\DM/\rdrag$ and $\rdrag H$ at $z=2.4$ lower by $0.25$ and
$0.3$ of \planck's $\sigma$, leaving the overall $2.3\,\sigma$ tension with these
results almost unchanged.  As shown by \citet{Aubourg:2014yra}, it is
difficult to construct well-motivated extensions to the base-\LCDM\
model that can resolve the tension with the Ly\,$\alpha$ BAOs. Further work is needed to assess whether the
discrepancy between \Planck\ and the Ly\,$\alpha$ BAO results is a statistical fluctuation,
is caused by small systematic errors, or is a signature of new physics.

\subsection{Type Ia supernovae}\label{sec:SN}

\begin{figure}[ht!]
\begin{center}
\includegraphics[width=\columnwidth,angle=0]{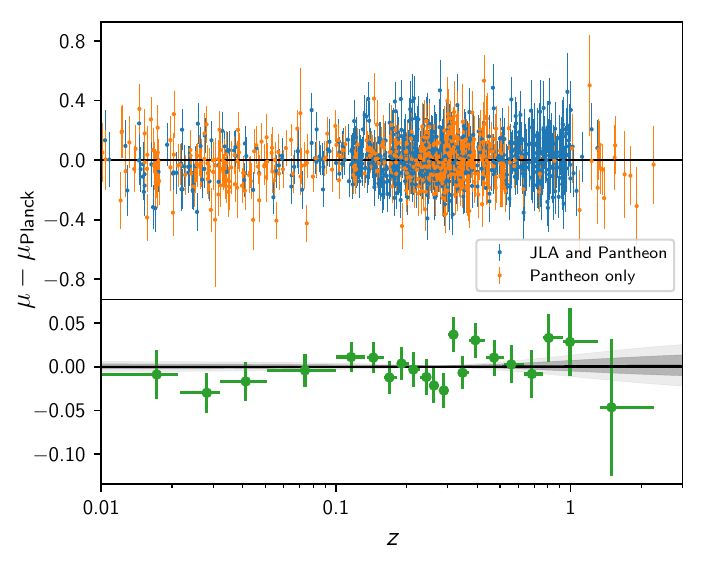}
\end{center}
\caption{\rep{Distance modulus $\mu=5\log_{10}(D_{\rm L}) + {\rm constant}$ (where $D_{\rm L}$ is the luminosity distance) for supernovae in the Pantheon sample \citep{Scolnic:2017caz} with $1\,\sigma$ errors, compared to the \planckall+lensing \lcdm\ best fit. Supernovae that were also in the older Joint Lightcurve Analysis \citep[JLA]{Betoule:2014}  sample are shown in blue. The peak absolute magnitudes of the SNe,  corrected for light-curve shape, colour, and host-galaxy mass correlations \citep[see equation~3 of][]{Scolnic:2017caz},  are fixed to an absolute distance
scale using the $H_0$ value from the \Planck\ best fit. The lower panel shows the binned errors, with equal numbers of supernovae per redshift bin (except for the two highest redshift bins). The grey bands show the $\pm1$ and $\pm2\,\sigma$ bounds from the \planckall+lensing chains, where each model is calibrated to the best fit, as for the data.}
}
\label{fig:Pantheon}
\end{figure}

The use of type Ia supernovae (SNe) as standard candles has been of
critical importance to cosmology, leading to the discovery of cosmic
acceleration \citep{Riess:1998,Perlmutter:1999}. For \LCDM\ models, however,
SNe data have little statistical power compared to \Planck\ and BAO and
in this paper they are used mainly to test models involving evolving
dark energy and modified gravity. For these extensions of the base cosmology,
 SNe data are useful in fixing the background cosmology at low redshifts,
where there is not enough volume to allow high precision constraints
from BAO.

 In \paramsII\ we used the ``Joint Light-curve Analysis'' (JLA) sample
 constructed from the SNLS and SDSS SNe plus several samples of
 low redshift SNe described in
 \citet{Betoule:2013,Betoule:2014} and \citet{Mosher:2014}.
 In this paper, we use the
 new ``Pantheon'' sample of \citet{Scolnic:2017caz}, which adds 276
 supernovae from the Pan-STARRS1 Medium Deep Survey at $0.03< z< 0.65$
 and various low-redshift and HST samples to give a total
of 1048 supernovae spanning the redshift range $0.01<z<2.3$. The Pantheon compilation
applies cross-calibrations of the photometric systems of all of the sub-samples
 used to construct the final catalogue \citep{Scolnic:2015},
reducing the impact of calibration systematics on cosmology\footnote{\rep{We use the November 2018 data file available from \url{https://github.com/dscolnic/Pantheon/}, which includes heliocentric redshifts and no bulk-flow corrections for $z>0.08$.}}. The Pantheon data are
compared to the predictions of the \planckall+lensing base-\LCDM\ model best fit in Fig.~\ref{fig:Pantheon}. The agreement is excellent.  The JLA and Pantheon samples are consistent with each other (with Pantheon providing tighter constraints on cosmological parameters) and there
would be no significant change to our science conclusions had we chosen to use the JLA sample in this paper. To illustrate this point we give results for a selection of models using both samples
in the parameter tables available in the \PLA; Fig.~\ref{fig:invladder}, illustrating inverse-distance-ladder constraints on $H_0$ (see Sect.~\ref{sec:hubble}),
shows a specific example.

\subsection{Redshift-space distortions}\label{sec:RSD}
\begin{figure}
\begin{center}
\includegraphics[width=85mm,angle=0]{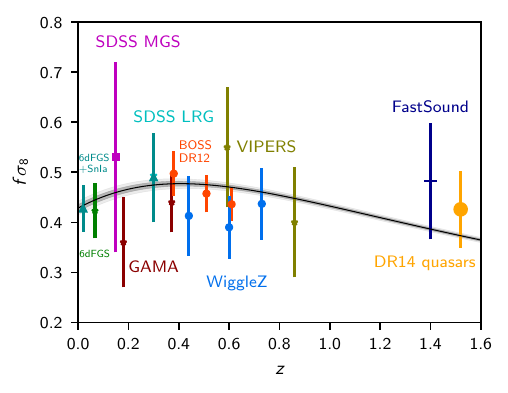}
\end{center}
\caption {
Constraints on the growth rate of fluctuations from various redshift
surveys in the base-\LCDM\ model:
dark cyan, 6dFGS and velocities from SNe Ia \citep{Huterer:2016uyq};
green, 6dFGRS \citep{Beutler:2012px};
purple square, SDSS MGS \citep{Howlett:2014opa};
cyan cross, SDSS LRG \citep{Oka:2013cba};
dark red, GAMA \citep{Blake:2013nif};
red, BOSS DR12 \citep{Alam:2016hwk};
blue, WiggleZ \citep{Blake:2012pj};
olive, VIPERS \citep{Pezzotta:2016gbo};
dark blue, FastSound \citep{Okumura:2015lvp};
and orange, BOSS DR14 quasars \citep{Zarrouk:2018vwy}.
Where measurements are reported in correlation with other variables, we here show the marginalized posterior means and errors.
Grey bands show the $68\,\%$ and $95\,\%$ confidence ranges allowed by \planckalllensing.
 }
\label{fig:RSD}
\end{figure}

The clustering of galaxies observed in a redshift survey exhibits anisotropies
induced by peculiar motions (known as redshift-space distortions, RSDs).
Measurement of RSDs can provide constraints on the growth rate of structure and
the amplitude of the matter power spectrum \citep[e.g.,][]{PerWhi09}.
Since it uses non-relativistic tracers, RSDs are sensitive to the time-time
component of the metric perturbation or the Newtonian potential.
A comparison of the amplitude inferred from RSDs with that inferred from
lensing (sensitive to the Weyl potential, see Sect. \ref{sec:results}).
provides a test of General
Relativity.

\begin{figure*}
\begin{center}
\includegraphics[width=165mm,angle=0]{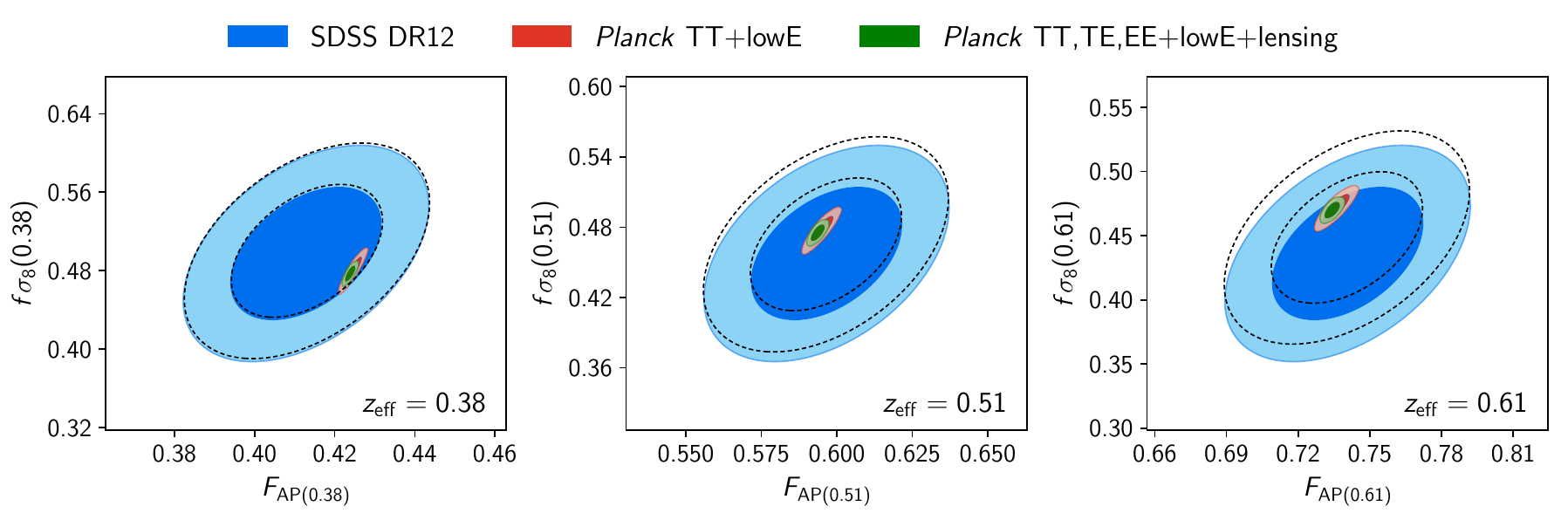}
\end{center}
\caption {Constraints on $f\, \sigma_8$ and $F_{\rm AP}$ (see Eqs.~\ref{eq:RSD1} and \ref{eq:RSD2}) from
analysis of redshift-space distortions.
The blue contours show $68\,\%$ and $95\,\%$ confidence ranges on $(f\,\sigma_8, F_{\rm AP})$ from BOSS-DR12,
marginalizing over $D_V$. Constraints from \Planck\ for the base-\LCDM\ cosmology are shown by the red and green contours.
The dashed lines are the $68\,\%$ and $95\,\%$ contours for BOSS-DR12, conditional on the \planckall+lensing constraints on
$D_V$. }
\label{fig:fsig8FAP}
\end{figure*}

Measurements of RSDs are usually quoted as constraints on $f\,\sigma_8$,
where for models with scale-independent growth
$f=d\ln D/d\ln a$. For \LCDM, $d\ln D/d\ln a\approx\Omega_{\rm m}^{0.55}(z)$.
 We follow \paramsII, defining
\begin{equation}
  f\,\sigma_8 \equiv \frac{\left[\sigma_8^{\rm (vd)}(z)\right]^2}
                          {      \sigma_8^{\rm (dd)}(z)},
\label{eq:RSD1}
\end{equation}
where $\sigma_8^{\rm (vd)}$ is the density-velocity correlation in spheres
of radius $8\,h^{-1}$Mpc in linear theory.

Measuring $f\,\sigma_8$ requires modelling nonlinearities and
scale-dependent bias and is considerably more complicated than
estimating the BAO scale from galaxy surveys. One key problem is
deciding on the precise range of scales that can be used in an RSD
analysis, since there is a need to balance potential systematic errors
associated with modelling nonlinearities against reducing statistical
errors by extending to smaller scales. In addition, there is a partial
degeneracy between distortions caused by peculiar motions and the
Alcock-Paczynski effect. Nevertheless, there have been substantial improvements in
modelling RSDs in the last few years, including extensive tests of systematic
errors using numerical simulations. Different techniques for measuring $f\,\sigma_8$
are now consistent to within a few percent \citep{Alam:2016hwk}.

Figure~\ref{fig:RSD}, showing $f\,\sigma_8$ as a function of redshift,
is an update of figure~16 from \paramsII. The most significant changes
from \paramsII\ are the new high precision measurements from BOSS DR12,
shown as the red points.  These points are the ``consensus'' BOSS D12
results from \citet{Alam:2016hwk}, which averages the results from
four different ways of analysing the DR12 data
\citep{Beutler:2017,Grieb:2017,Sanchez:2017, Satapathy:2017}. These
results are in excellent agreement with the \Planck\ base
\LCDM\ cosmology (see also Fig.~\ref{fig:fsig8FAP}) and provide
the tightest constraints to date on the growth rate of
fluctuations. We have updated the VIPERS constraints to those of the
second public data release \citep{Pezzotta:2016gbo} and added a data
point from the Galaxy and Mass Assembly (GAMA) redshift survey
\citep{Blake:2012pj}. Two new surveys have extended the reach of RSD
measurements (albeit with large errors) to redshifts greater than
unity: the deep FASTSOUND emission line redshift survey
\citep{Okumura:2015lvp}; and the BOSS DR14 quasar survey
\citep{Zarrouk:2018vwy}. We have also added a new low redshift
estimate of $f\,\sigma_8$ from \cite{Huterer:2016uyq} at an effective
redshift of $z_{\rm eff} = 0.023$, which is based on correlating deviations
from the mean magnitude-redshift relation of SNe in the Pantheon
sample with estimates of the nearby peculiar velocity field determined
from the 6dF Galaxy Survey \citep{Springob:2014}. As can be seen from Fig.~\ref{fig:RSD},
these growth rate measurements are consistent with the \Planck\ base-\LCDM\ cosmology over
the entire redshift range $0.023 < z_{\rm eff} < 1.52$.

Since the BOSS-DR12 estimates provide the strongest constraints on RSDs, it is worth comparing these results with
\Planck\ in greater detail. Here we use the ``full-shape consensus'' results\footnote{When using RSDs to constraint dark energy in Sect.~\ref{sec:darkenergy}, we use the alternative
$\DM$, $H$, and $f\,\sigma_8$ parameterization from \cite{Alam:2016hwk} for consistency with the
DR12 BAO-only likelihood that we use elsewhere.
} on $D_V$, $f\,\sigma_8$, and $F_{\rm AP}$ for each of the three
redshift bins from \cite{Alam:2016hwk} and the associated $9\times9$ covariance matrix, where $F_{\rm AP}$
is the Alcock-Paczinski parameter,
\begin{equation}
F_{\rm AP}(z) = \DM(z) {H(z) \over c}.
\label{eq:RSD2}
\end{equation}
Figure~\ref{fig:fsig8FAP} shows the constraints from BOSS-DR12 on $f\,\sigma_8$
and $F_{\rm AP}$ marginalized over $D_V$. \Planck\ base-\LCDM\ constraints are shown by the red and green contours. For each
redshift bin, the \Planck\ best-fit values of $f\,\sigma_8$ and $F_{\rm AP}$ lie within the $68\,\%$ contours from BOSS-DR12.
Figure~\ref{fig:fsig8FAP} highlights the impressive consistency of the base-\LCDM\ cosmology from the high redshifts probed
by the CMB to the low redshifts sampled by BOSS.

\subsection{The Hubble constant} \label{sec:hubble}

Perhaps the most controversial tension between the
\Planck\ \lcdm\ model and astrophysical data is the discrepancy with
\rep{traditional distance-ladder} measurements of the Hubble constant $H_0$.
\paramsI\ reported a value of $H_0=(67.3 \pm 1.2) \Hunit$ for the
base-\LCDM\ cosmology, substantially lower that the distance-ladder
estimate of $H_0 = (73.8 \pm 2.4) \Hunit$ from the SH0ES\footnote{SN,
 H0, for the Equation of State of dark energy.} project
\citep{Riess:2011yx} and other $H_0$ studies
\citep[e.g.,][]{Freedman:2001,Freedman:2012}. Since then, additional
data acquired as part of the SH0ES project \citep[][hereafter
 R18]{Riess:2016, Riess18} has exacerbated the tension. R18 conclude
that $H_0=(73.48 \pm 1.66) \Hunit$, compared to our \planckall+lensing
estimate from Table~1 of $H_0=(67.27 \pm 0.60) \Hunit$. Using Gaia
parallaxes \citet{Riess:2018byc} slightly tightened their measurement
to $H_0=(73.52 \pm 1.62)\Hunit$.  \rep{Recently \citet{Riess:2019cxk}
  then used improved measurements of LMC Cepheids to further
  tighten\footnote{By default in this paper (and in the \PLA) we use
    the \citet{Riess18} number (available at the time we ran our
    parameter chains) unless otherwise stated; using the updated
    number would make no significant difference to our conclusions.}
  the constraint to $H_0 = (74.03\pm 1.42)\Hunit$}.  Interestingly,
the central values of the SH0ES and \planck\ estimates have hardly
changed since the appearance of \paramsI, but the errors on both
estimates have shrunk so that the discrepancy has grown from around
$2.5\,\sigma$ in 2013 to $3.5\,\sigma$ today \citep[\rep{$4.4\,\sigma$
 using}][]{Riess:2019cxk}. This discrepancy has stimulated a number
of investigations of possible systematic errors in the either the
\planck\ or SH0ES data, which have failed to identify any obvious
problem with either analysis \citep[e.g.,][]{Spergel:2013rxa,
 Addison:2015wyg, planck2016-LI, Efstathiou:2014, Cardona:2017,
 Zhang:2017,Follin:2017ljs}. It has also been argued that the Gaussian
likelihood assumption used in the SH0ES analysis leads to an
overestimate of the statistical significance of the discrepancy
\citep{Feeney:2018a}.

\rep{Recently, \cite{Freedman:2019} have reported a determination of
$H_0$ using the tip of the red giant branch as a distance estimator.
This analysis gives $H_0 = (69.8\pm 1.9)\Hunit$, i.e., intermediate
between the SH0ES measurement and the \Planck\ base-\LCDM\ value.
There has been some controversy (see \citet{Yuan:2019}) concerning the calibration of the
tip of the red giant branch adopted in \citet{Freedman:2019}, though a recent reanalysis
by \citet{Freedman:2020} yields a value of $H_0$ that is almost identical to that reported in \citet{Freedman:2019}.}

\rep{Measurements of the Hubble constant using strong gravitational-lensing
 time delays are also higher than the \planck\ base-\lcdm\ value.
 The most recent results, based on six strongly lensed
  quasars, give $H_0 = 73.3^{+1.7}_{-1.8} \Hunit$ \citep{Wong:2019kwg},
  which is about $3.2\,\sigma$ higher than the \Planck\ value. A
  number of other techniques have been used to infer $H_0$, including
  stellar ages \citep[e.g.,][]{Jimenez:2002, Amendola:2018}, distant
  megamasers \citep{Reid:2013, Kuo:2013, Gao:2016} and gravitational-wave
  standard sirens \cite{Ligo:2017}. These measurements span a range of
values. Nevertheless, there is a tendency for local determinations to sit high compared to the
\Planck\ base-\LCDM\ value, with the SH0ES Cepheid-based measurement giving the most
statistically significant discrepancy. }

In this paper, we take the R18 estimate at face value and include it as a
prior in combination with \Planck\ in some of the parameter tables available on the \PLA.
The interested reader can then assess the impact of the R18 measurement on
a wide range of extensions to the base-\LCDM\ cosmology.

\begin{figure}
\begin{center}
\includegraphics[width=85mm,angle=0]{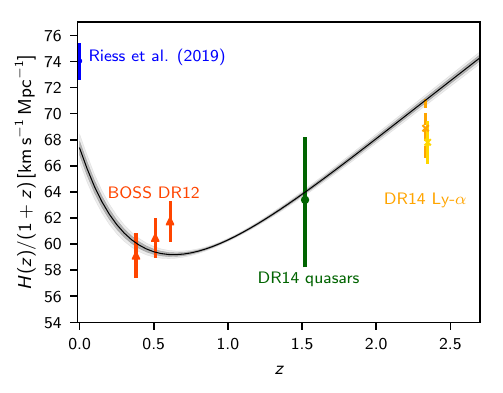}
\end{center}
\caption {
 Comoving Hubble parameter as a function of redshift.
 The grey bands show the $68\,\%$ and $95\,\%$ confidence ranges allowed by
 \planckalllensing\ in the base-\lcdm\ model, clearly showing the onset of acceleration around $z=0.6$.
 Red triangles show the BAO measurements from BOSS DR12
 \citep{Alam:2016hwk}, the green circle is from BOSS DR14 quasars \citep{Zarrouk:2018vwy},
 the orange dashed point is the constraint from the
 \rep{
  BOSS DR14 Ly\,$\alpha$ auto-correlation at $z=2.34$ \citep{Agathe:2019vsu}, and the solid gold point is the joint constraint from the Ly\,$\alpha$ auto-correlation and cross-correlation with quasars from \citet{Blomqvist:2019rah}.
 }
 All BOSS measurements are used in combination with the \planck\ base-model measurements of the sound horizon $\rdrag$, and the DR12 points are correlated. The blue point at redshift zero shows the inferred forward-distance-ladder Hubble measurement from \citet{Riess:2019cxk}.
 }
\label{fig:Hz}
\end{figure}

\begin{figure}
\begin{center}
\includegraphics[width=\columnwidth,angle=0]{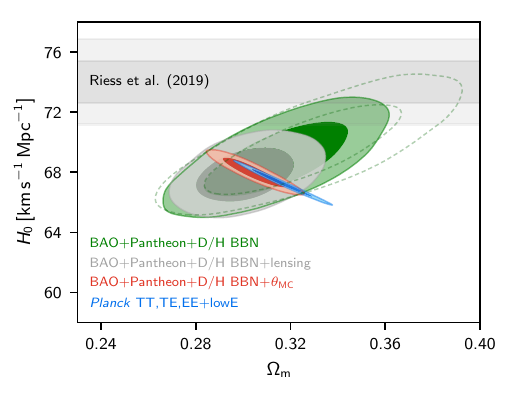}
\end{center}
\caption {\rep{
 Inverse-distance-ladder constraints on the Hubble parameter and $\Omm$ in the base-\lcdm\ model, compared to the result from the full \planck\ CMB
power-spectrum data.
 BAO data constrain the ratio of the sound horizon at the epoch of baryon drag and the distances; the sound horizon depends on the baryon density, which is constrained by the conservative prior of $\Omb h^2=0.0222\pm 0.0005$, based on the measurement of D/H by \citet{Cooke:2017cwo} and standard BBN with modelling uncertainties.
 Adding \planck\ CMB lensing constrains the matter density, or adding a conservative \planck\ CMB ``BAO'' measurement ($100\thetaMC = 1.0409 \pm 0.0006$) gives a tight constraint on $H_0$, comparable to that from the full CMB data set.
 Grey bands show the local distance-ladder measurement of \citet{Riess:2019cxk}.
 Contours contain $68\,\%$ and $95\,\%$ of the probability. Marginalizing over the neutrino masses or allowing dark energy equation of state parameters $\wzero >-1$ would only lower the inverse-distance-ladder constraints on $H_0$. The dashed contours show the constraints from the data combination
BAO+JLA+D/H BBN
}.
 }
\label{fig:invladder}
\end{figure}

We already mentioned in Sect.~\ref{sec:BAO} that BAO measurements along
the line of sight constrain $H(z)r_{\rm drag}$.  \planck\ constrains
$r_{\rm drag}$ to a precision of $0.2\,\%$ for the base-\LCDM\ model and
so the BAO measurements can be accurately converted into absolute
measurements of $H(z)$. This is illustrated by Fig.~\ref{fig:Hz}, which
shows clearly how well the \Planck\ base-\LCDM\ cosmology fits the
 BAO measurements of $H(z)$ over the redshift range $0.3$--$2.5$, yet fails
to match the R18 measurement of $H_0$ at $z=0$. \rep{The model is also consistent with
the most recent Ly\,$\alpha$ BAO measurements at $z \approx 2.3$.}

\paramsI\ and \paramsII\ emphasized that this mismatch between BAO
measurements and \rep{forward distance-ladder} measurements of $H_0$
is not sensitive to the \planck\ data at high multipoles. For example,
combining WMAP with BAO measurements leads to $H_0 = (68.14 \pm 0.73)
\Hunit$ for the base-\LCDM\ cosmology, which is discrepant with the
R18 value at the $2.9\,\sigma$ level.

\citet{Heavens:2014}, \citet{Cuesta:2014asa}, and
\citet{Aubourg:2014yra} showed that the combination of CMB, BAO, and
SNe data provides a powerful ``inverse-distance-ladder'' approach to
constructing a physically calibrated distance-redshift relation down
to very low redshift.  For the base-\lcdm\ model, this
inverse-distance-ladder approach can be used to constrain $H_0$
without using any CMB measurements at all, or by only using
constraints on the CMB parameter $\theta_{\rm MC}$ \citep[see
 also][]{Bernal:2016gxb,Addison:2018,
 Abbott:2017smn,Lemos:2018smw}. This is illustrated in
Fig.~\ref{fig:invladder}, which shows how the constraints on $H_0$ and
$\Omm$ converge to the \planck\ values as more data are included. The
green contours show the constraints from BAO and the Pantheon SNe
data, \rep{together with a BBN constraint on the baryon density ($\Omb h^2
= 0.0222 \pm 0.0005$)} based on the primordial deuterium abundance
measurements of \citet[][see Sect.~\ref{sec:BBN}]{Cooke:2017cwo}. The
dashed contours in this figure show how the green contours shift if
the Pantheon SNe data are replaced by the JLA SNe sample. Adding
\planck\ CMB lensing (grey contours) constrains $\Omm h^2$ and shifts
$H_0$ further away from the R18 measurement. Using a ``conservative''
\planck\ prior of $100\thetaMC = 1.0409 \pm 0.0006$ (which is
consistent with all of the variants of \LCDM\ considered in this paper
to within $1\,\sigma$, see Table~\ref{tab:base_extensions}) gives the
red contours, with $H_0=(67.9\pm 0.8)\Hunit$ and $\Omm=0.305\pm
0.001$, very close to the result using the full \planck\ likelihood
(blue contours). Evidently, there is a significant problem in matching
the base-\LCDM\ model to the R18 results and this tension is not
confined exclusively to the \Planck\ results.

The question then arises of whether there is a plausible extension to
the base-\LCDM\ model that can resolve the discrepancy.
Table~\ref{tab:base_extensions} summarizes the \planck\ constraints on $H_0$
for variants of \LCDM\ considered in this paper. $H_0$ remains
discrepant with R18 in all of these cases, with the exception of
models in which we allow the dark energy equation of state to
vary. For models with either a fixed dark energy equation-of-state
parameter, $\wzero$, or time-varying equation of state parameterized
by $\wzero$ and $w_a$ (see Sect.~\ref{sec:w0wa} for definitions and further
details), \planck\ data alone lead to poor constraints on $H_0$.
However, for most physical dark energy models where $p_{\rm de} \ge - \rho_{\rm de}$ (so $\wzero >-1$), and the density is only important after recombination, $H_0$ can only decrease with respect to \lcdm\ if the measured CMB acoustic scale is maintained, making the discrepancy with R18 worse.
If we allow for $\wzero <-1$, then adding BAO and SNe data is critical to obtain a useful constraint
\citep[as pointed out by][]{Aubourg:2014yra}, and we find
\beglet
\begin{eqnarray}
  H_0 &=& \rep{(68.34 \pm 0.81)} \Hunit, \quad (\wzero \ {\rm varying}), \quad \\
  H_0 &=& \rep{(68.31 \pm 0.82)} \Hunit, \quad ({\wzero, w_a} \ {\rm varying} ),
\end{eqnarray}
\endlet
for the parameter combination {\planckalllensing\dataplus\BAO{\dataplus}Pantheon}. Modifying the dark energy sector \rep{in the late universe} does not resolve the discrepancy with R18.

If the difference between base \LCDM\ and the R18 measurement of $H_0$ is caused by new physics,
then it is unlikely to be through some change to the late-time distance-redshift relationship. Another
possibility is a change in the sound horizon scale. If we use the R18 measurement of $H_0$, combined with Pantheon
supernovae and BAO, the acoustic scale is $\rdrag = (136.4\pm 3.5)\ \Mpc$. The difficulty is to find a model that can give this much smaller value of the sound horizon (compared to $\rdrag = (147.05\pm 0.3)\ \Mpc$ from \planckall\ in \lcdm), while preserving a good fit to the CMB power spectra and a baryon density consistent with BBN. We discuss some extensions to \lcdm\ in Sect.~\ref{sec:grid} that allow larger $H_0$ values (e.g., $\nnu > 3.046$); however, these models are not preferred by the \planck\ data, and tend to introduce other tensions, such as a higher value of $\sigma_8$.\footnote{To obtain simultaneously higher values of $H_0$, lower values of $\sigma_8$, and consistent values of $\Omm$ it is necessary to invoke
less common extensions of the \lcdm\ model,\rep{ such as models featuring non-standard interactions in the neutrino, dark-matter, dark-radiation, and/or dark-energy sector \rep{\citep[see e.g.,][]{Pettorino:2013, Lesgourgues:2015wza,planck2014-a16,Archidiacono:2016kkh,Lancaster:2017ksf,Oldengott:2017fhy,DiValentino:2017oaw,
Buen-Abad:2017gxg,Poulin:2018cxd,Kreisch:2019yzn,Agrawal:2019lmo,Lin:2019qug, Archidiacono:2019wdp}
}. Such models are likely to be highly constrained by the \Planck, BAO, and supernova data used in this paper and by future
CMB observations and surveys of large-scale structure.}}

\rep{ The tension between base \LCDM\ and the SH0ES $H_0$ measurement is intriguing and emphasizes the need for independent
measurements of the distance scale. It will be interesting in the future to compare the Cepheid distance scale in more
detail with other distance indicators, such as the tip of the red giant branch \citep{Freedman:2019}, and with completely
different techniques such as gravitational-lensing time delays \citep{Suyu:2013} and gravitational-wave standard sirens
\citep{Holz:2005df, Ligo:2017, Chen:2017, Feeney:2018b}.}

\subsection{Weak gravitational lensing of galaxies}
\label{sec:WL}

\begin{figure*}[h!]
\begin{center}
\includegraphics[width=90mm]{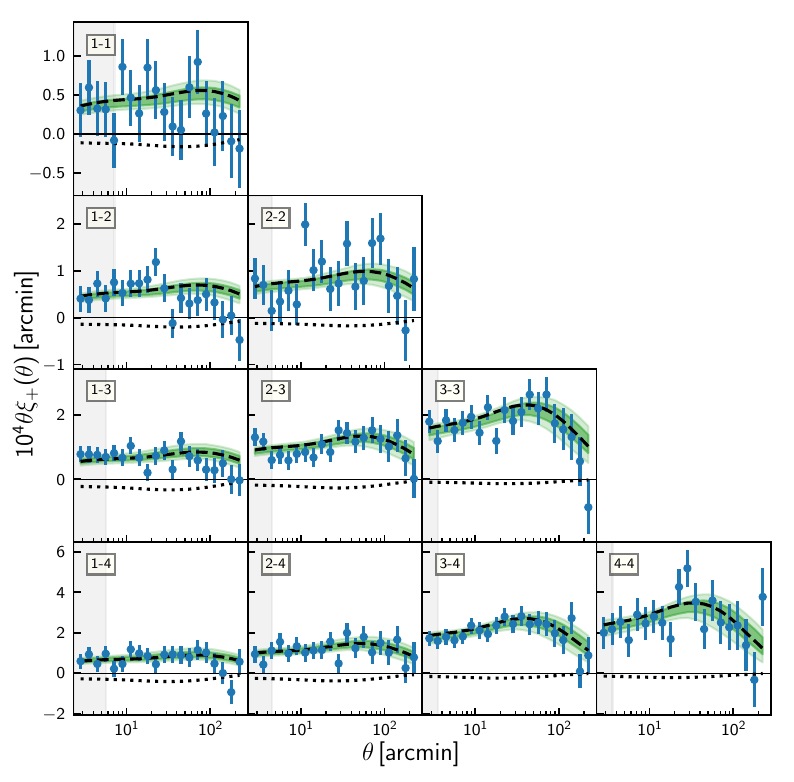}
\includegraphics[width=90mm]{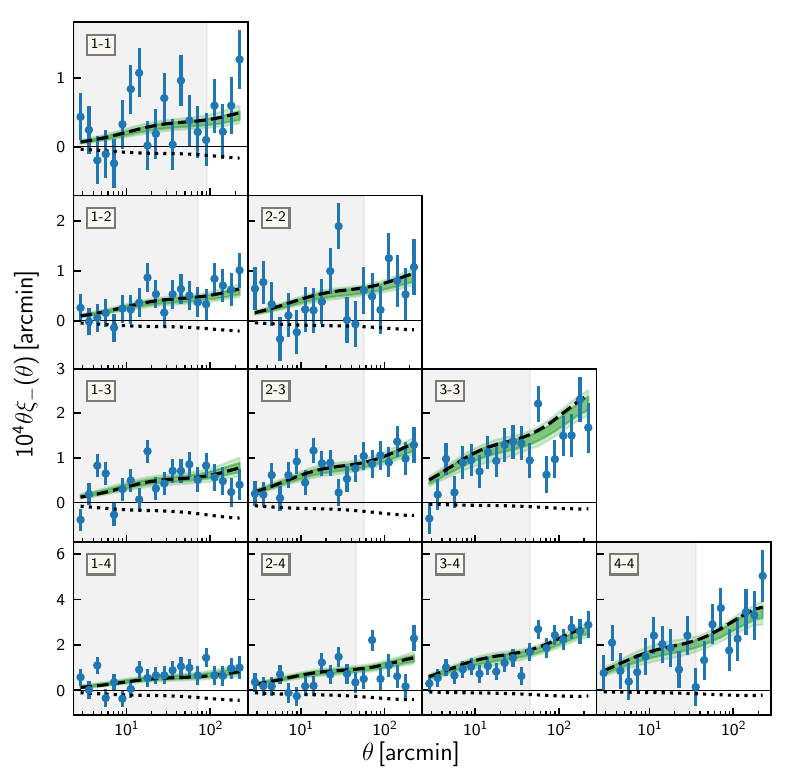}
\end{center}
\caption {
\rep{
Dark Energy Survey (DES) shear correlations functions, $\xi_+$ (left) and $\xi_-$ (right), for the auto- and cross-correlation between the four DES source redshift bins \citep{Troxel:2017xyo}. Green bands show the 68\,\% and 95\,\% distribution of model fits in the DES lensing-only base-\lcdm\ parameter fits. The dashed line shows the DES lensing parameter best fit when the cosmological parameters are fixed to the best fit model for \planckall\ only; dotted lines show the size of the contribution of intrinsic alignment terms to the dashed lines. Grey bands show the scales excluded from the DES analysis, in order to reduce sensitivity to nonlinear effects.}
 }
\label{fig:DESshearspec}
\end{figure*}

The distortion of the shapes of distant galaxies by lensing due to large-scale structure along the line of sight is known as galaxy lensing or cosmic
shear \citep[see e.g.,][for a review]{Bartelmann:1999yn}.
It constrains the gravitational potentials at lower redshift than CMB lensing, with tomographic information and completely different systematics, so the measurements are complementary.
Since the source galaxy shapes and orientations
are in general unknown, the lensing signal is a small effect that can only be detected statistically. If it can be measured robustly it is a relatively clean way of measuring the Weyl potential (and hence, in GR, the
total matter fluctuations); however, the bulk of the statistical power comes from scales where the signal is significantly nonlinear, complicating the cosmological interpretation.
The measurement is also complicated by several other issues.
Intrinsic alignment between the shape of lensed galaxies and their surrounding potentials means that the galaxy shape correlation functions actually measure a combination of lensing and intrinsic alignment effects \citep{Hirata:2004gc}. Furthermore, to get a strong statistical detection, a large sample of galaxies is needed, so most current results use samples that rely mainly on photometric redshifts; accurate calibration of  the photometric redshifts and modelling of the errors are required in order to use the observed lensing signal for cosmology.

Cosmic shear measurements are available from several collaborations, including \CFHTLENS\ \citep[][which we discussed in~\paramsII]{Heymans:2012gg,Erben:2012zw}, {\tt DLS}~\citep{Jee:2015jta}, \rep{ and more recently the Dark Energy Survey \citep[DES, ][]{Abbott:2017wau}, Hyper Suprime-Cam \citep[HSC, ][]{Hikage:2018qbn,Hamana:2019etx}, and
\KIDS~\citep{Hildebrandt:2016iqg,Kohlinger:2017sxk,Hildebrandt:2018yau}}. The \CFHTLENS\ and \KIDS\ results found a modest tension with the \planck\ \lcdm\ cosmology, preferring lower values of $\Omm$ or $\sigma_8$. A
 combined analysis of \KIDS\ with GAMA~\citep{vanUitert:2017ieu} galaxy clustering has found results consistent with \Planck,
whereas a similar analysis combining \KIDS\ lensing measurements with spectroscopic data from the 2-degree Field Lensing
Survey and BOSS claims a $2.6\,\sigma$ discrepancy with \Planck\ \citep{Joudaki:2018}.
\cite{Troxel:2018qll} have shown that a more
accurate treatment of the intrinsic galaxy shape noise, multiplicative shear calibration uncertainty, and angular scale of each bin can significantly change \rep{earlier} \KIDS\ results (by about $1\,\sigma$), making them more consistent with \planck.
\rep{At the time of running our chains the DES lensing results had been published and included this improved modelling, while an updated analysis from \KIDS\ was not yet available}; we therefore only consider the DES results in detail here. \cite{Troxel:2018qll} reports
consistent results from DES and their new analysis of \KIDS, and HSC also report results consistent with \planck.
\rep{However, the more recent \KIDS\ analysis by \cite{Hildebrandt:2018yau} still finds a  $2.3\,\sigma$ discrepancy with \planck, and
\cite{Joudaki:2019pmv} claim that a recalibration of the DES redshifts gives results compatible with \KIDS\ and a combined $2.5\,\sigma$ tension with \planck.}

The DES collaboration analysed 1321$\,{\rm deg}^2$ of imaging data from the first year of DES. They analysed the cosmic shear correlation functions of 26 million source galaxies in four redshift bins \citep{Troxel:2017xyo}, and also considered the auto- \citep{Elvin-Poole:2017xsf} and cross-spectrum \citep{Prat:2017goa} of 650\,000 lens galaxies in five redshift bins. To be conservative they restricted their parameter analysis to scales that are only weakly affected by nonlinear modelling (at the expense of substantially reducing the statistical power of the data). To account for modelling uncertainties, the cosmic shear analysis marginalizes over 10 nuisance parameters, describing uncertainties in the photometric redshift distributions, shear calibrations, and intrinsic alignments; the joint analysis adds an additional 10 nuisance parameters describing the bias and redshift uncertainty of the lens galaxies.

We use the first-year DES lensing (cosmic shear) likelihood, data cuts, nuisance parameters, and nuisance parameter priors, as described by \citet{Troxel:2017xyo, Abbott:2017wau,Krause:2017ekm}.
We implement the theory model code independently, but use the same physical model and assumptions as the DES analysis,\footnote{Except for the modified-gravity models in Sect.~\ref{sec:darkenergy} where we calculate the lensing spectrum directly from the power spectrum of the Weyl potential (rather than from the matter power spectrum assuming standard GR).} treating the nuisance parameters as fast parameters for sampling in \COSMOMC.
In this section we adopt the cosmological parameter priors assumed by \citet{Troxel:2017xyo}, but to be consistent with our other \lcdm\ analyses, we assume a single minimal-mass eigenstate rather than marginalizing over the neutrino mass, and use {\tt HMcode} for the nonlinear corrections.\footnote{The results are quite sensitive to the choice of cosmological parameter priors, see \PlanckLensThree\ for an analysis using the different priors assumed by the \Planck\ CMB lensing analysis. Here we assume consistent (DES) priors for DES and CMB lensing results; however, the $\planck$ power spectrum constraints are much less sensitive to priors and we use our default priors for those.} The shear correlation data points and parameter fits are shown in Fig.~\ref{fig:DESshearspec}. Note that intrinsic alignments contribute significantly to the observed
shear correlation functions (as shown by the dotted lines in the figure).
This introduces additional modelling uncertainty and a possible source of bias if the intrinsic alignment model is not correct. The DES model is validated in~\cite{Troxel:2017xyo,Krause:2017ekm}.

\begin{figure}
\begin{center}
\includegraphics[width=85mm,angle=0]{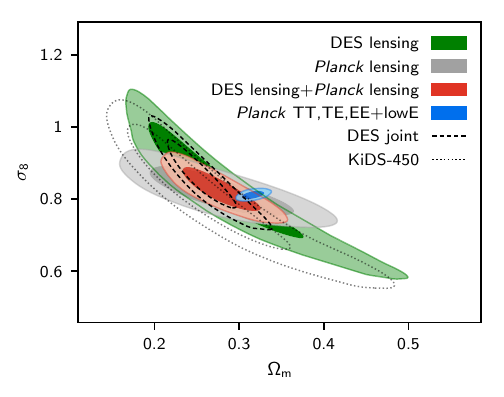}
\end{center}
\caption {\rep{
  Base-\lcdm\ model $68\,\%$ and $95\,\%$ constraint contours on the matter-density parameter $\Omm$ and fluctuation amplitude $\sigma_8$ from DES lensing \citep[][green]{Troxel:2017xyo}, \planck\ CMB lensing (grey), and the joint lensing constraint (red). For comparison, the dashed line shows the constraint from the DES cosmic shear plus galaxy-clustering joint analysis \citep{Abbott:2017wau}, the dotted line the constraint from the original \KIDS-450 analysis \citep[][without the corrections considered in \citealt{Troxel:2018qll}]{Hildebrandt:2016iqg}, and the blue filled contour shows the independent constraint from the \planck\ CMB power spectra.}
 }
\label{fig:DESlens}
\end{figure}

Figure~\ref{fig:DESlens} shows the constraints in the $\Omm$--$\sigma_8$ plane from DES lensing, compared to the constraints from the CMB power spectra and CMB lensing. The DES cosmic shear constraint is of comparable statistical power to CMB lensing, but due to the significantly lower mean source redshift, the degeneracy directions are different (with DES cosmic shear approximately constraining $\Omm\sigma_8^{0.5}$ and CMB lensing constraining $\Omm\sigma_8^{0.25})$.
The correlation between the DES cosmic shear and CMB lensing results is relatively small, since the sky area of the CMB reconstruction is much larger than that for DES, and it is also mostly not at high signal-to-noise ratio. Neglecting the cross-correlation,
we combine the DES and \planck\ lensing results to break a large part of the degeneracy, giving a substantially tighter constraint than either alone. The lensing results separately, and jointly, are both consistent with the main \planck\ power-spectrum results, although preferring $\sigma_8$ and $\Omm$ values at the lower end of those allowed by \planck. The DES joint analysis of lensing and clustering is also marginally consistent, but with posteriors preferring lower values of $\Omm$ (see the next subsection).
Overlap of contours in a marginalized 2D subspace does not of course guarantee consistency in the full parameter space. However, the values of the Hubble parameter in the region of $\Omm$--$\sigma_8$ parameter space consistent with \planck\ $\Omm$ and $\sigma_8$ are also consistent with \planck's value of $H_0$.
A joint analysis of DES with BAO and a BBN baryon-density constraint gives values of the Hubble parameter that are very consistent with the \planck\ power spectrum analysis \citep{Abbott:2017smn}.

\subsection{Galaxy clustering and cross-correlation}
\label{sec:galaxyclustering}

The power spectrum of tracers of large-scale structure can yield a biased estimate of the matter power spectrum, which can then be used as a probe of cosmology.
For adiabatic Gaussian initial perturbations the bias is expected to be constant on large scales where the tracers are out of causal contact with each other, and nearly constant on scales where nonlinear growth effects are small.
Much more information is available if small scales can also be used, but this requires detailed modelling of perturbative biases out to $k\approx 0.3$--$0.6\,\Mpc^{-1}$, and fully nonlinear predictions beyond that. Any violation of scale-invariant bias on super-horizon scales would be a robust test for non-Gaussian initial perturbations protected by causality \citep{Dalal:2007cu}. However, using the shape of the biased-tracer power spectrum on smaller scales to constrain cosmology requires at least a model of constant bias parameters for each population at each redshift, and, as precision is increased, or smaller scales probed, a model for the scale dependence of the bias.
Early galaxy surveys provided cosmology constraints that were competitive with those from CMB power spectrum measurements \citep[e.g.,][]{Percival:2001hw}, but as precision has improved, focus has mainly moved away to using the cleaner BAO and RSD measurements and, in parallel, developing ways to get the quasi-linear theoretical predictions under better control. Most recent studies of galaxy clustering have focussed on investigating bias rather than background cosmology, with the notable exception of WiggleZ \citep{Parkinson:2012vd}.

Here we focus on the first-year DES survey measurement of galaxy clustering \citep{Elvin-Poole:2017xsf} and the cross-correlation with galaxy lensing \citep[``galaxy-galaxy lensing'']{Prat:2017goa}.  By simultaneously fitting for the clustering, lensing, and cross-correlation, the bias parameters can be constrained empirically \citep{Abbott:2017wau}. Similar analyses using \KIDS\
lensing data combined with spectroscopic surveys have been performed by~\citet{vanUitert:2017ieu} and \citet{Joudaki:2018}.

To keep the theoretical model under control (nearly in the linear regime), DES exclude
all correlations on scales where modelling uncertainties in the nonlinear regime could begin to  bias parameter constraints (at the price of substantially reducing the total statistical power available in the data). %
Assuming a constant bias parameter for each of the given source redshift bins, parameter constraints are obtained after marginalizing over the bias, as well as a photometric redshift window mid-point shift parameter to account for redshift uncertainties. Together with galaxy lensing parameters, the full joint analysis has 20 nuisance parameters. Although this is a relatively complex nuisance-parameter model, it clearly does not fully model all possible sources of error: for example, correlations between redshift bins may depend on photometric redshift uncertainties that are not well captured by a single shift in the mean of each window's population. However, ~\citet{Troxel:2017xyo} estimate that the impact on parameters is below $0.5\,\sigma$ for all more complex models they considered.
The DES theoretical model for the correlation functions (which we follow) neglects redshift-space distortions, and assumes that the bias is constant in redshift and $k$ across each redshift bin; these may be adequate approximations for current noise levels and data cuts, but will likely need to be re-examined in the future as statistical errors improve.

\begin{figure}
\begin{center}
\includegraphics[width=\columnwidth]{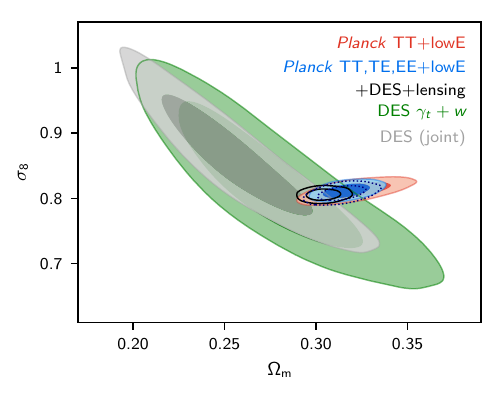}
\end{center}
\caption {\rep{
Base-\lcdm\ model constraints from the Dark Energy Survey (DES), using the shear-galaxy correlation and the galaxy auto-correlation data (green) and the joint result with DES lensing (grey), compared with \planck\ results using \shortTT\ and \shortall.
The black solid contours show the joint constraint from \planckalllensing+DES, assuming the difference between the data sets is purely statistical. The dotted line shows the \planckall\ result using the \camspec\ likelihood, which is slightly more consistent with the DES contours than using the default \plik\ likelihood. Contours contain $68\,\%$ and $95\,\%$ of the probability.
 }}
\label{fig:DESjoint}
\end{figure}

Using the full combined clustering and lensing DES likelihood, for a total of 457 data points \citep{Abbott:2017wau}, the best-fit \lcdm\ model has $\chi^2_{\rep{\rm eff}}\approx 500$ or \rep{513} with the \planck\ best-fit cosmology. Parameter constraints from the galaxy auto- and cross-correlation are shown in Fig.~\ref{fig:DESjoint}, together with the joint constraint with DES lensing (the comparison with DES galaxy lensing and CMB lensing alone is shown in Fig.~\ref{fig:DESlens}).

Using the joint DES likelihood in combination with DES cosmological parameter priors gives (for our base-\lcdm\ model with $\sumnu=0.06\eV$)
\twoonesig[2cm]{S_8 &\equiv \sigma_8(\Omm/0.3)^{0.5} = \rep{0.793}\pm 0.024, }{\Omm &= \rep{0.256}^{+0.023}_{-0.031},}{DES.}
\planckalllensing\ gives a higher value of $S_8=0.832\pm0.013$, as well as larger $\Omm=0.315\pm0.007$.
As shown in the previous section, the DES lensing results are quite compatible with \planck, although peaking at lower $\Omm$ and $\sigma_8$ values. The full joint DES likelihood, however, shrinks the error bars in the $\sigma_8$--$\Omm$ plane so that only 95\,\% confidence contours overlap with \planck\ CMB data, giving a moderate (roughly 2\,\% PTE) tension, as shown in Fig.~\ref{fig:DESjoint}. The dotted contour in Fig.~\ref{fig:DESjoint} shows the result using the \camspec\ \planck\ likelihood,  which gives results slightly more consistent with DES than the default \plik\ likelihood. The \planck\ result is therefore sensitive to the details of the polarization modelling at the $0.5\,\sigma$ level, and the tension cannot be quantified robustly beyond this level.

Combining DES with the baseline \planck\ likelihood pulls the \planck\ result to lower $\Omm$ and slightly lower $\sigma_8$, giving
\threeonesig[4cm]{S_8 &= 0.811\pm 0.011,}{\Omm &= \rep{0.3040\pm 0.0060},}{\sigma_8 &= \rep{0.8062\pm 0.0057},}{\planckall\dataplus\lensing\dataplus{DES}.}
A similar shift is seen without including \planck\ lensing, and is disfavoured by \planck\ CMB with a total $\dchisquare\approx 13$ for the CMB likelihoods (comparing the \planck-only best fit to the fit when combined with DES). The shift in parameters is also larger than would be expected for Gaussian distributions, given the small change in parameter covariance.
The corresponding change in $\chi^2_{\rm eff}$  for the DES likelihood is $\dchisquare\approx 10$, which is high,
 but less surprising given the $4\text{--}5$ contribution expected from the number of parameters that are much better constrained by \planck.
The summary consistency statistic $\chi^2_{\rm eff,joint} - \chi^2_{\rm eff, DES} - \chi^2_{\rm eff, Planck}  \approx 14$, which is high at the roughly $1\,\%$ PTE level, given the expected value of 4, assuming roughly Gaussian statistics~\citep{Raveri:2018wln}.

In summary, the DES combined probes of \lcdm\ parameters are in moderate percent-level tension with \planck.
Whether this is a statistical fluctuation, evidence for systematics, or new physics is currently unclear. In this paper, we follow the philosophy of \paramsI\ and \paramsII\ of making minimal use of other astrophysical data in combination with \planck, using BAO as our primary complementary data set. We therefore
do not include DES results in most of the parameter constraints discussed in
 this paper.
We do, however, consider the impact of the DES weak lensing results on dark-energy and modified-gravity constraints in Sect.~\ref{sec:darkenergy} and on neutrino masses in Sect.~\ref{subsec:mnu}. We also include DES for a wider
range of models in the \Planck\  parameter tables available on the \PLA.

\subsection{Cluster counts}
\label{sec:clusters}

Counts of clusters of galaxies provide an additional way of constraining the
amplitude of the power spectrum at low redshifts \citep[e.g.,][and references therein]
{Pierpaoli:2001,Komatsu:2002}.  \Planck\ clusters, selected via the thermal
Sunyaev-Zeldovich (tSZ) signature, were used to explore  cosmological parameters
in \cite{planck2013-p15}. This analysis was revisited using a deeper
sample of \Planck\ clusters in \cite{planck2014-a30}. We have not produced a new tSZ
cluster catalogue in the 2018 \Planck\ data release and so the results presented
in this section are based on the 439 clusters in the MMF3 cluster cosmology sample,
as analysed in \cite{planck2014-a30}.
 Comparison with the 2018 CMB \Planck\ power
spectrum results show differences primarily from changes to the base-\LCDM\ model
parameters caused by the tighter constraints on $\tau$. The impact of the lower value
of $\tau$ reported in \cite{planck2014-a10} on the interpretation of cluster counts
has been discussed by \cite{Salvati:2017rsn}.

We first review the main results from \cite{planck2014-a30}.
There has been increasing recognition that the calibration of cluster masses is the dominant uncertainty in
using cluster counts to estimate cosmological parameters. In the analysis of \Planck\
clusters, the cluster tSZ observable
was related to the cluster mass $M_{500}$\footnote{The mass contained
 within a sphere of radius $R_{500}$, centred on the cluster, where
 $R_{500}$ is the radius at which the mean density is 500 times the
 critical density at the redshift of the cluster.} using X-ray
scaling relations \citep{Arnaud:2010}, calibrated against a subsample
of the \Planck\ clusters. The X-ray masses are, however, derived
assuming hydrostatic equilibrium and are expected to be biased low
\citep[e.g.,][]{Nagai:2007}. This was accounted for by multiplying the
true masses by a so-called ``hydrostatic mass bias'' factor of $(1-b)$.
The strongest constraints on
this bias factor come from weak gravitational lensing estimates of
cluster masses. \citet{planck2014-a30} considered three lensing mass calibrations\footnote{\rep{See \cite{Sifon:2016} for a discussion of dynamical mass estimates for SZ-selected clusters.}}:
$(1-b) = 0.69 \pm 0.07$ from 22 \Planck\ clusters from the Weighing the Giants lensing
programme \citep{vonderLinden:14}; $(1-b) = 0.78 \pm 0.08$ from 37 \Planck\ clusters
calibrated by the Canadian Cluster Comparison Project \citep{Hoekstra:15}; and
$1/(1-b) = 0.99 \pm 0.19$ from \Planck\ CMB-lensing mass estimates of the MMF3
cluster sample \citep{planck2014-a30}. More recently, \citet{Sereno:2017zcn} have analysed 35 \Planck\ clusters with galaxy shear data from the CFHTLenS \citep{Heymans:2012} and RCSLenS \citep{Hildebrandt:2016} surveys, finding $(1-b) \approx 0.77 \pm 0.11$ for all clusters and $(1-b) = 0.68 \pm 0.11$ for the 15 clusters in the cosmological sample. Additionally, \citet{Penna-Lima:2016tvo}
use gravitational lensing measurements from HST images of 21 \Planck\ clusters finding $(1-b)=0.73 \pm 0.10$.

The determination of cosmological parameters such as $\sigma_8$ and $\Omm$ from \Planck\ cluster counts is
 strongly dependent on the prior adopted for the mass bias parameter.
In this paper, we use the \planckalllensing\ likelihood in combination with the \Planck\ cluster
counts to derive a constraint on $(1-b)$ (following similar analyses described in \citealt{planck2014-a24} and \citealt{Salvati:2017rsn}). This gives \oneonesig[5cm]{(1-b) = 0.62 \pm 0.03}{\planckalllensing}{, \label{equ:cluster1}}
compared to $0.58 \pm 0.04$ using the 2015 TT,TE,EE+lowP\ likelihood \citep{planck2014-a24}. The roughly $1\,\sigma$ upward shift in Eq.~\eqref{equ:cluster1} is mainly caused by the $2018$ change in the $\tau$ constraint. The mass bias of Eq.~\eqref{equ:cluster1} is at the lower end of the weak-lensing mass
estimates, but is about $2\,\sigma$ lower that the \Planck\ CMB-lensing mass calibration reported
in \cite{planck2014-a24}.

\rep{
\citet{Zubeldia:2019brr} have revisited the \Planck\ CMB-lensing mass calibration, incorporating the CMB-lensing mass estimates within a likelihood describing the \Planck\ cluster counts, together with a \Planck\ prior on $\theta_{\rm MC}$. This study
corrects for significant biases in the analysis reported in \cite{planck2014-a24}. \cite{Zubeldia:2019brr}
find $(1-b) = 0.71\pm 0.10$ and
$\sigma_8(\Omm/0.33)^{0.25} =  0.765\pm 0.035$.
These results, based entirely on \Planck\ data, are consistent with the base-\LCDM\ parameters from the \Planck\ power spectra and with the inferred mass bias of Eq.~\eqref{equ:cluster1}.}

Since \paramsII\ there have been a number of new analyses of cluster counts using
other surveys. Two recent studies \citep{Mantz:2015, deHaan:2016qvy}, with very different
selection criteria, use weak gravitational
lensing mass determinations from the Weighing the Giants programme to calibrate cluster scaling
relations. \cite{deHaan:2016qvy} analysed a sample of 377 clusters at $z>0.25$ identified with SPT,
finding $\sigma_8(\Omm/0.27)^{0.3} = 0.797 \pm 0.031$, while \citet{Mantz:2015} analysed an X-ray-selected sample of clusters from the ROSAT All-Sky survey, finding $\sigma_8(\Omm/0.3)^{0.17} = 0.81 \pm 0.03$.
These measurements can be compared to our baseline \LCDM\ constraints (\planckalllensing) of
$\sigma_8(\Omm/0.27)^{0.3} = 0.849 \pm 0.010$ and $\sigma_8(\Omm/0.3)^{0.17} = 0.817 \pm 0.076$.

\citet{Schellenberger:2017usb} have analysed a sample of 64 of the brightest X-ray clusters using a prior on the
hydrostatic mass bias from \cite{Biffi:2016}. These authors find $\Omega_{\rm m} = 0.303 \pm 0.009$, $\sigma_8 = 0.790^{+0.030}_{-0.028}$, and $S_8 = \sigma_8 (\Omm/0.3)^{1/2} = 0.792 \pm 0.054$. Each of these numbers is within about $1\,\sigma$ of the
\Planck\ base-\LCDM\ best-fit cosmology reported in this paper.
 Finally, we mention the analysis of ROSAT-observed X-ray clusters carried out by \citet{Bohringer:2014ooa,
Boehringer:2017wvr}. These authors choose a central value for the hydrostatic mass bias of $(1-b) = 0.9$, although they allow
for small variations in the slope ($7\,\%$) and normalization ($14\,\%$) of the X-ray luminosity-mass relation; they find constraints the $\sigma_8$--$\Omm$ plane in tension with \Planck\ at about $2.5\,\sigma$.

In summary, accurate calibrations of cluster masses are essential if
cluster counts are to be used as cosmological probes. Given the uncertainties in these
calibrations, we do not use cluster counts in our main parameter grid. Consistency of
cluster counts with the best-fit \Planck\ base-\LCDM\ cosmology
requires hydrostatic mass biases (Eq.~\ref{equ:cluster1}) that are at
the lower end, but within about $1\,\sigma$ of bias factors estimated from
weak-lensing cluster masses. The combined \Planck\ CMB-lensing and
cluster-count analysis reported by \cite{Zubeldia:2019brr} is in
good agreement with the \Planck\ base-\LCDM\ cosmology. At this time, there is no
compelling evidence for a discrepancy between \Planck-, SPT-, or X-ray-selected cluster
counts and the base-\LCDM\ model.

\section{Internal consistency of \boldlcdm\ model parameters}\label{sec:internalconsistency}

In this section we briefly discuss a couple of curious features of the \planck\ data that lead to moderate tensions in parameter consistency tests.
We first discuss how parameters vary between high and low multipoles, and the relevant features in the power spectra that may be responsible for these shifts.
We then discuss the related issue of how the full multipole range appears to prefer more lensing than predicted by \lcdm\ fits.
\rep{We end this section with a discussion of systematic uncertainties.}

\subsection{Consistency of high and low multipoles}\label{sec:highlow}

\begin{figure*}[htbp!]
\begin{center}
\includegraphics[width=\textwidth]{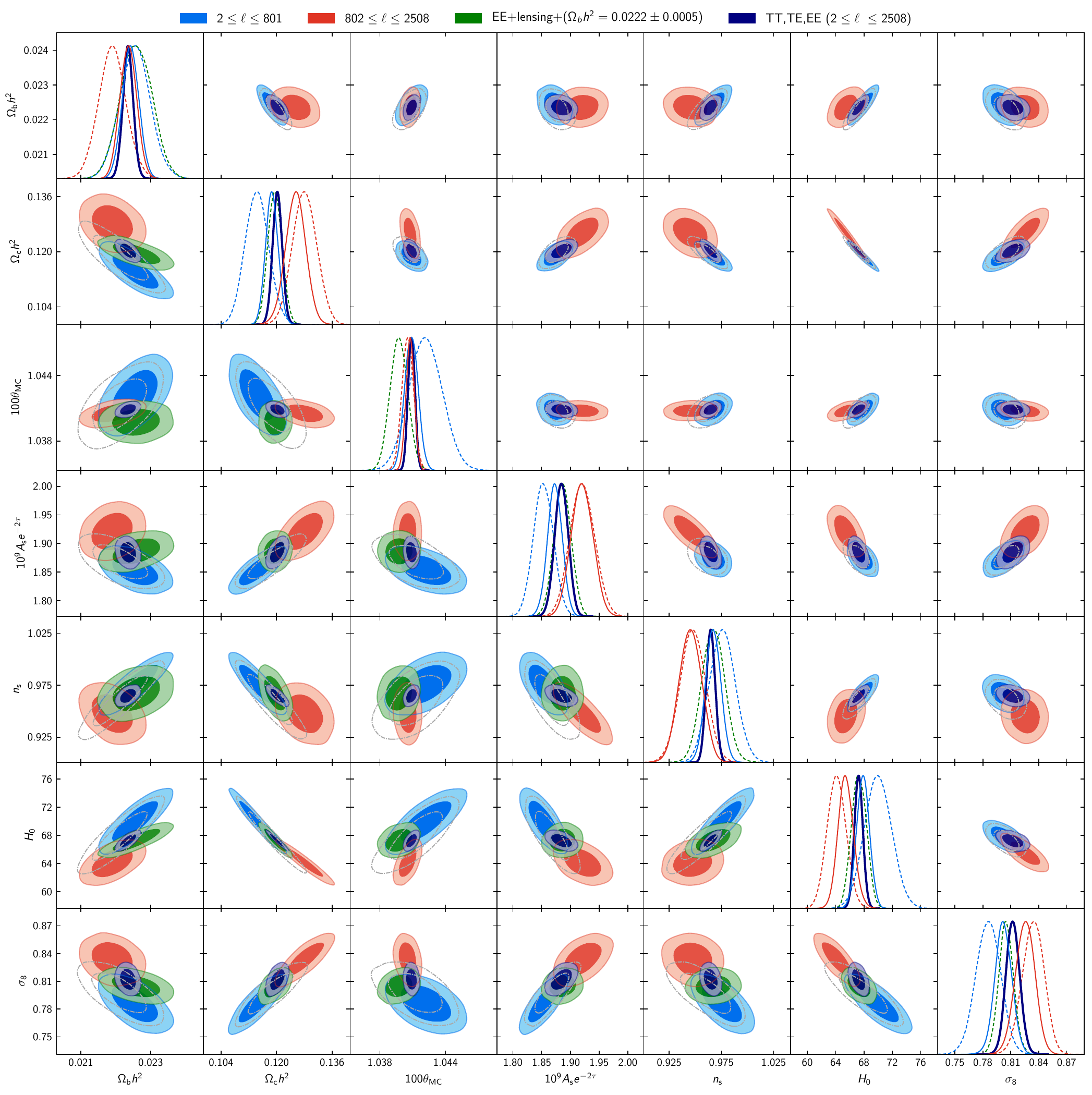}
\end{center}
\vspace{-3mm}
\caption{Base-\lcdm\ 68\,\% and 95\,\% parameter constraint contours from the CMB power spectra using $\ell \le 801$ (blue), compared to $\ell \ge 802$ (red). All results use the \pliklite\ \planck\ likelihood, and also include the low-$\ell$ \simall\ ``lowE'' likelihood to constrain the optical depth $\tau$; the \commander\ likelihood is used for temperature multipoles $\ell <30$.
The lower triangle contains the \planck\ temperature likelihoods, which show a moderate tension between high and low multipoles; however, they intersect in a region of parameter space consistent with the nearly-independent constraint from EE+lensing combined with a conservative prior $\Omb h^2 = 0.0222\pm 0.0005$, motivated by element-abundance observations (green).
The upper triangle shows the equivalent results from \planckall\ at low and high multipoles. The full combined result from \planckall\ is shown as the navy contours.
The unfilled grey contours show the result for multipoles
$30 \le \ell \le 801$ (that is, removing the low-$\ell$ \commander\ likelihood
that pulls parameters to give lower temperature power, due to the dip below
$\ell \approx 30$). The diagonal plots are the marginalized parameter constraints,
where results corresponding to the lower triangle are shown dashed, while the upper triangle are the solid curves.
}
\label{fig:lmaxlmin}
\end{figure*}

\begin{figure*}[]
\begin{center}
\includegraphics[width=\textwidth]{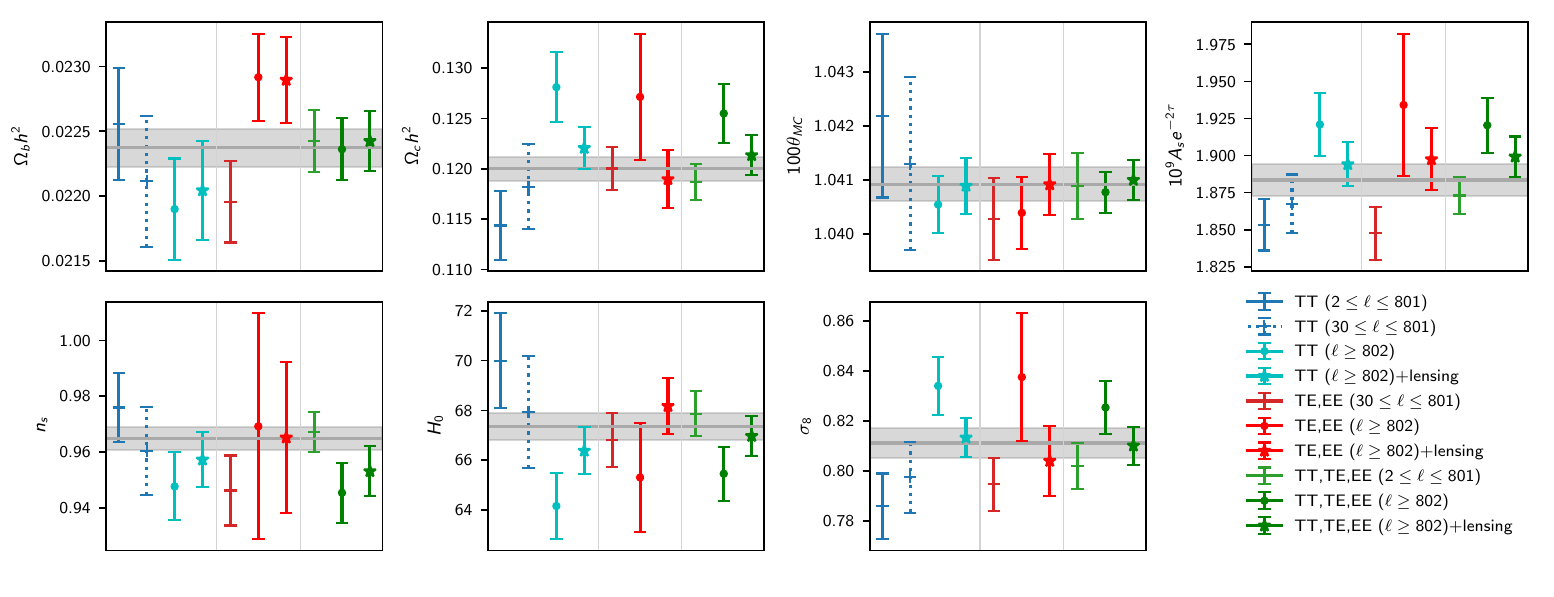}
\end{center}
\vspace{-3mm}
\caption{Base-\lcdm\ 68\,\% marginalized parameter constraints for various combinations of power spectrum ranges (all using \pliklite\ and also including low-$\ell$ polarization ``lowE'').
Points marked with a cross are from $2 \le \ell\le 801$, while points marked with a circle are from $\ell \ge 802$.
Dotted errors are the result from $30 \le \ell \le 801$, without the \commander\ large-scale temperature likelihood, showing that $\ell < 30$ pulls the low-multipole parameters further from the joint result. Points marked with a star
are from $\ell \ge 802$ combined with the lensing likelihood, showing that constraining the lensing amplitude pulls all the results from high multipoles towards better consistency with the results from lower multipoles. The grey horizontal band shows the combined $68\,\%$ constraint from \planckalllensing.
}
\label{fig:lmaxlminparams}
\end{figure*}

The \planck\ CMB temperature power spectrum shows a conspicuous dip over the multipole
range $20\la \ell\la 30$ compared to \lcdm\ fits, as can be seen in Fig.~\ref{fig:coadded}.
This feature was first observed by WMAP \citep{Bennett:2003bz}, and was discussed in detail in \paramsI.
Since it is detected consistently by both WMAP and \planck\ at multiple frequencies, it cannot plausibly be explained by an instrumental systematic or foreground.
The large-scale \planck\ temperature map is signal dominated, so the dip feature is almost identical in this final release.
\paramsI\ also noted an approximately $2.7\,\sigma$ mismatch between the best-fit \LCDM\ cosmology and the amplitude of the measured temperature power spectrum at $\ell \le 30$. However, with the tighter optical depth constraints used in this paper and improvements in the
high multipole likelihoods we find no strong evidence for an amplitude mismatch.
 The \planckalllensing\ best-fit \LCDM\ model provides a good overall fit to the temperature multipoles at $\ell < 30$ ($\effchisquare \approx 23$ for 28 data points), and because of the skewed $\chi^2$-like distribution of the CMB spectrum estimators, it is expected that typically more than half of the data points are below the theoretical model values (see Fig.~\ref{fig:coadded}). The statistical significance of the dip feature is hard to quantify, since it was identified a posteriori, but \likeII\ suggest a significance of about 2.8\,\% after maximizing over extremal $\ell$ ranges found in simulations.
This could be an indication of new physics at large scales, for example associated with a sharp feature in the
inflationary potential \citep[as considered by][and many subsequent researchers]{Peiris:2003}.
Alternatively, it could just be a statistical fluctuation, which is our baseline assumption. However,
since the dip is a relatively unusual fluctuation and it is near one end of the multipole range, it tends to pull cosmological parameters more than would
be expected in typical realizations of a \LCDM\ cosmology. This needs to be borne in mind in assessing parameter shifts
between low and high multipoles.

WMAP measured the CMB temperature fluctuations up to $\ell \approx 800$ \citep{Bennett:2012zja}. The higher-resolution data from \planck\ substantially increases the multipole range of the temperature power spectrum out to $\ell \approx 2500$.
Cosmological parameters are therefore expected to shift (usually towards the truth) from the mean posterior values measured by WMAP, together with a reduction in the error bars. This is what is seen, with the \planck\ values of $H_0$ and $\ns$ decreasing, and $\Omm$ and $\Omm h^2$ increasing, along with substantially smaller errors.
However as noted in \paramsI\ the magnitudes of the shifts appear to be slightly larger than might be expected
statistically, assuming the base-\LCDM\ cosmology. This stimulated additional work on the consistency of
the \Planck\ power spectra reported in \likeII\ and to further investigations of the consistency of cosmological parameters
measured from high and low multipoles from \planck\ \citep{Addison:2015wyg, planck2016-LI}.
As noted in the introduction, there is a very good agreement between \planck\ and WMAP temperature maps on the scales observed by WMAP \citep{planck2014-a01,Huang:2018xle},
but an inconsistency with high multipoles could indicate either new physics beyond \lcdm, or the presence of some unidentified systematics associated with the \Planck\ data and/or the foreground model. \citet{planck2016-LI} find that although some cosmological
parameters differ by more than $2\,\sigma$ between $\ell < 800$ and $\ell > 800$, accounting for the multi-dimensional parameter space including correlations between
parameters, the shifts are at the 10\,\% level and hence not especially unusual. Nonetheless, parameter shifts, particularly in the fluctuation amplitude and Hubble parameter (which are directly relevant for the \lcdm-comparison with external data, as discussed in Sect.~\ref{sec:datasets}) are worth a brief re-examination using the additional information provided by the \Planck\
polarization spectra.

Constraints on cosmological parameters from power spectra at high multipoles require
a foreground model. Previous studies have shown that results are not very sensitive to
the specific assumptions that are made within the broad context of slowly varying foreground spectra expected on physical grounds \citep{Addison:2015wyg,planck2016-LI}. In this section, we use the \pliklite\ \planck\ likelihood, described in detail in \likeIII, which has the standard \plik\ foreground and nuisance parameters marginalized out without further assumptions on the cosmology.\footnote{We do not attempt to quantify likelihood modelling differences in this section, but a \camspec-based likelihood gives slightly less tension between high and low multipoles (especially with polarization), associated with the weaker preference for $\Alens>1$, as discussed in more detail in Sect.~\ref{sec:Alens}.}  For standard model extensions \pliklite\
accurately reproduces
results from the full \plik\ likelihood. It allows us to explore the high-$\ell$ likelihood accounting for foreground uncertainties, but with the foregrounds constrained in a sensible way from their spectra over the full multipole range. We consider the multipole ranges $\ell \le 801$ and $\ell \ge 802$ (corresponding to the boundary of one of the \pliklite\ bins), so that the low-multipole range is roughly comparable to WMAP and the two ranges have similar statistical power on most parameters. Results splitting at $\ell\approx1000$ are similar, but with larger errors in the high multipole range.

Figure~\ref{fig:lmaxlmin} shows a comparison of the high and low
multipole ranges, both for temperature (lower triangle, as previously
discussed by \citealt{Addison:2015wyg} and \citealt{planck2016-LI}), and new results
for the combined temperature-polarization likelihood (upper triangle).
Part of the difference between the low- and high-multipole ranges is caused
by the large-scale temperature dip discussed above; if we exclude
multipoles $\ell <30$ (unfilled grey
contours), the contours from $\ell \le 801$ shift towards
the area of consistency with the high multipoles. This could indicate that the low-multipole results have
been pulled unusually far from the truth by the large-scale power
spectrum dip; if so, the WMAP temperature results would also have been
pulled at a similar (but not identical) level. The region of overlap
of the high- and low-multipole parameter constraints is consistent with
constraints from the nearly-independent combination of $EE$
polarization and lensing with a conservative $\Omb h^2$ prior (green
contours). This is consistent with a statistical fluctuation pulling
 the low and high multipoles in opposite directions, so that their
intersection is closer to the truth if \lcdm\ is correct.

Figure~\ref{fig:lmaxlminparams} shows marginalized individual
parameter constraints, and also a comparison with the results from the
polarization likelihoods at high and low multipoles.  The $\ell\ge
802$ temperature results pull parameters to a region of higher matter
density and fluctuation amplitude (and to lower $\ns$ and $H_0$) than
the lower multipole range, and predict a CMB lensing amplitude
parameter $\sigma_8 \Omm^{0.25}=0.649\pm 0.018$. This is in tension
with the CMB lensing-reconstruction measurement of $\sigma_8
\Omm^{0.25}=0.589\pm 0.020$ at $2.2\,\sigma$
(as pointed out by \citealt{Addison:2015wyg} with 2015 data; also see the closely-related
discussion in the next subsection).  As shown in
Fig.~\ref{fig:lmaxlminparams}, combining the $\ell\ge 802$ CMB likelihood
with the lensing reconstruction, all parameter results move back
towards the same region of parameter space as combining with $\ell\le
801$, consistent with the high-$\ell$ temperature result having
fluctuated high along the main degeneracy direction.
As discussed in Sects.~\ref{sec:lensing} and \ref{subsec:amplitudes}, the combined CMB power spectrum results over the full multipole range are consistent with the lensing likelihood.

It is also interesting to compare to parameter constraints from the CMB power spectrum multipoles $\ell \le 801$ combined with the lensing and BAO, which gives
\threeonesig[2.5cm]{H_0 &= (67.85\pm 0.52)\,\Hunit,}{\sigma_8 &= 0.8058\pm 0.0063,}{\Omm &= 0.3081\pm 0.0065.}
 {TT,TE,EE [$\ell\le 801$]\dataplus{lowE}\dataplus\lensing\dataplus\BAO\label{lensingBAOlowl}. }
These results are entirely independent of the cosmological parameter fit to the $\ell \ge 801$ power spectra, but agree
well at the $1\,\sigma$ level with the full joint results in Table~\ref{table:default} (which have similar errors on these parameters). An equivalent result could be obtained using WMAP data after replacing their low-$\ell$ polarization with the \planck\ HFI measurement (i.e., lowE).

For the temperature likelihoods, the difference between the low- and
high-multipole constraints remains evident, with
$\Omm h^2$ differing at the $2.8\,\sigma$ level.
 Adding polarization, the
results from the multipole ranges are more consistent, as shown in
Fig.~\ref{fig:lmaxlminparams}, though the difference in $\Omm h^2$ is
still unusual at the roughly $2\,\sigma$ level.  However, the shifts in the different
parameters are all highly correlated, due to partial parameter
degeneracies, so the significance of any individual large shift is
lower after accounting for the number of parameters
\citep{planck2016-LI}.  The internal tensions between multipole ranges
appear to be consistent with moderate statistical fluctuations,
related to the low-$\ell$ dip at large scales and correlated with the
lensing amplitude on small scales. The large-scale feature is well
determined by both WMAP and \planck\ and very robustly measured.
The internal consistency of the \Planck\ power spectra between different frequencies and
detectors (\likeII, \likeIII) argues against systematics driving large
parameter shifts at high multipoles.
Equation~\eqref{lensingBAOlowl} also demonstrates that any effect from the high-multipole spectra alone cannot be pulling our baseline parameters by more than about $1\,\sigma$.
In the next
subsection we describe in more detail the apparent preference for a higher
lensing amplitude, and the features in the observed spectrum that
could be responsible for it.

\subsection{Lensing smoothing and $\Alens$}\label{sec:Alens}

\begin{figure}[t]
\centering
\includegraphics[]{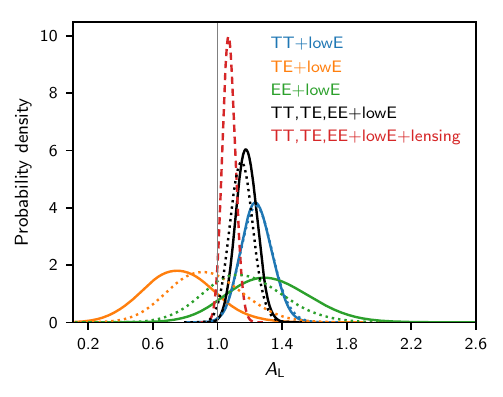}
\caption{
Constraints on the value of the consistency parameter $\Alens$, as a single-parameter extension to the base-\LCDM\ model, using various combinations of \planck\ data. When only power spectrum data are used, $\Alens>1$ is favoured at about $3\,\sigma$, but including the lensing reconstruction the result is consistent at $2\,\sigma$ with $\Alens=1$. The dotted lines show equivalent results for the \camspec\ likelihood, which peak slightly nearer to $\Alens=1$, indicating some sensitivity of the $\Alens$ results to choices made in constructing the high-multipole likelihoods.
\label{fig:Alens}
}
\end{figure}

\begin{figure}[t]
\centering
\includegraphics[width=\columnwidth]{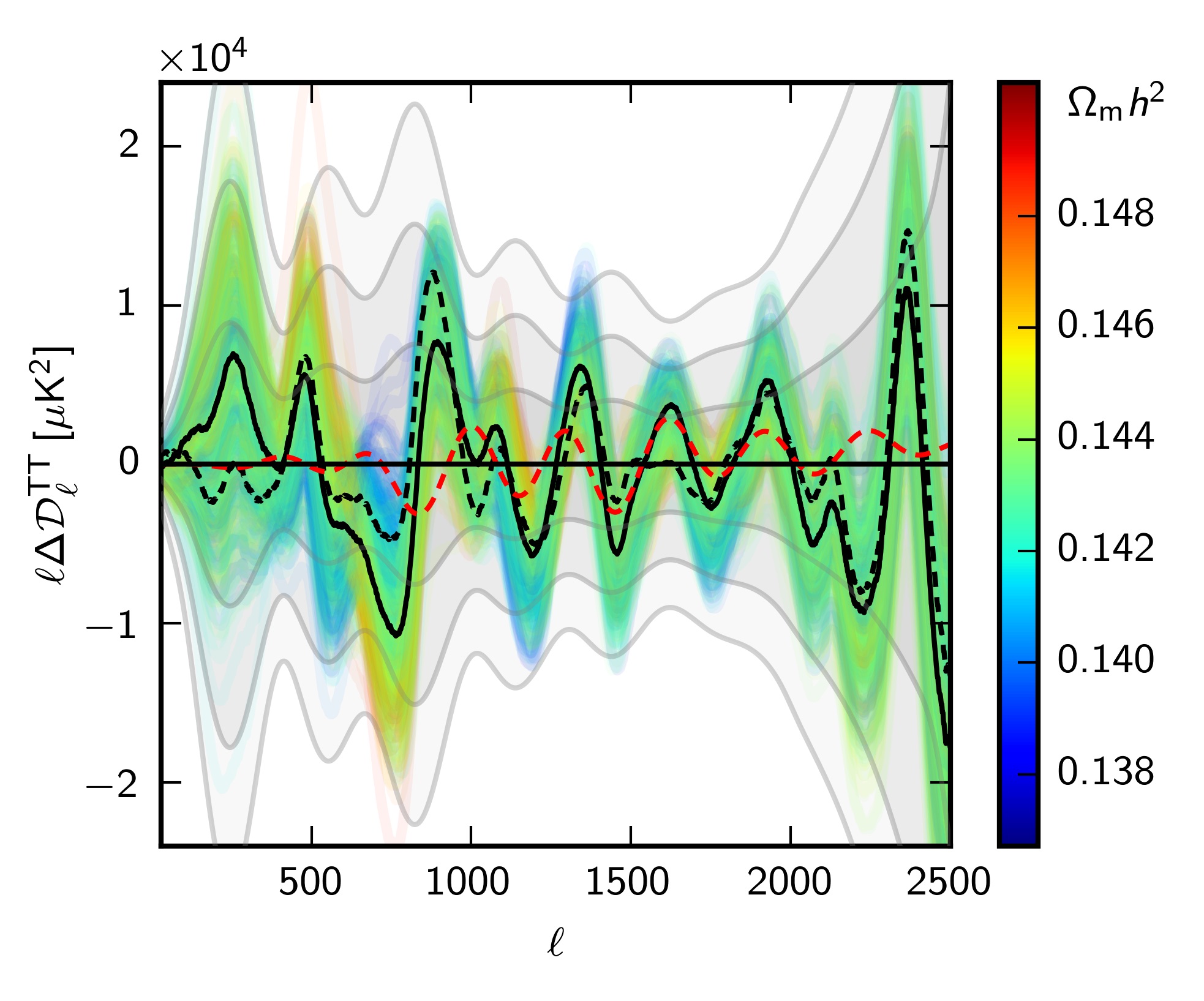}
\caption{
Base-\lcdm\ model ($\Alens=1$) $TT$ power spectrum residuals smoothed with a
Gaussian kernel of width $\sigma_\ell = 40$. The black line shows the smoothed
difference between the coadded data points and the theoretical model for the
\planckTT\ best-fit model, while coloured lines show the residuals for samples
over the allowed parameter space coloured by the value of $\Omm h^2$. Grey
bands show the 1, 2, and 3 $\sigma$ diagonal range expected for the smoothed
residuals in the best-fit model. The red dashed line shows 10\,\% of the
lensing-smoothing difference predicted in the best-fit model, displaying the
oscillatory signal expected if there were more lensing of the acoustic peaks.
The data residuals are not particularly anomalous, but the residuals have a
 similar pattern to the lensing smoothing difference over the
approximate range $\ell=1100$--2000,
giving a preference for around 10\,\% more lensing at fixed cosmological
parameters. Allowed models with lower $\Omm h^2$ (and hence higher $H_0$)
predict less lensing and give a larger oscillatory residual, preferring
relatively more lensing smoothing than models with high matter density.
The black dashed line shows the smoothed residual for the \planckTT\ best fit
to \lcdm+$\Alens$ (with $\Alens= 1.19$).
\label{fig:smoothresiduals}
}
\end{figure}

\begin{figure*}[t]
\centering
\includegraphics[width=0.95\textwidth]{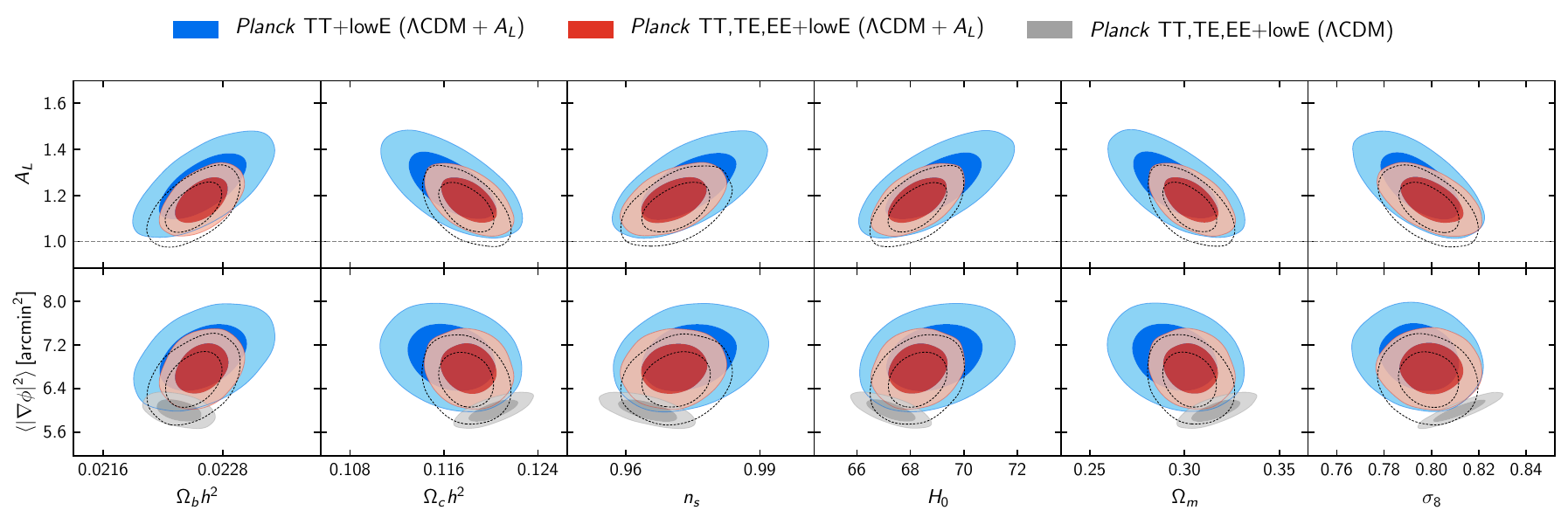}
\caption{
Marginalized $68\,\%$ and $95\,\%$ parameter constraint contours when adding $\Alens$ as a single-parameter extension to the base-\LCDM\ model, with (red) and without (blue) small-scale polarization,
 compared to the constraints in the base-\lcdm\ model (grey). The dashed contours show equivalent results for \planckall\ when using the \camspec\ likelihood, which gives results with $\Alens$ nearer unity and with slightly larger errors. The second row of subplots show, on the left axis, the predicted lensing deflection angle variance (from lensing multipoles $2\le L\le 2000$), which is a measure of the amount of actual lensing: the \shortall\ likelihood prefers about 10\,\% more actual lensing power (associated with lensing smoothing), but in the (unphysical) varying-$\Alens$ case this can be achieved using cosmological parameters that predict less lensing than in \lcdm\ but substantially larger $\Alens$, giving a preference for $\Alens\approx 1.2$.
\label{fig:Alens_degen}
}
\end{figure*}

In addition to the direct measurement of CMB lensing described in Sect.
\ref{subsec:CMBlensing} and \PlanckLensThree, lensing can be seen in the \planck\ CMB power spectra via the lensing-induced smoothing of the acoustic peaks
and transfer of power to the damping tail.
This effect is modelled in our main parameter analysis, and can be calculated accurately from the unlensed CMB power spectra and the CMB lensing potential power spectrum in each model \citep{Seljak:1995ve,Lewis:2006fu}. Interesting consistency checks include testing if the amplitude of the smoothing effect in the CMB power matches expectation and whether the amplitude of the smoothing is consistent with that measured by the lensing reconstruction. To do this,
 the theoretical prediction for the lensing spectrum in each model is often scaled by an ``$\Alens$'' consistency parameter, where the theoretical expectation is that $\Alens=1$ \citep{Calabrese:2008rt}.

As shown in Fig.~\ref{fig:lenspower}, the \planck\ lensing-reconstruction power spectrum is consistent with the amplitude expected for \lcdm\ models that fit the CMB spectra, so the \planck\ lensing measurement is compatible with $\Alens=1$.
However, the distributions of $\Alens$ inferred from the CMB power spectra alone are shown in Fig.~\ref{fig:Alens} for various different data combinations, and these indicate a preference for $\Alens>1$, with
\beglet
\begin{eqnarray}
\Alens &=& 1.243\pm 0.096 \quad\onesig{\planckTT},\\
\Alens &=& 1.180\pm 0.065 \quad\onesig{\planckall},
\end{eqnarray}
\endlet
assuming a \lcdm+$\Alens$ model.
The TE polarization data alone slightly prefer $\Alens<1$, with the EE data slightly preferring $\Alens>1$; however, both are consistent with $\Alens=1$ within $2\,\sigma$. The joint combined likelihood shifts the value preferred by the TT data downwards towards $\Alens=1$, but the error also shrinks, increasing the significance of $\Alens>1$ to $2.8\,\sigma$ ($99.8\,\%$ of parameter samples have $\Alens>0$, so the one-tailed limit is almost exactly $3\,\sigma$). Moreover, combining with the lensing likelihood further pulls the constraint towards $\Alens=1$, which is then consistent with the data to within $2\,\sigma$; we see that the preference for $\Alens>1$ is driven by the CMB power spectra alone.

The preference for high $\Alens$ is not just a volume effect in the full
parameter space (see \paramsI\ for discussion of such effects in multi-parameter fitting), with the best fit improved by $\dchisquare = -8.7$ when adding $\Alens$ for TT+\lowE\ and
$\dchisquare=-9.7$ for \TTTEEE+\lowE. The bulk of the $\dchisquare$ comes from the high-$\ell$ likelihood (mostly in the range $600< \ell<1500$); however, the low-$\ell$ temperature {\tt commander} likelihood fit is also improved if $\Alens$ is free, with $\dchisquare=-2.3$ and $\dchisquare=-1.3$ for the TT+\lowE\ and \TTTEEE+\lowE, respectively, due to the lower amplitude of the $\Alens$ fit on large scales. The change in fit to the low-$\ell$ polarization is not very significant ($\dchisquare=-0.2$ and $\dchisquare=-0.4$).

The determination of $\Alens$ from the high-$\ell$ polarization data and the \TTTEEE+\lowE\ joint combination depends on the calibration of the polarization channels, and is affected by different ways of modelling the polarization efficiencies, as discussed in Sect.~\ref{sec:likelihood}. The results from the \camspec\ likelihood (which uses spectrum-based rather than map-based calibrations for $TE$ and $EE$) are somewhat shifted with respect to the \plik\ likelihood, as shown by the dotted lines in Fig.~\ref{fig:Alens}, and have larger errors, giving
\beglet
\begin{eqnarray}
 \Alens \hspace{-1mm} &=& \hspace{-1mm} 1.246^{+0.092}_{-0.100} \quad\onesig{\shortTT~[\camspec]},\\
\Alens \hspace{-1mm} &=& \hspace{-1mm} 1.149 \pm 0.072 \ \onesig{\shortall~[\camspec]}.
\end{eqnarray}
\endlet
Using \camspec\ there is still a clear preference for $\Alens>1$, but the joint result with polarization is now only just over $2\,\sigma$ above $\Alens=1$.
\rep{The differences between these \plik\ and \camspec\ results arise from differences in the methodologies used to create the likelihoods. Although both likelihoods clearly show a
preference for $\Alens>1$, this cannot be claimed to be a robust detection at much over $2\,\sigma$ \citep[see also][]{Efstathiou:2019}.}

The preference for $\Alens>1$ within the \lcdm\ model is a curious feature of the \planck\ CMB power spectrum data, and has already been discussed extensively in \paramsI, \paramsII, and \citet{planck2016-LI}, although it is now slightly more significant. In temperature, over half of the small (approximately 0.02) upward shift in $\Alens$ compared to 2015 is explained by the lower optical depth from the 2018 low-$\ell$ likelihood: lower $\tau$ implies lower $\As$ to match the high-$\ell$ CMB fluctuation amplitude, and hence larger $\Alens$ to yield a lensing amplitude and hence amount of smoothing at the same level as 2015. In polarization about $40\,\%$ of the shift in $\Alens$ is explained by changes in $\tau$, with changes in the maps, modelling for beam leakage, and polarization efficiencies explaining the rest.

The high-$\ell$ temperature likelihood preference for more lensing smoothing than allowed by \lcdm\ can be seen by eye in the smoothed data residuals plotted in Fig.~\ref{fig:smoothresiduals}; over almost all the allowed \lcdm\ parameter space there is an oscillatory residual in the range $1100\la \ell \la 2000$ that matches the shape of the lensing smoothing\footnote{Although the oscillatory pattern looks most similar to lensing at high multipoles, an increase in the foreground model amplitude can decrease the oscillation amplitude in the theory contribution to the spectrum, and hence appear as an oscillatory difference. For example $\Delta \ns\approx -0.02$, combined with an implausibly large change in the foreground model, gives a difference in the predicted spectrum with an oscillatory component that has similar amplitude to $\Delta \Alens \approx 0.1$; see the related discussion in \citet{planck2016-LI}.
} (although in other multipole ranges it does not match at all). The residual is not obviously anomalous, with the TT \lcdm\ best fit improving by $\Delta\chi^2 \approx 4$ if a best-fit oscillatory residual (with $\Alens\approx 1.1$) is added to the best-fit \lcdm\ theory model.
The stronger preference for $\Alens>1$ when $\Alens$ varies arises because degeneracies between $\Alens$, cosmological parameters, and foregrounds improves the fit at both high and lower multipoles, as shown by the black dashed line in Fig.~\ref{fig:smoothresiduals}. In \lcdm\ the lensing amplitude can be increased by increasing $\Omm h^2$; however, the model then becomes a bad fit because of the poorer agreement at $\ell < 1000$).
Varying $\Alens$ allows a high $\Alens$ to remove the oscillatory residual at high multipoles that appears in \lcdm\ with lower $\Omm h^2$, giving best fits with lower $\Omm h^2$ and higher $H_0$ (by $1.5$--$2.0\,\sigma$, depending on the exact combination of data used) that are not favoured in the physical \lcdm\ model. Lower values of $\Omm h^2$ give higher values of $\ns$, lowering the theory prediction on large scales, so high $\Alens$ models are also slightly preferred by the dip in the $\ell<30$ \planck\ temperature data.
The parameter degeneracies are illustrated in Fig.~\ref{fig:Alens_degen}.

The $\Alens$ results appear to be robust to changes in foreground modelling in the baseline likelihood, with the \camspec\ 545 GHz cleaned likelihood (see Appendix~\ref{appendix:camspec}) giving very similar results. However, the dip in the residuals at $1420\la \ell\la 1480$, part of the oscillatory feature that looks like additional lensing, nearly coincides with an approximately $3\,\sigma$ discrepancy (for the best-fit foreground cosmology model) between the 143-GHz and 217-GHz power spectra at $1450\la\ell\la 1510$, with the 217-GHz spectrum pulling the coadded spectrum low compared to 143\,GHz by an amount comparable to the coadded residual (at $\ell = 1480$ the 217-GHz spectrum is $\mathcal{D}_\ell \approx 7 \muK^2$ lower than 143\,GHz with smoothing $\sigma_\ell=40$; see \likeIII).
This may be an indication that the preference for $\Alens$ at high multipoles is partly due to unknown systematics or foregrounds.
However, tightly cutting the $\ell$ range that contributes to the $3\,\sigma$ frequency difference does not in itself shift $\Alens$ to substantially lower values (though cutting all of $1420\la \ell\la 1480$ does), and the significance of the oscillatory feature in the \lcdm\ CMB residual is in any case not very high. If it is largely a statistical fluctuation, it would be expected to vary with changes in sky area; that is somewhat the case, with around $80\,\%$ sky area giving a substantially less oscillatory residual to the same best-fit \lcdm\ model at $\ell \la 1600$, but still favouring high $\Alens$. Different power spectrum analyses have also shown the preference for $\Alens$ \citep{Spergel:2013rxa,Couchot:2015eea}, though with varying significance, which could indicate that our roughly $3\,\sigma$ significance is partly an issue of analysis choices, e.g., the sky areas included and foreground priors chosen.

The dashed line in Fig.~\ref{fig:lenspower} shows the lensing power spectrum in the \planckall\ best-fit \lcdm+$\Alens$ model, which is clearly inconsistent with the lensing reconstruction, since it lies above almost all of the measured data points.
Because the amplitude of the lensing smoothing effect can be calculated from the lensing potential power spectrum alone, which we can also empirically measure, it is impossible to increase the lensing smoothing of the CMB peaks without also increasing the measured lensing reconstruction amplitude. This remains true if the lensing power spectrum is allowed to vary in shape \citep{Motloch:2018pjy}.
The actual lensing smoothing effect can also partly be removed by delensing, as shown by \cite{Larsen:2016wpa} and \cite{Carron:2017vfg}. In \PlanckLensThree\ we update these delensing analyses, and show (using the internal lensing reconstruction, a \planck\ CIB map as a tracer of the lensing potential, and a combined estimate) that the amount of peak sharpening observed after delensing is consistent with theoretical expectations (e.g., for the $TT$ spectrum, we measure a reduction in lensing smoothing of $0.411\pm 0.028$,
compared to the expected value of $0.375$ when using a combination of CIB and \planck\ lensing reconstruction).

Although the residuals shown in Fig.~\ref{fig:smoothresiduals} between the data and the \lcdm\ best fit temperature spectrum show what looks like an oscillatory lensing residual at high $\ell$, the fit itself is determined by the entire range of multipoles (and the low-$\ell$ polarization than constrains $\tau$). The preference for $\Alens>1$ could therefore be attributed to other scales when considering the CMB spectra alone. For example, after removing $\ell<30$ in both temperature and polarization, $\Alens$ from $TT$ is consistent with unity to within $1\,\sigma$. However, in this case the \lcdm\ lensing amplitudes are still large, giving a value of $\Omm \sigma_8^{0.25}$, in $2\,\sigma$ tension with the lensing reconstruction. This is another reflection of the tension noted in Sect.~\ref{sec:highlow} between the lensing reconstruction and the lensing amplitude predicted using temperature multipoles $\ell\ga 800$: the two tensions are therefore not independent and largely driven by the same features of the \lcdm\ fit to the temperature and low-$\ell$ polarization data.

If $\Alens>1$ is not just a statistical fluctuation, but comes from new physics changing the theoretical predictions, it could be something that mimics the smoothing effect in the CMB peaks.
The \emph{lensing} smoothing effect comes from averaging over the sky a spectrum that is locally varying (due to magnification and shear locally changing the scale and shape of the CMB peaks). Conceptually, the temperature lensing reconstruction works by looking for this spatial variation in scale and shear of the local power. Any non-lensing isotropic change in the amplitude of the small-scale peaks and troughs, either from new physics or random fluctuations, would therefore only have a small effect on the lensing reconstruction, which is sensitive to scale and shape, not amplitude.

One locally anisotropic physical effect that has been considered as a possible explanation is the presence of large-scale compensated isocurvature modes, discussed in detail in~\cite{planck2016-l10}.
Because the large-scale isocurvature modes locally vary the baryon-to-photon ratio, they can partially mimic the lensing smoothing effect by spatially varying the acoustic scale~\citep{Munoz:2015fdv,Valiviita:2017fbx}.
However, because they have a similar local effect to lensing, they also affect the large-scale lensing reconstruction~\citep{Smith:2017ndr}. Combining with the \planck\ 2018 lensing reconstruction, which now extends down to $L=8$, as shown in~\cite{planck2016-l10} this model therefore does not offer a significant improvement in overall fit ($\dchisquare = -3.3$ with \planckalllensing).

If the $\Alens>1$ preference is simply a statistical excursion (perhaps the most likely explanation), this indicates that there are random features in the spectrum that are pulling some parameters unusually far from
expected values.\footnote{It is not trivial to assess how unlikely a fluctuation in a consistency parameter like this is given the number of different cosmological and consistency test parameters we might have looked at. We are only discussing $\Alens$ in detail here because it comes out high; other consistency parameters, for example the relative amplitude of ISW, Doppler, and Sachs-Wolfe contributions to the temperature spectrum, come out perfectly consistent with expectations.}
There are several theoretical models that can fit the CMB power spectra
and also predict larger lensing amplitudes. These include \LCDM\ models
with spatial curvature, for which we find $\Omega_K<0$ at over $3\,\sigma$ (Sect.~\ref{sec:curv}) from the CMB power spectra, and some dark energy and modified gravity models (Sect.~\ref{sec:darkenergy}). For extensions to base-\LCDM,
parameters that decrease the lensing amplitude are more constrained by the \Planck\ power spectra
than might otherwise be expected; for example, higher neutrino masses lower the predicted lensing power compared to base \lcdm, leading to surprisingly tight constraints (Sect.~\ref{subsec:mnu}).
Adding the lensing-reconstruction information significantly reduces the parameter space of larger lensing amplitudes and partially mitigates these effects.
However, the statistical power of the \planck\ power spectra is sufficiently high that the joint constraints prefer lensing amplitudes
in the higher range allowed by the lensing data.

Even within \lcdm, the fact that the data prefer more lensing leads to a preference for higher fluctuation amplitudes,
hence the high-$\ell$ data yield higher $\As$ and higher $\tau$ than we infer in combination with large-scale $E$-mode polarization (Sect.~\ref{sec:lowl}) or lensing reconstruction.
Since these preferences are degenerate with $\Omm h^2$, $\ns$, and $H_0$ (see \paramsI, \paramsII), these parameters are also pulled ($\Omm h^2$ higher, $\ns$ and $H_0$ lower). Our baseline best-fit results include both the ``lowE'' data and the lensing reconstruction, each of which restrict the range of allowed variation, so the remaining pulls should be modest; however, it should not perhaps be too much of a surprise if the central values of the parameters inferred from \planck\ turn out to be slightly more shifted than typical with respect to the ultimate truth if the base-\lcdm\ model is correct.

\section{Extensions to the base-\boldlcdm\ model}\label{sec:maingrid}

\subsection{Grid of extended models}\label{sec:grid}
\begin{figure*}[h!]
\centering
\includegraphics[width=18cm]{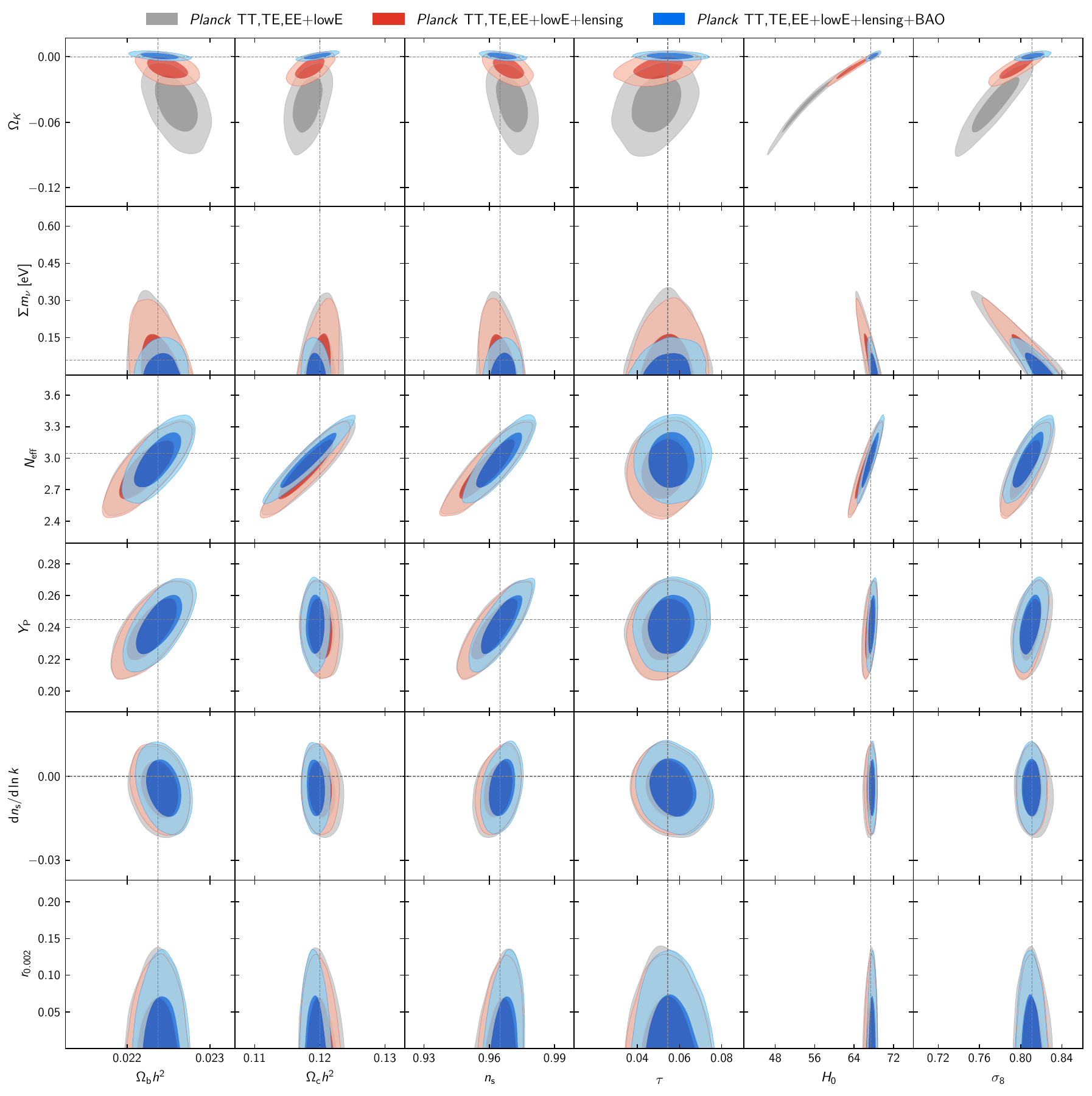}
\caption {Constraints on 1-parameter extensions to the base-\lcdm\ model.  Contours show 68\,\% and 95\,\% confidence regions
for \planckall\ (grey), \planckall+lensing (red),
and \planckall+lensing+BAO (blue).
Horizontal dashed lines correspond to the parameter values assumed in the
base-\LCDM\ cosmology,
while vertical dashed lines show the mean posterior values in the base model
for \planckall+lensing.}
\label{fig:grid_1paramext}
\end{figure*}

We have studied a range of extension to the base \lcdm\ model. A
full grid of results from standard parameter extensions is available online through the \PLA.\footnote{Chains are available at \url{https://pla.esac.esa.int}, with description and parameter tables in \citet{planck2016-ES}.}
Figure~\ref{fig:grid_1paramext} and Table~\ref{tab:grid_1paramext}
summarize the constraints on 1-parameter extensions to the base-$\Lambda$CDM
model.  As in 2013 and 2015 we find no strong evidence in favour of any of
these extensions, using either the \Planck\ data alone or \Planck\ combined
with BAO. We also find that constraints on the base-\lcdm\ parameters are remarkably robust to a variety of possible extensions to the \lcdm\ model, as shown in Table~\ref{tab:base_extensions}: many of these parameters are constrained to high precision
in a nearly model-independent way.

We now discuss some specific extensions in more detail.

\begin{table*}
\begin{center}
\caption{Constraints on 1-parameter extensions to the base-\lcdm\ model for
combinations of \planck\ power spectra, \planck\ lensing, and BAO (equivalent results using the \camspec\ likelihood are given in Table~\ref{tab:grid_1paramext_camspec}). Note that we quote 95\,\% limits here.
\label{tab:grid_1paramext}
}
\begingroup
\newdimen\tblskip \tblskip=5pt
\nointerlineskip
\vskip -3mm
\footnotesize
\setbox\tablebox=\vbox{
    \newdimen\digitwidth
    \setbox0=\hbox{\rm 0}
    \digitwidth=\wd0
    \catcode`"=\active
    \def"{\kern\digitwidth}
    \newdimen\signwidth
    \setbox0=\hbox{+}
    \signwidth=\wd0
    \catcode`!=\active
    \def!{\kern\signwidth}
\halign{
\hbox to 1.2in{$#$\leaderfil}\tabskip=1.5em&
\hfil$#$\hfil\tabskip=1.5em&
\hfil$#$\hfil\tabskip=1.5em&
\hfil$#$\hfil\tabskip=1.5em&
\hfil$#$\hfil\tabskip=0pt\cr
\noalign{\doubleline}
\omit\hfil\text{Parameter}
 \hfil& \hfil \quad \TT{+}\lowE\quad\hfil& \hfil \TTTEEE{+}\lowE \hfil& \hfil \TTTEEE{+}\lowE{+}\lensing \hfil& \hfil \TTTEEE{+}\lowE{+}\lensing{+}\BAO \cr
\noalign{\vskip 3pt\hrule\vskip 5pt}

\Omega_K & -0.056^{+0.044}_{-0.050} & -0.044^{+0.033}_{-0.034} & -0.011^{+0.013}_{-0.012} & 0.0007^{+0.0037}_{-0.0037}\cr
\Sigma m_\nu\,[\mathrm{eV}] & < 0.537 & < 0.257 & < 0.241 & < 0.120\cr
N_{\mathrm{eff}} & 3.00^{+0.57}_{-0.53} & 2.92^{+0.36}_{-0.37} & 2.89^{+0.36}_{-0.38} & 2.99^{+0.34}_{-0.33}\cr
Y_{\mathrm{P}} & 0.246^{+0.039}_{-0.041} & 0.240^{+0.024}_{-0.025} & 0.239^{+0.024}_{-0.025} & 0.242^{+0.023}_{-0.024}\cr
\mathrm{d}n_{\mathrm{s}}/\mathrm{d}\ln k & -0.004^{+0.015}_{-0.015} & -0.006^{+0.013}_{-0.013} & -0.005^{+0.013}_{-0.013} & -0.004^{+0.013}_{-0.013}\cr
r_{0.002} & < 0.102 & < 0.107 & < 0.101 & < 0.106\cr
w_0 & -1.56^{+0.60}_{-0.48} & -1.58^{+0.52}_{-0.41} & -1.57^{+0.50}_{-0.40} & -1.04^{+0.10}_{-0.10}\cr \noalign{\vskip 5pt\hrule\vskip 3pt}
} 
} 
\endPlancktable
\endgroup
\end{center}
\end{table*}

\begin{table*}
\begin{center}
\caption{Constraints on standard cosmological parameters from \planckalllensing\ when the base-\lcdm\ model is extended by varying additional parameters.
The constraint on $\tau$ is also stable but not shown for brevity; however,
we include $H_0$ (in $\Hunit$) as a derived parameter (which is very poorly constrained from \planck\ alone in the \lcdm+$\wzero$ extension).
Here $\alpha_{-1}$ is a matter isocurvature amplitude parameter, following \paramsII.  All limits are 68\,\% in this table.
The results assume standard BBN except when varying $\yhe$ independently (which requires non-standard BBN). Varying $\Alens$ is not a physical model (see \Alenssec).
\label{tab:base_extensions}
}
\begingroup
\newdimen\tblskip \tblskip=5pt
\nointerlineskip
\vskip -4mm
\setbox\tablebox=\vbox{
    \newdimen\digitwidth
    \setbox0=\hbox{\rm 0}
    \digitwidth=\wd0
    \catcode`"=\active
    \def"{\kern\digitwidth}
    \newdimen\signwidth
    \setbox0=\hbox{+}
    \signwidth=\wd0
    \catcode`!=\active
    \def!{\kern\signwidth}
\halign{
\hbox to 1.2in{$#$\leaderfil}\tabskip=1.5em&
\hfil$#$\hfil\tabskip=1.5em&
\hfil$#$\hfil\tabskip=1.5em&
\hfil$#$\hfil\tabskip=1.5em&
\hfil$#$\hfil\tabskip=1.5em&
\hfil$#$\hfil\tabskip=1.0em&
\hfil$#$\hfil\tabskip=0pt\cr
\noalign{\doubleline}
\omit\hfil Parameter(s)\hfil
& \Omega_{\mathrm{b}} h^2& \Omega_{\mathrm{c}} h^2& 100\theta_{\mathrm{MC}}& H_0& n_\mathrm{s}& \ln(10^{10} A_\mathrm{s})\cr
\noalign{\vskip 5pt\hrule\vskip 5pt}
\mathrm{Base}\ \Lambda\mathrm{CDM}& 0.02237\pm 0.00015& 0.1200\pm 0.0012& 1.04092\pm 0.00031& 67.36\pm 0.54& 0.9649\pm 0.0042& 3.044\pm 0.014\cr
r& 0.02237\pm 0.00014& 0.1199\pm 0.0012& 1.04092\pm 0.00031& 67.40\pm 0.54& 0.9659\pm 0.0041& 3.044\pm 0.014\cr
\mathrm{d}n_{\mathrm{s}}/\mathrm{d}\ln k& 0.02240\pm 0.00015& 0.1200\pm 0.0012& 1.04092\pm 0.00031& 67.36\pm 0.53& 0.9641\pm 0.0044& 3.047\pm 0.015\cr
\mathrm{d}n_{\mathrm{s}}/\mathrm{d}\ln k, r& 0.02243\pm 0.00015& 0.1199\pm 0.0012& 1.04093\pm 0.00030& 67.44\pm 0.54& 0.9647\pm 0.0044& 3.049\pm 0.015\cr
\mathrm{d}^2n_{\mathrm{s}}/\mathrm{d}\ln k^2, \mathrm{d}n_{\mathrm{s}}/\mathrm{d}\ln k& 0.02237\pm 0.00016& 0.1202\pm 0.0012& 1.04090\pm 0.00030& 67.28\pm 0.56& 0.9625\pm 0.0048& 3.049\pm 0.015\cr
N_{\mathrm{eff}}& 0.02224\pm 0.00022& 0.1179\pm 0.0028& 1.04116\pm 0.00043& 66.3\pm 1.4& 0.9589\pm 0.0084& 3.036\pm 0.017\cr
N_{\mathrm{eff}}, \mathrm{d}n_{\mathrm{s}}/\mathrm{d}\ln k& 0.02216\pm 0.00022& 0.1157\pm 0.0032& 1.04144\pm 0.00048& 65.2\pm 1.6& 0.950\pm 0.011& 3.034\pm 0.017\cr
\Sigma m_\nu& 0.02236\pm 0.00015& 0.1201\pm 0.0013& 1.04088\pm 0.00032& 67.1^{+1.2}_{-0.67}& 0.9647\pm 0.0043& 3.046\pm 0.015\cr
\rep{\Sigma m_\nu, N_{\mathrm{eff}}} & 0.02221\pm 0.00022 & 0.1179^{+0.0027}_{-0.0030} & 1.04116\pm 0.00044 & 65.9^{+1.8}_{-1.6} & 0.9582\pm 0.0086 & 3.037\pm 0.017\cr
m_{\nu,\,\mathrm{sterile}}^{\mathrm{eff}}, N_{\mathrm{eff}}& 0.02242^{+0.00014}_{-0.00016}& 0.1200^{+0.0032}_{-0.0020}& 1.04074^{+0.00033}_{-0.00029}& 67.11^{+0.63}_{-0.79}& 0.9652^{+0.0045}_{-0.0056}& 3.050^{+0.014}_{-0.016}\cr
\alpha_{-1}& 0.02238\pm 0.00015& 0.1201\pm 0.0015& 1.04087\pm 0.00043& 67.30\pm 0.67& 0.9645\pm 0.0061& 3.045\pm 0.014\cr
w_0& 0.02243\pm 0.00015& 0.1193\pm 0.0012& 1.04099\pm 0.00031& \dots& 0.9666\pm 0.0041& 3.038\pm 0.014\cr
\Omega_K& 0.02249\pm 0.00016& 0.1185\pm 0.0015& 1.04107\pm 0.00032& 63.6^{+2.1}_{-2.3}& 0.9688\pm 0.0047& 3.030^{+0.017}_{-0.015}\cr
Y_{\mathrm{P}}& 0.02230\pm 0.00020& 0.1201\pm 0.0012& 1.04067\pm 0.00055& 67.19\pm 0.63& 0.9621\pm 0.0070& 3.042\pm 0.016\cr
Y_{\mathrm{P}}, N_{\mathrm{eff}}& 0.02224\pm 0.00022& 0.1171^{+0.0042}_{-0.0049}& 1.0415\pm 0.0012& 66.0^{+1.7}_{-1.9}& 0.9589\pm 0.0085& 3.036\pm 0.018\cr
A_{\mathrm{L}}& 0.02251\pm 0.00017& 0.1182\pm 0.0015& 1.04110\pm 0.00032& 68.16\pm 0.70& 0.9696\pm 0.0048& 3.029^{+0.018}_{-0.016}\cr
\noalign{\vskip 5pt\hrule\vskip 3pt}
} 
} 
\endPlancktable
\endgroup
\end{center}
\end{table*}

\subsection{Early Universe}\label{sec:early}
CMB observations probe the state of the universe at the earliest
time that is directly observable with the electromagnetic spectrum.
The physics of the anisotropies is well understood, and can be predicted accurately with linear theory given a set of initial conditions. \planck\ observations can therefore be used to give powerful constraints on the initial conditions, i.e., the perturbations present at the start of the hot big bang. We discuss in turn constraints on the scalar and tensor perturbations, allowing for deviations from a purely power-law scalar spectrum, and discuss the interpretation within the context of the most popular inflationary models.
\subsubsection{Primordial scalar power spectrum}

\begin{figure}
\centering
\includegraphics[]{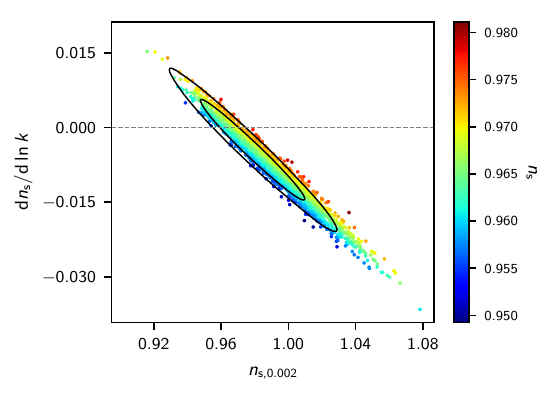}
\caption{
Constraints on the running of the scalar spectral index
in the \lcdm\ model, using \planckall+lensing when marginalizing over $r$ (samples,
coloured by the spectral index at $k = 0.05\Mpc^{-1}$), and the equivalent result
when $r=0$ (black contours). The \planck\ data are consistent
with zero running, but also allow for significant negative
running, which gives a positive tilt $n_{{\rm s},0.002}$, and hence less power,
 on large scales ($k \approx 0.002\Mpc^{-1}$).
\label{fig:nrun_ns}
}
\end{figure}

The \planck\ data are consistent with purely adiabatic primordial scalar curvature perturbations, with no evidence for isocurvature modes \citep[see][]{planck2016-l10}, as predicted by the simplest single-field inflation models. The primordial power spectrum is then just a function of scale. In this
section, we characterize the scalar fluctuation spectrum in terms of a spectral
index $\ns$ and its first two derivatives with respect to $\ln k$ (the ``running'' and ``running of the running'' of the spectral index):
\beglet
\begin{eqnarray}
  \clp_\clr(k) &=& \As \left(\frac{k}{k_0}\right)^{n(k)}, \\
  n(k) & = & \ns-1+(1/2)(\nrun) \ln(k/k_0) \nonumber \\
       &   &  \hspace{10mm} + (1/6)(\nrunrun) (\ln(k/k_0))^2. \label{PS1}
\end{eqnarray}
\endlet

In the absence of any running of the spectral index, our constraint on $\ns$
for the base-\LCDM\ model (Eq.~\ref{equ:n_s}) shows an $8\,\sigma$ tilt away from scale invariance. Adding BAO tightens the constraint to nearly $9\,\sigma$:
\oneonesig{\ns = 0.9665\pm 0.0038}{\shortall\dataplus\lensing\dataplus\BAO}{.}
The need for a red-tilted scalar spectrum is quite robust to
extensions to base \lcdm, as summarized in Table~\ref{tab:base_extensions}.
In all cases, we find $\ns <1$ at $\geq3\,\sigma$.

Adding running of the spectral index, $\nrun$, as a single additional parameter
to base \lcdm, we find
\beglet
\threeonesig[3.0cm]{\nrun &= -0.0045\pm 0.0067,}{\ns &= 0.9641\pm 0.0044,}{\nszerotwo &=0.979\pm 0.021,}{\shortall\dataplus\lensing,}
\threeonesig[3.0cm]{\nrun &= -0.0041\pm 0.0067,}{\ns &= 0.9659\pm 0.0040,}{\nszerotwo&=0.979\pm
0.021,}{\shortall\dataplus\lensing\dataplus\BAO,}
\endlet
where $\ns$ is defined by default at $k=0.05\,\Mpc^{-1}$ and $\nszerotwo$ is the corresponding tilt at
$k=0.002\,\Mpc^{-1}$.
The slight preference for negative running is driven by \rep{the mild tension between the
CMB temperature power spectrum at high and low multipoles discussed in Sect. \ref{sec:highlow}}, with negative running allowing higher large-scale tilt, giving less power on large scales (see Fig.~\ref{fig:nrun_ns} and the extensive discussions in \paramsI\ and \paramsII).
The measurements of the tilt and running around the pivot scale of $k \simeq 0.05\,\Mpc^{-1}$ are robust to allowing even more freedom for the spectrum to vary with
scale. For example, allowing for running of the running we find
\threeonesig[3.0cm]{\nrunrun &= 0.009\pm 0.012,}{\nrun &= 0.0011\pm 0.0099,}{\ns &= 0.9647\pm 0.0043,}{\shortall\dataplus\lensing\dataplus\BAO.}
Here the slight preference for negative running has almost disappeared, and there is instead a slight preference for lower large-scale power by having positive running of the running, leaving a near power-law solution on small scales. There is no evidence for any significant deviation from a power law on small scales. This is consistent with the simplest slow-roll inflation models where the running (and higher derivatives of the spectral index) are higher order in slow-roll (so that $\nrun = \clo(|\ns-1|^2)$, $\nrunrun = \clo(|\ns-1|^3)$)
and all deviations from a constant spectral index can be neglected at \planck\ sensitivity.

An analysis of more general parameterizations of the primordial power spectrum are presented in section~6 of \citet{planck2016-l10}, including various specific physically motivated models, as well as general parametric reconstructions.
Models with many more free parameters can provide better fits to the data, but none are favoured; in all cases the small-scale spectrum is found to be consistent with a power law over the range $0.008\,\Mpc^{-1}\la k\la 0.1\,\Mpc^{-1}$, with low-significance hints of larger-scale features corresponding to the dip in the low-$\ell$ temperature power spectrum.
The introduction of the additional degrees of freedom in the initial power spectrum had no significant impact on the determination of the main cosmological parameters for the parameterizations considered.

\subsubsection{Tensor modes}

\begin{figure}[t]
\centering
\includegraphics[width=88mm]{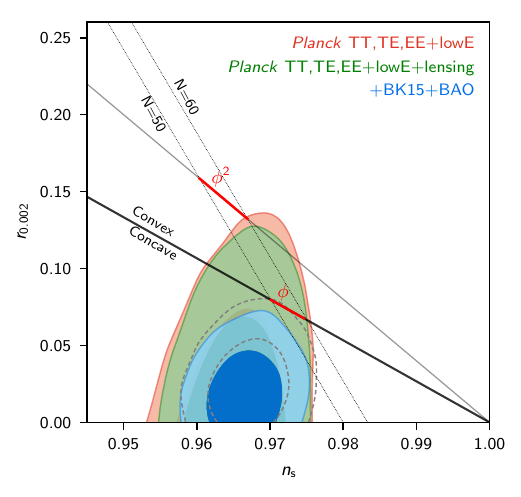}
\caption {
\rep{
Constraints on the tensor-to-scalar ratio $\rzerotwo$ in the
\lcdm\ model, using \planckall\ and \planckalllensing\,
(red and green, respectively), and joint constraints with BAO and BICEP2/Keck \citep[blue, including \planck\ polarization to determine the foreground components,][]{Ade:2018gkx}.
This assumes the inflationary consistency relation and negligible running.
Dashed grey contours show the joint constraint when using \camspec\ instead of \plik\ as the high-$\ell$ \planck\ likelihood, indicating the level of modelling uncertainty in the polarization results.
Dotted lines show the loci of approximately constant $e$-folding number $N$,
assuming simple $V\propto (\phi/\mpl)^p$ single-field inflation. Solid lines
show the approximate $\ns$--$r$ relation for locally quadratic and linear
potentials to first order in slow roll;
red lines show the approximate allowed range assuming $50<N<60$ and a power-law
potential for the duration of inflation.
The solid black line (corresponding to a linear potential) separates concave
and convex potentials.
\label{fig:ns_r_inflation}
}}
\end{figure}

Primordial gravitational waves\footnote{\rep{The polarization anisotropies generated by gravitational waves was discussed
first by \cite{Polnarev:1985}.}}, or tensor modes, source a distinctive curl-like (``$B$-mode'') pattern in the CMB polarization and add additional power to the large-scale temperature power spectrum \citep{Kamionkowski:1996ks,Seljak:1997gy}.
\planck's $B$-mode measurement is noise and systematics limited and provides a relative weak constraint
on the tensor-to-scalar ratio $r_{0.002}\,{<}\,0.41$ (95\,\% CL, \citealt{planck2016-l05}).
As with the 2013 and 2015 releases, the strongest constraint on tensor
modes from the \Planck\ data alone comes from the $TT$ spectrum at $\ell\la 100$.

The precision of the \Planck\ temperature constraint remains limited by cosmic
variance from the scalar component and is model dependent. The tightest and least model-dependent
constraints on the tensor amplitude come from the \citet[][BK15]{Ade:2018gkx} analysis of the
BICEP2/Keck field, in combination with \planck\ and WMAP maps to remove polarized
Galactic dust emission.
The \BK\ observations measure the $B$-mode polarization power spectrum in nine bins at $\ell\la 300$, with the tensor amplitude information coming mainly from scales $\ell\simeq 100$, where the $B$-mode spectrum from scattering at recombination is expected to peak. The \planck\ CMB power spectrum measurements use a much larger sky area, and are useful to convert this measurement into a constraint on the tensor-to-scalar ratio $r$ at a given scale with little additional cosmic variance error. To relate the tensor measurement to constraints on specific inflation models (which usually predict a region in the $\ns$--$r$ plane), combining with the \planck\ data is also essential, although model dependent.

Figure~\ref{fig:ns_r_inflation} shows the constraints in the $\ns$--$r$ plane, with $r$ added as
a single additional parameter to the base model and plotted at pivot scale $0.002\,\Mpc^{-1}$.
We assume the tensor-mode spectrum is close to scale invariant, with spectral index given by the inflation consistency relation to second order in slow-roll parameters.
\planck\ alone gives
\begin{equation}
  r_{0.002} < 0.10, \quad \twosig{\shortall+lensing},
\end{equation}
with $\ns = 0.9659\pm 0.0041$ at $1\,\sigma$.
Adding \BK\ to directly measure the tensor amplitude significantly tightens the $r$ constraint, and
adding BAO data tightens (slightly) the $\ns$ constraint.
Using the \planck\ temperature likelihoods we find
\onetwosig[6cm]{r_{0.002} < 0.055}{\shortTT\dataplus\lensing\dataplus\BK\dataplus\BAO}{,}
with $\ns = 0.9661\pm 0.0040$ at $1\,\sigma$, or adding polarization
\onetwosig{r_{0.002} < 0.058}{\shortall\dataplus\lensing\dataplus\BK\dataplus\BAO}{,\label{equ:tensorconstraint2}}
with $\ns = 0.9668\pm 0.0037$ at $1\,\sigma$. However, the small change when adding polarization is not stable to the choice of polarization likelihood; when using the \camspec\ \shortall\ likelihood in place of \plik, we find the weaker constraint $r_{0.002} < 0.065$ for the same data combination as that used in Eq.~\eqref{equ:tensorconstraint2}.

All the \rep{combined} $\ns$--$r$ contours exclude convex potentials at about the $95\,\%$ confidence (marginally less if we use the \camspec\ likelihood, see Fig. \ref{fig:ns_r_inflation}), which substantially restricts the range of allowed inflation models and disfavours all simple integer power law potentials. More generally, since $r$
depends on the slope of the potential, the smallness of the empirical
upper limit on $r$ implies that the inflationary potential must have been nearly flat when modes exited the horizon. The measured $\ns$ must then be determined largely by the second derivative of the potential, suggesting a hierarchy in the magnitudes of the slow-roll parameters, favouring hilltop-like potentials. For a detailed discussion of the implications for specific inflation models see \cite{planck2016-l10}.

If we allow running of the spectral index in addition to tensor modes, the
constraint on $r_{0.002}$ weakens if we use only the \Planck\ likelihood;
a negative running allows $\ns$ at large scales to shift to higher values,
lowering the large-scale scalar amplitude, and hence allowing a larger tensor
contribution.  Inclusion of the \BK\ likelihood significantly reduces the
extent of this degeneracy by constraining the tensor amplitude more directly, giving
\beglet
\twotwosig{r_{0.002} &< 0.16,}{\nrun &= -0.008^{+0.014}_{-0.015},}{\shortall\\+lensing,}
\twotwosig[3.5cm]{r_{0.002} &< 0.066,}{\nrun &=-0.006\pm 0.013,}{\shortall\\+lensing+\BK+BAO.}
\endlet

The combination of \planck\ and \BK\ robustly constrain the tensor ratio to be small, with $r_{0.002}\la 0.06$.
The implications for inflation are slightly more model dependent as a result of
degeneracies between $\ns$ and additional parameters in extended \lcdm\
models. However, as shown in Table~\ref{tab:base_extensions},
the extensions of \lcdm\ that we consider in this paper cannot substantially shift the value of the spectral index when the tensor amplitude is small, so the overall conclusions are unlikely to change substantially in extended models.

\subsection{Spatial curvature}
\label{sec:curv}

\begin{figure}[t]
\centering
\includegraphics[width=88mm]{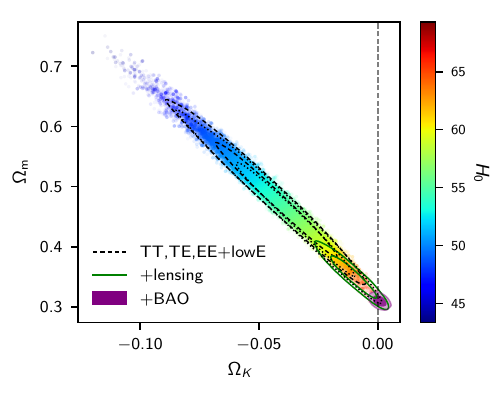}
\caption {
Constraints on a non-flat universe as a minimal extension to the base-\lcdm\ model.
Points show samples from the \planckall\ chains coloured by the value of the Hubble parameter and with transparency proportional to the sample weight.
Dashed lines show the corresponding $68\,\%$ and $95\,\%$ confidence contours that close away from the flat model (vertical line), while
dotted lines are the equivalent contours from the alternative \camspec\ likelihood.
The solid dashed line shows the constraint from adding \planck\ lensing, which pulls the result back towards consistency with flat (within $2\,\sigma$). The filled contour shows the result of also adding BAO data, which makes the full joint constraint very consistent with a flat universe.
\label{fig:omegak_omegam}
}
\end{figure}

The base-$\Lambda$CDM model assumes that the spatial hypersurfaces are flat,
such as would be predicted (to within measurable precision) by the simplest inflationary models.
This is a prediction that can be tested to high accuracy by the combination
of CMB and BAO data (the CMB alone suffers from a geometric degeneracy, which is
weakly broken with the addition of CMB lensing). This is illustrated in Fig.~\ref{fig:omegak_omegam}.

The combination of the \Planck\ temperature and polarization power spectra
give
\beglet
\oneonesig{  \Omega_K = -0.056^{+0.028}_{-0.018}}{\planckTT}{,}
\oneonesig{  \Omega_K = -0.044^{+0.018}_{-0.015}}{\planckall}{,}
\endlet
an apparent detection of curvature at well over $2\,\sigma$.
The $99\,\%$ probability region for the \shortall\ result is $-0.095<\Omega_K < -0.007$, with only about 1/10000 samples at $\Omega_K\ge 0$. This is not entirely a volume effect, since the best-fit $\chi^2$ changes by $\Delta \chi^2_{\rm eff} = -11$ compared to base \lcdm\ when adding the one additional curvature parameter. The reasons for the pull towards negative values of $\Omega_K$ are
 discussed at length in \paramsII\ and \Alenssec. They are essentially the
same as those that lead to the preference for $\Alens>1$,
although slightly exacerbated in the case of curvature, since the low multipoles also fit the low-$\ell$ temperature likelihood slightly better if $\Omega_K < 0$.
As with the $\Alens>1$ preference, the joint \planck\ polarization result is not robust at the approximately $0.5\,\sigma$ level to modelling of the polarization likelihoods, with the \camspec\ \shortall\ likelihood giving $\Omega_K = -0.037^{+0.019}_{-0.014}$.

Closed models predict substantially higher lensing amplitudes than in \lcdm,
so combining with the lensing reconstruction (which is
 consistent with a flat model) pulls
parameters back into consistency with a spatially flat universe to well within $2\,\sigma$:
\beglet
\oneonesig[4cm]{\Omega_K = -0.0106\pm 0.0065}{\shortall\dataplus\lensing}{.}
The constraint can be further sharpened by combining the \Planck\ data with
BAO data; this convincingly breaks the geometric degeneracy to give
\oneonesig[4cm]{\Omega_K = 0.0007\pm 0.0019}{\shortall\dataplus\lensing\dataplus\BAO}{.}
\endlet
The joint results suggests our Universe is spatially flat to a $1\,\sigma$ accuracy of
$0.2\,\%$.

\subsection{Dark energy and modified gravity}\label{sec:darkenergy}
\label{sec:results}

The late-time accelerated expansion of the Universe \citep{Riess:1998,
Perlmutter:1999} is still considered one of the most mysterious
aspects of the standard cosmology.  In the base \lcdm\ model the
acceleration is driven by a cosmological constant, added into the
Einstein equations of General Relativity \citep[GR,][]{Einstein_1917}.
Although \lcdm\ fits the data well, $\Lambda$ is a phenomenological
parameter without an underlying theoretical basis to explain its
value \citep[though see][]{Weinberg:1987}.  In addition, the
empirically required value of $\Lambda$ marks our epoch as a special
time in the evolution of the Universe.  Attempts have therefore been
made to find a dynamical mechanism that leads to cosmic acceleration,
with evolving background energy densities close to \lcdm. Such dynamics is
usually associated with a fluid (a scalar field) which we refer to as
``dark energy'' (DE), or with modifications of GR, which we refer to
as ``modified gravity'' (MG).

A detailed analysis of the impact of \Planck\ data on dark energy and
modified gravity was presented in a dedicated paper that accompanied
the 2015 \Planck\ release, \cite[hereafter \PDEII]{planck2014-a16}.
We refer the reader to this paper for a review of different
cosmological models, and for constraints
from \Planck\ on its own and in combination with galaxy weak lensing
(WL) and redshift-space distortions (RSDs).  In \PDEII\ it was shown
that although the base-\lcdm\ model fits \Planck\ data, there were some
tensions (at levels as high as $3\,\sigma$) when \Planck\ was
combined with RSD and WL data, even when conservative cuts were applied to
exclude nonlinear scales. However, the addition
of \Planck\ lensing data was found to reduce these tensions. Updated
constraints on a few specific models, using more recent WL data, are
presented in \cite{Abbott:2017wau}.

In this paper, we follow a similar methodology to \PDEII,
distinguishing between models that directly affect only the background
(and impact perturbations predominantly through changes in the
expansion rate) and those that directly affect perturbations.
However, we restrict the analysis to a smaller range of models here. As in
the rest of this paper, we show results for the
baseline \planckalllensing\ data set and for combinations with other
relevant data sets. Such external data are particularly useful for
constraining DE and MG models because the largest deviations
from \lcdm\ are usually at late times, which are not well constrained
by the CMB power-spectra and CMB lensing.  However, CMB lensing
provides important information that mitigates the preference for
$\Alens\,{>}\,1$ seen in the \Planck\ temperature power spectra
(Sect.~\ref{sec:Alens}), so we explicitly comment on the impact of CMB lensing
wherever relevant.  We recall here that the lensing likelihood assumes
a fiducial \lcdm\ model, but linear corrections to the fiducial mode
are accounted for self-consistently. \PlanckLensThree\ explicitly tested
that this procedure is unbiased, even when the lensing spectrum differs
from the fiducial spectrum by as much as 20\,\% (which is much larger
than differences allowed by the CMB lensing data).

We consider the following external data sets:
\begin{itemize}
\item SNe + BAO (see Sects.~\ref{sec:BAO}, \ref{sec:SN}, and \ref{sec:hubble}
 for discussions of the data sets and
comments on why we do not combine \Planck\ data with direct measurements of
$H_0$);
\item RSDs (as described in Sect.~\ref{sec:RSD}), where we specifically use
BOSS-DR12 data from \cite{Alam:2016hwk}, adopting the $f \sigma_8$--$H$--$\DM$
parameterization;
\item WL data from DES (as described in Sect.~\ref{sec:WL}),
except that here we use the Weyl potential to obtain theoretical predictions for the lensing correlation functions, rather than assuming the matter-sourced Poisson equation to relate the
lensing potential power spectrum to the matter power spectrum.
\end{itemize}

We calculate all results both fixing and varying the neutrino mass.
Neutrino masses are known
to be degenerate with DE and MG and should be varied consistently
when testing such models \citep[as discussed in][]{Dirian:2017pwp}; fixing the
neutrino mass to the minimal value of $0.06\,{\rm eV}$ (as for our baseline \LCDM\ results)
gives tighter constraints than allowing the neutrino mass
to vary and partly shifts
results towards $\Lambda$CDM. These shifts are usually small, often
negligible, and always less than $1\,\sigma$ for marginalized
results.
\rep{We model the small-scale nonlinear power spectrum using {\tt HMcode} \citep{Mead:2015yca,Mead:2016zqy} as in the main parameter grid of extensions to base-\LCDM, neglecting any differences arising from modified gravity. In using the DES weak-lensing correlation functions, we exclude scales where nonlinear modelling uncertainties are important, but since the modified gravity models introduce an additional level of uncertainty, we also marginalize over the feedback amplitude $B$ with a flat prior, $2\le B \le 4$. This parameter is used by {\tt HMcode} to introduce an additional uncertainty in the nonlinear correction due to the modelling of the baryonic effects on the matter power spectrum at small scales, modifying the halo mass-concentration relation and the shape of the halo density profile. In this context, however, we marginalize over this parameter in order to reduce the residual sensitivity of our results on the nonlinear modelling in modified gravity theories; marginalizing over $B$ reduces the constraining power coming from nonlinear scales, where the correction recipe used by {\tt HMcode} may not correctly reproduce the perturbation evolution for all the models included in our parameterization.}

Throughout this section we will adopt the metric given by the line
element
\begin{equation}
ds^2 = a^2 \left[ - (1+2 \Psi) d\tau^2 + (1-2 \Phi) dx^2 \right] \, ,
 \label{eq:metric}
\end{equation}
with the speed of light $c$ set to 1.
The functions $\Phi(\tau,x)$ and $\Psi(\tau,x)$ are the gauge-invariant gravitational potentials, which are very nearly equal at late times in \lcdm.
For the background parameterization we use the standard {\tt CAMB} code,
while for the perturbation parameterization we use the publicly available code {\tt MGCAMB}\footnote{Available at
\url{http://www.sfu.ca/\textasciitilde aha25/MGCAMB.html} (February 2014 version, but updated to correctly output the power spectrum of the Weyl potential).}
\citep{Zhao:2008bn,Hojjati:2011ix} integrated into the latest version of \COSMOMC.
For the effective field theory (EFT) models of Sect.~\ref{sec:eft} we use {\tt EFTCAMB}\footnote{Available at
\url{http://eftcamb.org/} (version 2.0).} \citep{Hu:2013twa,Raveri:2014cka}.

\subsubsection{Background parameterization: \texorpdfstring{$w_0$, $w_a$}{w₀,wₐ}}
\label{sec:w0wa}

If the DE is a generic dynamical fluid, its equation of state
parameter $w \equiv p/\rho$ will in general be a function of time.
 Here $p$ and $\rho$ are the
spatially-averaged (background) DE pressure and density.

To test a time-varying equation of state we adopt the functional form
\begin{equation} \label{w0wapar}
w(a) = w_0 + (1-a) w_a \, ,
\end{equation}
where $w_0$ and $w_a$ are assumed to be constants. In \LCDM, $w_0 = -1$ and $w_a=0$.
We use the parameterized post-Friedmann (PPF) model of \cite{Fang:2008sn}
to explore expansion histories where $w$ crosses $-1$.
The PPF equations are modelled on the perturbations of quintessence dark energy, i.e., they correspond
to a fluid with vanishing anisotropic stress and a rest-frame speed of sound approximately equal to the speed of light.
Because of the high sound speed, dark-energy density perturbations are suppressed inside the horizon and are irrelevant
compared to the matter perturbations, except on the very largest scales.
\rep{While this is the standard procedure adopted in the literature, we should emphasize that a single minimally-coupled canonical scalar field (quintessence) {\it cannot\/} cross $w=-1$ \rep{\citep{Vikman:2004dc}}.} Such a crossing could happen in models with two scalar fields (one of which would have to be a phantom field with the opposite sign of the kinetic term); in such models
the perturbations remain close to the quintessence case \citep[see e.g.,][]{Kunz:2006wc}. Alternatively,
the phantom ``barrier'' can be crossed with a sound speed that vanishes in the phantom domain \citep{Creminelli:2008wc}
or in models with additional terms in the action, such as in kinetic-gravity-braiding \citep{Deffayet:2010qz}, or with non-minimal couplings \citep{Amendola:1999er, Pettorino:2008ez}. These and other modified gravity models, typically also change the behaviour of
the perturbations.

\begin{figure}[!t]
\begin{center}
\includegraphics[width=0.455\textwidth]{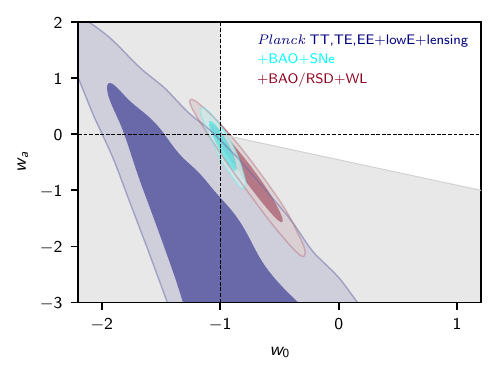}
\caption{Marginalized posterior distributions of the ($w_0, w_a$)
parameters for various data combinations.
The tightest constraints come from the combination \planckall\dataplus\lensing+SNe+BAO and are compatible
with \lcdm. Using \planckall\dataplus\lensing\ alone is considerably less constraining and allows for an area in parameter space that corresponds to large values of the Hubble constant (as already discussed in \citealt{planck2014-a15} and \PDEII).
The dashed lines indicate the point corresponding to the \lcdm\ model. The parametric equation of state given by Eq.~\eqref{w0wapar} stays out of the phantom regime (i.e., has $w\geq-1$) at all times only in the (upper-right) unshaded region.
}
\label{fig:w2D_wwa}
\end{center}
\end{figure}

Marginalized contours of the posterior distributions for $w_0$ and
$w_a$ are shown in Fig.~\ref{fig:w2D_wwa}.  Note that CMB lensing has
only a small effect on the constraints from \planck\ alone (see the
parameter grid tables in the \PLA).  Using \planck\ data alone, a wide volume
of dynamical dark-energy parameter space is allowed, with contours cut
off by our priors ($-3<w_0<1$, $-5<w_a<5$, and $0.4< h < 1$; note that
Fig.~\ref{fig:w2D_wwa} does not show the complete prior range).
However, most of the allowed region of parameter space corresponds to
phantom models with very high values of $H_0$ (as discussed
in \PDEII); such models are inconsistent with the late-time evolution
constrained by SNe and BAO data.  This is illustrated in Fig.~\ref{fig:w2D_wwa}
which also shows constraints if we add BAO/RSD+WL and BAO+SNe to the \planckalllensing\
likelihood. The addition of external data sets narrows the constraints towards the
\LCDM\ values of $w_0=-1$, $w_a=0$. The tightest constraints are found for the
data combination \planckalllensing+BAO+SNe; the difference in $\chi^2$
between the best-fit DE and \lcdm\ models for this data combination is
only $\Delta\chi^2=-1.4$ (which is not significant given the two
additional parameters).  Numerical constraints for these data
combinations, as well as $\chi^2$ differences, are presented in
Table~\ref{tab:resw0wa}.  It is also apparent that for the
simple $w_0$, $w_a$ parameterization of evolving DE, \Planck\ combined with
external data sets does not allow significantly lower values of $S_8$ or
higher values of $H_0$ compared to the base-\LCDM\ cosmology.

Fixing the evolution parameter $w_a=0$, we obtain the tight constraint
\oneonesig{\rep{w_0 = -1.028\pm0.031}}{\planckalllensing\dataplus{SNe}\dataplus\BAO}{,}
and restricting to $w_0>-1$ (i.e., not allowing phantom equations of state), we find
\onetwosig[5.5cm]{w_0 < -0.95}{\planckalllensing\dataplus{SNe}\dataplus\BAO}{.}
Here we only quote two significant figures, so that the result is robust to differences between
the \plik\ and \camspec\ likelihoods.

For the remainder of this section, we assume \lcdm\ at the background level (i.e., $w = -1$ at all times), but instead turn our attention to constraining the behaviour of the dark sector perturbations.

\begin{table}
\begin{center}
\caption{Marginalized values and 68\,\% confidence limits for cosmological parameters obtained by combining \planckalllensing\ with other data sets, assuming the $(w_0,w_a)$ parameterization of $w(a)$ given by Eq.~\eqref{w0wapar}. The $\Delta \chi^2$
values for best fits are computed with respect to the \lcdm\ best fits computed from the corresponding data set combination.
\label{tab:resw0wa}
}
\begingroup
\newdimen\tblskip \tblskip=5pt
\nointerlineskip
\vskip -3mm
\footnotesize
\setbox\tablebox=\vbox{
    \newdimen\digitwidth
    \setbox0=\hbox{\rm 0}
    \digitwidth=\wd0
    \catcode`"=\active
    \def"{\kern\digitwidth}
    \newdimen\signwidth
    \setbox0=\hbox{+}
    \signwidth=\wd0
    \catcode`!=\active
    \def!{\kern\signwidth}
\halign{
\hbox to 1.0in{#\leaderfil}\tabskip=1.5em&
\hfil#\hfil\tabskip=1.5em&
\hfil#\hfil\tabskip=0pt\cr
\noalign{\doubleline}
\noalign{\vskip -1pt}
\omit\hfil\text{Parameter}\hfil& \rep{\Planck+SNe+BAO}& \Planck+BAO/RSD+WL\cr
\noalign{\vskip 3pt\hrule\vskip 4pt}
$w_0$& $-0.957\pm0.080$& $-0.76\pm0.20$\cr
\noalign{\vskip 2pt}
$w_a$& $-0.29^{+0.32}_{-0.26}$& $-0.72^{+0.62}_{-0.54}$\cr
\noalign{\vskip 2pt}
$H_0\,[\Hunit]$& $68.31\pm0.82$& $66.3\pm{1.8}$\cr
\noalign{\vskip 2pt}
$\sigma_8$& $0.820\pm0.011$& $0.800^{+0.015}_{-0.017}$\cr
\noalign{\vskip 2pt}
$S_8$& $0.829\pm0.011$& $0.832\pm0.013$\cr
\noalign{\vskip 4pt\hrule\vskip 3pt}
$\Delta\chi^2$& $-1.4$& $-1.4$\cr
\noalign{\vskip 3pt\hrule\vskip 5pt}}}
\endPlancktablewide
\endgroup
\end{center}
\end{table}

\subsubsection{Perturbation parameterization: \texorpdfstring{$\mu$, $\eta$}{μ, η}} \label{sec:mueta}

\begin{figure}[!t]
\begin{center}
\includegraphics[width=0.48\textwidth]{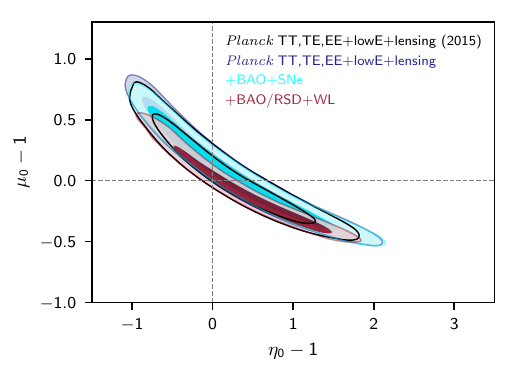}
\includegraphics[width=0.48\textwidth]{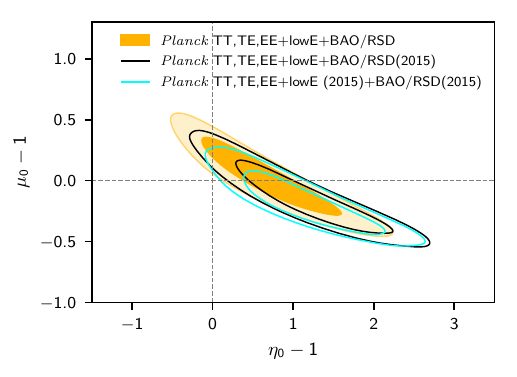}
\caption{{\it Top}: Marginalized posterior distributions of the MG parameters $\mu$ and $\eta$ for \planckall\dataplus\lensing\ data alone and in combination with external data (as indicated in the legend), using the late-time parameterization and neglecting any scale dependence. The dashed lines show the standard $\Lambda$CDM model. {\it Bottom}: Impact of the BAO/RSD and \planckall\ data, compared to the 2015 results.  For the 2018 \Planck\ data, the contours shift towards lower values of $\eta_0 -1$, along the maximum degeneracy line (black versus cyan contours) and shift in the same direction when using the BAO/RSD data (yellow versus black contours).
}
\label{fig:mu_eta}
\end{center}
\end{figure}

In the types of DE or MG models considered here, changes to observables only arise via the impact on the geometry of the Universe.
At the level of perturbations,
it is then sufficient to model the impact on the gravitational potentials $\Phi$ and
$\Psi$, or, equivalently, on two independent
combinations of these potentials
\citep[e.g.,][]{Zhang:2007nk,Amendola:2007rr}.
Following \PDEII\ we consider two phenomenological functions, $\mu$ and $\eta$, defined as follows.
\begin{enumerate}
\item $\mu(a, k)$: a modification of the Poisson equation for $\Psi$,
\begin{equation} \label{eq:mudef}
 k^2\Psi=-\mu(a,k) \, 4\pi G a^2\left[\rho\Delta+3(\rho+P)\sigma\right],
\end{equation}
where $\rho\Delta = \rho_{\rm m} \Delta_{\rm m} + \rho_{\rm r} \Delta_{\rm r}$, using comoving fractional density perturbations $\Delta$, and where $\sigma$ is the anisotropic stress from relativistic species (photons and neutrinos).
\item $\eta(a,k)$: an effective additional anisotropic
stress, leading to a difference between the gravitational potentials $\Phi$ and $\Psi$, defined implicitly through
\begin{equation}\label{eq:neweta}
 k^2\left[\Phi-\eta(a,k)\Psi\right]=\mu(a,k) \, 12\pi G a^2(\rho+P)\sigma.
\end{equation}
At late times, $\sigma$ from standard particles is negligible and we find
\begin{equation} \label{eq:etadef}
 \eta(a,k) \approx \Phi/\Psi.
\end{equation}
\end{enumerate}

\vskip 0.1in

These definitions are phenomenological, in the sense that they are not
derived from a theoretical action. However, they are able to capture a
generic deviation of the perturbation evolution from \LCDM\ that does
not need to correspond to a known model.  This approach is
complementary to constraints on action-based models, which are the
topic of the next subsection.  When $\eta=\mu=1$ we recover GR at all
times, including when there are non-zero contribution from photons and
neutrinos to the density perturbation or anisotropic stress. In the
parameterization adopted here (described further below), the MG
contribution to $\eta$ is only relevant at late times, when the
anisotropic stress from relativistic particles is negligible.

In this section we fix the background evolution to that of \LCDM\ ($w = -1$ at all times), so
that any significant deviation of $\mu$ or $\eta$ from unity would indicate a deviation from \LCDM.
We also consider constraints on the derived quantity $\Sigma$, defined as
\begin{equation} \label{eq:sigmadef}
 k^2\left[\Phi+\Psi\right]=-\Sigma(a,k)4\pi G a^2\left[2\rho\Delta-3(\rho+P)\sigma\right].
\end{equation}
Since $\Sigma$ measures deviations of the lensing potential from the GR prediction, it is better constrained by WL data than $\mu$ and $\eta$ separately.

For simplicity we only allow $\mu$ and $\eta$ to vary with time (as in \PDEII). Scale dependence
increases the number of degeneracies in parameter space and may require, for example, higher-order statistics in WL observables \citep{2018arXiv180505146P} to break the degeneracies.
We use the late-time DE parameterization of \PDEII\ and \citet{Casas:2017eob}, where the time evolution of all quantities is assumed to be proportional to the relative dark-energy density:
\beglet
\begin{equation}
\mu(z)=1+E_{11}\Omega_{\rm DE}(z); \label{equ:DE1}
\end{equation}
\begin{equation}
\eta(z)=1+E_{21}\Omega_{\rm DE}(z). \label{equ:DE2}
\end{equation}
\endlet
This defines the constants $E_{11}$ and $E_{21}$. We report results in terms of $\mu_0\equiv \mu(z=0)$ and $\eta_0\equiv \eta(z=0)$, which are determined from $E_{11}$ and $E_{21}$, given the dark-energy density parameter today.
This parameterization is motivated by the assumption that the impact of dark energy depends on its density and
therefore allows for more deviation of $\mu$ and $\eta$ from \lcdm\ at late times. The alternative early-time parameterization included in \PDEII\ led to similar results and is not discussed here for brevity. Our choice of parameterization, of course,
limits the nature of possible deviations from \LCDM; however, the choices of Eqs.~\eqref{equ:DE1} and \eqref{equ:DE2}
allow us to compare our results directly with those of \PDEII.

\begin{table*}
\begin{center}
\caption{Marginalized values and 68\,\% confidence regions for cosmological parameters obtained combining \planckall\ with other data sets, assuming the ($\mu$,\,$\eta$) parameterization of modified gravity. The $\Delta \chi^2$ values are computed with respect to the best-fit \LCDM\ cosmology, using the same data combination.
The quantity $\langle d^2 \rangle^{1/2}$ is the root-mean-square CMB lensing deflection angle, which is pulled high by the CMB data unless galaxy lensing (WL) or CMB lensing are included. The combination $\Sigma_0 S_8$ is approximately the
lensing amplitude parameter best constrained by the DES WL data at lower redshift.
}
\label{tab:resMGde}
\begingroup
\newdimen\tblskip \tblskip=5pt
\nointerlineskip
\vskip -3mm
\footnotesize
\setbox\tablebox=\vbox{
    \newdimen\digitwidth
    \setbox0=\hbox{\rm 0}
    \digitwidth=\wd0
    \catcode`"=\active
    \def"{\kern\digitwidth}
    \newdimen\signwidth
    \setbox0=\hbox{+}
    \signwidth=\wd0
    \catcode`!=\active
    \def!{\kern\signwidth}
\halign{
\hbox to 1.2in{#\leaderfil}\tabskip=1.5em&
\hfil#\hfil\tabskip=1.0em&
\hfil#\hfil\tabskip=1.0em&
\hfil#\hfil\tabskip=2.0em&
\hfil#\hfil\tabskip=1.0em&
\hfil#\hfil\tabskip=1.0em&
\hfil#\hfil\tabskip=0pt\cr
\noalign{\doubleline}
\omit& \multispan3\hfil With CMB lensing\hfil& \multispan3\hfil Without CMB lensing\hfil\cr
\noalign{\vskip -5pt}
\omit& \multispan3\hrulefill& \multispan3\hrulefill\cr
\omit& \Planck& \Planck& \Planck& \Planck& \Planck& \Planck\cr
\omit\hfil\text{Parameter}\hfil& & +SNe+BAO& +BAO/RSD+WL& & +SNe+BAO& +BAO/RSD+WL\cr
\noalign{\vskip 3pt\hrule\vskip 4pt}
$\mu_0-1$& $0.10^{+0.30}_{-0.42}$& $0.05^{+0.26}_{-0.39}$& $-0.07^{+0.19}_{-0.32}$& $0.12^{+0.29}_{-0.51}$& $0.10^{+0.30}_{-0.50}$& $-0.12^{+0.17}_{-0.32}$\cr
$\eta_0-1$& $0.22^{+0.55}_{-1.0}$& $0.32^{+0.63}_{-0.89}$& $0.32^{+0.63}_{-0.89}$& $0.55^{+0.78}_{-1.2}$& $0.62^{+0.79}_{-1.2}$& $0.52^{+0.67}_{-0.86}$\cr
$\Sigma_0-1$& $0.100\pm0.093$& $0.106\pm0.086$& $0.018^{+0.059}_{-0.048}$& $0.27^{+0.15}_{-0.13}$& $0.27^{+0.15}_{-0.13}$& $0.017^{+0.058}_{-0.050}$\cr
$\tau$& $0.0481^{+0.0087}_{-0.0072}$& $0.0487^{+0.0088}_{-0.0074}$& $0.0524\pm 0.0075$& $0.0504\pm0.0080$& $0.0505\pm0.0080$& $0.0526\pm 0.0079$\cr
$H_0\,[\Hunit]$& $68.20\pm0.63$& $68.19\pm0.45$& $68.09\pm 0.45$& $68.23\pm0.71$& $68.26\pm0.48$& $68.09\pm 0.46$\cr
$\sigma_8$& $0.812^{+0.034}_{-0.040}$& $0.807^{+0.029}_{-0.039}$& $0.799^{+0.023}_{-0.033}$& $0.817^{+0.032}_{-0.053}$& $0.814^{+0.033}_{-0.052}$& $0.794^{+0.020}_{-0.032}$\cr
$S_8$& $0.817\pm0.037$& $0.812^{+0.033}_{-0.038}$& $0.806^{+0.027}_{-0.034}$& $0.822^{+0.040}_{-0.051}$& $0.819^{+0.037}_{-0.052}$& $0.801^{+0.025}_{-0.034}$\cr
$\langle d^2\rangle^{1/2}$\,[{\rm arcmin}]& $2.531^{+0.046}_{-0.052}$& $2.529\pm 0.049$& $2.453\pm 0.032$& $2.697^{+0.095}_{-0.082}$& $2.695^{+0.099}_{-0.080}$& $2.456\pm 0.043$ \cr
$\Sigma_0 S_8$& $0.898\pm 0.067$& $0.897^{+0.068}_{-0.061}$& $0.820^{+0.043}_{-0.035}$& $1.04^{+0.12}_{-0.099}$& $1.04^{+0.12}_{-0.098}$& $0.814^{+0.044}_{-0.038}$ \cr
\noalign{\vskip 3pt\hrule\vskip 4pt}
$\Delta \chi^2$& $ -4.6$& $ -5.5$& $-1.2 $& $ -10.2$& $-11.0 $& $-0.7$\cr
\noalign{\vskip 3pt\hrule\vskip 5pt}}}
\endPlancktablewide
\endgroup
\end{center}
\end{table*}

\begin{table*}[h!]
\begin{center}
\caption{Marginalized values and 68\,\% confidence regions for cosmological parameters obtained by combining \planckall\ with other data sets, assuming the EFT parameterization $\Omega_0^\mathrm{EFT}(a)$. The $\Delta \chi^2$ values are computed with respect
to the best-fit \lcdm\ model using the same data combination. Values in brackets give the significance of the deviation from zero assuming a Gaussian posterior distribution.
}
\label{tab:reseft}
\begingroup
\newdimen\tblskip \tblskip=5pt
\nointerlineskip
\vskip -3mm
\footnotesize
\setbox\tablebox=\vbox{
    \newdimen\digitwidth
    \setbox0=\hbox{\rm 0}
    \digitwidth=\wd0
    \catcode`"=\active
    \def"{\kern\digitwidth}
    \newdimen\signwidth
    \setbox0=\hbox{+}
    \signwidth=\wd0
    \catcode`!=\active
    \def!{\kern\signwidth}
\halign{
\hbox to 1.2in{#\leaderfil}\tabskip=1.5em&
\hfil#\hfil\tabskip=1.0em&
\hfil#\hfil\tabskip=2.0em&
\hfil#\hfil\tabskip=1.0em&
\hfil#\hfil\tabskip=0pt\cr
\noalign{\doubleline}
\omit& \multispan2\hfil With CMB lensing\hfil& \multispan2\hfil Without CMB lensing\hfil\cr
\noalign{\vskip -5pt}
\omit& \multispan2\hrulefill& \multispan2\hrulefill\cr
\omit& \Planck& \Planck& \Planck& \Planck\cr
\omit\hfil\text{Parameter}\hfil& & +BAO/RSD+WL& & +BAO/RSD+WL\cr
\noalign{\vskip 3pt\hrule\vskip 4pt}
$\Omega_0^\mathrm{EFT}$& $-0.049^{+0.037}_{-0.024}\ (1.6\,\sigma)$& $-0.019^{+0.024}_{-0.019}\ (0.8\,\sigma)$& $-0.101^{+0.059}_{-0.038}\ (2.1\,\sigma)$& $-0.021\pm0.025\ (0.9\,\sigma)$\cr
$\alpha_{M0}$& $-0.040^{+0.041}_{-0.016}$& $-0.015^{+0.019}_{-0.017}$& $-0.075^{+0.073}_{-0.028}$& $-0.014^{+0.017}_{-0.014}$\cr
$\beta$& $0.72^{+0.38}_{-0.14}$& $0.66^{+0.44}_{-0.21}$& $0.66^{+0.38}_{-0.16}$& $0.62^{+0.45}_{-0.24}$\cr
$\tau$& $0.0489^{+0.0083}_{-0.0072}$& $0.0549^{+0.0096}_{-0.011}$& $0.0497\pm0.0082$& $0.0528\pm0.0086$\cr
$H_0\,[\Hunit]$& $68.19\pm0.67$& $68.22\pm0.46$& $68.30\pm0.71$& $68.16\pm0.46$\cr
$\sigma_8$& $0.8198\pm0.0074$& $0.8151\pm0.0067$& $0.845^{+0.013}_{-0.015}$& $0.8164^{+0.0087}_{-0.010}$\cr
$S_8$& $0.826\pm0.013$& $0.8205\pm0.0098$& $0.849\pm0.017$& $0.823\pm0.011$\cr
\noalign{\vskip 3pt\hrule\vskip 4pt}
$\Delta \chi^2$& $-4.3$& $-2.1$& $-9.7$& $-2.9$\cr
\noalign{\vskip 3pt\hrule\vskip 5pt}}}
\endPlancktablewide
\endgroup
\end{center}
\end{table*}

\begin{figure}[!t]
\begin{center}
\includegraphics[width=0.48\textwidth]{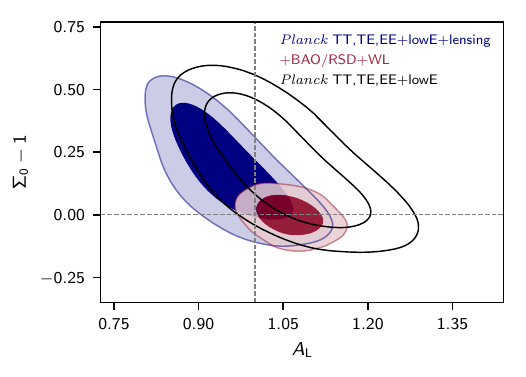}
\caption{Degeneracy between $\Alens$ and $\Sigma_0-1$, computed as
a derived parameter in our ($\mu,\eta$) parameterization.  The horizontal dashed line includes $\Lambda$CDM (but is also marginalized over one of the two degrees of freedom in the $\mu$--$\eta$ space). The vertical dashed line shows $\Alens=1$.
The filled contours use the \planckalllensing\ likelihood, alone and in combination with WL+BAO/RSD data. The
unfilled contours show the constraints from \planckall. Note that $\Alens$ only affects CMB lensing of the \Planck\ power spectra
by definition, as discussed in Sect.~\ref{sec:Alens}.}
\label{fig:mueta_alens}
\end{center}
\end{figure}

Figure~\ref{fig:mu_eta} shows the marginalized constraints on $\mu_0$
and $\eta_0$ from different combinations of data, and also compares
with the results from \PDEII. Marginalized mean values and errors for
cosmological parameters are presented in Table~\ref{tab:resMGde}. This table also lists
results for $\langle d^2 \rangle^{1/2}$, the root-mean-square CMB lensing deflection angle, and the parameter combination
$\Sigma_0 S_8$ that is well-constrained by the DES WL data. These quantities allow the reader to assess the impact of
lensing data on the parameter constraints.
The $\mu$ parameter affects the growth of structure, so, for example, higher $\mu$ gives larger values of $\sigma_8$. The CMB only constrains MG via the integrated Sachs-Wolfe effect (on large scales, where there is large cosmic variance) and CMB lensing. Lensing observations do not constrain the fluctuation amplitude directly, but the amplitude scaled by $\Sigma$ (as defined in Eq.~\ref{eq:sigmadef}). The degeneracy direction shown in Fig.~\ref{fig:mu_eta} corresponds to approximately constant lensing amplitude, with higher $\mu_0$ requiring lower $\Sigma_0$ and hence lower $\eta_0$. The thickness of the degeneracy contour and its location depends on the constraint on lensing. With
\planck\ data alone, or \planck+SNe+BAO, the lensing amplitude is pulled to high values by the preference for more lensing discussed in Sect.~\ref{sec:Alens}, so the contours are slightly shifted with respect to \LCDM.  The inclusion of
WL\footnote{Tests during the writing of this paper revealed a bug in
{\tt MGCAMB} that was also present in 2015. This bug reduced the
constraining power of WL data for the ($\mu,\eta$) parameterization
(which in 2015 was suppressed by the very conservative excision of
nonlinear scales). The CMB and BAO/RSD constraints and other
cosmological models were not affected by this bug.}
data shrinks the contour, and reduces the offset with respect to \LCDM;
DES WL data disfavour higher lensing amplitudes than predicted by the \planck\ \lcdm\ cosmology. DES also measures lensing at much lower redshift than CMB lensing, so it is a more powerful probe of MG models where changes to GR only appear at late times (as we have assumed).

The BAO/RSD data constrain $\mu_0$ directly, since redshift distortions are a probe of structure growth. The lower panel of
Fig.~\ref{fig:mu_eta} shows constraints with BAO/RSD alone, and also demonstrates that removing the CMB lensing reconstruction data shifts the contour further from \lcdm; this is consistent with the pull away from \lcdm\ being driven by the preference for more lensing in the high-$\ell$ CMB power spectra.

We can further demonstrate the effect of CMB lensing by varying the consistency parameter $\Alens$ within MG models.
Figure~\ref{fig:mueta_alens} shows the degeneracy between $\Alens$ and $\Sigma_0-1$, which is computed as
a derived parameter in our ($\mu,\eta$) parameterization. Here $\Alens$ affects lensing of the CMB power spectra
only, while $\Sigma_0$ encodes modifications to the lensing
amplitude caused by modifications of gravity. The contours show that
 MG models ($\Sigma_0 \ne 1$) are preferred by the \Planck\ power spectra (although not strongly) if $\Alens = 1$.
The preference for higher $\Sigma_0$ values is reduced by allowing
larger $\Alens$; the preference for MG that we find is therefore largely another
reflection of the preference for $\Alens > 1$ discussed in Sect.~\ref{sec:Alens}. Adding \Planck\
CMB lensing measurements shifts the contours back into consistency
with \lcdm\ (blue contours). Adding BAO/RSD + WL tightens the constrains (red contours)
which remain consistent with \lcdm.
Using the \camspec\ likelihood gives slightly less preference for high $\Alens$, and the results for \planckalllensing\ shift by about $0.2\,\sigma$ towards better consistency with \lcdm.

\subsubsection{Effective field theory description of dark energy\label{sec:eft}}

To investigate action-based models that can give interesting values of
$\mu$ and $\eta$, we limit ourselves to a sub-class of effective field
theories
\citep[EFTs,][]{2008JHEP...03..014C,Creminelli:2008wc,Gubitosi:2012hu}. The
EFTs we consider contain models with a single scalar field and at most
second-order equations of motion, a restriction that is in general
necessary to avoid the so-called Ostrogradski instability. In addition,
EFTs typically assume a universal coupling to gravity; models with
non-universal couplings \citep{Amendola:1999er, Pettorino:2008ez},
multiple scalar fields, additional vector \citep{Hellings:1973zz} or
tensor fields \citep{Hassan:2011tf}, or non-local
models \citep{Belgacem:2017cqo} do not fall into this class and are
not considered here. Nevertheless EFTs provide a general set of models
for which we can, in principle, compute all quantities of interest,
including $\mu$ and $\eta$ (which will span a restricted part of the
 $\mu$--$\eta$ space considered in the previous section.)

As described in section~5.2.1 of \PDEII, the degrees of freedom in
actions of this class of models can be reduced to the expansion rate
$H$ and five additional functions of time~\citep{2013JCAP...08..025G,
2014JCAP...07..050B}
\{$\alpha_{\rm M}, \alpha_{\rm K}, \alpha_{\rm B}, \alpha_{\rm T}, \alpha_{\rm H}$\}.
However, measurements of the speed of gravitational waves \citep{TheLIGOScientific:2017qsa} imply that $\alpha_T (z\,{=}\,0) \simeq 0$, which reduces the space of acceptable models \rep{\citep{Lombriser:2015sxa,McManus:2016kxu,Creminelli:2017sry,Ezquiaga:2017ekz,Sakstein:2017xjx,Baker:2017hug}}.

Apart from models where gravitational wave propagation is not modified
at all, which would necessarily limit us to
$\eta=1$ \citep{2014PhRvL.113s1101S}, only conformally (non-minimally)
coupled models\footnote{Recently the EFT action has been extended
to include degenerate higher-order theories \citep[DHOST,
][]{Zumalacarregui:2013pma,Gleyzes:2014dya,BenAchour:2016fzp}, which
feature an additional parameter called $\beta_{\rm
1}$ \citep{Langlois:2017mxy}. DHOST models can also give
$\alpha_{\rm T}\approx 0$, but $\alpha_{\rm H}$ and $\beta_{\rm 1}$ are
constrained to be small from astrophysical tests of
gravity \citep{2018PhRvD..97h4004C,Langlois:2017dyl,Dima:2017pwp,Saltas:2018mxc}.}
with $\alpha_{\rm B} = -\alpha_{\rm M}$ (and $\alpha_{\rm H}=0$) naturally lead to
$\alpha_{\rm T}=0$.  For these reasons we focus on this latter class of
models (and for simplicity we assume $\alpha_{\rm T} (z) \simeq 0$ at all
times), and in addition choose the kinetic terms of the scalar (set by
$\alpha_{\rm K}$) to keep the scalar sound speed equal to the speed of light
(current observational data are not able to constrain the sound speed
significantly, see e.g., the $k$-essence model constraints
in \PDEII). We finally end up with a non-minimally coupled $k$-essence
model described by the single function $\alpha_{\rm M}$ that determines the
running of the Planck mass. The background expansion is chosen to be
the same as in \lcdm, as in our analysis of the $\mu$--$\eta$ parameterization described in the previous
section. The
main difference relative to \PDEII\ is that we now allow for
$\alpha_{\rm M}<0$, which corresponds to a Planck mass decreasing with time.

\begin{figure}[!t]
\begin{center}
\includegraphics[width=0.48\textwidth]{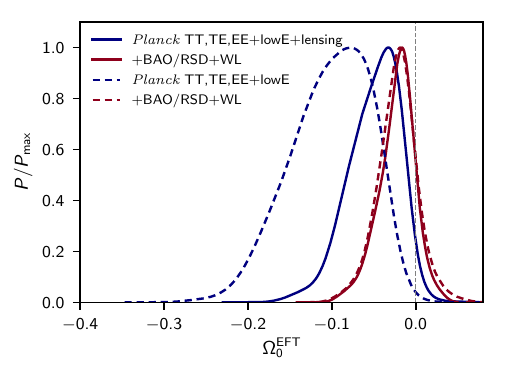}
\caption{Marginalized posterior distribution of $\Omega_0^\mathrm{EFT}$ that parameterizes the evolution of the Planck mass according to Eq.~\eqref{eq:planckmass} in the EFT model. We show constraints
for \planckall\dataplus\lensing\ data (solid lines), as well as \planckall\ data without CMB lensing (dashed lines), both alone and in combination with WL+BAO/RSD data. The $\Lambda$CDM limit lies at $\Omega_0^\mathrm{EFT}=0$ (vertical dashed line).
}
\label{fig:eft}
\end{center}
\end{figure}

As in \PDEII\ we adopt the parameterization $\alpha_M = \alpha_{M0}
a^{\,\beta}$, where $\alpha_{M0}$ is the value of $\alpha_M$ today and
where $\beta>0$ determines how quickly the absolute value of
$\alpha_M$ decreases at high redshift. In terms of the non-minimal
coupling function $\Omega$ multiplying the Ricci scalar $R$ in the
action, this corresponds to\footnote{
This notation is conventional; note that $\Omega^\mathrm{EFT}$ here is not the contribution to the critical density, and $\Omega_0^\mathrm{EFT}$ is not the value of $\Omega^\mathrm{EFT}(a)$ at $a=1$.
}
\begin{equation}
\Omega^\mathrm{EFT}(a) = \exp\left\{ \frac{\alpha_{M0}}{\beta} a^{\,\beta}\right\} -1 = \exp\left\{ \Omega_0^\mathrm{EFT} a^{\,\beta} \right\} - 1 \, ,
\label{eq:planckmass}
\end{equation}
which agrees with the exponential model built-in to {\tt
EFTCAMB} \citep{Raveri:2014cka} (which we use to compute the model
predictions presented here). The resulting posterior distribution on
$\Omega_0^\mathrm{EFT}$, marginalized over $\beta$ and other
parameters, is shown in Fig.~\ref{fig:eft}. The $\Lambda$CDM limit
lies at $\Omega_0^\mathrm{EFT}=0$ (vertical dashed line). We see that
the posterior distribution prefers negative values of
$\Omega_0^\mathrm{EFT}$, with a shift of $1.6\,\sigma$ for the
baseline \planckalllensing\ likelihood and $2.1\,\sigma$ if CMB lensing
is excluded. These shifts are reduced to $0.8\,\sigma$ with the
addition of BAO/RSD+WL to the \planckalllensing\ likelihood and to $0.9\,\sigma$
if \planck\ lensing is excluded.
 Table~\ref{tab:reseft} gives the parameter constraints for these data combinations and lists
the changes in $\chi^2$ of the best fits relative to base \LCDM.
As was the case for the ($\mu,\eta$) parameterization, DES WL measurements
pull the contours towards \lcdm. If we determine $\mu_0$ and $\eta_0$ that
correspond to the mean values of the EFT parameters for a specific
scale choice, we find that the parameters lie in the top-left quadrant of
the ($\mu,\eta$) parameter space shown in Fig.~\ref{fig:mu_eta}.
 Another class of models that predicts values of ($\mu,\eta$) in the top-left quadrant of Fig.~\ref{fig:mu_eta} are the non-local models, specifically the RR model of \cite{Dirian:2016puz}; these models are not discussed here.

Overall, the EFT sub-class of non-minimally coupled $k$-essence models considered here is
not preferred by current data. Without using CMB and galaxy WL lensing, \planck\ gives a
moderate preference for models that predict more lensing compared to \LCDM\ (as found in
our investigation of the ($\mu,\eta$) parameterization). However, combining \planck\ with CMB and DES WL
lensing measurements disfavours high lensing amplitudes and pulls the parameters towards \lcdm.

\subsubsection{General remarks}

\planck\ alone provides relatively weak constraints on dark energy and modified gravity, but \Planck\ does constrain other cosmological parameters extremely well. By combining \Planck\ with external data we then obtain tight constraints on these
models. We find no strong evidence for deviations from \lcdm, either at the background level or when allowing for changes to the perturbations. At the background level, \lcdm\ is close to the best fit. In the simple $\mu$--$\eta$ and EFT parameterizations of perturbation-level deviations from GR, we do find better fits to the \planckall\ data compared to \lcdm,
but this is largely associated with the preference in the CMB power spectra for higher lensing amplitudes (as discussed in Sect.~\ref{sec:Alens}), rather than a distinctive preference for modified gravity. Adding weak lensing data disfavours the large lensing amplitudes and our results are consistent with \lcdm\ to within  $1\,\sigma$.
Since neutrino masses are in general degenerate with DE and MG parameters, it is also worth testing the impact of varying neutrino masses versus fixing them to our base-\lcdm\ value of $m_\nu = 0.06$\,eV. We find similar trends, with slightly
larger posteriors when varying the neutrino mass.

\subsection{Neutrinos and extra relativistic species}\label{sec:neutrinos}
\newcommand{\JL}[1]{\textcolor{Brown}{$\triangleleft$ JL: #1 $\triangleright$}}

\subsubsection{Neutrino masses}
\label{subsec:mnu}

The \Planck\ base-\lcdm\ model assumes a normal mass hierarchy
with the minimal mass $\sum m_\nu= 0.06\,$eV allowed by neutrino flavour oscillation experiments.  However, current observations are consistent with many neutrino mass models, and there are no compelling theoretical reasons to
strongly prefer any one of them.  Since the masses are already known to be non-zero, allowing for larger $\sum m_\nu$ is one of the most well-motivated extensions of the base model.
The normal hierarchy, in which the lowest two mass eigenstates have the smallest mass splitting, can give any $\sum m_\nu \ga 0.06\eV$; an inverted hierarchy, in which the two most massive eigenstates have the smallest mass separation, requires $\sum m_\nu \ga 0.1\eV$.  A constraint that $\sum m_\nu <0.1\eV$ would therefore rule out the inverted hierarchy. For a review of neutrino physics and the impact on cosmology see e.g.,~\citet{LesgourgueBook}.

As in \paramsI\ and \paramsII, we quote constraints assuming
three species of neutrino with degenerate mass, a Fermi-Dirac distribution,
and zero chemical potential.  At \planck\ sensitivity the small mass
splittings can be neglected to good accuracy \citep[see e.g.,][]{Lesgourgues:2006nd}.
Neutrinos that become non-relativistic around recombination produce distinctive signals in the CMB power spectra,
which \planck\ and other experiments have already ruled out. If the neutrino mass is low enough that they became non-relativistic after recombination ($m_\nu \ll 1\eV$), the main effect on the CMB power spectra is a change in the angular diameter distance that is degenerate with decreasing $H_0$. The \planck\ data then mainly constrain lower masses via the lensing power spectrum and the impact of lensing on the CMB power spectra. Since the CMB power spectra prefer slightly more lensing than in the base-\lcdm\ model, and neutrino mass can only suppress the power, we obtain somewhat stronger constraints than might be expected in typical realizations of a minimal-mass neutrino model.

In \paramsII\ no preference for higher neutrino masses was found, but a tail to high neutrino masses was still allowed, with relatively high primordial amplitudes $\As$ combining with high neutrino mass to give acceptable lensing power.
The tighter 2018 constraint on the optical depth from polarization at low multipoles restricts the primordial $\As$ to be smaller, to match the same observed high-$\ell$ power ($C_\ell \propto \As e^{-2\tau}$); this reduces the parameter space with larger neutrino masses, giving tighter constraints on the mass.
With only temperature information at high $\ell$, the 95\,\% CL upper bound moved from 0.72\,eV (\paramsII\ TT+lowP) to 0.59\,eV \citep[using the {\tt SimLow} polarization likelihood of][at low $\ell$]{planck2014-a10}. This now further tightens to
\beglet
\onetwosig{\sumnu < 0.54\,\eV}{\planckTT}{.}
Adding high-$\ell$ polarization further restricts residual parameter degeneracies, and the limit improves to
\onetwosig{\sumnu < 0.26\,\eV}{\planckall}{.}
\endlet
Although the high-$\ell$ TT spectrum prefers more lensing than in base \lcdm, the lensing reconstruction is very consistent with expected amplitudes.
In \paramsII, the 2015 lensing likelihood weakened joint neutrino mass constraints because it preferred substantially less lensing than the temperature power spectrum. The 2018 lensing construction gives a slightly (1--2\,\%) higher lensing power spectrum amplitude than in 2015, which, combined with the decrease in the range of higher lensing amplitudes allowed by the new TT+\lowE\ likelihood, means that the constraints are more consistent.
Adding lensing therefore now slightly tightens the constraints to

\beglet
\begin{eqnarray}
\sumnu &<& 0.44\,\eV \quad\twosig{\shortTT\dataplus\lensing},\\
\sumnu &<& 0.24\,\eV \quad\twosig{\shortall\dataplus\lensing}.
\end{eqnarray}
\endlet

The joint constraints using polarization are however sensitive to the details of the high-$\ell$ polarization likelihoods, with the \camspec\ likelihood giving significantly weaker constraints with polarization:
\beglet
\begin{eqnarray}
  \sumnu &< 0.38\,\eV \quad &\mbox{\text{\leftparbox{5.7cm}{(95\,\%,~\shortall\,[\camspec])}}}
  \\
  \sumnu &< 0.27\,\eV \quad&\mbox{\text{\leftparbox{5.7cm}{(95\,\%,~\shortall\dataplus\lensing\,[\camspec]).}}}
\end{eqnarray}
\endlet
As discussed in \Alenssec, the \camspec\ \shortall\ likelihood shows a weaker
preference for higher lensing amplitude $\Alens$ than the default \plik\ likelihood, and this propagates directly into a weaker constraint on the neutrino mass, since for small masses the constraint is largely determined by the lensing effect. The differences between \plik\ and \camspec\ are much smaller if we add
CMB lensing, since the lensing measurements restrict the lensing amplitude to
values closer to those expected in base \LCDM.

The combination of the acoustic scale measured by the CMB ($\thetaMC$) and BAO data is sufficient to largely determine the background geometry in the \lcdm+$\sumnu$ model, since the lower-redshift BAO data break the geometric degeneracy. Combining BAO data with the CMB lensing reconstruction power spectrum (with priors on $\Omb h^2$ and $\ns$, following \PlanckLensTwo), the neutrino mass can also be constrained to be
\onetwosig{\sumnu < 0.60\,\eV}{\planck\ \lensing\dataplus\BAO\dataplus$\thetaMC$}{.}
This number is consistent with the tighter constraints using the CMB power spectra, and almost independent of lensing effects in the CMB spectra; it would hold even if the $\Alens$ tension discussed in \Alenssec\ were interpreted as a sign of unknown residual systematics.
 Since the constraint from the CMB power spectra is strongly limited by the geometrical degeneracy,
adding BAO data to the \planck\ likelihood significantly tightens the neutrino mass constraints. Without CMB lensing we find
\beglet
\onetwosig{\sumnu < 0.16\,\eV}{\planckTTBAO}{,}
\vspace{-2.0\baselineskip}
\onetwosig{\sumnu < 0.13\,\eV}{\planckallBAO}{,}
\endlet
and combining with lensing the limits further tighten to
\beglet
\onetwosig{\sumnu < 0.13\,\eV}{\planckTTlensing\dataplus\BAO}{,}
\vspace{-2.0\baselineskip}
\onetwosig{\sumnu < 0.12\,\eV}{\planckalllensing\dataplus\BAO}{. \label{equ:mnu1}}
\endlet
These combined constraints are almost immune to high-$\ell$ polarization modelling uncertainties, with the \camspec\ likelihood giving the $95\,\%$ limit $\sumnu < 0.13\,\eV$ for \planckalllensing\dataplus\BAO.

Adding the Pantheon SNe data marginally tightens the bound to
$\sumnu\,{<}\,0.11\,\eV$ (95\,\%,~\planckalllensing\dataplus\BAO\dataplus\mksym{Pantheon}).
In contrast the full DES 1-year data prefer a slightly lower $\sigma_8$ value than the \Planck\ \lcdm{} best fit, so DES slightly favours higher neutrino masses, relaxing the bound to
$\sumnu\,{<}\,0.14\,\eV$ (95\,\%,~\planckalllensing\dataplus\BAO\dataplus\mksym{DES}).

\begin{figure}[tbp!]
\centering
\includegraphics[]{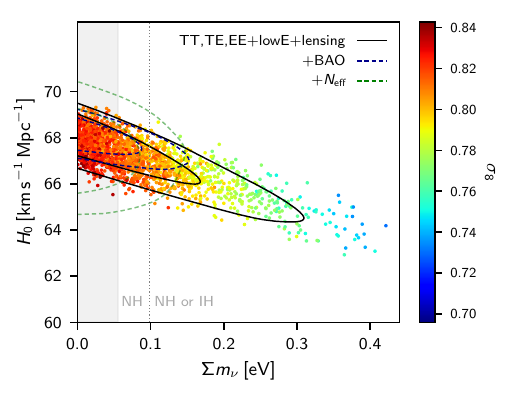}
\caption{Samples from \planckall\ chains in the
$\mnu$--$H_0$ plane, colour-coded by $\sigma_8$.
Solid black contours show the constraints from \planckall+lensing, while dashed
blue lines show the joint constraint from \planckall+lensing+BAO, and
the dashed green lines additionally marginalize over $\nnu$.
The grey band on the left shows the region with $\sumnu < 0.056\eV$ ruled out by neutrino oscillation experiments.
Mass splittings observed in neutrino oscillation experiments also imply that the region left of the dotted vertical line can only be
a normal hierarchy (NH), while the region to the right could be either the normal hierarchy or an inverted hierarchy (IH).
\label{fig:mnu-H-sigma}
}
\end{figure}

Increasing the neutrino mass leads to lower values of $H_0$, and hence aggravates the tension with the distance-ladder determination
of \citet[][see Fig.~\ref{fig:mnu-H-sigma}]{Riess18}. Adding the \citet{Riess18}
$H_0$ measurement to \Planck\ will therefore give even tighter neutrino mass constraints (see the parameter tables in the \PLA),
but such constraints should be interpreted cautiously until the Hubble tension is better understood.

The remarkably tight constraints using CMB and BAO data are comparable with the latest bounds from combining with Ly\,$\alpha$ forest data \citep{Palanque-Delabrouille:2015pga,Yeche:2017upn}. Although Ly\,$\alpha$ is a more direct probe of the neutrino mass (in the sense that it is sensitive to the matter power spectrum on scales where the suppression caused by neutrinos is expected to be significant)
the measurements are substantially more difficult to model and interpret than the CMB and BAO data.
Our $95\,\%$ limit of $\sumnu < 0.12\,\eV$ starts to put pressure on the inverted mass hierarchy (which requires $\sumnu\ga 0.1\eV$)
independently of Ly\,$\alpha$ data. This is consistent with constraints from neutrino laboratory experiments which also slightly prefer the normal hierarchy at $2$--$3\,\sigma$ \citep{Adamson:2017gxd,Abe:2017aap,Capozzi:2018ubv,deSalas:2017kay,deSalas:2018bym}.

\subsubsection{Effective number of relativistic species}

\begin{figure}[htbp!]
\centering
\includegraphics[]{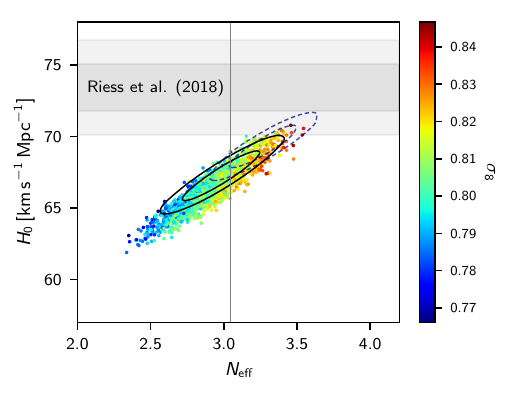}
\caption{Samples from \planckall\ chains in the
$\neff$--$H_0$ plane, colour-coded by $\sigma_8$.  The grey bands show the
local Hubble parameter measurement $H_0=(73.45\pm 1.66)\,\Hunit$ from \citet{Riess18}.
Solid black contours show the constraints from \planckall+lensing+BAO,
while dashed lines the joint constraint also including \citet{Riess18}.
Models with $\neff < 3.046$ (left of the solid vertical line) require photon
heating after neutrino decoupling or incomplete thermalization.
\label{fig:nnu-H-sigma}
}
\end{figure}

\begin{figure*}[htbp!]
\centering
\includegraphics[width =0.7\textwidth]{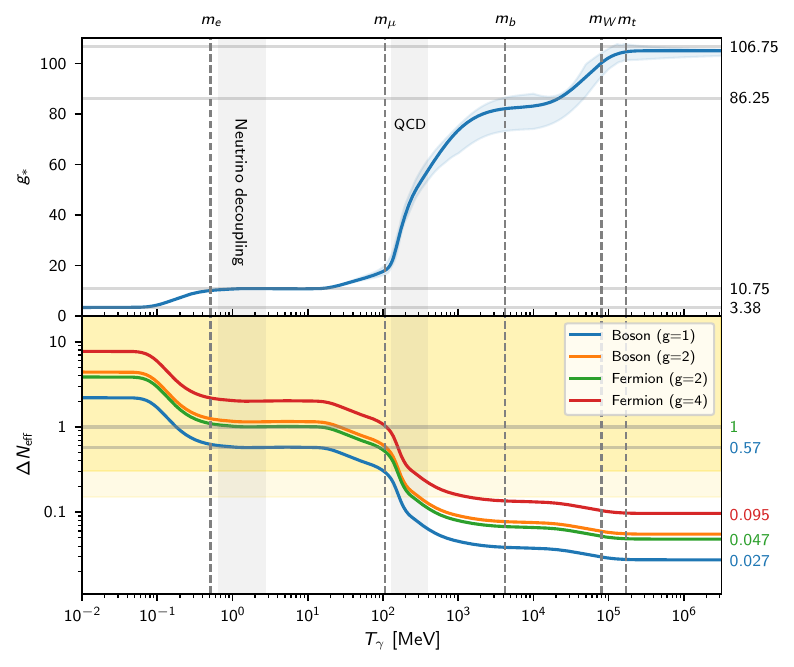}
\caption{
Constraints on additional relativistic particles.
{\it Top}:
Evolution of the effective degrees of freedom for Standard Model particle density, $g_*$, as a function of photon temperature in the early Universe. Vertical bands show the approximate temperature of neutrino decoupling and the QCD phase transition, and dashed vertical lines denote some mass scales at which corresponding particles annihilate with their antiparticles, reducing $g_*$. The solid line shows the fit of \citet{Borsanyi:2016ksw} plus standard evolution at $T_\gamma<1 \MeV$, and the pale blue bands the estimated $\pm1\,\sigma$ error region from \citet{Saikawa:2018rcs}. Numbers on the right indicate specific values of $g_*$ expected from simple degrees of freedom counting.
{\it Bottom}:
Expected $\Delta \nnu$ today for species decoupling from thermal equilibrium as a function of the decoupling temperature, where lines show the prediction from the \citet{Borsanyi:2016ksw} fit assuming a single scalar boson ($g=1$, blue), bosons with $g=2$ (e.g., a massless gauge vector boson, orange), a Weyl fermion with $g=2$ (green), or fermions with $g=4$ (red).
One-tailed $68\,\%$ and $95\,\%$ regions excluded by \planckall+lensing+BAO are shown in gold; this rules out at 95\,\% significance light thermal relics decoupling after the QCD phase transition (where the theoretical uncertainty on $g_*$ is negligible), including specific values indicated on the right axis of $\Delta\nnu = 0.57$ and 1 for particles decoupling between muon and positron annihilation. At temperatures well above the top quark mass and electroweak phase transition, $g_*$ remains somewhat below the naive 106.75 value expected for all the particles in the Standard Model, giving interesting targets for $\Delta \nnu$ that may be detectable in future CMB experiments \citep[see e.g.][]{Baumann:2017}.
\label{fig:Neff-Temp}
}
\end{figure*}

New light particles appear in many extensions of the Standard Model of
particle physics.
Additional dark relativistic degrees of freedom are usually parameterized
by $\nnu$, defined so that the total relativistic energy density well after electron-positron annihilation
is given by
\begin{equation}
  \rho_{\rm rad} = \nnu\ \frac{7}{8}\left(\frac{4}{11}\right)^{4/3}
  \ \rho_\gamma.
\end{equation}
The standard cosmological model has $\nnu\approx 3.046$, slightly larger than
$3$ since the three standard model neutrinos were not completely decoupled at
electron-positron annihilation \citep{Gnedin:1998, Mangano2005,deSalas:2016ztq}.

We can treat any additional massless particles produced well before
recombination (that neither interact nor decay) as simply an additional
contribution to $\nnu$.  Any species that was initially in thermal equilibrium
with the Standard Model particles produces a $\Delta \nnu
\,(\equiv \nnu - 3.046)$ that depends only
on the number of degrees of freedom and decoupling temperature.
Using conservation of entropy, fully thermalized relics with $g$ degrees of freedom contribute
\begin{equation}
  \Delta \nnu = g\,\left[\frac{43}{4\,g_s}\right]^{4/3}\times
  \left\{ \begin{array}{cl} 4/7 & {\rm boson}, \\
                            1/2 & {\rm fermion}, \end{array} \right.
\end{equation}
where $g_s$ is the effective degrees of freedom for the entropy of the other thermalized relativistic species that are present when they decouple.\footnote{For most of the thermal history $g_s\approx g_*$, where $g_*$ is the effective degrees of freedom for density, but they can differ slightly, for example during the QCD phase transition~\citep{Borsanyi:2016ksw} .}
Examples range from a fully thermalized sterile neutrino decoupling at
$1 \simlt T \simlt 100\,$MeV, which produces $\Delta \nnu=1$, to a thermalized boson
decoupling before top quark freeze-out, which produces
$\Delta \nnu\approx 0.027$.

Additional radiation does not need to be fully thermalized, in which case
$\Delta \nnu$ must be computed on a model-by-model basis.
We follow a phenomenological approach in which we treat $\nnu$
as a free parameter.  We allow $\nnu<3.046$ for completeness, corresponding to
standard neutrinos having a lower temperature than expected,
even though such models are less well motivated theoretically.

The 2018 \Planck\ data are still entirely consistent
with $\nnu\approx 3.046$, with the new low-$\ell$ polarization constraint lowering the 2015 central value slightly and with a corresponding
10\,\% reduction in the error bar, giving
\beglet
\onetwosig{\neff = 3.00^{+0.57}_{-0.53}}{\planckTT}{,}
\vspace{-2.0\baselineskip}
\onetwosig{\neff = 2.92^{+0.36}_{-0.37}}{\planckall}{,}
\endlet
with similar results including lensing.
Modifying the relativistic energy density before recombination changes the sound horizon, which is partly degenerate with changes in the late-time geometry.
Although the physical acoustic scale measured by BAO data changes in the same way,
the low-redshift BAO geometry helps to partially break the degeneracies.
Despite improvements in both BAO data and \Planck\ polarization measurements,
the joint \Planck+BAO constraints remain similar to \paramsII:
\beglet
\onetwosig[5.5cm]{\neff = 3.11^{+0.44}_{-0.43}}{ \shortTT\dataplus\lensing\dataplus\BAO}{;}
\vspace{-2.0\baselineskip}
\onetwosig[5.5cm]{\neff = 2.99^{+0.34}_{-0.33}}{ \shortall\dataplus\lensing\dataplus\BAO}{. \label{equ:neff1}}
\endlet
For $\nnu>3.046$ the \Planck\ data prefer higher values of the
Hubble constant and fluctuation amplitude, $\sigma_8$, than for the
base-\lcdm\ model.  This is because higher $\nnu$ leads to a
smaller sound horizon at recombination and $H_0$ must rise to keep the
acoustic scale, $\theta_*=\rstar/\DM$, fixed at the observed value. Since the change in the allowed Hubble constant with $\nnu$ is associated with a change in the sound horizon, BAO data do not help to strongly exclude larger values of $\nnu$.
Thus varying $\nnu$ allows the tension with \citet[][R18]{Riess18} to be somewhat eased, as illustrated in Fig.~\ref{fig:nnu-H-sigma}. However, although the 68\,\% error from \planckalllensing+BAO on the Hubble parameter is weakened when allowing varying $\nnu$, it is still discrepant with R18 at just over $3\,\sigma$, giving $H_0 = (67.3\pm 1.1)\Hunit$. Interpreting this discrepancy
as a moderate statistical fluctuation, the combined result is
\twoonesig[3cm]{\neff &= 3.27\pm 0.15}{H_0 &=(69.32\pm0.97)\Hunit}{\shortall\dataplus\lensing\dataplus\BAO\dataplus{R18.}}
However, as explained in \paramsII, this set of parameters requires an increase in $\sigma_8$ and a decrease in $\Omm$, potentially
increasing tensions with weak galaxy lensing and (possibly) cluster count data. Higher values for $\nnu$ also start to come into tension with
observational constraints on primordial light element abundances (see Sect.~\ref{sec:BBN}).

Restricting ourselves to the more physically motivated models with $\Delta\nnu>0$, the one-tailed \planckall+lensing+BAO constraint is $\Delta \nnu < 0.30$ at $95\,\%$.  This rules out light thermal relics that
decoupled after the QCD phase transition (although new species are still allowed if they decoupled at higher temperatures and with $g$ not too large).
Figure~\ref{fig:Neff-Temp} shows the detailed constraint as a function of decoupling temperature, assuming only light thermal relics and other Standard Model particles.

\subsubsection{Joint constraints on neutrino mass and $\nnu$}

\begin{figure}[t]
\centering
\includegraphics[]{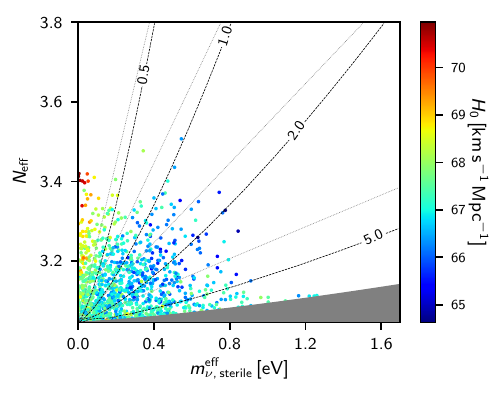}
\caption{Samples from \planckall+lensing, colour coded by the value of the Hubble parameter $H_0$,
for a model with minimal-mass active neutrinos and one additional sterile neutrino with mass parameterized by $\meffsterile$.
The physical mass for thermally-produced sterile neutrinos, $\msthermal$, is constant along the grey lines labelled by the mass in $\eV$;
the equivalent result for sterile neutrinos produced via the Dodelson-Widrow mechanism \citep{Dodelson:1993je} is shown by the adjacent thinner lines.
The dark grey shaded region shows the part of parameter space excluded by our default prior $\msthermal <10\,\eV$, where the
sterile neutrinos would start to behave like dark matter for CMB constraints.
\label{fig:meffsterile}
}
\end{figure}

There are various theoretical scenarios in which it is possible to have both sterile neutrinos and neutrino
mass. We first consider the case of massless relics combined with the three
standard degenerate active neutrinos, varying $\neff$ and $\sumnu$ together.
The parameters are not very correlated, so the mass constraint is similar
to that obtained when not also varying $\neff$.
We find:
\twotwosig[5cm]{\neff &= 2.96^{+0.34}_{-0.33},}{ \sumnu &< 0.12\,\eV,}{\planckalllensing\dataplus\BAO.}
The bounds remain very close to the bounds on either $\neff$ (Eq.~\ref{equ:neff1}) or $\sumnu$ (Eq.~\ref{equ:mnu1})
in 7-parameter models, showing that the data clearly differentiate between
the physical effects generated by the addition of these
 two parameters. Similar results are found without
lensing and BAO data. Although the mass constraint is almost unchanged,
varying $\nnu$ does allow for larger Hubble parameters, as shown in
Fig.~\ref{fig:mnu-H-sigma}. However, as discussed in \paramsII\ and the previous section,
 this does not
substantially help to resolve possible tensions with $\sigma_8$ measurements from other astrophysical
data.

The second case that we consider is massive sterile neutrinos combined with standard
active neutrinos having a minimal-mass hierarchy, parameterizing the sterile
mass by $\meffsterile\equiv \Omega_{\nu,\rm{sterile}}h^2 (94.1 \eV)$
as in \paramsI\ and \paramsII. The physical mass of the sterile neutrino in this
case is $\msthermal = (\Delta \nnu)^{-3/4}\meffsterile$ assuming a thermal
sterile neutrino, or $m_{\rm sterile}^{\rm DW} =
(\Delta\nnu)^{-1}\meffsterile$ in the case of production via the mechanism
described by \citet{Dodelson:1993je}. For low $\Delta \nnu$ the physical mass
can therefore become large, in which case the particles behave in the same way as cold
dark matter. In our grid of parameter chains we adopt a prior that $\msthermal <
10 \eV$ (and necessarily $\Delta \nnu \geq 0$) to exclude parameter space that is
degenerate with a change in the cold dark matter density; as we show in
Fig.~\ref{fig:meffsterile}, detailed constraints will depend on this choice of
prior. Assuming $\msthermal < 10 \eV$ we find
\beglet
\twotwosig[5cm]{\neff &< 3.29,}{\meffsterile &<0.65\,\eV,}{\planckalllensing\dataplus\BAO,\label{mefften}}
or adopting a stronger prior of $\msthermal < 2\eV$, we obtain the stronger constraint
\twotwosig[5cm]{\neff &< 3.34,}{\meffsterile &<0.23\,\eV,}{\planckalllensing\dataplus\BAO.\label{mefftwo}}
\endlet
The mass constraint in Eq.~\eqref{mefften} actually appears weaker than in
\paramsII; this is because the change in optical depth reduces the high-$\nnu$
parameter space, and the remaining lower-$\nnu$ parameter space has significant
volume associated with models having relatively large $\meffsterile$
(close to the $\msthermal$ prior cut). Removing this high-physical-mass
parameter space by tightening the prior to $2\,\eV$ gives the mass constraint
in Eq.~\eqref{mefftwo}, which is substantially tighter than the result quoted in
\paramsII\ without high-$\ell$ polarization.

One thermalized sterile neutrino with $\Delta \nnu=1$ is excluded at about
$6\,\sigma$ irrespective of its mass, or at about $7\,\sigma$ when
assuming a mass $m_\mathrm{sterile}^\mathrm{DW} \approx 1\,\eV$.
This is especially interesting in the context of the controversial evidence for
light sterile neutrinos, invoked to explain the neutrino short baseline
(SBL) anomaly.
The latest MiniBooNE data on electron-neutrino appearance \citep{Aguilar-Arevalo:2018gpe} support previous anomalous results by LSND \citep{Aguilar:2001ty}, with a combined significance of $6.1\,\sigma$ in favour of electron-neutrino appearance.
However, this contradicts recent muon-neutrino
disappearance data from MINOS+ and IceCube \citep{Dentler:2018sju}, \rep{when considered along with electron-antineutrino disappearance results \citep{Dentler:2017tkw,Gariazzo:2018mwd}, and also appears to be excluded by OPERA \citep{Agafonova:2018dkb}}. The
long-standing evidence for electron-neutrino disappearance in reactor
experiments has also recently been challenged by new data from STEREO
\citep{Almazan:2018wln} and PROSPECT \citep{Ashenfelter:2018iov}. It is worth
noting, however, that removing any \emph{individual} experiment does not relieve
the tension between the remaining experiments, and mild tension still persists
if all electron (anti-)neutrino appearance (disappearance) data are removed
\citep[see][for a detailed summary]{MaltoniTalk}.
Several analyses have shown that in order to fit the anomalous data sets with one sterile neutrino, one needs an active-sterile neutrino mixing angle such that the fourth neutrino mass eigenstate would acquire a thermal distribution in the early Universe \citep[see e.g.,][]{Hannestad13,Bridle17,Knee18}, thus contributing as $\Delta N_\mathrm{eff} \approx 1$.\,\footnote{Note that $\Delta N_\mathrm{eff}$ could in principle be reduced if there was a small amount of lepton asymmetry in the early Universe; however, this would raise other types of problems \citep{Saviano:2013ktj}.} Our
\Planck\ results confirm that the presence of a light thermalized sterile
neutrino is in strong contradiction with cosmological data, and that the
production of sterile neutrinos possibly explaining the SBL anomaly would need to be suppressed by some
non-standard interactions \citep{Archidiacono:2016kkh,Chu:2015ipa},
low-temperature reheating \citep{deSalas:2015glj}, or another special mechanism.

\subsection{Big-bang nucleosynthesis}\label{sec:BBN}

\subsubsection{Primordial element abundances}
\label{subsec:bbn}

{\it Primordial helium.} The latest estimates of the primordial helium abundance come from the data compilations of \citet{Aver:2015iza}, giving $\ypbbn \equiv 4 n_{\rm He}/n_{\rm b} = 0.2449 \pm 0.0040$ (68\,\% CL) and
\citet{Peimbert:2016bdg}, giving a slightly tighter constraint $\ypbbn = 0.2446 \pm 0.0029$ (68\,\% CL). These two estimates are consistent with each other. \citet{Izotov:2014fga} find a higher value, $\ypbbn = 0.2551 \pm 0.0022$ (68\,\% CL) in moderate (2.2$\,\sigma$ to 2.9$\,\sigma$) tension with the previous two.
\citet{Aver:2015iza} discuss the differences between their results and \citet{Izotov:2014fga}, which are caused by modelling differences involving neutral hydrogen
collisional emission, corrections for dust absorption, and helium emissivities, amongst other effects. This raises the issue, which has long-plagued helium abundance measurements, of whether
the systematic errors are accurately incorporated in the quoted uncertainties.  In this paper, we will use the more conservative \citet{Aver:2015iza} results as the baseline; however,
we will occasionally quote bounds based on the combined \citet{Aver:2015iza} and \citet{Peimbert:2016bdg} results ($\ypbbn = 0.2447 \pm 0.0023$ (68\,\% CL)) and for
the \citet{Izotov:2014fga} results.

\begin{figure}[htbp!]
\includegraphics[width=9.0cm]{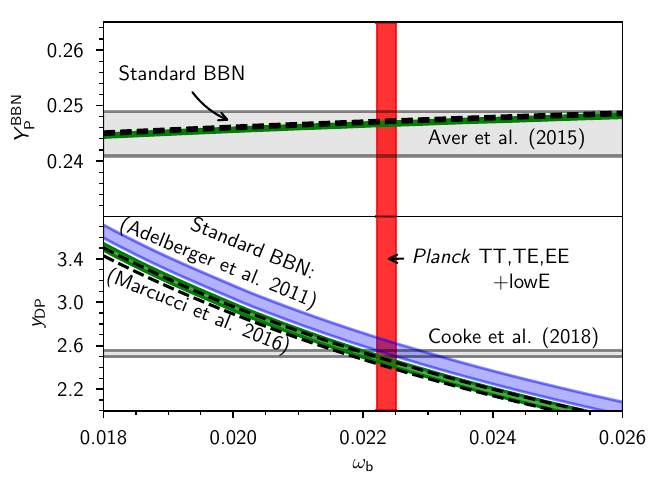}
\caption{Summary of BBN results with $\nnu=3.046$, using \planckall. All bands are 68\,\% credible intervals. The standard BBN predictions computed with \parthenope{} are shown in green (case (b) in the text), while those from {\tt PRIMAT} are in black dashed lines (case (c)). The blue lines show the \parthenope{} results based on the experimental determination of nuclear rates by \cite{Adelberger:2010qa}, instead of the theoretical rate of \citet[][case (a)]{Marcucci:2015yla}.}
\label{fig:bbn_a}
\end{figure}
\vspace{0.5cm}

Compared to the measurement used in \paramsII, the \citet{Aver:2015iza} error bar has decreased by a factor of 2.4.
To relate the primordial helium abundance to early Universe parameters under the assumption of standard BBN, we use two public BBN codes: first, version 1.10 of
\parthenope\footnote{\url{http://parthenope.na.infn.it}.
Note that \parthenope\ already exists in version~2.0, but the
difference with respect to 1.10 is only at the level of numerical
methods and performance. The physical input data and results are
identical.} \citep{Pisanti:2007hk}; and second, the recently released
{\tt PRIMAT}
code\footnote{\url{http://www2.iap.fr/users/pitrou/primat.htm}} \citep{Pitrou:2018cgg}. The
most relevant particle physics parameter for helium-abundance
calculations is the neutron lifetime. \parthenope\ 1.10 uses the
average value $\taun=(880.2\pm 1.0)\,$s (68\,\% CL) taken from the
Particle Data Group summary \citep{Patrignani:2016xqp}. This is a very
small shift with respect to the value of $\taun=(880.3\pm 1.1)\,$s
used in \paramsII. The PRIMAT code uses instead an average over
post-2000 measurements only, $\taun=(879.5\pm 0.8)\,$s \citep[68\,\%
CL,][]{Serebrov:2017bzo}. The two codes find (consistently) that
uncertainties of $\sigma(\taun)=1.0\,$s and $0.8\,$s correspond to
theoretical errors for the helium fraction of
$\sigma(\ypbbn)=3.0 \times 10^{-4}$ and $2.4 \times 10^{-4}$,
respectively.  Given the \Planck{} result for the baryon density in
the base-\lcdm\ model,
\onetwosig[4cm]{\omb = 0.02236 \pm 0.00029}{\shortall}{,\label{bbn:omegab}}
\parthenope{} predicts
\beglet
\onetwosig[3.6cm]{\ypbbn=
0.24672^{+(0.00011)0.00061}_{-(0.00012)0.00061}\hspace{-1mm}}{\shortall}{,}
while {\tt PRIMAT} gives
\onetwosig[3.6cm]{\ypbbn=0.24714^{+(0.00012)0.00049}_{-(0.00013)0.00049}\hspace{-1mm} }{\shortall}{.}
\endlet
The first set of error bars (in parentheses)
reflects only the uncertainty on $\omb$, while the second set
includes the theoretical uncertainty $\sigma(\ypbbn)$ added in
quadrature. The two mean values are shifted by $\Delta \ypbbn \approx 4.2 \times 10^{-4}$ because of differences in the adopted
neutron lifetime and because {\tt PRIMAT} includes a more elaborate treatment of weak interaction rates. However, this shift is quite close to the theoretical
errors estimated from both codes, and about an order of magnitude smaller that the observational error quoted by \cite{Aver:2015iza}.  As shown in Fig.~\ref{fig:bbn_a},
 the results from both codes
lie well within the region favoured by the \citet{Aver:2015iza} observations. They are also compatible at the 1$\,\sigma$ level with the combined \citet{Aver:2015iza} and \citet{Peimbert:2016bdg} results, but in 3.6--$3.8\,\sigma$ tension with
the \citet{Izotov:2014fga} results. Evidently, there is an urgent need to resolve the differences between
the helium abundance measurements and this tension should be borne in mind when we use the \citet{Aver:2015iza} measurements
 below.

\noindent {\it Primordial deuterium.} There has been significant progress related to deuterium abundance determination since the completion of \paramsII.
On the observational side, \cite{Cooke:2017cwo} have published a new estimate based on their best seven measurements in metal-poor damped Ly\,$\alpha$ systems,
$y_\mathrm{DP} \equiv 10^5 n_{\rm D}/n_{\rm H} = 2.527 \pm 0.030$ (68\,\% CL). On the calculational side, the value of the nuclear reaction rate $d(p,\gamma)^3 \mbox{He}$,
which has a major impact on BBN computations of the primordial deuterium calculation, has now been calculated ab initio.
The most recent theoretical calculation is presented in \citet[][leading to a smaller value of $y_\mathrm{DP}$]{Marcucci:2015yla} and differs significantly
from previous predictions extrapolated from laboratory experiments by \cite{Adelberger:2010qa}. This issue should be settled by forthcoming precise measurements by the LUNA experiment \citep{2017EPJWC.13601009G}. In this paper we will compare the results obtained when the deuterium fraction is computed in three different ways:
\begin{description}
\item[(a)] with \parthenope{}, assuming the experimental rate from \cite{Adelberger:2010qa};
\item[(b)] with \parthenope{}, using the theoretical rate of \cite{Marcucci:2015yla};
\item[(c)] with {\tt PRIMAT}, using the rate from \cite{Iliadis:2016vkw}, based on a hybrid method
 that consists of assuming the energy dependence of the rate computed ab initio by \cite{Marcucci:2005zc}
 and normalizing it with a fit to a selection of laboratory measurements.

\end{description}
In addition to the $d(p,\gamma)^3 \mbox{He}$ reaction rates, the current
versions of \parthenope{}, {\tt PRIMAT}, and other codes
\citep[such as that developed by][]{Nollett:2000fh,Nollett:2011aa} make
different assumptions on other rates, in particular those of the
deuterium fusion reactions $d(d,n)^3$He and $d(d,p)^3$H, which also
contribute significantly to the error budget of the primordial
deuterium fraction. \parthenope{} estimates these rates by averaging
over all existing measurements, while {\tt PRIMAT} again uses a hybrid
method based on a subset of the existing data. When using one of approaches
(a), (b), or (c), we adopt different theoretical errors. For
(a), \cite{Adelberger:2010qa} estimate that the error in their
extrapolated rate propagates to $\sigma(y_\mathrm{DP})=0.06$. For (b),
we rely on the claim by \cite{Marcucci:2015yla} that the error is now
dominated by uncertainties on deuterium fusion and propagates to
$\sigma(y_\mathrm{DP})=0.03$. For (c), the error computed by {\tt
PRIMAT} (close to the best-fit value of $\omb$) is similar,
$\sigma(y_\mathrm{DP})=0.032$.

These systematic error estimates are consistent with the differences
between different BBN codes. Taking $d(p,\gamma)^3 \mbox{He}$
from \cite{Marcucci:2015yla}, the
prediction of \parthenope\ 1.10 is higher than that of the code by
\cite{Nollett:2011aa} by about $\Delta y_\mathrm{DP}=0.04$, which is
comparable to the theoretical error adopted in this paper. \cite{Nollett:2011aa}
attribute this shift to their different assumptions on the deuterium
fusion rates. The shift between cases (b) and (c) is smaller, $\Delta
y_\mathrm{DP}=0.015$, suggesting that differences in
$d(p,\gamma)^3 \mbox{He}$ and in the deuterium fusion rates nearly
compensate each other in the final result.

Nuclear rate uncertainties are critically important in the discussion
of the compatibility between deuterium measurements and CMB
data. \cite{Cooke:2017cwo} reported that their measurement of
primordial deuterium was in moderate 2.0$\,\sigma$ tension with
the \Planck\ baryon density from \paramsII{}. This is based on the
predictions of the code of \cite{Nollett:2011aa} with the nuclear rate
of \citet{Marcucci:2015yla}.
\rep{Switching to \parthenope\ (b) and including the theoretical error
$\sigma(y_\mathrm{DP})=0.03$, we find consistency to
1.1$\,\sigma$. With our three BBN calculation pipelines, the deuterium abundance measurement of \cite{Cooke:2017cwo} translates into the following bounds on $\omb$,
\threetwosig
{&\mathrm{(a)\ } \,\, \omb =  0.02270 \pm 0.00075}
{&\mathrm{(b)\ } \,\, \omb =  0.02198 \pm 0.00044}
{&\mathrm{(c)\ } \,\, \omb =  0.02189 \pm 0.00046}
{\\Cooke~(2018),}
including theoretical errors. In several places in this work and in \PlanckLensThree, we refer to a ``conservative BBN prior,'' $\omb =  0.0222 \pm 0.0005$ (68\% CL), set to be compatible with each of these three predictions.}

We now update this discussion using the latest \Planck\ results. With our three assumptions (a), (b), and (c) on standard BBN, the determination of $\omb$ by \planck\ 2018 for the base-$\Lambda$CDM model (see Eq.~\ref{bbn:omegab}) implies
\threetwosig{&\mathrm{(a)\ } \,\, y_\mathrm{DP} =2.587^{+(0.055)0.13}_{-(0.052)0.13}}{&\mathrm{(b)\ } \,\, y_\mathrm{DP} =2.455^{+(0.054)0.081}_{-(0.053)0.080}}{&\mathrm{(c)\ } \,\, y_\mathrm{DP} =2.439^{+(0.053)0.082}_{-(0.051)0.081}}{\shortall,}
with the $\omb$-only error between parentheses, followed by the total error including the theoretical uncertainty.
These results are in agreement with the \cite{Cooke:2017cwo} measurement to within 0.8$\,\sigma$, 1.4$\,\sigma$, and 1.7$\,\sigma$, respectively.
Thus no significant tensions are found in any of these cases.
\vspace{0.5cm}

\noindent
\rep{{\it Other light elements.} We do not discuss other light elements, such as tritium and lithium, since
the observed abundance measurements and their interpretation in terms of the standard models of BBN are more
controversial \citep[see][for reviews]{Fields:2011, Fields:2014uja}. The \planck\ results
do not shed any further light on these problems compared to earlier CMB experiments.}

\begin{figure}[]
\includegraphics[width=9.0cm]{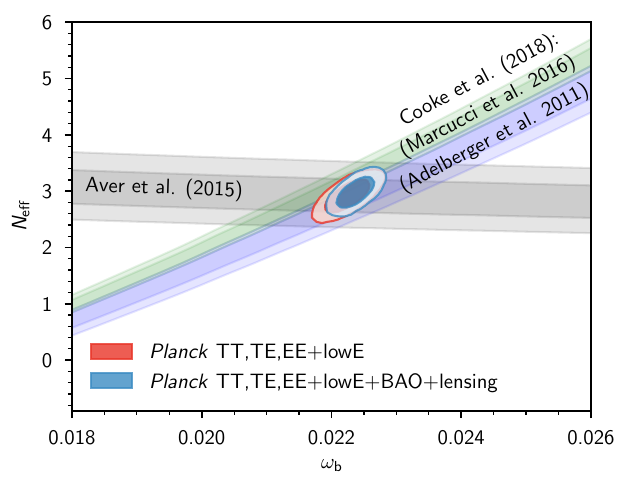}
\caption{
Constraints in the $\omb$--$\neff$ plane from \planckall{} and \planckallBAO\dataplus\lensing{} data
(68\,\% and 95\,\% contours) compared to the predictions of BBN combined with
primordial abundance measurements of helium \citep[][in grey]{Aver:2015iza} and deuterium \citep[][in green and blue, depending on which reaction rates are assumed]{Cooke:2017cwo}.
In the CMB analysis, $\nnu$ is allowed to vary as an additional parameter to the base-\LCDM{} model, while $Y_{\rm P}$ is inferred
from $\omb$ and $\nnu$ according to BBN
predictions. For clarity we only show the deuterium predictions based on the \parthenope code with two assumptions on the nuclear rate $d(p,\gamma)^3 \mbox{He}$ (case (a) in blue, case (b) in green). These constraints assume no significant
lepton asymmetry.}
\label{fig:bbn_b}
\end{figure}

\vspace{0.5cm}

\noindent
{\it Nuclear rates from bounds from Planck.} The previous paragraphs highlighted the importance of assumptions on the radiative-capture process ${\rm d}({\rm p},\gamma)^3 \mbox{He}$ for deuterium abundance predictions. It is worth checking whether the comparison of CMB and deuterium abundance data provides an indirect estimate of this rate. This approach was suggested in \cite{Cooke:2013cba} and implemented in \cite{DiValentino:2014cta} and \paramsII. We can now update it using the latest {\it Planck} and deuterium data.

We parameterize the thermal rate $R_2(T)$ of the ${\rm d}({\rm p},\gamma)^3 \mbox{He}$ process in the \texttt{PArthENoPE} code by rescaling the rate $R_2^{\rm ex}(T)$ fitted to experimental data by \cite{Adelberger:2010qa} with a factor $A_2$:
\begin{equation}
R_2 (T) = A_2 \, R_2^{\rm ex} (T)\, .
\end{equation}
This factor does not account in an exact way for the differences between
the experimental fit and the theoretical predictions; it should instead be seen as a consistency parameter, very much like $\Alens$ for CMB lensing in \Alenssec. The rate $R_2^{\rm th}(T)$ predicted by \cite{Marcucci:2005zc} has a temperature dependence that is close to what is measured experimentally, and can be very well approximated by a rescaling factor $A_2 = 1.055$. The new theoretical rate obtained by \cite{Marcucci:2015yla} has a slightly different temperature dependence but is well approximated
by an effective rescaling factor $A_2^{\rm th}=1.16$ (Mangano \& Pisanti, private communication).

Assuming the base-\lcdm\ model, we then constrain $A_2$ using \planck\ data combined with the latest deuterium abundance measurements from \citet{Cooke:2017cwo}. We still need to take into account theoretical errors on deuterium predictions arising from uncertainties on other rates, and from the difference between various codes. According to \citet{Marcucci:2015yla} and \citet{Pitrou:2018cgg}, the deuterium fusion uncertainties propagate to an error $\sigma(y_{\rm DP})=0.03$, which encompasses the difference on deuterium predictions between \parthenope\ versus {\tt PRIMAT}. Thus we adopt $\sigma(y_{\rm DP})=0.03$ as the theoretical error on deuterium predictions in this analysis. Adding the theoretical error in quadrature to the observational error of \cite{Cooke:2017cwo}, we obtain a total error of $\sigma(y_{\rm DP})=0.042$ on deuterium, which we use in our joint fits of \planck+deutrium (D) data. We find
\beglet
\oneonesig[4.2cm]{A_2 = 1.138 \pm 0.072}{\planckTT\dataplus{\rm D}}{,}
\vspace{-10mm}
\oneonesig[4.2cm]{A_2 = 1.080 \pm 0.061}{\planckall\dataplus{\rm D}}{.}
\vspace{-4mm}
\endlet
If we compare these results with those from \paramsII, the tension between the \planckTT\dataplus{\rm D} prediction and the experimental rate slightly increases to 1.9$\,\sigma$. However the inclusion of polarization brings the \planckall\dataplus{\rm D} prediction half-way between the experimental value and the theoretical rate of \cite{Marcucci:2015yla}, in agreement with both at the 1.3$\,\sigma$ level. The situation is thus inconclusive and highlights the need for a precise experimental determination of the ${\rm d}({\rm p},\gamma)^3 \mbox{He}$ rate with LUNA \citep{2017EPJWC.13601009G}.

\vspace{0.5cm}

\noindent
{\it Varying the density of relic radiation.} We can also relax the assumption that $\neff=3.046$ to check the agreement between CMB and primordial element abundances in the $\omb$--$\neff$ plane. Figure~\ref{fig:bbn_b} shows that this agreement is very good, with a clear overlap of the 95\,\% preferred regions of \Planck\ and of the helium+deuterium measurements. This is true with any of our assumptions on the nuclear rates. For clarity in the plot, we only include the predictions of \parthenope{} (cases (a) and (b)), but those of {\tt PRIMAT} are very close to case (b). Since all these data sets are compatible with each other, we can combine them to obtain marginalized bounds on $\neff$, valid in the 7-parameter \lcdm+$\neff$ model, with an error bar reduced by up to 30\,\% compared to the \planck+BAO bounds of Eq.~\eqref{equ:neff1}:
\threetwosig{&\mathrm{(a)\ }\,\, \neff = 2.89_{-0.29}^{+0.29}}{&\mathrm{(b)\ }\,\, \neff = 3.05_{-0.27}^{+0.27}}{&\mathrm{(c)\ }\,\, \neff = 3.06_{-0.28}^{+0.26}}{\planckall \dataplus Aver~(2015) \dataplus Cooke~(2018);}
\threetwosig{&\mathrm{(a)\ }\,\, \neff = 2.94_{-0.27}^{+0.27}}{&\mathrm{(b)\ }\,\, \neff = 3.10_{-0.25}^{+0.26}}{&\mathrm{(c)\ }\,\, \neff = 3.12_{-0.26}^{+0.25}}{\planckallBAO \dataplus Aver~(2015) \dataplus Cooke~(2018).}
The bounds become even stronger if we combine the helium measurements of \citet{Aver:2015iza} and \citet{Peimbert:2016bdg}:
\threetwosig{&\mathrm{(a)\ }\,\, \neff = 2.93_{-0.23}^{+0.23}}{&\mathrm{(b)\ }\,\, \neff = 3.04_{-0.22}^{+0.22}}{&\mathrm{(c)\ }\,\, \neff = 3.06_{-0.22}^{+0.22}}{\planckallBAO \dataplus Aver~(2015) \dataplus Peimbert~(2016) \dataplus Cooke~(2018).}
\rep{
However, as noted in the previous section, there is some inconsistency between the helium abundance measurements reported by different authors. If we use the helium abundance measurement of \citet{Izotov:2014fga} in place of \citet{Aver:2015iza} and \citet{Peimbert:2016bdg}, the mean value of $\neff$ shifts by about $0.35$ (e.g., for case (b), $\neff = 3.37\pm 0.22$ at the 95\% level), in 2.9$\,\sigma$ tension with the standard model value of $3.046$.}

Note finally that one can obtain $\neff$ bounds independently of the details of
the CMB spectra at high multipoles by combining the helium, deuterium, and BAO data sets with a nearly model-independent prior on the scale of the sound horizon at decoupling inferred from \planck{} data, $100\thetaMC=1.0409\pm 0.0006$ (68\,\%).
This gives a very conservative bound, $\neff = 2.95^{+0.56}_{-0.52}$ (95\,\%), when BBN is modelled as in case (b), along with a
$68\,\%$ bound on the Hubble rate, $H_0=(67.2\pm 1.7)\Hunit$.

\subsubsection{CMB constraints on the helium fraction} \label{sec:yhebbn}

\begin{figure}[]
\centering
\includegraphics[width=9.0cm]{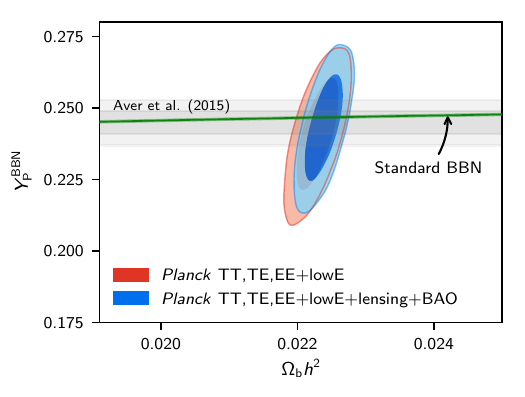}
\caption{Constraints on the helium abundance $\ypbbn$ from \planck, assuming the standard value of $\nnu=3.046$.
Results are consistent with the predictions of standard BBN (green line), and also the observed helium abundance \citep[68\,\% and 95\,\% grey bands from][]{Aver:2015iza}.
\label{fig:bbnyhe}
}
\end{figure}

\begin{figure}[htbp!]
\centering
\includegraphics[width=9.0cm]{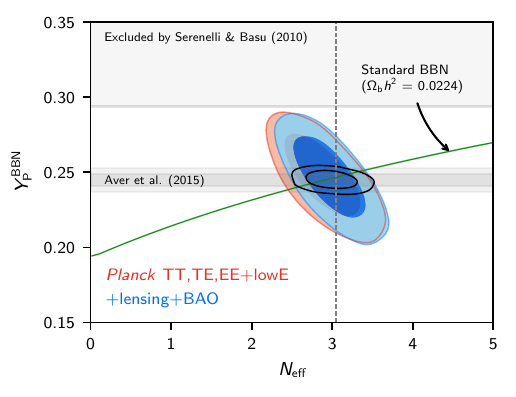}
\caption{Constraints on the helium abundance $\ypbbn$ and number of effective neutrino species $\nnu$ from \planckall\ and in combination with lensing and BAO.
Results are consistent with the predictions of standard BBN (green line), and also the observed helium abundance \citep[68\,\% and 95\,\% grey bands from][]{Aver:2015iza}.
The grey band at the top shows a conservative 95\,\% upper bound inferred from the Solar helium abundance \citep{Serenelli:2010fk}.
The black contours show the joint BBN-independent constraint from combining \planckall+lensing+BAO and \cite{Aver:2015iza}.
\label{fig:bbnyheneff}
}
\end{figure}

We now allow the helium fraction to vary independently of BBN, and compare \planck\ constraints with expectations.
In the parameter chains we vary the mass fraction $\yhe$ and compute the nucleon fraction $\ypbbn$ as a derived parameter, obtaining
\beglet
\onetwosig{\ypbbn = 0.241\pm0.025 \hspace{-2.2mm}}{\planckall}{,}
with similar results combined with lensing and BAO,
\onetwosig{\ypbbn = 0.243^{+0.023}_{-0.024}}{\planckalllensing\dataplus\BAO}{.}
\endlet
The \planck\ constraints on $\yhe$ and $\Omb h^2$ are shown in Fig.~\ref{fig:bbnyhe}, and are in good agreement with standard BBN predictions and
the helium abundance measurement of \citet{Aver:2015iza}.

Since both helium abundance and relativistic degrees of freedom affect
the CMB damping tail, they are partially degenerate.  Allowing $\nnu$
to also vary in addition to $\yhe$, we obtain the somewhat weaker
constraints:
\twotwosig[4.5cm]{\ypbbn &= 0.247^{+0.034}_{-0.036},}{\nnu &= 2.89^{+0.63}_{-0.57,}}{\planckall,}
\twotwosig[4.5cm]{\ypbbn &= 0.246\pm 0.035,}{\nnu &= 2.97^{+0.58}_{-0.54},}{\planckalllensing\dataplus\BAO.}
These constraints are shown in Fig.~\ref{fig:bbnyheneff}, and are again entirely consistent with standard assumptions.
The direct helium abundance measurement of \citet{Aver:2015iza}
provides significantly tighter constraints than those from \planck\ CMB measurements. By combining \planck\ with \citet{Aver:2015iza} we obtain a slightly tighter BBN-independent constraint on $\nnu$, while substantially improving the $\ypbbn$ result:
\twotwosig[4.3cm]{\ypbbn &=0.2437^{+0.0077}_{-0.0080},}{\nnu &=2.99^{+0.43}_{-0.40},}{\planckalllensing\dataplus\BAO\dataplus{\rm Aver (2015).}}
In our main grid results we assume that $\yhe$ can be determined accurately using standard BBN predictions from \parthenope\ based on a neutron lifetime $\taun = (880.2\pm 1.0)\,$s.
This uncertainty on $\taun$ is sufficiently small that it has negligible impact on constraints for non-BBN parameters.

If the $\taun$ constraint is relaxed, for example to allow a systematic shift towards the beam
measurement $\taun = [887 \pm 1.2 (\rm stat.) \pm 1.9 (\rm sys,)]\,$s of
\citet{Yue:2013qrc}, there would be a slight shift in cosmological parameters; however, taking the central value of $\taun \approx 887{\rm s}$ would shift \lcdm\ parameters by at most $0.2\,\sigma$ (for $\thetaMC$). As shown in Table~\ref{tab:base_extensions} the base-\lcdm\ parameters are very stable to marginalization over $\yhe$ with no constraint, at the expense of only
modest increase in uncertainties. There is therefore only very limited
scope for shifting the main \planck\ parameters by changing the BBN
model, especially given the BBN-independent requirement of consistency
with the observed helium abundances of \citet{Aver:2015iza}.

Finally, we can assume that standard BBN is an accurate theory, but
take $\taun$ as a free parameter to obtain an indirect constraint on
the neutron lifetime from CMB or CMB+helium data. This is potentially
interesting in the context of the long-standing difference between
neutrino lifetime measurements performed by beam and bottle
experiments. The PDG result, $\taun = (880.2\pm 1.0)\,$s, is based on
an average over two beam and five bottle
experiments~\citep{Patrignani:2016xqp}. The beam-only average gives
$\taun = (888.0\pm 2.0)\,$s, while the bottle-only average yields
$\taun = (879.2\pm 0.6)\,$s; these determinations are in 4.0$\,\sigma$
tension. To derive an independent prediction, following the lines
of \cite{Salvati:2015wxa}, we combine our \lcdm+$\yhe$ chains with the
function $\ypbbn(\omb, \taun)$ predicted by \parthenope\ or {\tt
PRIMAT} to obtain a posterior probability distribution in
$(\omb, \taun)$ space.\footnote{For simplicity, here we fix the extra
relativistic degrees of freedom to the standard value $\nnu=3.046$;
see \cite{Salvati:2015wxa} for discussion.} After marginalizing over
$\omb$, for CMB-only data, we find
\beglet
\oneonesig{\taun = (851 \pm 60)\,\mathrm{s}}{\planckall}{,}
using {\tt PRIMAT} (or, with \parthenope, $\taun = (855\pm62)\,$s).
Adding helium measurements from \citet{Aver:2015iza}, we find
\oneonesig{\taun = (867 \pm 18)\,\mathrm{s}}{\planckall\dataplus{\rm Aver (2015)}}{,}
\endlet
using {\tt PRIMAT} (or, with \parthenope, $\taun = (870\pm18)\,$s).
These results do not provide a statistically significant preference for either the
beam or bottle values.
If we make a similar prediction by combining \Planck\ with the helium measurement of \cite{Izotov:2014fga}, we obtain a range, $\taun = (920\pm11)\mathrm{s}$ (68\,\%CL), in 3.6$\,\sigma$ tension with all direct measurements of the neutron lifetime; this is a potentially interesting result, emphasizing again the need to
resolve tensions between different analyses of the primordial helium abundance.

\subsection{Recombination history}\label{sec:recombination}
\label{sec:projections_mu}
The cosmological recombination era marks an important phase in the
history of the Universe, determining precisely how CMB photons
decoupled from baryons around redshift
$z\,{\approx}\,10^3$ \citep{Sunyaev1970, Peebles1970}. With precision
data from \Planck, we can test physical assumptions of the
recombination process \citep{Hu1995, Seljak2003}, studying both
standard and non-standard physics.

The \Planck\ data are sensitive to several subtle atomic physics and
radiative-transfer effects \citep[see e.g.,][]{Chluba2006, Kholupenko2007,
Switzer2007I} that were omitted in earlier calculations of the
recombination history \citep{Zeldovich68, Peebles68,
Seager2000}. These effects can lead to significant biases to several
cosmological parameters \citep[e.g.,][]{Jose2010, Shaw2011}; however,
as the \Planck\ 2015 analysis confirmed, at the present level of
precision these can be reliably incorporated within the advanced
recombination codes \COSMOREC\ \citep{Chluba2010b}
and \HYREC\ \citep{Yacine2010}, as well as the \RECFAST\
code \citep{Seager:1999bc, Wong2008}, modified using corrections
calculated with the more precise codes.

In this section, we update the \paramsII\ search for deviations from
the standard recombination history. In particular, improved
polarization data provide additional constraining power that warrants
revisiting this question. As in 2015, we find no significant
indication for departures of the recombination history from the
standard prediction.

We use a semi-blind eigen-analysis (often referred to as a
principal-component analysis) of deviations of the free-electron
fraction, $x_{\rm e}(z)=n_{\rm e}/n_{\rm H}$, where $n_{\rm H}$
denotes the number density of hydrogen nuclei, away from the standard
recombination history \citep{Farhang2011, Farhang2013}. Specifically,
a perturbation, $\delta x_{\rm e}(z)/x_{\rm e}^{\rm fid}(z)$, is
expanded in $N_z=80$ bands of $\delta z$, spanning redshifts from well
before helium recombination to well past hydrogen recombination (taken
to be $200\le z \le 3500$).  Here, $x_{\rm e}^{\rm fid}(z)$ describes
the ionization history, assuming the standard recombination physics
and using the best-fitting cosmological parameters from \planckall.

\begin{figure}
\begin{center}
\includegraphics[width=\columnwidth]{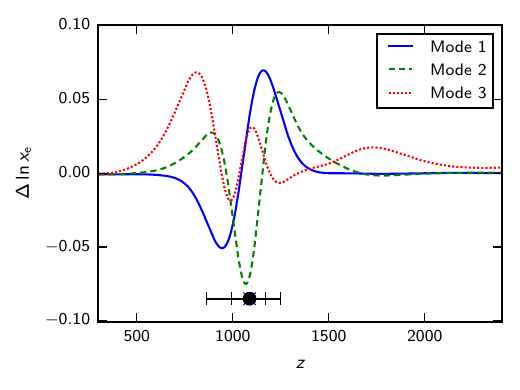}
\end{center}
\caption{First three normalized $x_{\rm e}$ modes constructed using the \planckall\ likelihood. The modes are marginalized over standard and
nuisance parameters. The forecast measurement uncertainties for the mode amplitudes are $\sigma_{\mu_1}=0.16$, $\sigma_{\mu_2}=0.23$, and $\sigma_{\mu_3}=0.73$. The position and width of the Thomson visibility function are indicated by the error bars at the bottom of the figure.}
\label{fig:eXeM}
\end{figure}

 We then form the Fisher information matrix for the $N_z+N_{\rm
std}+N_{\rm nuis}$ parameters, corresponding to the
$x_{\rm e}$-perturbation, standard cosmological, and nuisance
parameters, respectively. The Fisher matrix is then inverted to obtain
the parameter-parameter correlation matrix. Our focus is on the
$N_z \times N_z$ block of this Fisher inverse, containing the
marginalized errors and correlations of the $x_{\rm e}$ parameters.
The $x_{\rm e}$ block is diagonalized, and the corresponding diagonal
variances are rank-ordered from the lowest to highest fluctuation
variance (i.e., from the best to worst constrained mode). The rotation
diagonalizing the Fisher inverse defines the $x_{\rm e}$
eigenmodes. Truncation of the eigenmode hierarchy to determine the
number of $x_{\rm e}$ modes used for parameter estimation is performed
according to some suitably chosen selection criterion.  We refer to
these modes as ``eXeMs'', the first three of which are shown in
Fig.~\ref{fig:eXeM}. Only a small number are probed by \Planck\ 2018
data, even with the addition of the higher quality polarization
information. If instead we diagonalized the $N_z$ block of the Fisher
matrix before inverting, the modes would be characterized by the fixed
best-fitting cosmological and nuisance parameter values, i.e., they
would not be marginalized. Those $x_{\rm e}$ modes differ from the
eXeMs, but would give similar results \citep[as discussed
in][]{Farhang2011}.

For our analysis, we use the eXeMs, applying them to
the \planckalllensing\dataplus\BAO\ data combination. By construction,
these modes are orthogonal to each other;\footnote{In practice, our
mode generation method gives slight mode correlations at the level of
3--9\,\% due to the numerical procedure and smoothing of the
mode-functions \citep[see][for details]{Farhang2011}.} however,
correlations arise once the standard and nuisance parameters are
varied. This slightly modifies the errors and can also cause small
parameter biases \citep{Farhang2011}. Although the lowest order
$x_{\rm e}$ modes given in \paramsII\ look similar to those for the
2018 data, the precision of the \planck\ data requires the
eigen-analysis to be updated around the new fiducial point in
parameter space; indeed, we find subtle differences, e.g., a small
shift in the position of the first mode, to which the data are
sensitive.

 \begin{figure}
\begin{center}
\includegraphics[width=\columnwidth]{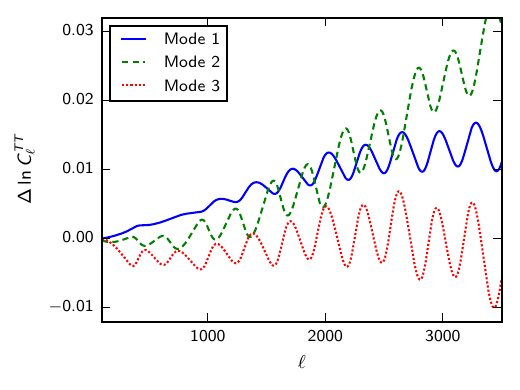}
\\
\includegraphics[width=\columnwidth]{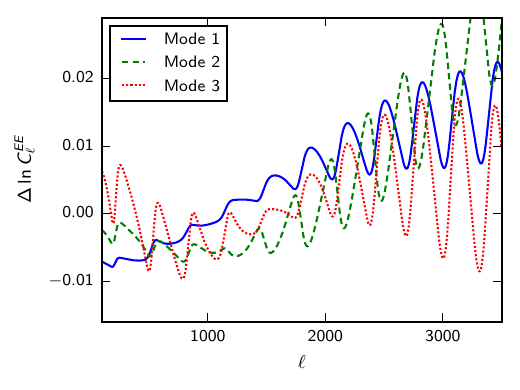}
\end{center}
\caption{Power spectrum responses to the first three $x_{\rm e}$ modes, constructed using \planckall, shown in Fig.~\ref{fig:eXeM}. For each curve, the corresponding $x_{\rm e}$ mode was added to the standard recombination history with an amplitude corresponding to their predicted $1\,\sigma$ uncertainties (i.e., $\sigma_{\mu_1}=0.16$, $\sigma_{\mu_2}=0.23$, and $\sigma_{\mu_3}=0.73$ for the first three eXeMs).
}
\label{fig:Cls}
\end{figure}

\begin{table}[tb]
\begingroup
\newdimen\tblskip \tblskip=5pt
\caption{%
Standard cosmological parameters, along with the first three $x_{\rm e}$-mode amplitudes, as determined using \planckalllensing\dataplus\BAO\ (all errors are $68\,\%$ CL).
}
\label{xe-mcmc-2}
\nointerlineskip
\vskip -3mm
\small
\setbox\tablebox=\vbox{
   \newdimen\digitwidth
   \setbox0=\hbox{\rm 0}
   \digitwidth=\wd0
   \catcode`*=\active
   \def*{\kern\digitwidth}
   \newdimen\signwidth
   \setbox0=\hbox{+}
   \signwidth=\wd0
   \catcode`!=\active
   \def!{\kern\signwidth}
\halign{
\hbox to 0.7in{#\leaderfil} \tabskip 0.1em&
\hfil#\hfil\tabskip 0.25em&
\hfil#\hfil\tabskip 0.1em&
\hfil#\hfil\tabskip 0pt\cr
\noalign{\doubleline}
\noalign{\vskip -2pt}
\omit \hfil Parameter\hfil& *+ 1 mode& *+ 2 modes& *+ 3 modes\cr
\noalign{\vskip 3pt\hrule\vskip 5pt}
$100\Omb h^2$&
 $*2.241\pm0.016$& $*2.241\pm0.018$& $!*2.239\pm0.018$\cr
$\Omc h^2$&
 $*0.1191\pm0.0009$& $*0.1192\pm0.0010$& $!*0.1192\pm0.0010$\cr
$H_0$&
 $67.72\pm0.43$& $67.72\pm0.44$& $!67.84\pm0.45$\cr
$\tau$&
 $*0.054\pm0.007$& $*0.055\pm0.007$& $!*0.055\pm0.007$\cr
$\ns$&
 $*0.9667\pm0.0051$& $*0.9668\pm0.0050$& $!*0.9657\pm0.0051$\cr
$\ln(10^{10}A_{\rm s})$&
 $*3.042\pm0.015$& $*3.042\pm0.014$& $!*3.040\pm0.015$\cr
$\mu_1$&
 $*0.02\pm0.12$& $*0.01\pm0.12$& $!*0.03\pm0.13$\cr
$\mu_2$&
               *\dots& $*0.01\pm0.17$& $!*0.05\pm0.17$\cr
$\mu_3$&
               *\dots&               *\dots& $*-0.84\pm0.69$\cr
\noalign{\vskip 3pt\hrule\vskip 3pt}}}
\endPlancktablewide
\endgroup
\end{table}

As discussed in \paramsII, the first mode corresponds mainly to a
change in the width and height of the Thomson visibility function,
while the second mode leads to a change in the position of the
visibility peak. The third mode introduces a superposition of the
change in the width, height, and position of the visibility peak. Each
mode causes a response in $\delta C_\ell /C_\ell^{\rm fid}$, as illustrated
in Fig.~\ref{fig:Cls}.

In the eigen-analysis, each eXeM is multiplied by an amplitude,
$\mu_i$, which is determined by MCMC sampling along with all of
the other standard cosmological and nuisance parameters. These
amplitudes and their errors are summarized in Table~\ref{xe-mcmc-2}
for the data combination \planckalllensing\dataplus\BAO. There is
stability in the amplitudes as the mode number is increased, and all
are consistent with no deviation from standard recombination within
the errors. We also find that cosmological parameters do not shift
with the inclusion of these modes, agreeing well (though with slightly
larger errors) with the \lcdm\ values computed assuming the standard
recombination history.  The four-mode case (not reported here) gives
similar results, but with slightly larger errors.

In \paramsII, an equivalent exercise also showed no evidence for
deviations from the standard recombination history. Using the
2015 \Planck\ high-multipole temperature power spectra, only two modes
were well-constrained; however, adding the preliminary high-multipole
polarization data in \paramsII\ allowed a third mode to be
constrained. The 2018 \Planck\ temperature and improved polarization
data used in this paper provide a more robust analysis.  Relative to
2015, we find comparable errors on the first and second mode
amplitudes and a small decrease in the uncertainty of the
third mode amplitude.

\subsection{Reionization}\label{sec:reionization}
At scales smaller than the horizon size at reionization ($\ell\,{\ga}\,10$),
free electrons generated during reionization can scatter and partially
damp the CMB anisotropies. This leads to a mostly scale-independent
suppression of power above $\ell\,{\approx}\,10$ by a factor of $e^{2\tau}$, where
$\tau$ is the total integrated optical depth to reionization, related to the
free electron fraction $x_{\rm e}(z)\equiv n^{\rm reion}_{\rm e}(z)/n_{\rm H}(z)$ by
\begin{equation}
    \tau = n_{\rm H}(0) c \sigma_{\rm T} \int_0^{z_{\rm max}} dz \, x_{\rm e}(z) \frac{(1+z)^2}{H(z)}.
\end{equation}
Here $n^{\rm reion}_{\rm e}(z)$ is the number density of free electrons from reionization,
$n_{\rm H}(z)$  is the  total number of hydrogen nuclei, and $\sigma_{\rm T}$ is the Thomson scattering
cross-section. We set $z_{\rm max}\,{=}\,50$, which is early enough to capture
the entirety of the expected contribution from reionization. We assume that the first reionization of helium
happens at the same time as the reionization of hydrogen, so complete first reionization corresponds to $x_{\rm e}>1$. There is an additional increase in $x_{\rm e}$ at $z\la 3.5$ when the helium is fully ionized;
this only has a small contribution to $\tau$ and in all cases we model it with a simple smooth transition at $z=3.5$.

At large scales in polarization ($\ell\,{\la}\,30$), anisotropies are instead
created by the rescattering of the local temperature quadrupole, which
varies maximally across Hubble-sized patches. This leads to a ``bump'' today in
the large-scale polarization power spectrum at the Hubble scale during
reionization. The amplitude of the bump scales like $\tau^2$, but the exact
shape encodes information on the detailed evolution of the ionization fraction
and can therefore constrain $x_{\rm e}(z)$ \citep{zaldarriaga1997, Kaplinghat:2002vt}.
Conversely, the inferred value of $\tau$ depends on the model assumed for
$x_{\rm e}(z)$, thus the reionization history has implications for other
cosmological parameters, which are important to quantify. Throughout the 2018
papers, we use the simple TANH model for reionization (described below and in Sect. \ref{subsec:amplitudes}). In this
section, we augment this with two other models to check whether our choice has
any impact on the $\tau$ constraints, and to assess  the extent to which \Planck\
data can place  model-independent bounds on
reionization. The three models we use are the
following.

\begin{itemize}
    \item TANH, which assumes a smooth transition from a neutral to ionized
    Universe, with a parametric form for $x_{\rm e}(z)$ based on a hyperbolic
    tangent (see footnote~\ref{footnote:tanh}). This model is not physically motivated, but makes the optical depth
    approximately independent of the transition width \citep{Lewis:2008wr}. It
    has been used previously in \paramsI\ and \paramsII, and is the default
    model in these 2018 papers.
    \item PCA (principle-component analysis), which decomposes the reionization history into eigenmodes that
    form a complete basis for any observable history \citep{hu2003}. In general,
    one must also specify a set of bounds to prevent the
    reconstruction from giving unphysical (e.g., negative) ionization fractions,
    and for this we use the optimal bounds given in \cite{Millea:2018bko}. The PCA model
    has some deficiencies: firstly, model parameters (the eigenmode
    amplitudes) do not have a straightforward  physical interpretation;
    secondly,
    even with the optimal physicality bounds, physicality cannot be
    enforced exactly \citep{mortonson2008,Millea:2018bko}. Nevertheless, the PCA
    approach serves as a useful alternative for comparison, and although we do not do so here, it can be
    used to construct an approximate likelihood that can be convenient way of
    exploring other models \citep{Heinrich:2018btc,miranda2017}.
    \item FlexKnot, which reconstructs any arbitrary reionization history using
    an interpolating function between a varying number of knots, with marginalization over the number
    of knots  \citep{Millea:2018bko}. Here,
    the model parameters are directly tied to the physical quantity of the
    ionization fraction, and as such physicality can be enforced by design. This
    model is the exact analogue of the  model used in reconstructing the
    primordial power spectrum from \Planck\ data
    \citep{vazquez2012,planck2014-a24}.
\end{itemize}

For each of these models, we must also specify the prior on the model
parameters, which in turn corresponds to some particular prior on $\tau$.
Previous analyses of \Planck\ data such as \cite{heinrich2017},
\cite{Obied:2018qdr}, \cite{hazra2017}, or \cite{Villanueva-Domingo:2017ahx},
have not considered the impact of these (sometimes implicit) priors, which
differed among the different analyses and consequently caused some partial
disagreement between results.
To allow direct comparison of $\tau$ values, unless otherwise stated we will use a prior that is
uniform on $\tau$. \cite{Heinrich:2018btc} construct a prior that is uniform on
$\tau$, but which increases the allowed unphysical parameter space and is chosen
a posteriori. Here we instead use the flat prior constructed by the procedure
described in \citet{Millea:2018bko} and \citet{Handley:2018gel}, which does not
admit extra unphysical models and gives the most generic prior that leaves the
prior on $\tau$ uniform.

Evidence based on observations of the Gunn-Peterson trough in the spectra of
high-redshift quasars show that  the inter-galactic medium is highly ionized by
$z\,{\approx}\,6$ \citep[see e.g.,][]{bouwens2015}. We enforce this bound in the
case of the TANH model by requiring that the central redshift of reionization be
greater than $z\,{=}\,6.5$; since the assumed duration in the TANH model is
$\Delta z\,{=}\,0.5$, this ensures that reionization is nearly complete by
$z\,{\approx}\,6$. The corresponding lower limit for the optical depth is
$\tau\,{\ga}\,0.0430$, modulo some small dependence on other cosmological
parameters. In the case of the FlexKnot model, the Gunn-Peterson bounds are
enforced by constraining the knot redshifts to be at $z\,{>}\,6$. Here, because the
duration of reionization is not specified and can effectively be instantaneous,
the optical depth can be as low as $\tau\,{=}\,0.0385$. The PCA model also
implicitly includes the Gunn-Peterson bounds, since the eigenmodes only have
support within the range $z\,{\in}\,[6,30]$, although the imperfect physicality bounds do
allow values of $\tau$ slightly below $0.0385$.

\begin{figure}[t!]
\begin{center}
\resizebox{\columnwidth}{!}{\includegraphics{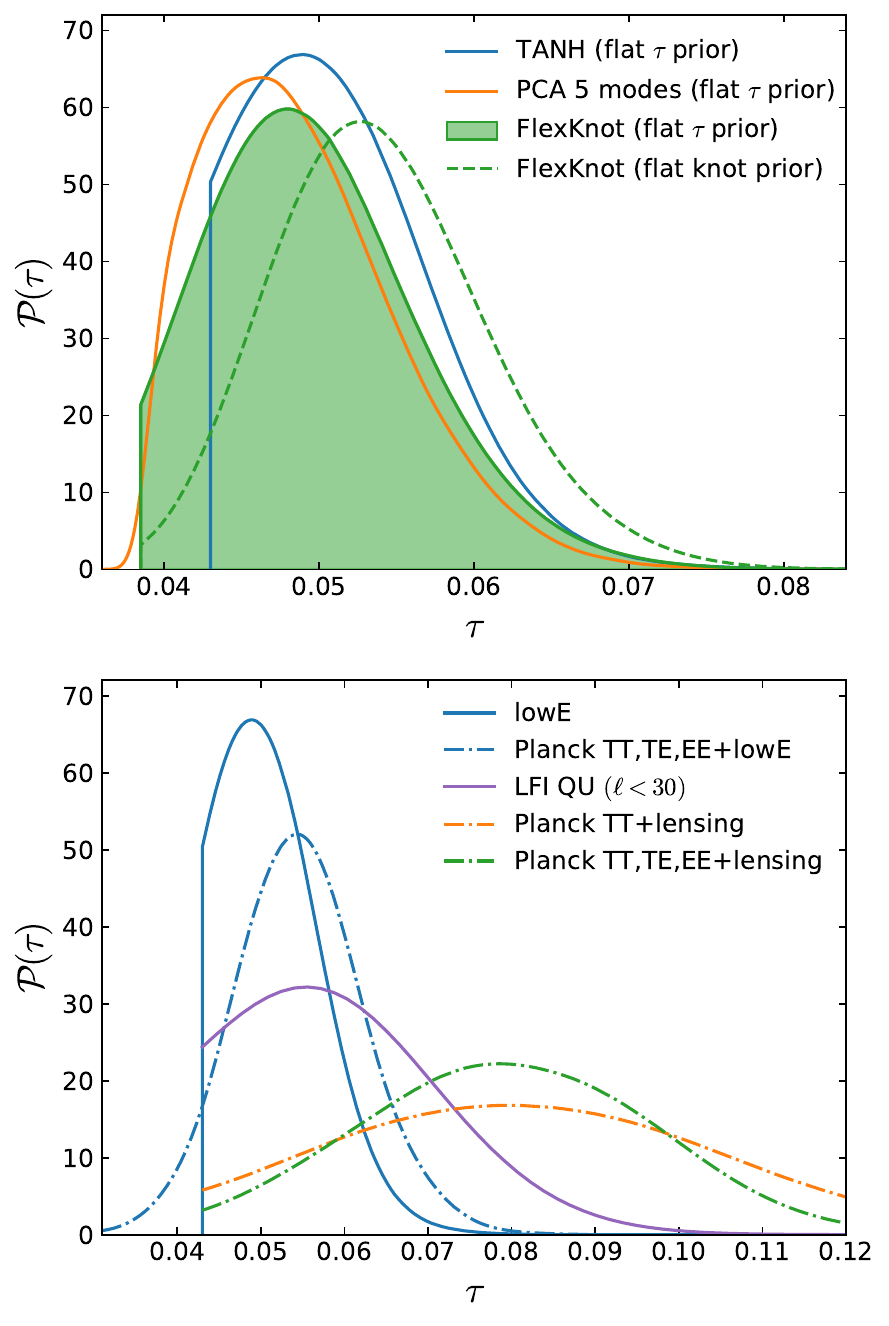}}
\end{center}
\caption{{\it Top}: Marginalized constraints on the optical depth to
reionization from \lowE\ alone, assuming different models of
reionization and different priors over the model parameters. Only reionization
parameters are varied here, with $\As e^{-2\tau}$ and other cosmological and
instrumental parameters held fixed at their best-fit values from
\planckTTTEEEonly. The solid lines use a flat prior on $\tau$, while the dashed
line uses a flat prior on the knot amplitudes; the difference between the green
lines is an example of the level to which these constraints depend on the choice
of prior. {\it Bottom}: Constraints from different data sets on the optical
depth assuming the TANH model and a flat $\tau$ prior (the cases that include
high-$\ell$ data are indicated by dot-dashed lines and also marginalize over
\LCDM\ parameters, as opposed to fixing them). The \planckTTTEEEonly\dataplus\lowE\
line is shown without the lower bound due to measurements of the Gunn-Peterson
trough, as a reminder that this bound is applied only in this section,
 resulting in some small extra shifts in the
central values of quoted constraints between this section and the remainder of
the paper. }
\label{fig:tau_simall}
\end{figure}

\begin{figure}[bhtp!]
\begin{center}
\resizebox{\columnwidth}{!}{\includegraphics{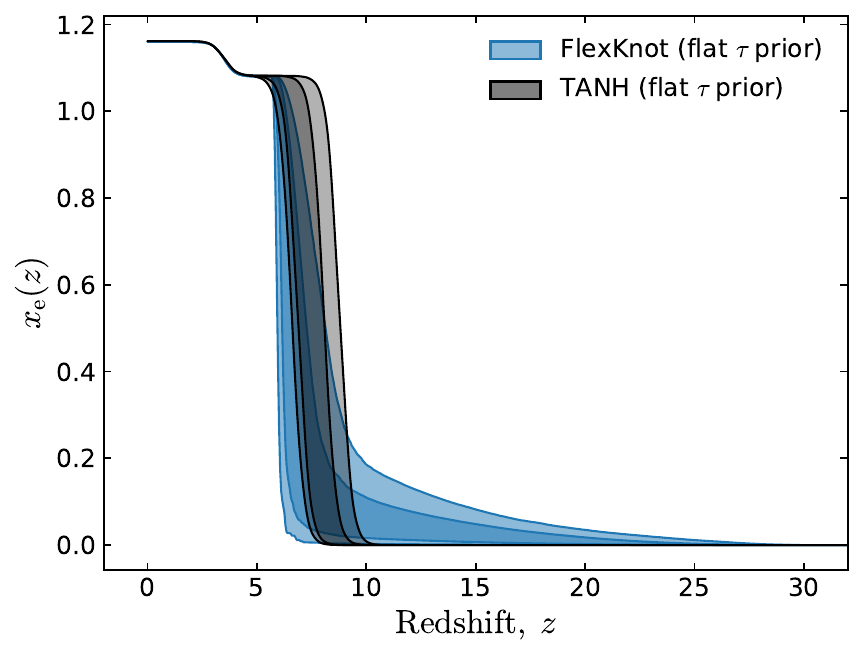}}
\end{center}
\caption{Constraints on the free electron fraction, $x_{\rm e}(z)$, from
\lowE\ alone, with $\As e^{-2\tau}$ and other cosmological and
instrumental parameters held fixed to their best-fit values from \planckTTTEEEonly, and with a
flat prior on $\tau$. The shaded bands are middle 68th and 95th percentiles
(note that this does not correspond exactly to confidence intervals). The FlexKnot
constraints show that any non-zero component of reionization above a redshift
of about 15 is highly disfavoured.
}
\label{fig:reio_hists}
\end{figure}

We begin by giving results using only the \lowE\ large-scale polarization
likelihood. As discussed in Sect.~\ref{sec:lowl}, this likelihood uses only $EE$
information, and  is restricted to $\ell \le 29$; we assume that the
reionization information in the polarization spectrum at $\ell\ge 30$ is
negligible, which is a good approximation for most models that can fit the
low-$\ell$ data. The \lowE\ data provide constraints on reionization that are largely
model independent, i.e., insensitive to other cosmological parameters. For
definiteness, we fix these other cosmological parameters to their best-fit
values from \planckTTTEEEonly, in particular holding $\As e^{-2\tau}$ rather
than $A_{\rm s}$ fixed, which better reflects the impact that the
$\ell\,{\ga}\,10$ data would have (we will comment at the end of this section on
how the high-$\ell$ data affect $\tau$). We plot posterior constraints from
\lowE\ in the top panel of Fig.~\ref{fig:tau_simall}. One can see the moderate
extent to which the hard cutoff of the Gunn-Peterson bound informs the posterior
in the TANH and FlexKnot cases (it of course also impacts the PCA case, although
the imperfect physicality priors in this case lead to the more gradual cutoff
visible in the figure). We find in the three cases the 68\,\% constraints:
\beglet
\begin{eqnarray}
    \tau &=& 0.0519^{+0.0030}_{-0.0079} \;\;(\rm \lowE; flat\,\tau\,prior; TANH); \\
    \tau &=& 0.0504^{+0.0050}_{-0.0079} \;\;(\rm \lowE; flat\,\tau\,prior; FlexKnot); \\
    \tau &=& 0.0487^{+0.0038}_{-0.0081} \;\;(\rm \lowE; flat\,\tau\,prior; PCA).
\end{eqnarray}
\endlet
The three results are in good agreement, showing that the \Planck\ data
prefer a late and fast transition from a
neutral to an ionized universe, which all models can capture equally well. The
TANH result gives slightly higher optical depth than the others, which is primarily
driven by the fixed duration of reionization assumed. The PCA result is
slightly lower, and is partly affected by the imperfect physicality priors that
allow unphysical negative ionization fractions. The FlexKnot result represents
our best model-independent estimate of the optical depth. Nevertheless, the
differences between this and the TANH result, or between the FlexKnot result
using either a flat prior on $\tau$ or on the knot positions and amplitudes (the
dashed line in Fig.~\ref{fig:tau_simall}), are small. For example, these
differences correspond to shifts in $\sigma_8$ of ${<}\,0.1\,\sigma$ when used
in conjunction with \planckall\ data. Thus, although future cosmological
inferences will depend somewhat on the details of reionization
\citep{Allison:2015qca, Millea:2018bko}, current \Planck\ data are
quite robust to how reionization is modelled.

The FlexKnot approach provides a model-independent reconstruction of the entire
reionization history, with physicality enforced exactly. This reconstruction is
presented in Fig.~\ref{fig:reio_hists}. A comparison against the TANH model is
also shown; although this imposes a fixed shape on the evolution, it
nevertheless matches the FlexKnot constraint fairly well. We find no preference
for any significant high-redshift contribution to the optical depth. \rep{This
conclusion does not depend qualitatively on our choice of prior either; we have
checked both a prior that is uniform on the knot positions and amplitudes, and
one that is uniform on the contribution to $\tau$ between redshifts 15 and 30,
$\tau(15,30)$.  We find\footnote{
\emph{Corrigendum:} In the original published version of this paper, the bounds given in Eqs.~(87a) and (87b) on the contribution to the early-time optical depth, $\tau(15,30)$, contained a numerical error in deriving the 95th percentile from the Monte Carlo samples. These corrected bounds are a factor of ${\sim}\,3$ larger than the originally reported results. Consequently, the new bounds do not significantly improve upon previous results from {\it Planck} data presented in \cite{Millea:2018bko} as was stated, but are instead comparable. Equations.~\eqref{eq:tauhigh_simall} and \eqref{eq:tauhigh_simallb} give results that are now similar to those of \cite{heinrich2021}, who used the same {\it Planck} 2018 data to derive a 95\,\% upper bound of 0.020 using the principal component analysis (PCA) model and uniform priors on the PCA mode amplitudes.
}:
\beglet
\begin{eqnarray}
    \label{eq:tauhigh_simall}
    \tau(15,30) &<& 0.018 \;\; (\rm lowE,\, flat \, \tau(15,30), FlexKnot); \\
    \label{eq:tauhigh_simallb} \tau(15,30) &<& 0.023 \;\; (\rm lowE,\, flat \, knot,\, FlexKnot).
\end{eqnarray}
\endlet}

This can be compared with the results of \cite{heinrich2017} \rep{and \cite{Obied:2018qdr}}, who
found a roughly $2\,\sigma$ preference for non-zero
$\tau(15,30)$ using \Planck\ 2015 data (which included a large-scale polarization
likelihood from the LFI instrument). \cite{Millea:2018bko} showed that the
majority of this preference disappeared when using the lower-noise
\planck\ HFI {\tt SimLow} likelihood \citep{planck2014-a10}, with an additional
sub-dominant effect due to the choice of prior. 

The upper bound on the contribution from $z\,{>}\,15$ to the total optical depth
limits some candidate explanations of the anomalously large 21-cm signal from
the EDGES experiment \citep{bowman2018}. Some otherwise plausible explanations
also lead, as a side-effect, to a significant number of ionizing photons being
generated at high redshift, enough to contribute significantly to $\tau(15,30)$.
These models are now highly disfavoured by the \Planck\ bound in their simplest
forms \citep[see e.g.,][]{Ewall-Wice:2018bzf}.

CMB data also probe high-redshift reionization via the patchy kinetic
Sunyaev-Zeldovich (kSZ) effect \citep{gruzinov1998, Knox:1998fp}. \Planck\ data,
together with smaller-scale ACT and SPT data (which are even more sensitive to this
effect), give upper bounds on the amplitude of the patchy kSZ power spectrum and
thus on the duration of reionization \citep{zahn2012, Sievers:2013ica,
planck2014-a25}. We do not attempt to derive new constraints here,
since it is not completely straightforward to turn a limit on the amplitude
of the patchy kSZ signal into one on the duration of reionization, especially
given the generic non-physical models for the ionization fraction
that we use here. However, in
the future kSZ should be a powerful probe of the details of reionization,
in particular with low-noise small-scale temperature measurements over large
fractions of the sky \citep{Smith:2016lnt,Ferraro:2018izc}.

The lower panel of Fig.~\ref{fig:tau_simall} compares the optical depth posteriors from different likelihoods.
Results from
the large-scale LFI polarization \citep{planck2016-l05} are in broad
agreement with \lowE, although with larger errors.
The \Planck\ lensing reconstruction data described in Sect.~\ref{sec:lensing}
can also provide a completely separate (although more
model-dependent) determination of the value of $\tau$; lensing is directly
sensitive to $\As$, and hence can partially break the $\As e^{-2\tau}$
degeneracy. By using the high-$\ell$ data in conjunction with the reconstructed
lensing-potential power spectrum, both of which are sensitive to lensing, we can
infer comparable constraints on $\tau$. These are shown in the bottom
panel of Fig.~\ref{fig:tau_simall}. Although the peak of the $\tau$ posterior
lies at higher values in this case, the difference between the $\tau$ estimates
from e.g., \lowE\ and \planckTTTEEEonly\dataplus\lensing\ is only
$1.4\,\sigma$ (where we compute the difference in posterior mean with respect to
the Gaussian error bars combined in quadrature). The preference for higher $\tau$
is driven by the same features in the CMB power spectrum data that prefer more lensing in \lcdm\ (giving $A_{\rm L}\,{>}\,1$; see Sect.~\ref{sec:Alens}):
the lensing amplitude can increased by increasing $A_{\rm s}$, which at constant
$A_{\rm s}e^{-2\tau}$ also increases $\tau$. Marginalizing over non-\LCDM\
parameters, for example $\Omega_K$ (which can also increase lensing by having $\Omega_K<0$),
can reduce the pull to higher $\tau$, but does not
change $A_{\rm s}e^{-2\tau}$ or the shape of the reionization bump significantly enough to affect the large-scale polarization result. This
type of model-independence has motivated our focus on only large-scale
polarization data in this section, although of course constraints on $\tau$
including higher-$\ell$ data (as are presented throughout the rest of this paper)
are equally valid, bearing in mind which model is assumed. Also, results in other sections do not apply the lower bound from the Gunn-Peterson constraint, which reduces the posterior mean values, somewhat disguising the larger peak values of the optical depth.

Overall, the results in this section leave us with a picture of reionization
that happened late and fast, and are consistent with reionization being driven by photons from
massive stars in low mass galaxies~\cite[see
e.g.][]{Robertson:2015, Parsa:2017tpd}. Our results are also consistent
with observations suggesting that the Universe is substantially neutral at
redshift $z\ga7.5$ \citep{Banados:2017unc,Davies:2018pdw,Mason:2017eqr}. The
low value of the optical depth makes the \planck\ constraints very robust to the
details of reionization modelling, with the simple TANH model adopted in this
paper causing no significant  biases in other parameters.

\subsection{Dark-matter annihilation}\label{sec:ann}
\begin{figure*}[htbp!]
\includegraphics[width=18cm]{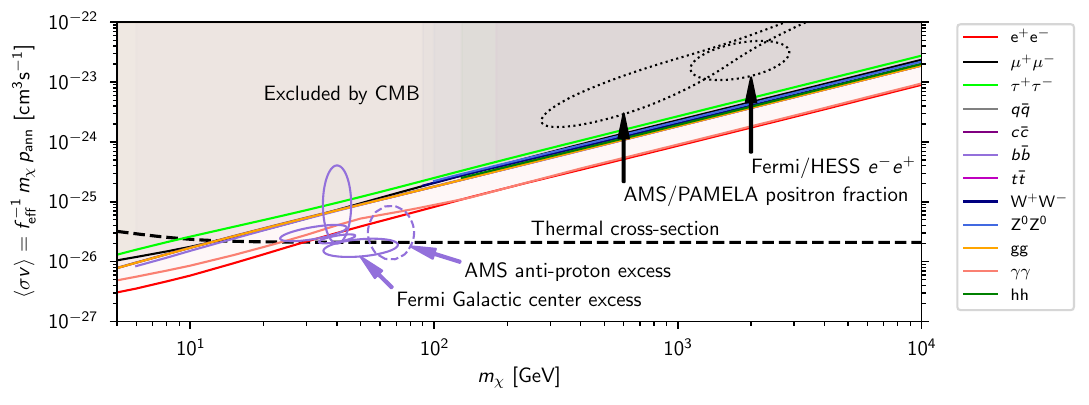}
\caption{\Planck\ 2018 constraints on DM mass and annihilation cross-section.
Solid straight lines show joint CMB constraints on several annihilation channels (plotted using different colours), based on $p_{\rm ann} < 3.2 \times 10^{-28} \,{\rm cm}^3\,{\rm s}^{-1}\,{\rm GeV}^{-1}$.
We also show the 2$\,\sigma$ preferred region suggested by the AMS proton excess (dashed ellipse) and the {\it Fermi\/} Galactic centre excess according to four possible models with references given in the text (solid ellipses), all of them computed under the assumption of annihilation into $b\bar{b}$ (for other channels the ellipses would move almost tangentially to the CMB bounds). We additionally show the 2$\,\sigma$ preferred region suggested by the AMS/PAMELA positron fraction and {\it Fermi}/H.E.S.S.\ electron and positron fluxes for the leptophilic $\mu^+\mu^-$ channel (dotted contours). Assuming a standard WIMP-decoupling scenario, the correct value of the relic DM abundance is obtained for a ``thermal cross-section'' given as a function of the mass by the black dashed line.}
\label{fig:ann}
\end{figure*}

CMB anisotropies are sensitive to energy injection in the
inter-galactic medium that could be a consequence, for example, of
dark-matter (DM) annihilation (see discussion in section
6.6. of \paramsII\ and references therein).
The current CMB sensitivity to the annihilation cross section of weakly-interactive massive particles (WIMPs) is competitive with and complementary to that of indirect DM search experiments.
The effective parameter constrained by CMB anisotropies is
\begin{equation}
p_{\rm ann} \equiv f_\mathrm{eff} \frac{\langle \sigma v \rangle}{m_\chi},
\end{equation}
where $m_\chi$ is the DM particle mass, $\langle \sigma v \rangle$ its
thermally averaged annihilation cross-section (assumed here to be
independent of temperature and redshift, as predicted for WIMPs
annihilating in s-wave channels), and $f_\mathrm{eff}$ is the fraction
of the energy released by the annihilation process that is transferred
to the intergalactic medium (IGM) around the redshifts to which the
CMB anisotropy data are most sensitive, namely $z \simeq
600$ \citep{Finkbeiner:2011dx}.

For each value of $p_{\rm ann}$, we compute CMB anisotropies using the
{\tt ExoClass} branch~\citep{Stocker:2018avm} of {\tt class} v2.6.3,
with recombination solved by {\tt HyRec} v2017 \citep{AliHaimoud:2010dx}. We assume that the energy injected
by DM annihilation is immediately transferred to the IGM
(the ``on-the-spot'' approximation), and splits between gas heating and
hydrogen excitation/ionization, according to the calculations
summarized in Table~V of \cite{Galli:2013dna}. Helium ionization and
beyond on-the-spot effects can be safely neglected here. Since CMB
anisotropies are very weakly sensitive to the redshift dependence of
the transferred energy fraction $f(z)$, we assume a constant fraction
$f(z)=f_\mathrm{eff}$.

We quote constraints on
$p_\mathrm{ann}^{28} \equiv  p_\mathrm{ann} / [10^{28}\,\mathrm{cm}^3\,\mathrm{s}^{-1}\,\mathrm{GeV}^{-1}] = 17.8 p_\mathrm{ann} / [10^{6}\,\mathrm{m}^3\,\mathrm{s}^{-1}\,\mathrm{kg}^{-1}]$:
\beglet
\vspace{-0.5\baselineskip}
\onetwosig{p_\mathrm{ann}^{28} < 3.5}{\planckall}{;}
\vspace{-2\baselineskip}
\onetwosig{p_\mathrm{ann}^{28} < 3.3}{\planckalllensing}{;}
\vspace{-2\baselineskip}
\onetwosig{p_\mathrm{ann}^{28} < 3.2}{\planckalllensing\dataplus\BAO}{.}
\vspace{-0.5\baselineskip}
\endlet
The bound based on CMB temperature and polarization data improves by
17\,\% compared to \paramsII. The difference is driven by
the new high-$\ell$ TT,TE,EE likelihood. This is consistent with the
fact that in addition to changing the physics of recombination, and
thus the scale and height of the acoustic peaks, DM annihilation enhances
the freeze-out value of the ionization fraction of the Universe after
recombination, and introduces a distinctive signature in the
polarization spectrum for $\ell \leq 200$. The new bounds are not only
stronger but also more robust, since polarization systematics in
the \Planck\ polarization spectra are now better understood. Adding
\Planck\ lensing and BAO further tightens the constraints.

In the baseline version of the \plik\ likelihood, the calibration
parameters of the polarization data, which correct for polarization
efficiencies, are fixed to the values computed assuming the base-\LCDM\
model, as described in Sect.~\ref{sec:plik}. This is
not necessarily consistent when the ionization history is
substantially modified by energy injection from DM or other
mechanisms. We thus performed further analyses in which the
polarization calibration parameters are varied, with a flat prior
within the range 0.8--1.2.  We found that our bounds remain
unaffected by floating these additional nuisance parameters, which are
not correlated with $p_\mathrm{ann}$.

Figure~\ref{fig:ann} translates the bounds on $p_\mathrm{ann}$ into joint limits on the mass $m_\chi$ and annihilation cross-section $\langle \sigma v \rangle$ of DM, assuming twelve plausible WIMP s-wave annihilation channels. The value of $f_\mathrm{eff}$ for each mass and channel was computed\footnote{Courtesy of P. St\"ocker.} using the public {\tt DarkAges} module of~ \cite{Stocker:2018avm}, which relies on the energy transfer functions presented by~\cite{Slatyer:2015kla}. We consistently account for corrections related to low-energy photons in the manner described in section V.B. of \cite{Slatyer:2015kla}. Finally, the {\tt DarkAges} module defines $f_\mathrm{eff}$ by convolving $f(z)$ in redshift space with the weighting function recommended by~\cite{Slatyer:2015jla}.
Note that for the $W^+W^-$ and $Z^0Z^0$ channels, the bounds assume on-shell 2-body processes and are cut sharply at the mass of the daughter particle, while in reality they would extend further to the left in Fig.~\ref{fig:ann}.

As usual the strongest bounds are obtained assuming annihilation into electron-positron pairs. The case of annihilation purely into neutrinos is not shown here, since the constraints are orders of magnitude weaker in that case. Assuming a thermal cross-section (shown in Fig.~\ref{fig:ann}), the 95\,\% CL lower bounds on the DM mass range from $m_\chi \geq 9$\,GeV for annihilation into tau/anti-tau, up to $m_\chi \geq 30$\,GeV for annihilation in electron/positron.
To compare with hints of DM annihilation in indirect DM search data, we first show the regions preferred by the AMS/PAMELA positron fraction and {\it Fermi}/H.E.S.S.\ electron-positron flux, assuming s-wave annihilation into muons and standard halo profiles. These regions, taken from \citet{Cirelli:2008pk}, have long been known to be in strong tension with CMB data.

We also indicate the regions suggested by the possible DM interpretation of several anomalies in indirect DM search data. The 95\,\% CL preferred region for the AMS anti-proton excess is extracted from \cite{Cuoco:2016eej,Cuoco:2017rxb}. The DM interpretation of the {\it Fermi\/} Galactic centre excess is very model-dependent and, as in figure~9 of \citet{Charles:2016pgz}, we choose to show four results from the analyses of \citet{Gordon:2013vta}, \citet{Abazajian:2014fta}, \citet{Calore:2014nla}, and \citet{Daylan:2014rsa}. For the {\it Fermi\/} Galactic centre excess and the AMS anti-proton excess, we only show results assuming annihilation into $b\bar{b}$, in order to keep the figure readable. About 50\,\% of the region found by \cite{Abazajian:2014fta} is excluded by CMB bounds, while other regions are still compatible. The 95\,\% CL preferred region for the AMS anti-proton excess is still compatible with CMB bounds for the $b\bar{b}$ channel shown in the figure, and we checked that this is also the case for other channels.

\section{Conclusions} \label{sec:conclusions}

This is the final \planck\ collaboration paper on cosmological
parameters and presents our best estimates of parameters defining the
base-\LCDM\ cosmology and a wide range of extended models.  As
in \paramsI\ and \paramsII\ we find that the base-\lcdm\ model
provides a remarkably good fit to the \Planck\ power spectra and
lensing measurements, with no compelling evidence to favour any of the
extended models considered in this paper.

Compared to \paramsII\ the main changes in this analysis come from
improvements in the \Planck\ polarization analysis, both at low and
high multipoles. The new \Planck\ polarization maps provide a tight
constraint on the reionization optical depth, $\tau$, from large-scale
polarization (and are consistent with the preliminary HFI polarization results presented in \citet{planck2014-a10}). This revision to the constraint on $\tau$ accounts for
most of the (small) changes in parameters determined from the
temperature power spectra in this paper compared to \paramsII. We have
characterized a number of systematic effects, neglected in \paramsII, which
affect the polarization spectra at high multipoles. Applying
corrections for these systematics (principally arising from errors in
polarization efficiencies and temperature-to-polarization leakage) we
have produced high multipole \TTTEEE\ likelihoods that provide
substantially tighter constraints than using temperature alone. We
have compared two \TTTEEE\ likelihoods that use different
assumptions to correct for polarization systematics and find
consistency at the $\la 0.5\,\sigma$ level. Although the \TTTEEE\
likelihoods are not perfect, the \planck\ parameter results presented
in this paper can be considered accurate to within their error bars.

\noindent Our main conclusions include the following.
\begin{unindentedlist}

\item The 6-parameter base-\lcdm\ model provides a good fit to the \planck\ \TT, \TE, and \EE\
 power spectra and to the \Planck\ CMB lensing measurements, either individually or in combination with each
other.

\item The CMB angular acoustic scale is measured robustly at
 $0.03\,\%$ precision to be $\theta_*=(0\pdeg5965 \pm 0\pdeg0002)$, and is
 one of the most accurately measured parameters in cosmology, of
 comparable precision to the measurement of the background CMB
 temperature \citep{Fixsen:2009ug}.

\item The \planck\ best fit base-\lcdm\ cosmology is in very good agreement with
 BAO, supernovae, redshift-space distortion measurements and BBN
 predictions for element abundance observations.  There is some
 tension (at about $2.5\,\sigma$) with high-redshift BAO measurements
 from quasar Ly\,$\alpha$ observations, but no standard extension of the base-\lcdm\
cosmology improves the fit
 to these data.

\item The new low-$\ell$ polarization likelihood tightens the reionization optical depth significantly compared to the 2015 analysis, giving $\tau = 0.054\pm 0.007$, suggesting a mid-point reionization redshift of $\zre = 7.7 \pm 0.7$. This is consistent with astrophysical observations of quasar absorption lines and
     models in which reionization happened relatively fast
    and late. We investigated more general models of reionization and
    demonstrated that our cosmological parameter results are
    insensitive to residual uncertainties in the reionization history.

\item The primordial fluctuations are consistent with Gaussian purely adiabatic scalar perturbations characterized by a
 power spectrum with a spectral index $\ns = 0.965\pm 0.004$,
    consistent with the predictions of slow-roll, single-field,
    inflation. Combined with BAO, we find that the Universe is spatially flat to
    high accuracy ($\Omk = 0.0007\pm 0.0019$), consistent the predictions of simple
    inflationary models. Combining with BICEP-Keck 2015 data
    on $B$-mode polarization we find a 95\,\% upper limit on the tensor-to-scalar ratio
    $r_{0.002} < 0.06$. Together with our measurement of $\ns$,
    these results favour concave over convex inflation potentials,
    suggesting a hierarchy between the slow-roll parameters measuring
    the slope and curvature of the potential.

\item The \planck\ base-\lcdm\ cosmology predicts a late-time clustering amplitude
$\sigma_8 = 0.811\pm 0.006$, and matter density parameter $\Omm =
0.315\pm 0.007$.  The parameter $S_8\equiv \sigma_8 \Omm^{0.5} =
0.831\pm 0.013$ is compatible with DES galaxy lensing, and
joint \planck-DES lensing results, although in modest tension with DES
results that also include galaxy clustering, which prefer a roughly $2.5\,\sigma$ lower
value of $S_8$. There is no obvious inconsistency
between the \planck\ base-\lcdm\ cosmology and counts of clusters (selected either
through the thermal Sunyaev-Zeldovich effect or via X-ray luminosity) because
of large uncertainties in the calibrations of cluster masses.

\item The \planck\ base-\lcdm\ cosmology requires a Hubble constant
$H_0 = (67.4\pm 0.5)\Hunit$, in substantial \rep{$4.4\,\sigma$} tension with the latest local determination by \citet{Riess:2019cxk}. The \planck\ measurement is in excellent agreement with independent inverse-distance-ladder measurements using BAO, supernovae, and element abundance results. None of the extended models that we have studied in this paper convincingly resolves the tension
with the \citet{Riess:2019cxk} value of $H_0$.

\item Allowing for extra relativistic degrees of freedom,
we measure the effective number of degrees of freedom in non-photon radiation density to be $\nnu = 2.89\pm 0.19$ ($\nnu=2.99\pm 0.17$ including BAO data), consistent with the value 3.046 expected in the standard model.
Light thermal relics that decoupled after the QCD phase transition are ruled out at the $2\,\sigma$ level. Allowing for larger $\nnu$ can slightly reduce tension with the local $H_0$ measurement and be consistent with BAO; however, the marginalized constraint on $H_0$ remains in tension with \citet{Riess:2019cxk} at over $3\,\sigma$ and higher values of $\nnu$ are not favoured by element abundance observations.

\item Combining \planck\ data with Pantheon supernovae and BAO data, the equation of state
of dark energy is tightly constrained to $\wzero = -1.03\pm 0.03$, consistent with a cosmological constant. We have also investigated a variety of modified-gravity models, finding no significant evidence for deviations from \lcdm.

\item Allowing for a free degenerate active neutrino mass, and combining with BAO measurements, we obtain the tight $95\,\%$ constraint on the sum of the masses $\mnu < 0.12\,\eV$.

\item
We find good agreement between the predictions of BBN for the \planck\
base-\lcdm\ parameters and element abundance
observations. Uncertainties in nuclear rates currently dominate the
error budget for the interpretation of deuterium abundances.

\item We have investigated a number of models for massive sterile neutrinos and dark-matter annihilation, finding no evidence for deviations from base \lcdm.

\end{unindentedlist}
\vskip 1pt

The overall picture from \Planck, since our first results were
presented in \paramsI, is one of remarkable consistency with the
6-parameter \LCDM\ cosmology. This consistency is strengthened with
the addition of the polarization spectra presented in this
paper. Nevertheless, there are a number of curious ``tensions,'' both
internal to the \Planck\ data (the tendency for \Planck\ to favour
$\Alens >1$, discussed in \Alenssec, is an example) and with some
external data sets.  Some of these tensions may reflect small systematic
errors in the \Planck\ data (though we have not found any evidence for
errors that could significantly change our results) and/or systematic
errors in external data. However, none of these, with the exception of the discrepancy
with direct measurements of $H_0$,  is significant at more than the
2--$3\,\sigma$ level. Such relatively modest discrepancies generate interest, in part,
because of the high precision of the \Planck\ data set.
We could, therefore, disregard these tensions
and conclude that the 6-parameter
\LCDM\ model provides an astonishingly accurate description of the Universe
from times prior to 380\,000 years after the Big Bang, defining the last-scattering
surface observed via the CMB, to the present day at an age of $13.8$ billion years.

Nevertheless, it is important to bear in mind that the main ingredients
of \LCDM, namely inflation, dark energy, and dark matter are not
understood at any fundamental level. There is, therefore, a natural
tendency to speculate that ``tensions'' may be hints of new physics,
especially given that the landscape of possible new physics is
immense.  In the post-\Planck\ era, the CMB provides enormous
potential for further discovery via high-sensitivity ground-based
polarization experiments and possibly a fourth-generation CMB
satellite. The next decade will see an ambitious programme of large
BAO and weak lensing surveys, and new techniques such as deep 21-cm
surveys and gravitational wave experiments. Uncovering evidence for new physics
is therefore a realistic possibility.  What we have learned, and the
legacy from \Planck, is that any signatures of new physics in the CMB
must be small.

\paragraph{Acknowledgements}
The Planck Collaboration acknowledges the support of: ESA; CNES, and
CNRS/INSU-IN2P3-INP (France); ASI, CNR, and INAF (Italy); NASA and DoE
(USA); STFC and UKSA (UK); CSIC, MINECO, JA, and RES (Spain); Tekes, AoF,
and CSC (Finland); DLR and MPG (Germany); CSA (Canada); DTU Space
(Denmark); SER/SSO (Switzerland); RCN (Norway); SFI (Ireland);
FCT/MCTES (Portugal); ERC and PRACE (EU). A description of the Planck
Collaboration and a list of its members, indicating which technical
or scientific activities they have been involved in, can be found at
\href{https://www.cosmos.esa.int/web/planck/planck-collaboration}{https://www.cosmos.esa.int/web/planck/planck-collaboration}.
We additionally acknowledge support from the European Research Council under
the European Union's Seventh Framework Programme (FP/2007-2013) / ERC Grant Agreement No. [616170]. This project has received funding from the European Research Council (ERC) under the European Union’s
Horizon 2020 research and innovation programme (grant agreement No 725456, CMBSPEC).
We thank Ofelia Pisanti for providing updated numerical BBN results from the \parthenope\ code, Cyril Pitrou for producing some results from the {\tt PRIMAT} code, and the DES
team for sharing their likelihoods. We also thank Marco Crisostomi, Ignacy Sawicky, Alessandra Silvestri, and Filippo Vernizzi for discussions on the dark-energy and modified-gravity models.
Some of the results in this paper have been derived using the {\tt HEALPix}
package.

\bibliographystyle{aat}

\providecommand{\aj}{Astron. J. }\providecommand{\apj}{ApJ
  }\providecommand{\apjl}{ApJ
  }\providecommand{\mnras}{MNRAS}\providecommand{\prl}{PRL}\providecommand{\prd}{PRD}\providecommand{\jcap}{JCAP}\providecommand{\aap}{A\&A}
\def\eprinttmppp@#1arXiv:@{#1}
\providecommand{\arxivlink[1]}{\href{http://arxiv.org/abs/#1}{arXiv:#1}}
\def\eprinttmp@#1arXiv:#2 [#3]#4@{\ifthenelse{\equal{#3}{x}}{\ifthenelse{
\equal{#1}{}}{\arxivlink{\eprinttmppp@#2@}}{\arxivlink{#1}}}{\arxivlink{#2}
  [#3]}}
\providecommand{\eprintlink}[1]{\eprinttmp@#1arXiv: [x]@}
\providecommand{\eprint}[1]{\eprintlink{#1}}
\providecommand{\adsurl}[1]{\href{#1}{ADS}}

\appendix
\section{Cosmological parameters from \texttt{\textbf{CamSpec}}}
\label{appendix:camspec}

\begin{figure*}[htbp!]
\begin{center}
\includegraphics[width=18cm]{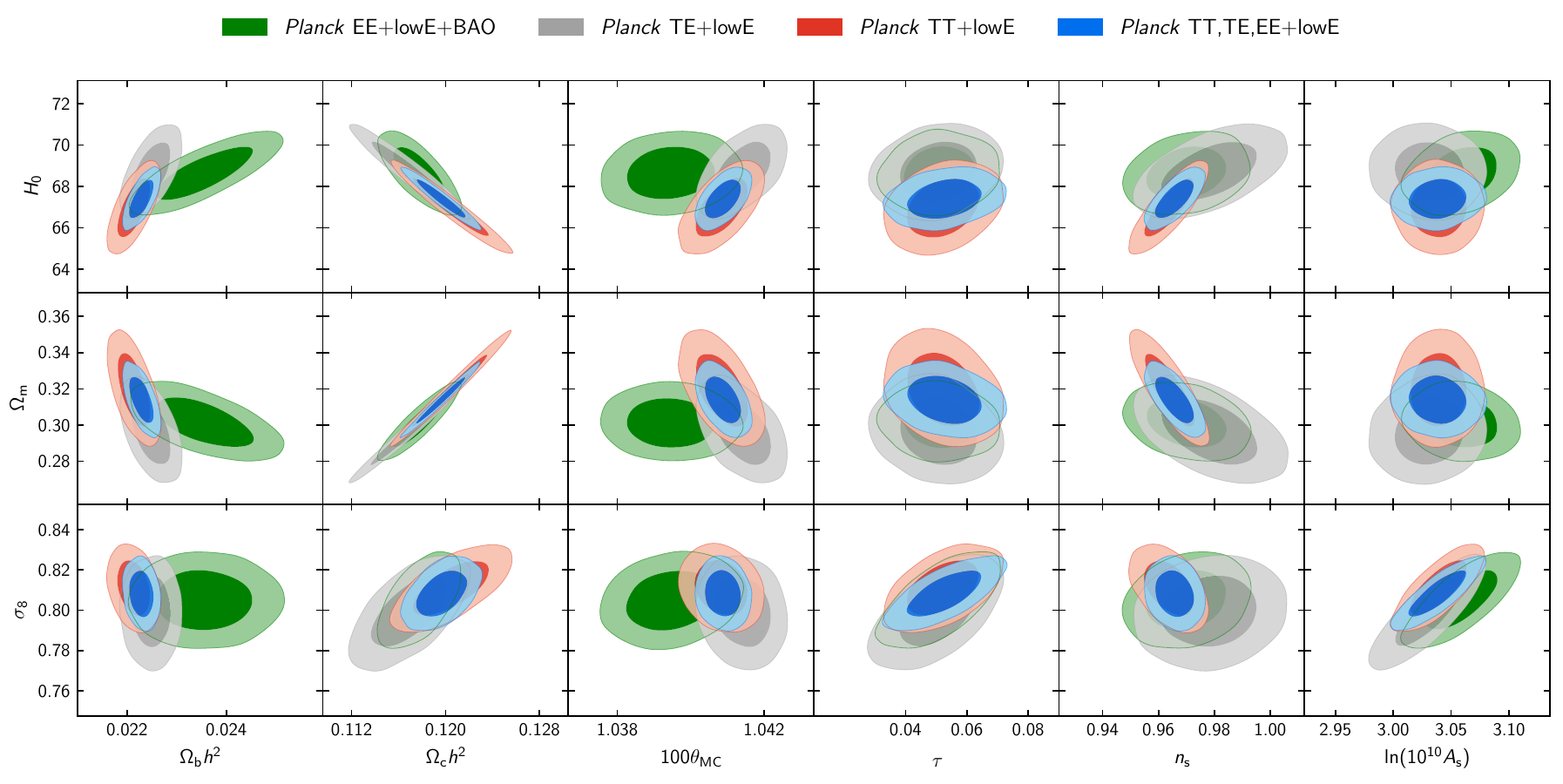}
\end{center}
\vspace{-3mm}
\caption {The equivalent of Fig.~\ref{fig:polrectangle} using \camspec\ in place of \plik, showing
constraints on parameters of the base-\lcdm\ model using the high-$\ell$ TT, TE, and EE separately (with the EE results also including BAO), and the combined result from the \TTTEEE\ likelihood.}
\label{fig:camspecpolrectangle}
\end{figure*}

Section~\ref{sec:likelihood} summarized the two high-multipole
likelihoods used in this paper. We stated that the two codes give very
similar answers in $TT$ but show some differences in $TE$ and $EE$. We
discuss these differences in more detail in this Appendix. Table
\ref{table:default} compares the base-\LCDM\ parameters from
\plik\ and \camspec\ for the \planckalllensing\ likelihood
combinations, showing that the two codes return cosmological parameters
that agree to within a fraction of a standard deviation.
Table~\ref{LCDMcompare_camspec} is the equivalent of Table~\ref{LCDMcompare},
but using the \camspec\ likelihood in place of \plik. In $TT$, the
parameters determined from the two codes agree to $0.2\,\sigma$
or better.  The agreement is less good in $TE$; the most discrepant
parameters are $\ns$, which is $1\,\sigma$ higher in
\camspec, and $\As e^{-2\tau}$, which is $1.2\,\sigma$ higher in \camspec.
Both of these parameters are sensitive to the calibration
of the polarization spectra which differ in the two codes. The other
cosmological parameters agree
to better than $0.5\,\sigma$ and the shifts in parameters between $TE$
and $TT$ are similar in both codes. In $EE$, the parameter shifts
compared to $TT$ are similar in both codes, although the $EE$ parameters
from the two codes typically differ by almost $1\,\sigma$. Since $EE$
from \planck\ is so noisy, the $EE$ differences have little impact on
the combined \TTTEEE\ parameters.

\begin{table*}
\begin{center}

\caption{The equivalent of Table~\ref{LCDMcompare}, but using the \camspec\ likelihood in place of \plik.
}
\label{LCDMcompare_camspec}

\begingroup
\openup 5pt
\newdimen\tblskip \tblskip=5pt
\nointerlineskip
\vskip -3mm
\scriptsize
\setbox\tablebox=\vbox{
    \newdimen\digitwidth
    \setbox0=\hbox{\rm 0}
    \digitwidth=\wd0
    \catcode`"=\active
    \def"{\kern\digitwidth}
    \newdimen\signwidth
    \setbox0=\hbox{+}
    \signwidth=\wd0
    \catcode`!=\active
    \def!{\kern\signwidth}
\halign{
\hbox to 0.9in{$#$\leaderfil}\tabskip=1.5em&$#$\hfil&$#$\hfil&$#$\hfil&$#$\hfil&$#$\hfil&\hfil$#$\hfil\tabskip=0pt\cr
\noalign{\doubleline}
\multispan1\hfil \hfil&\multispan1\hfil \shortTT\hfil&\multispan1\hfil TE+\lowE\hfil&\multispan1\hfil EE+\lowE\hfil&\multispan1\hfil \shortall\hfil&\multispan1\hfil \shortall+\lensing\hfil&\multispan1\hfil \shortall+\lensing+BAO\hfil\cr
\noalign{\vskip -3pt}
\omit\hfil Parameter\hfil&\omit\hfil 68\% limits\hfil&\omit\hfil 68\% limits\hfil&\omit\hfil 68\% limits\hfil&\omit\hfil 68\% limits\hfil&\omit\hfil 68\% limits\hfil&\omit\hfil 68\% limits\hfil\cr
\noalign{\vskip 3pt\hrule\vskip 5pt}
\Omega_{\mathrm{b}} h^2&0.02214\pm 0.00022&0.02248\pm 0.00026&0.0233\pm 0.0012&0.02229\pm 0.00016&0.02229\pm 0.00015&0.02234\pm 0.00014\cr
\Omega_{\mathrm{c}} h^2&0.1205\pm 0.0021&0.1169\pm 0.0021&0.1192\pm 0.0047&0.1196\pm 0.0014&0.1197\pm 0.0012&0.11907\pm 0.00094\cr
100\theta_{\mathrm{MC}}&1.04084\pm 0.00048&1.04141\pm 0.00051&1.03928\pm 0.00087&1.04088\pm 0.00032&1.04087\pm 0.00031&1.04095\pm 0.00030\cr
\tau&0.0521\pm 0.0080&0.0504\pm 0.0088&0.0504\pm 0.0088&0.0528\pm 0.0080&0.0536^{+0.0069}_{-0.0077}&0.0552^{+0.0067}_{-0.0076}\cr
\ln(10^{10} A_\mathrm{s})&3.039\pm 0.016&3.031\pm 0.021&3.058\pm 0.022&3.039\pm 0.016&3.041\pm 0.015&3.043^{+0.013}_{-0.015}\cr
n_\mathrm{s}&0.9638\pm 0.0058&0.978\pm 0.011&0.967\pm 0.014&0.9658\pm 0.0045&0.9656\pm 0.0042&0.9671\pm 0.0038\cr
\noalign{\vskip 5pt\hrule\vskip 3pt}
H_0\,[{\rm km}\,{\rm s}^{-1}\,{\rm Mpc}^{-1}]&66.98\pm 0.92&68.72\pm 0.93&67.9\pm 2.6&67.41\pm 0.62&67.39\pm 0.54&67.66\pm 0.42\cr
\Omega_\Lambda&0.680\pm 0.013&0.703\pm 0.012&0.687^{+0.035}_{-0.028}&0.6861\pm 0.0085&0.6858\pm 0.0074&0.6897\pm 0.0057\cr
\Omega_{\mathrm{m}}&0.320\pm 0.013&0.297\pm 0.012&0.313^{+0.028}_{-0.035}&0.3139\pm 0.0085&0.3142\pm 0.0074&0.3103\pm 0.0057\cr
\Omega_{\mathrm{m}} h^2&0.1432\pm 0.0020&0.1400\pm 0.0020&0.1431\pm 0.0038&0.1426\pm 0.0013&0.1426\pm 0.0011&0.14205\pm 0.00090\cr
\Omega_{\mathrm{m}} h^3&0.09593\pm 0.00045&0.09622\pm 0.00054&0.0971^{+0.0015}_{-0.0017}&0.09610\pm 0.00031&0.09610\pm 0.00031&0.09611\pm 0.00031\cr
\sigma_8&0.8110\pm 0.0089&0.799\pm 0.012&0.809^{+0.019}_{-0.017}&0.8083\pm 0.0076&0.8091\pm 0.0060&0.8083\pm 0.0060\cr
\sigma_8(\Omega_{\rm m}/0.3)^{0.5}&0.837\pm 0.024&0.795\pm 0.025&0.825\pm 0.058&0.827\pm 0.016&0.828\pm 0.013&0.822\pm 0.011\cr
\sigma_8 \Omega_{\mathrm{m}}^{0.25}&0.610\pm 0.012&0.590\pm 0.013&0.604\pm 0.028&0.6050\pm 0.0083&0.6058\pm 0.0064&0.6033\pm 0.0057\cr
z_{\mathrm{re}}&7.49^{+0.83}_{-0.75}&7.18^{+0.93}_{-0.75}&7.06^{+0.90}_{-0.76}&7.52^{+0.83}_{-0.75}&7.61\pm 0.75&7.75\pm 0.73\cr
10^9 A_{\mathrm{s}}&2.089\pm 0.034&2.072\pm 0.042&2.130\pm 0.046&2.088\pm 0.034&2.092^{+0.028}_{-0.031}&2.097^{+0.028}_{-0.032}\cr
10^9 A_{\mathrm{s}} e^{-2\tau}&1.882\pm 0.014&1.873\pm 0.019&1.925\pm 0.024&1.879\pm 0.011&1.879\pm 0.011&1.877\pm 0.011\cr
\mathrm{Age}\,[\mathrm{Gyr}]&13.825\pm 0.037&13.757\pm 0.039&13.75\pm 0.14&13.805\pm 0.025&13.805\pm 0.023&13.796\pm 0.020\cr
z_\ast&1090.26\pm 0.41&1089.51\pm 0.42&1088.8^{+1.6}_{-1.8}&1089.99\pm 0.28&1089.99\pm 0.26&1089.88\pm 0.22\cr
r_\ast\,[\mathrm{Mpc}]&144.49\pm 0.48&145.15\pm 0.50&143.94\pm 0.66&144.58\pm 0.31&144.57\pm 0.28&144.70\pm 0.23\cr
100\theta_\ast&1.04105\pm 0.00047&1.04158\pm 0.00050&1.03937\pm 0.00084&1.04107\pm 0.00031&1.04106\pm 0.00031&1.04114\pm 0.00030\cr
z_{\mathrm{drag}}&1059.43\pm 0.45&1059.98\pm 0.55&1061.9\pm 2.3&1059.73\pm 0.33&1059.74\pm 0.32&1059.79\pm 0.32\cr
r_{\mathrm{drag}}\,[\mathrm{Mpc}]&147.23\pm 0.48&147.79\pm 0.52&146.31\pm 0.69&147.27\pm 0.31&147.26\pm 0.28&147.38\pm 0.25\cr
k_{\mathrm{D}}\,[\mathrm{Mpc}^{-1}]&0.14054\pm 0.00052&0.14021\pm 0.00060&0.1423\pm 0.0012&0.14061\pm 0.00034&0.14063\pm 0.00033&0.14054\pm 0.00031\cr
z_{\mathrm{eq}}&3408\pm 48&3331\pm 48&3405\pm 90&3392\pm 31&3393\pm 27&3379\pm 22\cr
k_{\mathrm{eq}}\,[\mathrm{Mpc}^{-1}]&0.01040\pm 0.00015&0.01017\pm 0.00014&0.01039\pm 0.00027&0.010352\pm 0.000095&0.010355\pm 0.000083&0.010314\pm 0.000066\cr
100\theta_{\rm{s,eq}}&0.4487\pm 0.0046&0.4565\pm 0.0047&0.4492\pm 0.0091&0.4503\pm 0.0030&0.4502\pm 0.0026&0.4515\pm 0.0021\cr
\noalign{\vskip 5pt\hrule\vskip 3pt}
f_{2000}^{143}&30.8\pm 3.0&&&29.8\pm 2.8&29.7\pm 2.8&29.5\pm 2.8\cr
f_{2000}^{217}&107.6\pm 2.0&&&106.9\pm 1.9&106.9\pm 1.9&106.8\pm 1.9\cr
f_{2000}^{143\times217}&33.0\pm 2.1&&&32.2\pm 2.0&32.2\pm 2.0&32.0\pm 2.0\cr
\noalign{\vskip 5pt\hrule\vskip 3pt}
} 
} 
\endPlancktable
\endgroup
\end{center}
\end{table*}

The differences listed in Table~\ref{table:default} can be seen visually in Fig.~\ref{fig:camspecpolrectangle}, which
is the equivalent of Fig.~\ref{fig:polrectangle} in Sect.~\ref{sec:lcdm},
comparing base-\LCDM\ parameters determined separately
from TT, TE, and EE, and the combined result from the
\TTTEEE\ \camspec\ likelihood. The two figures are remarkably similar,
given the different methodologies and choices (e.g., polarization
masks, multipole cuts, etc.) used to construct the polarization blocks
of the likelihoods. For the base-\LCDM\ cosmology, the two likelihoods
are in such close agreement that it would make no difference to any of
the science conclusions in this paper if we used \camspec\ in place of
\plik.

\begin{figure*}
\begin{center}
\includegraphics[height=65mm,angle=0]{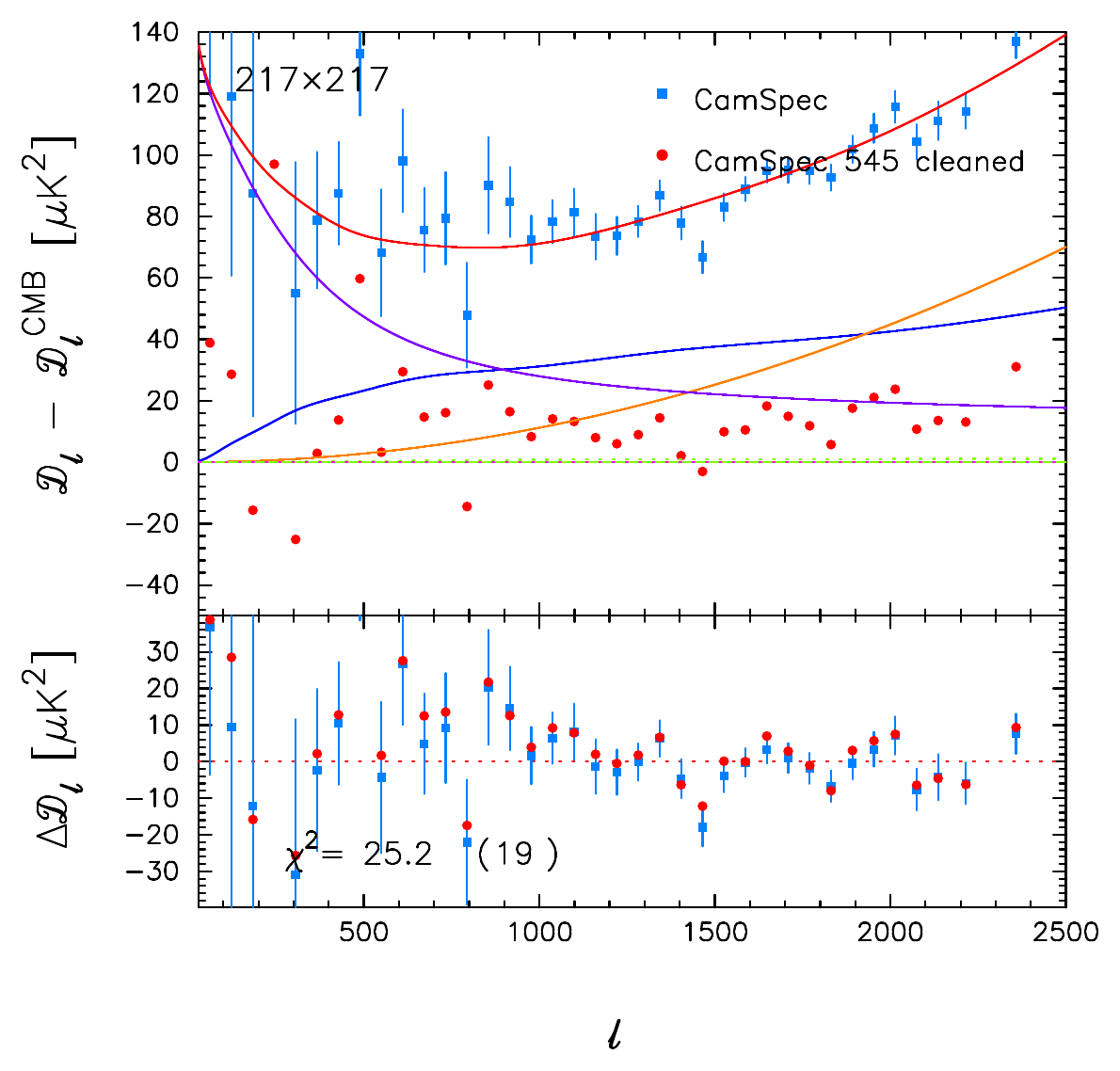}\hspace*{-2mm}
\includegraphics[height=65mm,angle=0]{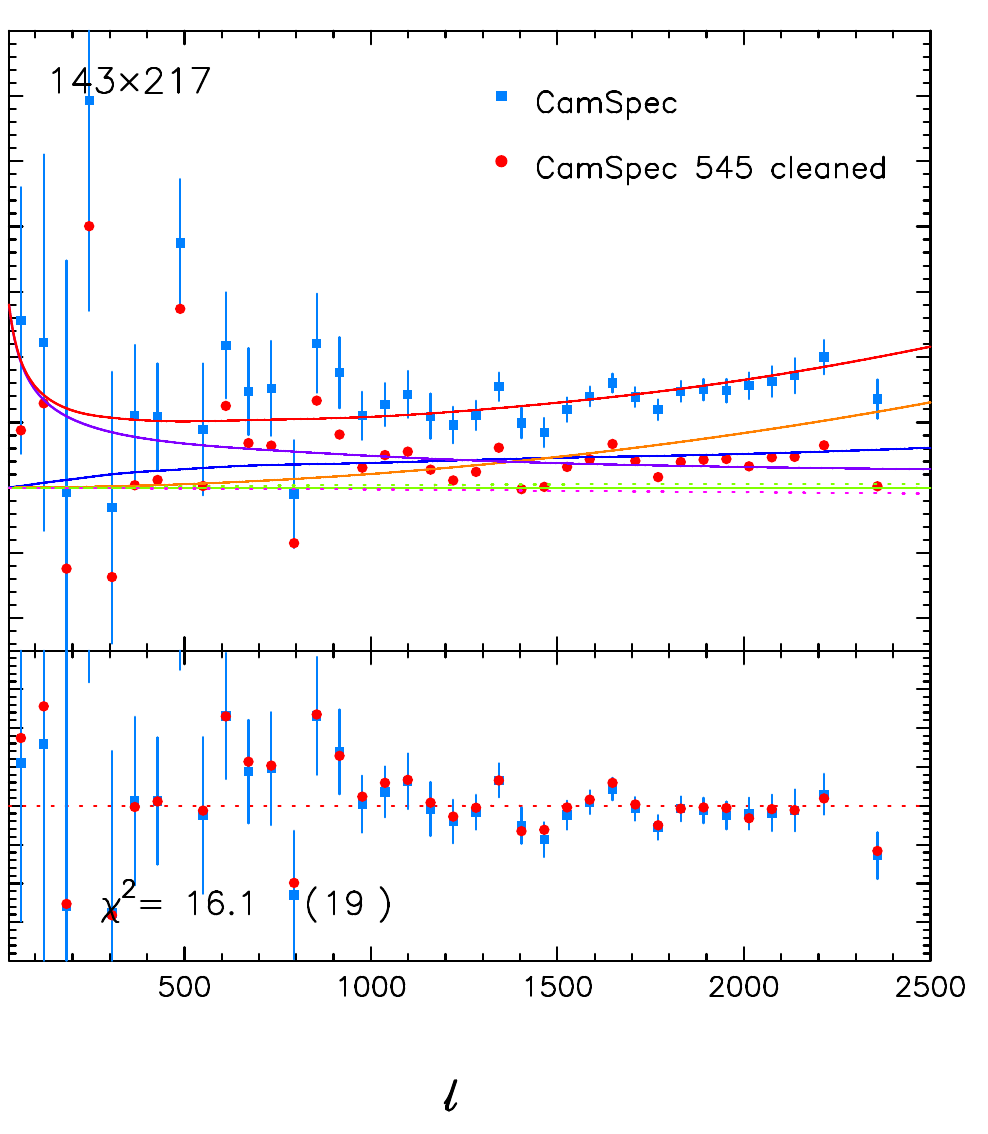}\hspace*{-2mm}
\includegraphics[height=65mm,angle=0]{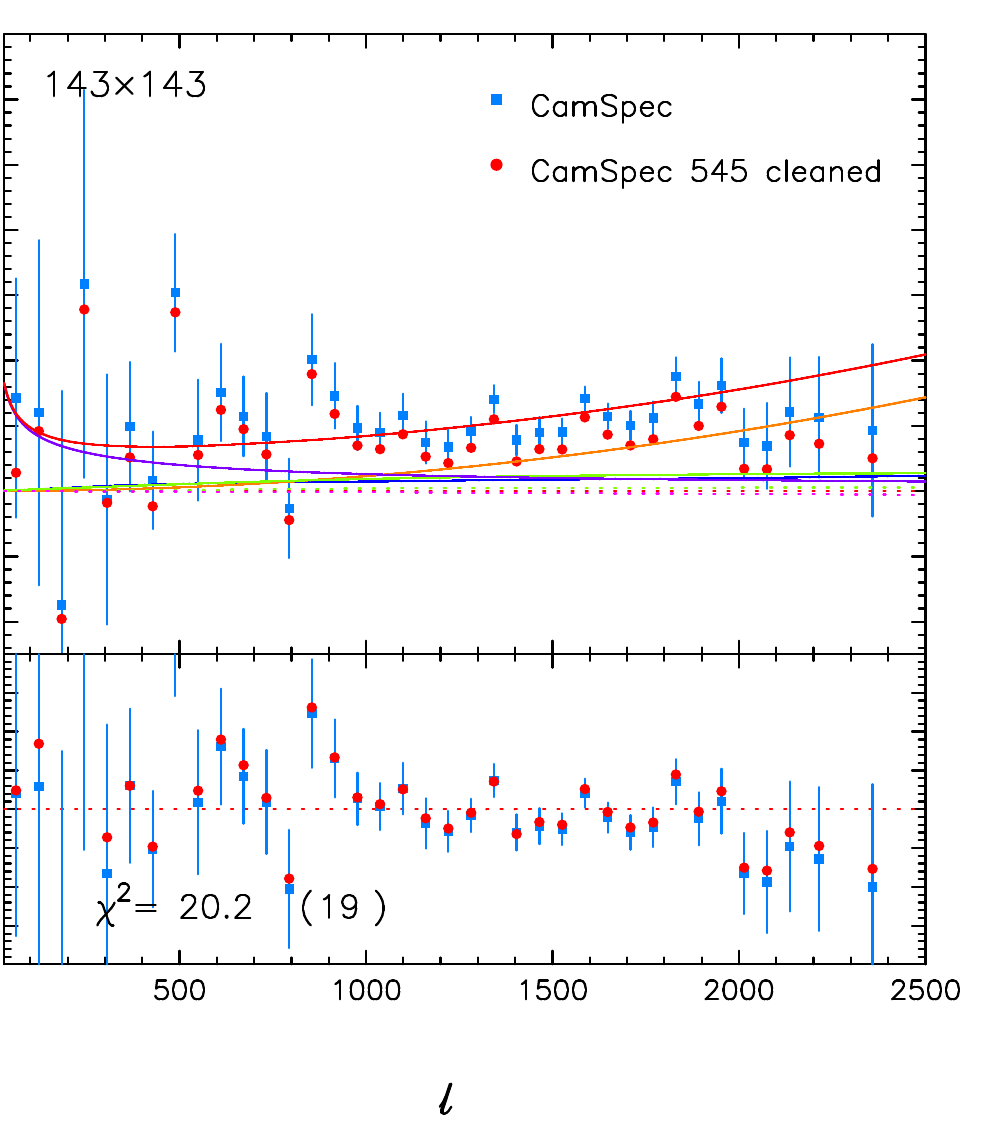}
\end{center}
\vspace{-3mm}
\caption {Residual plots illustrating the sensitivity of the $TT$ spectra to foreground modelling.
The blue points in the upper panels show the \camspec\ half-mission cross-spectra
after subtraction of the best-fit \LCDM\ spectrum fit to \shortTT.  The residuals in the
upper panel should be accurately described by the foreground model. Major
foreground components are shown by the solid lines, colour coded as follows:
total foreground spectrum (red); Poisson point sources (orange);
clustered CIB (blue); thermal SZ (green); and Galactic dust
(purple). Minor foreground components are shown by the dotted lines,
colour coded as follows: kinetic SZ (green); and tSZ$\times$CIB
cross-correlation (purple). The red points in the upper panel panels show
the 545\,GHz-cleaned spectra (minus best-fit CMB, as subtracted from the uncleaned spectra)
that are fit to a power-law residual foreground model, as discussed in the text.
The lower panels show the spectra after subtraction of the best-fit
foreground models.
The $\chi^2$ values of the residuals of the blue points, and the number of
band powers, are listed in the lower panels.
}
\label{fig:camspecclean}
\end{figure*}

The small differences between the \plik\ and \camspec\ TT likelihoods
are probably due to underestimates of the modelling uncertainties because the foreground
models in the two codes are almost identical. A more accurate impression of
foreground-modelling uncertainties can be gleaned by comparing
the default model with a heuristic foreground model applied to
spectra, cleaned using the 545-GHz temperature maps, as described in section~3.2 of \paramsII.
Since the low-frequency and high-frequency maps have different beams, the
subtraction is actually done in the power spectrum domain:
\begin{eqnarray}
\hat C^{T_{\nu_1}T_{\nu_2} {\rm clean}} &=&
 (1 + \alpha^{T_{\nu_1}})(1 + \alpha^{T_{\nu_2}})
 \hat C^{T_{\nu_1} T_{\nu_2}} \nonumber \\
 & & - (1 + \alpha^{T_{\nu_1}}) \alpha^{T_{\nu_2}}
 \hat C^{T_{\nu_1}T_{\nu_{\rm t}}} \nonumber \\
 & & - (1 + \alpha^{T_{\nu_2}}) \alpha^{T_{\nu_1}}
 \hat C^{T_{\nu_2}T_{\nu_{\rm t}}}
 + \alpha^{T_{\nu_1}} \alpha^{T_{\nu_2}}
 \hat C^{T_{\nu_{\rm t}}T_{\nu_{\rm t}}}, \label{MC2}
\end{eqnarray}
\citep[e.g.,][]{Spergel:2013rxa}, where $\hat C^{T_{\nu_1}T_{\nu_2}}$ etc.\ are the mask-deconvolved
beam-corrected power spectra at low frequencies and $\nu_{\rm t}$ is the frequency of the
template map. The coefficients $\alpha^{T_{\nu_i}}$
are determined by minimizing
\begin{equation}
\sum_{\ell =\ell_{\rm min}}^{\ell_{\rm max}}
 \sum_{\ell^\prime =\ell_{\rm min}}^{\ell_{\rm max}}
 \hat C^{T_{\nu_i}T_{\nu_i}{\rm clean}}_\ell
 \left(
 \hat{\tens{M}}^{T_{\nu_i}T_{\nu_i}}_{\ell \ell^\prime}
 \right)^{-1}
 \hat C^{T_{\nu_i}T_{\nu_i} {\rm clean}}_{\ell^\prime},
\end{equation}
where $\hat{\tens{M}}^{T_{\nu_i}T_{\nu_i}}$ is the covariance matrix of
the estimates $\hat C^{T_{\nu_i}T_{\nu_i}}$.  As in \paramsII\ we
choose $\ell_{\rm min}\,{=}\,100$ and $\ell_{\rm max}\,{=}\,500$ and
compute the spectra in Eq.~\eqref{MC2} by cross-correlating
half-mission maps using the 60\,\% mask used to compute the
$217\times217$ spectrum.  The resulting cleaning coefficients are
$\alpha^T_{143} = 0.00198$ and $\alpha^T_{217} = 0.00763$ (very close
to the coefficients adopted in \paramsII); note that all of the input
maps here are in units of thermodynamic temperature. The cleaned half-mission
$143\times143$, $143\times 217$ and $217\times 217$ spectra are
compared with uncleaned \camspec\ spectra (using the same masks for both sets of spectra)
and default foreground model
in Fig.~\ref{fig:camspecclean}.

To model residual foregrounds in the cleaned spectra, we assume that they
follow power-laws, $A_{f} (\ell/1500)^{\gamma_f}$ characterized by an amplitudes
and spectral indices for each of the $143\times143$, $143\times 217$ and $217\times217$ spectra
together with kinetic and thermal Sunyaev-Zeldovich templates, as in the default
foreground model. We assume the default foreground model for the $100\times 100$ spectrum.
We then form a \camspec\ ``cleaned'' likelihood (used in several places in the main body of
this paper) using the same covariance matrices as those computed for the uncleaned likelihood.
Comparing parameters for the \shortTT\ likelihood combination, the cleaned and uncleaned
cosmological parameters agree to a fraction of a standard deviation, with $\thetaMC$ and
$\ns$ from the cleaned likelihood each lower by $0.3\,\sigma$. We conclude
that systematics associated with modelling foregrounds do not introduce significant biases
in cosmological parameter determinations.

\begin{table*}
\begin{center}
\caption{Constraints on 1-parameter extensions to the base-\lcdm\ model using \camspec\ at high $\ell$, and also including \planck\ lensing and BAO.
 This is equivalent to Table~\ref{tab:grid_1paramext} for \plik, except that we have
added results for the cleaned TT \camspec\ likelihood in the third column. Note that we quote 95\,\% limits here.
}

\label{tab:grid_1paramext_camspec}

\begingroup
\newdimen\tblskip \tblskip=5pt
\nointerlineskip
\vskip -3mm
\footnotesize
\setbox\tablebox=\vbox{
    \newdimen\digitwidth
    \setbox0=\hbox{\rm 0}
    \digitwidth=\wd0
    \catcode`"=\active
    \def"{\kern\digitwidth}
    \newdimen\signwidth
    \setbox0=\hbox{+}
    \signwidth=\wd0
    \catcode`!=\active
    \def!{\kern\signwidth}
\halign{
\hbox to 0.7in{$#$\leaderfil}\tabskip=1.5em&
\hfil$#$\hfil\tabskip=1.5em&
\hfil$#$\hfil\tabskip=1.5em&
\hfil$#$\hfil\tabskip=1.5em&
\hfil$#$\hfil\tabskip=1.5em&
\hfil$#$\hfil\tabskip=0pt\cr
\noalign{\doubleline}
\noalign{\vskip -1pt}
\omit\hfil\text{Parameter}
 \hfil& \hfil \quad \TT{+}\lowE\quad\hfil& \hfil \quad \TT^{\rm clean}{+}\lowE\quad\hfil& \hfil \TTTEEE{+}\lowE \hfil& \hfil \TTTEEE{+}\lowE{+}\lensing \hfil& \hfil \TTTEEE{+}\lowE{+}\lensing{+}\BAO \cr
\noalign{\vskip 3pt\hrule\vskip 5pt}
\Omega_K & -0.058^{+0.046}_{-0.051} & -0.057^{+0.045}_{-0.051} & -0.037^{+0.032}_{-0.034} & -0.011^{+0.012}_{-0.013} & 0.0005^{+0.0038}_{-0.0040}\cr
\Sigma m_\nu\,[\mathrm{eV}] & < 0.569 & < 0.578 & < 0.379 & < 0.273 & < 0.131\cr
N_{\mathrm{eff}} & 2.94^{+0.59}_{-0.56} & 2.99^{+0.59}_{-0.57} & 2.92^{+0.45}_{-0.43} & 2.88^{+0.44}_{-0.42} & 2.98^{+0.39}_{-0.38}\cr
Y_{\mathrm{P}} & 0.242^{+0.040}_{-0.042} & 0.246^{+0.042}_{-0.042} & 0.246^{+0.035}_{-0.035} & 0.244^{+0.034}_{-0.035} & 0.248^{+0.032}_{-0.032}\cr
\mathrm{d}n_{\mathrm{s}}/\mathrm{d}\ln k & -0.003^{+0.015}_{-0.015} & -0.005^{+0.015}_{-0.015} & -0.001^{+0.013}_{-0.013} & -0.001^{+0.013}_{-0.013} & 0.000^{+0.013}_{-0.013}\cr
r_{0.002} & < 0.106 & < 0.105 & < 0.141 & < 0.136 & < 0.141\cr
w_0 & -1.54^{+0.59}_{-0.48} & -1.55^{+0.60}_{-0.48} & -1.52^{+0.56}_{-0.45} & -1.54^{+0.51}_{-0.41} & -1.03^{+0.10}_{-0.11}\cr
\noalign{\vskip 5pt \hrule \vskip 3pt}
} 
} 
\endPlancktable
\endgroup
\end{center}
\end{table*}

Table~\ref{tab:grid_1paramext_camspec} gives \camspec\ results for
extensions to the base-\LCDM\ cosmology, which is the equivalent to
Table~\ref{tab:grid_1paramext}, but with the addition of results for
the \camspec\ cleaned \shortTT\ likelihood.  The \camspec\ and
\plik\ likelihoods give closely similar results for these
extensions. The only noteworthy differences are the \TTTEEE\ results
for $\Omk$, this being slightly closer to zero in the
\camspec\ likelihood (we find similar behaviour for the lensing consistency parameter $\Alens$, as
discussed in \Alenssec) and somewhat weaker constraints on $\Sigma m_\nu$ and $r$.  These differences give an indication of the
sensitivity of our results to different methods and choices made in
constructing the \TTTEEE\ likelihoods and, in particular, to the schemes
used to calibrate effective polarization efficiencies. \rep{A detailed description
of \camspec, including further justification of the methodology used to construct
the polarization blocks of the likelihood, is given in \cite{Efstathiou:2019}.}

\end{document}